 \documentstyle [eqsecnum,preprint,aps,epsf,fixes]{revtex}

\ifpreprintsty\advance\textwidth by -0.1 in\fi

\begin{document}

\def\lsim{ \,\, \vcenter{\hbox{$\buildrel{\displaystyle <}\over\sim$}}
 \,\,}
\def\gsim{ \,\, \vcenter{\hbox{$\buildrel{\displaystyle >}\over\sim$}}
 \,\,}

 \newcommand{\eqntap}{& \displaystyle}
 \newcommand{\dum}{\vphantom{x}}
 \newcommand{\ud}{\underline}
 \newcommand{\Order}{{\cal O}}
 \newcommand{\mth}{m_{\rm th}}
 \newcommand{\mphys}{m_{\rm phys}}

 \setcounter{page}{1}

 \preprint{UW/PT-94}

 \title{Hydrodynamic Transport Coefficients
 	in Relativistic Scalar Field Theory}

 \author{Sangyong Jeon}

 \address{Department of Physics FM-15\\
	 University of Washington\\
	 Seattle, WA 98195}
 \date{\today}

 \maketitle

 \begin{abstract}
 {%
 {%
 \advance\leftskip -2pt
 Hydrodynamic transport coefficients may be evaluated from first
 principles in a weakly coupled scalar field theory at arbitrary
 temperature.
 In a theory with cubic and quartic interactions,
 the infinite class of diagrams which contribute
 to the leading weak coupling behavior are identified and summed.
 The resulting expression may be reduced to a single linear integral
 equation, which is shown to be identical to the corresponding result
 obtained from a linearized Boltzmann equation describing effective
 thermal excitations with temperature dependent masses and scattering
 amplitudes.  The effective Boltzmann equation is valid even at very
 high temperature where the thermal lifetime and mean free path are
 short compared to the Compton wavelength of the fundamental
 particles.  Numerical results for the shear and the bulk viscosities
 are presented.
 }%

 \bigskip
 \bigskip

 Submitted to Physical Review D.

 \ifpreprintsty
 \thispagestyle {empty}
 \newpage
 \thispagestyle {empty}
 \vbox to \vsize
 {}
 \fi
 }%
 \end{abstract}

 \setcounter{page}{1}

 \section{Introduction}
 \label{sec:introduction}

 Linear response theory provides a framework for calculating
 transport coefficients starting from first principles
 in a finite temperature quantum field theory.
 As reviewed below,
 the resulting ``Kubo'' formulae
 express the hydrodynamic transport coefficients in terms of
 the zero momentum, small frequency limit
 of stress tensor--stress tensor correlation
 functions\cite{MartinK}.
 One-loop calculations of transport coefficients
 using these Kubo formulae in a relativistic scalar $\phi^4$
 theory
 have appeared previously\cite{Hosoya,Horsley}.
 However, those calculations are wrong even in the weak coupling
 limit; they fail to include an infinite class of diagrams
 which contribute at the same order as the one-loop
 diagram.
 These multi-loop diagrams
 are not suppressed because powers of the single particle thermal
 lifetime compensate
 the explicit coupling constants provided by the interaction
 vertices.%
\footnote{%
     Similar phenomena occur in the calculation of transport
 coefficients
     in non-relativistic fluids\cite{Mahan}.
}

 In this paper,
 all diagrams which make leading order contribution to
 the viscosities in a weakly coupled relativistic scalar field
 theory
 with cubic and quartic interactions are identified.
 The diagrammatic rules needed to calculate the required finite
 temperature spectral densities of composite operator correlation
 functions were derived in a previous paper\cite{Jeon}
 (and are summarized below).
 The dominant diagrams are identified by
 counting the powers of the coupling constants
 which result from a given diagram,
 including those generated by near ``on-shell'' singularities
 which are cut-off by the single particle thermal lifetime.

 For the calculation of the shear viscosity,
 certain cut ``ladder'' diagrams,
 corresponding to the contribution of elastic scatterings only,
 are found to make the leading order contributions.
 The geometric series of cut ladder diagrams is then summed
 by introducing a set of effective vertices which satisfy coupled
 linear integral equations.
 The resulting expression is then shown to reduce to a single integral
 equation, which is solved numerically.

 For the calculation of the bulk viscosity,
 in addition to the leading order ladder diagrams,
 contributions from the next order diagrams containing
 inelastic scattering processes must also be summed.
 In general, the bulk viscosity is proportional to the relaxation
 time of the processes which restore
 equilibrium when the volume of a system changes\cite{Landau}.
 For a system of a single component real scalar field,
 such processes involve inelastic scatterings
 which change the number of particles.
 Hence, diagrams corresponding to such processes
 must be included.

 Boltzmann equations based on kinetic theory have traditionally
 been
 used to calculate transport properties of dilute weakly interacting
 systems.
 However,
 the validity of kinetic theory is restricted by the condition
 that the mean free path of the particles must be much larger
 than any other microscopic length scale.
 In particular, the mean free path must be large compared to
 the Compton wavelength of the underlying particle
 in order for the classical picture of particle propagation to
 be valid.
 A Boltzmann equation describing the fundamental particles cannot
 be
 justified when this condition fails to hold.
 Such is the case at extremely high temperature,
 where the mean free path scales as $1/T$.

 No such limitation exists when starting from the fundamental
 quantum field theory.
 Nevertheless,
 it will be shown that the correct transport coefficients, in
 a weakly
 coupled theory, may also be obtained by starting from a Boltzmann
 equation describing effective single particle thermal excitations
 with
 temperature dependent masses and scattering amplitudes.%
\footnote{%
     The Boltzmann equation
     derived by Calzetta and Hu\cite{Hu} via relativistic
     Wigner function is also expressed in terms of the thermal
 mass.
}
 This equivalence holds even for asymptotically large temperatures
 where
 both the thermal lifetime and mean free path are tiny compared
 to the
 zero temperature Compton wavelength.
 Hence, in a weakly coupled theory,
 although a kinetic theory description in terms of fundamental
 particles
 is only valid at low temperatures, a kinetic theory description
 of
 effective thermal excitations remains valid at all
 temperatures.%
\footnote{%
     In a strongly coupled theory,
     the mean free path can be comparable to the
     scattering time, or other microscopic scales, and no kinetic
     description is justified.
}

 The effective kinetic theory result
 presented in this paper is valid through all temperatures
 in the weak coupling limit.
 At low temperatures, $T \ll m_{\rm phys}$,
 where $m_{\rm phys}$ is the physical mass of the underlying
 particles at zero temperature,
 most particles are non-relativistic.
 Hence, the effective theory reduces
 to non-relativistic kinetic theory at low temperatures.
 If the temperature is in the range
 $m_{\rm phys}
 \,\,\vcenter {\hbox{$\buildrel{\displaystyle <}\over\sim$}} \,\,
 T \ll m_{\rm phys}/\sqrt{\lambda}$,
 where $\lambda$ is the quartic coupling constant,
 most particles are relativistic,
 but the thermal corrections to the mass and the scattering
 amplitude are negligible.
 Consequently, the viscosities at these temperatures can be calculated
 by a kinetic theory of relativistic particles with temperature
 independent mass and scattering amplitudes.

 The most interesting temperatures are those where
 $T={\cal O}(m_{\rm phys}/\sqrt{\lambda})$.
 At these temperatures,
 the thermal correction to the mass
 is comparable to the zero temperature mass.
 For weak coupling, this temperature is also large enough
 that most excitations are highly relativistic.
 One might expect that
 the thermal correction to the mass would then be irrelevant.
 This is true for some physical quantities
 which are insensitive to
 soft momentum contributions, such as the shear viscosity.
 However, other quantities, such as the bulk viscosity,
 are sensitive to soft momenta.
 For such quantities, including
 the thermal correction to the mass
 and the scattering amplitude will be shown to be essential.

 At very high temperature,
 $T
 \,\, \vcenter{\hbox{$\buildrel{\displaystyle >}\over\sim$}} \,\,
 m_{\rm phys}/\lambda$,
 all mass scales, including the cubic coupling constant,
 other than the temperature are completely negligible,
 and consequently the theory reduces to the massless scalar
 theory
 with only a quartic interaction.

 Through out this paper, we work with the Lagrangian
 \begin{equation}
 -{\cal L} =
 {1 \over 2}
 \phi ( -\partial_{\tau}^2 - \nabla^2 + m_0^2) \phi
 +
 {g\over 3!} \phi^3
 +
 {\lambda \over 4!} \phi^4
 \;.
 \label{eq:full_Lagrangian}
 \end{equation}
 It is assumed that $\lambda \ll 1$ and
 $g^2 = {\cal O}(\lambda m_0^2)$,
 so that the theory is always weakly coupled.  For simplicity,
 we also
 take $m_0^2 > 0$.  Note that at tree level, $m_0$ can be
 regarded
 as the physical mass $m_{\rm phys}$.
 Portions of the analysis will begin by assuming pure quartic
 interactions, after which the additional contribution arising from
 cubic interactions will be considered.
 The remainder of the paper is organized as follows.
 A brief review of various background material
 is presented in section~\ref{sec:background}.
 This material includes the definition of transport coefficients,
 basic linear response theory,
 diagrammatic ``cutting'' rules for the evaluation of spectral
 densities,
 and a summary of the behavior of self-energies at high temperature.
 Section~\ref{sec:classification} deals with the problem of identifying
 the leading order diagrams.
 By counting powers of coupling constants, including those from
 near on-shell singularities,
 ladder diagrams are identified as the leading order diagrams.
 The summation of these diagrams is discussed in
 section~\ref{sec:summation}.
 Section~\ref{sec:Boltzmann} contains a brief review of
 the computations of viscosities starting from the Boltzmann
 equation,
 and then discusses the relation between the resulting formulae
 and those in section~\ref{sec:summation}.
 Using the results of the previous sections,
 the final calculation of viscosities is discussed
 in section~\ref{sec:viscosities}, and numerical results presented in
 section~\ref{sec:numeric}.

 Several appendices contain technical details.
 Explicit forms of the imaginary-time
 and real-time propagators used in the main body of the paper
 are
 summarized in
 appendix~\ref{app:propagators}.
 Appendices~\ref{app:ladder} and \ref{app:zero_modes}
 present explicit forms of the ``ladder'' kernels discussed in
 section~\ref{sec:summation}.
 In appendix~\ref{app:temperature_integrals}, the first order
 correction
 to the equilibrium stress-energy tensor needed in
 sections~\ref{sec:summation} and \ref{sec:viscosities} are calculated.
 Appendix~\ref{app:soft_contributions} discusses the soft momentum
 and
 collinear contributions to finite temperature cut diagrams,
 and shows
 that they does not upset the estimates used in
 section~\ref{sec:classification}.
 Appendix~\ref{app:chains} contains technical details of summing
 up
 the ``chain'' diagrams appearing in section~\ref{subsec:chains}.

 \section{Background material}
 \label{sec:background}

 \subsection{Hydrodynamic transport coefficients}
 \label{subsec:transport_coeff}

 In a single component real scalar field theory,
 the only locally conserved quantities are energy and momentum.
 The transport coefficients associated with energy and momentum
 flow, known as the shear and bulk viscosities,
 may be defined by the following constitutive relation,
 \begin{equation}
 \langle T_{ij}\rangle
 \simeq
 -{\eta \over \langle \varepsilon {+} {\cal P} \rangle}
     \left(
 	 \nabla_i\langle T_{j}^0 \rangle
       + \nabla_j \langle T_{i}^0 \rangle
       - {\textstyle {2 \over 3}} \delta_{ij}
 	\nabla^l \langle T_{l}^0 \rangle
 	\right)
       -{\zeta \over \langle \varepsilon {+} {\cal P} \rangle}
        \delta_{ij}  \nabla^l \langle T_{l}^0 \rangle
 	+ \delta_{ij} \langle {\cal P} \rangle
 \;,
 \label{eq:constitutive_rel}
 \end{equation}
 valid when the length scale
 of energy and momentum fluctuations is much longer than the
 mean free path.
 Here,
 $T_{ij}$ is the spatial part of the stress-energy tensor
 $T_{\mu\nu}$,
 $\varepsilon \equiv T_{00}$ is the energy density,
 ${\cal P} = {1\over 3}T_i^i$ is the pressure,
 and
 $\eta$ and $\zeta$
 are the shear and the bulk viscosities, respectively.
 Also, $\langle \cdots \rangle$ denotes
 the expectation in a non-equilibrium thermal ensemble describing
 a system slightly out of equilibrium.
 Since there are no additional conserved charges, thermal conductivity
 is not an independent transport coefficient.%
\footnote{%
      Calculation of the thermal conductivity in a scalar
      $\lambda\phi^4$ theory in Ref.\cite{Hosoya} is in this
 sense
      misleading.
}

 The above constitutive relation,
 combined with the exact conservation equation
 \begin{equation}
 \partial_{\mu} T^{\mu\nu}(x) = 0
 \;,
 \label{eq:em_conservation}
 \end{equation}
 constitute linearized hydrodynamic equations for a relativistic
 fluid.
 With the help of the equilibrium thermodynamic relation
 \begin{equation}
 {\partial \langle {\cal P} \rangle_{\rm eq}
 \over
 \partial \langle\varepsilon\rangle_{\rm eq}}
 = v_{\rm s}^2
 \;,
 \label{eq:sound_speed}
 \end{equation}
 where $v_{\rm s}$ is the speed of sound,
 the linearized hydrodynamic equations can be reduced to two
 equations
 for the transverse part of the momentum density
 $\langle \pi_T^{\vphantom{x}} \rangle$
 and the pressure $\langle {\cal P} \rangle$,
 \begin{mathletters}
 \begin{eqnarray}
 & \displaystyle
 (\partial_t - D\nabla^2)\,
 \langle \pi_T^{\vphantom{x}}(x) \rangle = 0
 \;,
 &
 \\
 \noalign{\hbox{and}}
 & \displaystyle
 \left(
       \partial_t^2 - v_{\rm s}^2 \nabla^2 - \Gamma\nabla^2\partial_t
 \right)
 \langle {\cal P}(x) \rangle = 0
 &
 \;.
 \end{eqnarray}
 \label{eq:linear_hydrodynamic_eq}
 \end{mathletters}%
 \hspace{-1ex}%
 Here the diffusion constant $D$ is
 proportional to the shear viscosity
 \begin{eqnarray}
 D
 & \equiv & \displaystyle
 \eta/\langle\varepsilon{+}{\cal P}\rangle_{\rm eq}
 \;,
 \label{eq:def_D}
 \\
 \noalign{\hbox{and the sound attenuation constant $\Gamma$
 equals
 a linear combination of the viscosities}}
 \Gamma
 & \equiv & \displaystyle
 ({\textstyle{4\over 3}}\eta{+}\zeta)/
 \langle\varepsilon{+}{\cal P}\rangle_{\rm eq}
 \;.
 \end{eqnarray}

 Using the basic linear response result,
 one may express the viscosities in terms of
 the stress tensor-stress tensor
 correlation functions\cite{MartinK,Forster}.
 One finds the ``Kubo'' formulae,
 \begin{mathletters}
 \begin{eqnarray}
 & \displaystyle
 \eta
 =
 {\beta\over 20}
 	     \lim_{\omega \to 0}
 		\lim_{ {\bf q} \to 0 }
 	          \sigma_{\pi\pi}^{\vphantom{x}} (\omega,{\bf q})
 \;,
 &
 \label{eq:shear_viscosity}
 \\
 & \displaystyle
 \zeta
 =
 {\beta\over 2}
 	     \lim_{\omega \to 0}
 		\lim_{ {\bf q} \to 0 }
 	           \sigma_{\bar{\cal P}\bar{\cal P}}^{\vphantom{x}}
		      (\omega,{\bf q})
 \;.
 &
 \label{eq:bulk_viscosity}
 \end{eqnarray}
 \label{eq:viscosities}
 \end{mathletters}%
 \hspace{-1ex}%
 Here
 $\sigma_{\pi\pi}^{\vphantom{x}} (\omega, {\bf q})$
 is the Fourier transformed
 traceless stress-stress Wightman function,
 \begin{eqnarray}
 & \displaystyle
 \sigma_{\pi\pi}^{\vphantom{x}} (\omega, {\bf q})
 \equiv
 \int d^3 {\bf x}\, dt\, e^{-i{\bf q}{\cdot}{\bf x} +i\omega
 t}\,
     \langle \pi_{lm}(t,{\bf x}) \pi^{lm}(0) \rangle_{\rm eq}
 \;,
 &
 \\
 \noalign{\hbox{where}}
 & \displaystyle
 \pi_{lm}(x) \equiv T_{lm}(x)
 - {\textstyle{1\over 3}} \delta_{lm} T_i^i(x)
 \;
 &
 \label{eq:pi_lm}
 \end{eqnarray}
 is the traceless stress tensor.
 Similarly,
 \begin{eqnarray}
 & \displaystyle
 \sigma_{\bar{\cal P}\bar{\cal P}}^{\vphantom{x}} (\omega, {\bf q})
 \equiv
 \int d^3 {\bf x}\, dt\, e^{-i{\bf q}{\cdot}{\bf x} +i\omega
 t}\,
     \langle \bar{\cal P}(t,{\bf x})\bar{\cal P}(0) \rangle_{\rm
 eq}
 \;,
 &
 \\
 \noalign{\hbox{where}}
 & \displaystyle
 \bar{\cal P}(t,{\bf x})
 \equiv
 {\cal P}(t,{\bf x}) - v^2 {\varepsilon}(t,{\bf x})
 = {\textstyle{1\over 3}}T_i^i(x) - v^2 T_{00}(x)
 \;,
 &
 \label{eq:modified_pressure}
 \end{eqnarray}
 is a linear combination of pressure and energy density.
 The constant $v^2$ in this combination
 is arbitrary;
 due to energy-momentum conservation,
 Wightman functions
 involving the energy density vanish
 (at non-zero frequency) in the zero spatial momentum limit.
 However, as will be discussed in section~\ref{sec:summation},
 $v$ will eventually be chosen to equal the speed of sound.
 This will be necessary in order to make the final
 integral equation for the transport coefficient well defined.
 Note that, whereas the approximate constitutive relation
 (\ref{eq:constitutive_rel}) involves a non-equilibrium thermal
 expectation, the Kubo formulae (\ref{eq:viscosities}) express
 the
 transport coefficients solely
 in terms of equilibrium expectation values.

 \subsection{Qualitative behavior of the viscosities}
 \label{subsec:qualitative}

 In general, a transport coefficient is roughly
 proportional to the mean free path, or equivalently the relaxation
 time, of the processes responsible for the particular
 transport\cite{Landau}.

 This behavior is most easily seen in a diffusion constant
 (or in the shear viscosity).
 Consider a system with a conserved charge.
 In such a system, the diffusion of a charge density
 fluctuation may be modeled by a random walk\cite{Forster}.
 The rate of the diffusion then depends on two parameters;
 the step size (the mean free path)
 and the number of steps per time (the mean speed).
 A longer step size or a larger number of steps per time
 implies faster diffusion of the excess charge, {\it i.e.,} a larger
 diffusion constant $D$.
 Since the diffusion constant has the dimension of a length, one finds
 \begin{equation}
 D \sim l_{\rm free}\bar{v}
 \;.
 \end{equation}
 Recall that the diffusion constant in Eq.~(\ref{eq:def_D}) is given by
 $D = \eta/\langle\varepsilon{+}{\cal P}\rangle_{\rm eq}$.
 Applying the above estimate of $D$ yields
 \begin{equation}
 \eta \sim l_{\rm free}\bar{v}
      \langle \varepsilon{+}{\cal P} \rangle_{\rm eq}
 \;.
 \end{equation}

 Given the scattering cross-section $\sigma$ and the density of the
 particles $n$, the mean free path can be estimated as
 $l_{\rm free} \sim 1/n\sigma$.
 Consider a weakly coupled scalar $\lambda\phi^4$ theory.
 The lowest order scattering cross section in the $\lambda\phi^4$ theory
 is $\sigma \sim \lambda^2/s$ where $s$ is the square of the
 center of mass energy.
 At high temperature, $T \gg m_{\rm phys}$, the only relevant mass scale
 is the temperature.  (Here $m_{\rm phys}$ denotes the physical mass.)
 Hence, $\sigma \sim \lambda^2 /T^2$, $n \sim T^3$, and
 \begin{equation}
 \eta \sim T^3/\lambda^2
 \qquad
 ( T \gg m_{\rm phys} )
 \;.
 \end{equation}
 At low temperature, $T \ll m_{\rm phys}$,
 $\sigma \sim\lambda^2/m_{\rm phys}^2$
 and $\bar{v} \sim (T/m_{\rm phys})^{1/2}$.
 At these temperatures, the energy density
 $\langle \varepsilon \rangle_{\rm eq}
 \sim m_{\rm phys} n$, dominates over the pressure.
 Canceling two density factors in $l_{\rm free}$ and
 $\langle \varepsilon \rangle_{\rm eq}$,
 the shear viscosity can be estimated as
 \begin{equation}
 \eta
 \sim m_{\rm phys}^3 (T/m_{\rm phys})^{1/2}/\lambda^2
 \qquad
 ( T \ll m_{\rm phys} )
 \;.
 \end{equation}
 Note that the shear viscosity is not analytic in the weak coupling
 constant.
 This may be taken as an indication that
 the first few terms in the usual Feynman diagram expansion
 cannot produce the correct value of the leading order shear viscosity.

 For the bulk viscosity, the situation is more complicated
 than the simple picture given above.
 The bulk viscosity does not have an interpretation as a diffusion
 constant.  Hence,
 the random walk model cannot be directly applied.
 The bulk viscosity is still proportional to the mean free time
 (inverse transition rate) of a scattering process
 since the viscosities govern relaxation of a system towards equilibrium.
 However, the factors multiplying the mean free time cannot simply
 be $\langle \varepsilon{+}{\cal P} \rangle_{\rm eq}$ since
 the bulk viscosity $\zeta$ vanishes
 in a scale invariant system\cite{Weinberg}.

 To understand this,
 consider a slow uniform expansion of the volume of a system.
 In such an expansion, there can be no shear flow\cite{Smith}.
 Hence, the relaxation of disturbances caused by the expansion
 depends only on the bulk viscosity.
 For scale invariant systems,
 the restoration of local equilibrium does not require any relaxation
 process.
 A suitable scaling of the temperature alone
 can maintain local equilibrium at all times.
 Hence, for such systems,
 including the non-relativistic monatomic ideal gas\footnote{%
	The non-relativistic monatomic ideal gas is {\em not} equivalent
	to the low temperature limit of the single component real scalar
	field theory.  The number of particles in the non-relativistic
	monatomic ideal gas is conserved whereas the number of particles
	in the low temperature limit of the scalar field vanishes as the
	temperature goes to zero.  Hence, the low temperature limit of
	the scalar theory bulk viscosity need not vanish.
}
 and the ideal gas of massless particles,
 the bulk viscosity vanishes since the relaxation time vanishes.

 When the system is not scale invariant,
 the bulk viscosity must be proportional to
 a measure of the violation of scale invariance, or the mass
 $m_{\rm phys}$.
 In section~\ref{sec:summation}, the formula for the leading order
 bulk viscosity at high temperature is shown to be
 \begin{equation}
 \zeta
 \sim
 m_{\rm phys}^4 \tau_{\rm free}
 \;.
 \end{equation}
 The mean free time $\tau_{\rm free}$ here is given by the inverse of
 the transition rate per particle,
 \begin{equation}
 \tau_{\rm free} \equiv n/ (dW / dV dt)
 \;,
 \end{equation}
 where $dW/dVdt$
 the transition rate per volume.
 corresponding to the relaxation of the uniformly expanding system.
 In a number non-conserving system with broken scale invariance
 (such as a massive scalar theory),
 the number-changing inelastic scattering processes
 are ultimately responsible for relaxation towards equilibrium.
 As the system expands, the temperature must decrease
 since the system loses energy pushing the boundary.
 Decreasing energy implies decreasing particle number
 in a number non-conserving system with broken scale invariance.
 Hence, the relaxation toward equilibrium
 must involve number-changing scatterings.
 In the $g\phi^3{+}\lambda\phi^4$ theory,
 the lowest order number-changing process is
 2--3 scatterings involving 3 cubic vertices or
 1 cubic and 1 quartic vertices.
 In pure $\lambda\phi^4$ theory,
 the lowest order number changing process is
 2--4 scatterings involving 2 quartic vertices.

 At high temperature,
 $m_{\rm phys} \ll T
\,\, \vcenter{\hbox{$\buildrel{\displaystyle <}\over\sim$}} \,\,
 m_{\rm phys}/\sqrt{\lambda}$,
 most particles have momentum of ${\cal O}(T)$.
 However, due to the Bose-Einstein enhancement,
 the transition rate per volume will be shown to be dominated by the
 ${\cal O}(m_{\rm th})$ momentum components in the system
 where $m_{\rm th}$ is the thermal mass containing
 ${\cal O}(\sqrt{\lambda}T)$ thermal corrections.
 In the $g\phi^3{+}\lambda\phi^4$ theory,
 with the statistical factors for 5 particles involved in
 the scattering, the transition rate per volume is
 ${\cal O}(\lambda^2 g^2 T^5 / m_{\rm th}^3)$ which is
 ${\cal O}(T^3 /m_{\rm th}^3)$
 larger than the transition rate of the particles with ${\cal O}(T)$
 momentum.
 Hence at temperatures in the range
 $m_{\rm phys} \ll T
\,\, \vcenter{\hbox{$\buildrel{\displaystyle <}\over\sim$}} \,\,
 m_{\rm phys}/\sqrt{\lambda}$,
 the bulk viscosity will be
 \begin{equation}
 \zeta
 \sim m_{\rm phys}^4 m_{\rm th}^3  / \lambda^2 g^2 T^2
 \;.
 \end{equation}
 When $T = {\cal O}(m_{\rm phys}/\sqrt{\lambda})$,
 $\zeta = {\cal O}(T^3 / \sqrt{\lambda})$ which is ${\cal O}(\lambda^{3/2})$
 smaller than the shear viscosity.
 At the same temperature, the $\lambda\phi^4$ theory transition
 rate per volume is ${\cal O}(\lambda^3 T^4)$ again due to the
 Bose-Einstein enhancement.
 In this case,
 $\zeta = {\cal O}(\beta m_{\rm phys}^4 / \lambda^3) =
 {\cal O}(T^3/\lambda)$ which is ${\cal O}(\lambda)$
 smaller than the shear viscosity.

 At very high temperature $T \gg m_{\rm phys}/\lambda$,
 all mass scales, including the cubic coupling constant,
 other than the temperature are completely negligible,
 and consequently the theory reduces to the massless scalar
 theory with only a quartic interaction.
 The massless scalar
 theory is classically scale invariant.
 However, quantum mechanics breaks the scale invariance.
 The measure of the violation of scale invariance in this case is the
 renormalization group $\beta$-function.
 Since the transition rate per volume must still be
 ${\cal O}(\lambda^3 T^4)$ due to the thermally generated
 ${\cal O}(\sqrt{\lambda} T)$ mass,
 the bulk viscosity in this case is
 ${\cal O}( T^3 \beta(\lambda)^2/\lambda^3 ) = {\cal O}(\lambda T^3)$.

 At low temperature, $T \ll m_{\rm phys}$,
 the bulk viscosity is proportional to the mean free time and
 the number density.  At these temperatures,
 the transition rate per volume in the $g\phi^3{+}\lambda\phi^4$ theory
 is ${\cal O}(e^{- 3 \beta m_{\rm phys}})$, since
 the center of mass energy must exceed $3m_{\rm phys}$ for a 2--3 process
 to occur.  Since the density
 $n = {\cal O}(e^{-\beta m_{\rm phys}})$ at low temperature,
 the bulk viscosity is then ${\cal O}(e^{\beta m_{\rm phys}}/\lambda^3)$.
 The bulk viscosity at low temperature is hence much larger than
 the ${\cal O}(1/\lambda^2)$ shear viscosity.
 In the $\lambda\phi^4$ theory, the transition rate is
 ${\cal O}(e^{-4\beta m_{\rm phys}})$ since the center of mass energy in
 this case must exceed $4m_{\rm phys}$ for a 2--4 process to occur.
 The bulk viscosity is then ${\cal O}(e^{2\beta m_{\rm phys}}/\lambda^4)$,
 and again much larger than the shear viscosity.

 \subsection{Linear response theory}
 \label{subsec:linear_response}

 Linear response theory describes the behavior of a many-body
 system
 which is slightly displaced from equilibrium.
 First order time dependent perturbation theory implies that\cite{Forster}
 \begin{equation}
 \delta\langle \hat{A}_{a}(t, {\bf x}) \rangle
 =
 i\int d^3{\bf x}' \!
 \int_{-\infty}^t dt' \,
 \langle
 	[ \hat{A}_{a}(t, {\bf x}), \hat{A}_{b}(t',{\bf x}')]
 \rangle_{\rm eq} \,
 F_{b} (t', {\bf x}')
 \;,
 \label{eq:linear_response}
 \end{equation}
 where
 $\{ F_{b}(t,{\bf x}) \}$
 is some set of generalized external forces coupled to
 the interaction picture charge density operators
 $\{ \hat{A}_b(t,{\bf x}) \}$ so that
 \begin{equation}
 \delta\hat{H}(t) = - \int d^3{\bf x}\,
 	F_{b}(t,{\bf x}) \,
 	\hat{A}_b(t,{\bf x})
 \;,
 \end{equation}
 and $\langle \cdots \rangle_{\rm eq}$
 denotes an equilibrium thermal expectation.
 (Summation over the repeated index $b$ should be understood.)

 To examine transport properties, it is convenient to consider a
 relaxation process in which the external field is held constant
 for a long time (allowing the system to re-equilibrate in the presence of
 the external field), and then suddenly switched off,
 \begin{equation}
 F_b(t, {\bf x}) \equiv F_b({\bf x})\, e^{\epsilon t} \, \theta(-t)
 \;,
 \end{equation}
 where $\epsilon$ is a positive infinitesimal number.
 Once the field is switched off, the system will relax back towards
 the
 original unperturbed equilibrium state.
 Spatial translational invariance implies that
 Fourier components of the initial values
 $\delta\langle \hat{A}_{a}(0, {\bf x}) \rangle$
 are linearly related to the Fourier components of
 $F_b({\bf x})$.
 Hence, after a Fourier transform in space
 and a Laplace transform in time,
 Eq.~(\ref{eq:linear_response}) turns into
 an algebraic relation\cite{Forster},
 \begin{equation}
 \delta\tilde{A}_{a} (z, {\bf k})
 =
 {1\over i z}
 [
 \chi_{ab}^{\vphantom{x}}(z, {\bf k})\,
 \chi_{bc}^{-1}(i\epsilon, {\bf k})
 -
 \delta_{ac}
 ]
 \delta\tilde{A}_{c} (t{=}0, {\bf k})
 \;,
 \label{eq:initial_value_prob}
 \end{equation}
 where
 $\delta\tilde{A}_a (z, {\bf k})$ are Laplace and Fourier transformed
 deviations from equilibrium values
 $\delta\langle \hat{A}_{a}(t, {\bf x}) \rangle$,
 and
 $\delta\tilde{A}_c (t{=}0, {\bf k})$ are
 Fourier transformed initial values
 $\delta\langle \hat{A}_{c}(t{=}0, {\bf x}) \rangle$.
 Here,
 $\chi_{ab}^{\vphantom{x}}(z, {\bf k})$ is
 the retarded correlation function with complex frequency $z$;
 it has the spectral representation
 \begin{equation}
 \chi_{ab}^{\vphantom{x}}(z, {\bf k})
 =
 \int
 {d\omega\over 2\pi}\,
 {
 	\rho_{ab}^{\vphantom{x}}(\omega, {\bf k})
 		\over
        	     \omega - z
 }
 \;,
 \label{eq:chi_ab}
 \end{equation}
 where the spectral density is
 \begin{equation}
 \rho_{ab}^{\vphantom{x}}(\omega, {\bf k})
 \equiv
 \int d^4 x\, e^{-i{\bf k}{\cdot}{\bf x} + i\omega t}\,
 \langle [\hat{A}_{a}(t,{\bf x}), \hat{A}_{b}(0)] \rangle_{\rm
 eq}
 \;.
 \end{equation}

 If $\hat{A}_a$'s are conserved charge densities, then Ward identities
 can be shown to imply that the response functions have hydrodynamic
 poles (poles in the frequency plain which vanish as the spatial
 momentum
 goes to zero)\cite{Yaffe}.
 In the case of the conserved energy and momentum densities,
 the response functions in Eq.~(\ref{eq:initial_value_prob})
 can be shown to have a pole at $z = -iD{\bf k}^2$ when
 the disturbed charge is the transverse part of the momentum
 density
 $\pi_T^{\vphantom{x}}$, and poles at
 $z^2 = v^2{\bf k}^2 - i\Gamma z {\bf k}^2$
 when the disturbed charge is the energy density or the longitudinal
 part of the momentum density.

 Eq.~(\ref{eq:initial_value_prob}) solves the initial value problem
 in
 terms of the response function.
 When the conserved quantities are energy and momentum densities,
 the time evolution of the initial values can be also described
 (for low frequency and momentum)
 by the phenomenological hydrodynamic equations
 (\ref{eq:linear_hydrodynamic_eq}).
 When Fourier transformed in space and Laplace transformed in
 time,
 Eqs.~(\ref{eq:linear_hydrodynamic_eq})
 yield response functions with exactly the same
 diffusion and sound poles.
 Hence, by extracting the diffusion constant $D$
 and the sound attenuation constant $\Gamma$ from
 the pole positions in the correlation functions,
 one may derive the Kubo formulae (\ref{eq:viscosities})
 for the viscosities.

 The Wightman functions appearing in formulae (\ref{eq:viscosities})
 for
 the viscosities
 are trivially related to the corresponding spectral densities:
 \begin{eqnarray}
 \rho_{\pi\pi}^{\vphantom{x}}(\omega, {\bf q})
 & \displaystyle = & \displaystyle
 (1 - e^{-\beta\omega})\, \sigma_{\pi\pi}(\omega, {\bf q})
 \\
 \rho_{\bar{\cal P}\bar{\cal P}}^{\vphantom{x}}(\omega, {\bf q})
 & \displaystyle = & \displaystyle
 (1 - e^{-\beta\omega})\,
 \sigma_{\bar{\cal P}\bar{\cal P}}(\omega, {\bf q})
 \;.
 \end{eqnarray}
 Hence, the viscosities can equivalently be written as
 zero frequency derivatives of spectral densities,
 \begin{mathletters}
 \begin{eqnarray}
 & \displaystyle
 \eta =
	 {1\over 20}\,
	 \lim_{\omega\to 0}\,\lim_{{\bf q}\to 0}\,
         {\partial \over \partial \omega}
	 \rho_{\pi\pi}(\omega, {\bf q})
 \;,
 &
 \\
 \noalign{\hbox{and}}
  & \displaystyle
  \zeta =
	 {1 \over 2}\,
	 \lim_{\omega\to 0}\,\lim_{{\bf q}\to 0}\,
         {\partial \over \partial \omega}
	 \rho_{\bar{\cal P}\bar{\cal P}}(\omega, {\bf q})
 \;.
 &
 \end{eqnarray}
 \end{mathletters}

 \subsection{Cutting rules}
 \label{subsec:cutting_rules}

 In Ref.\cite{Jeon}, diagrammatic cutting rules for the perturbative
 calculation of the spectral density of an arbitrary
 two-point correlation function were derived starting from imaginary
 time
 finite temperature perturbation theory.
 These rules are a generalization of the
 standard zero-temperature Cutkosky
 rules, to which they reduce as temperature goes to zero.

 To calculate the perturbative expansion of a finite temperature
 spectral density,
 one should draw all cut Feynman diagrams for the two-point
 correlation function of interest.%
\footnote{%
	Only half the cut diagrams, those in which the external momentum
	flows into the shaded region, need to be considered if one
	includes an additional overall factor of
	$(1{-}e^{-q^0\beta})$, where $q^0$ is the external frequency.
	Omitting this factor, the same rules generate the Wightman
	function instead of the spectral density.
}
\begin{figure}
 \setlength {\unitlength}{1cm}
\vbox
    {%
    \begin {center}
 \begin{picture}(0,0)
 \put(1.8,4){A}
 \end{picture}
	\leavevmode
	\def\epsfsize	#1#2{0.4#1}
	\epsfbox {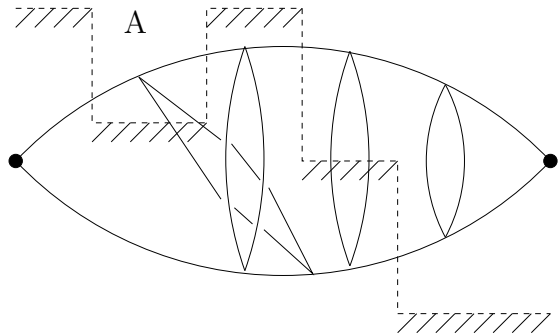}
    \end {center}
    \caption{%
    A typical cut diagram in a scalar $\lambda\phi^4$ theory.
 }
 \label{fig:typical_cut}
    }%
\end {figure}
 \noindent
 All cuts that separate the two external operators
 are allowed at non-zero temperature.
 Each line corresponds to either
 a cut or uncut thermal propagator, as described below.

 A typical example of a finite temperature cut diagram is shown
 in Fig.~\ref{fig:typical_cut}.
 Note that a cut at finite temperature can
 separate a diagram into multiple connected pieces, some of which
 are
 disconnected from the external operators.
 A disconnected piece, such as the portion labeled
 A in Fig.~\ref{fig:typical_cut},
 cannot contribute in a zero temperature cut diagram because
 of energy
 momentum conservation.
 For example,
 at zero temperature the piece labeled A would
 represent an impossible event of
 four incoming on-shell physical particles
 scattering and disappearing altogether.
 However, at finite temperature there
 exist physical thermal excitations in the medium.
 Thus,
 the above disconnected piece also represents
 the elastic scattering of a particle off of a thermal excitation
 already present in the medium.
 This scattering process is clearly possible; the amplitude
 is proportional to the density of the thermal
 particles (as the form of the cut propagator shown below clearly
 indicates).

 An uncut line in the unshaded region
 corresponds to a real-time time-ordered propagator
 $\langle {\cal T} (\phi(x)\phi(0)) \rangle$,
 an uncut propagator in the shaded region is
 $\langle {\cal T} (\phi(x)\phi(0)) \rangle^*$,
 and a cut line corresponds to the
 Wightman function $\langle \phi(x)\phi(0) \rangle$.
 In momentum space, the uncut propagator
 has the following spectral representation
 \begin{equation}
 \tilde{G} (k) \equiv
 \int {d\omega \over 2\pi} \,
 [ 1 {+} n(\omega) ] \,
 \rho(\omega, |{\bf k}|)
 \left(
 {
        2i\omega
         \over
 (k^0)^2 - (\omega {-} i\epsilon)^2
 }
 \right)
 \;,
 \label{eq:realtimeprop}
 \end{equation}
 where $\rho(\omega,|{\bf k}|)$ is the single particle spectral
 density.
 The cut propagator is proportional to the single particle spectral
 density
 \begin{equation}
 S(k) \equiv [1{+}n(k^0)]\, \rho(k)
 \;,
 \end{equation}
 where $n(k^0)$ is the Bose statistical factor $1/(e^{k^0\beta}{-}1)$.
 In more physical terms,
 \begin{equation}
 dV(k) \equiv
 \theta(k^0)\,S(k)\,{d^4 k \over (2\pi)^4}
 =
 \theta(k^0)\,[1{+}n(k^0)]\,\rho(k)\,{d^4 k \over (2\pi)^4}
 \;
 \end{equation}
 is the thermal phase space volume available to a final state
 particle in a scattering process, and
 \begin{equation}
 dN(k) \equiv
 \theta(-k^0)\,S(k)\,{d^4 k \over (2\pi)^4}
 =
 \theta(k^0)\,n(k^0)\,\rho(k)\,{d^4 k \over (2\pi)^4}
 \;
 \end{equation}
 is the number of thermal excitations within the 4-momentum range
 $(k_\mu, k_\mu{+}dk_\mu)$.

 If the single particle spectral density is approximated by a
 delta
 function (to which it reduces at zero temperature),
 {\it i.e.},
 \begin{equation}
 \rho_{\rm free}^{\vphantom{x}}(k) = {\rm sgn}(k^0)\,2\pi\delta(k^2+m_0^2)
 \;,
 \label{eq:free_sd}
 \end{equation}
 then self-energy insertions on any line
 generate ill-defined products of on-shell delta functions.
 Although these on-shell singularities disappear
 when all cut diagrams are summed, it is far more convenient
 to first resum single particle self-energy insertions.
 The resummed single particle spectral density $\rho(k)$ will
 then
 include the thermal lifetime of single particle
 excitations,
 which will smear the $\delta$-function peaks and produce a smooth
 spectral density.
 Henceforth,
 all single particle propagators will include the thermal self-energy
 and
 no self-energy insertions will appear explicitly in any cut
 diagram.

 \subsection{Propagators and self-energies at high temperature}
 \label{subsec:high_T_self}

 To analyze near infrared singularities, the explicit forms
 of single particle propagators will be needed.
 The resummed single particle spectral density can be calculated as
 \begin{figure}
 \setlength {\unitlength}{1cm}
\vbox
    {%
    \begin {center}
 \begin{picture}(0,0)
 \put(1.6,-0.5){(a)}
 \put(6.0,-0.5){(b)}
 \put(10.3,-0.5){(c)}
 \end{picture}
	\leavevmode
	\def\epsfsize	#1#2{0.4#1}
	\epsfbox {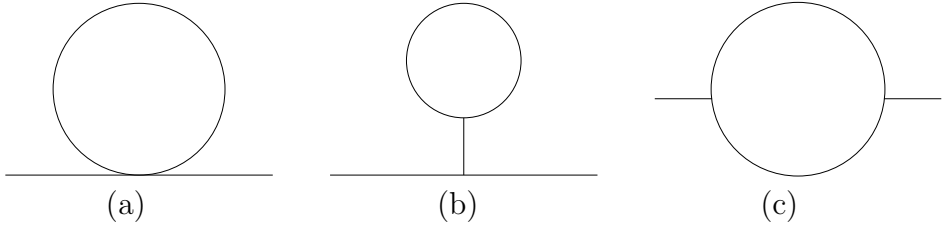}
    \end {center}
 \vspace{0.5cm}
 \caption{%
 One-loop self-energy diagrams in a scalar
 $g\phi^3{+}\lambda\phi^4$ theory.
 At high temperatures,
 the diagrams (a) and (b)
 produce a thermal mass squared of order $\lambda T^2$.
 The contribution of the diagram (c) to the thermal mass is
 $\Order(g^2 T/\mth)=\Order(\lambda^{3/2} T^2)$.
 }
 \label{fig:self_one_loop}
 }
 \end{figure}
 \noindent
 the discontinuity of the analytically continued imaginary-time
 propagator across the real frequency axis,
 \begin{eqnarray}
 \rho(k)
 & = & \displaystyle
 -i\left(
 \tilde{G}_E(k^0{+}i\epsilon, {\bf k})
 -
 \tilde{G}_E(k^0{-}i\epsilon, {\bf k})
 \right)
 \nonumber\\
 & = & \displaystyle
 {-i\over k^2 + m_{\rm th}^2 + \Sigma(k) }
 +
 {i\over k^2 + m_{\rm th}^2 + \Sigma(k)^* }
 \nonumber\\
 & = & \displaystyle
 -iG_R(k) + iG_A(k)
 =
 {2\Sigma_I(k)\over |k^2 + m_{\rm th}^2 +\Sigma(k)|^2}
 \;,
 \label{eq:single_ptl_sd}
 \end{eqnarray}
 where
 the subscript $R$ indicates the retarded propagator
 given by the analytic continuation of the Euclidean propagator
 \begin{equation}
 G_R(k) = \tilde{G}_E(k^0{+}i\epsilon,{\bf k})
 \;,
 \end{equation}
 and the subscript $A$ indicates the advanced propagator
 defined similarly, but with $k^0{-}i\epsilon$
 instead of $k^0{+}i\epsilon$\cite{Fetter,Brown}.
 Here, the thermal mass $m_{\rm th}$
 includes the ${\cal O}(\lambda T^2)$ one-loop
 corrections shown in Fig.~\ref{fig:self_one_loop}, and
 $\Sigma(k)$ is the analytically continued single particle self-energy
 \begin{equation}
 \Sigma(k) \equiv
 \Sigma_E(k^0{+}i\epsilon,{\bf k})
 =
 \Sigma_R(k){-}i\Sigma_I(k)
 \label{eq:retarded_prop}
 \;.
 \end{equation}
 The thermal mass squared $m_{\rm th}^2$ may be
 defined by the (off-shell) condition $\Sigma_R(0)\equiv 0$.
 Also, note that since $\rho(k)$ is the spectral density of a
 correlation function of the CPT even Hermitian operators $\hat{\phi}$,
 $\rho(k)$ must be
 an odd function of the frequency $k^0$\cite{Jeon,Forster}.
 This implies that $\Sigma_I(k)$ is also an odd function of
 $k^0$.

 As will be reviewed below, the imaginary part of the self-energy
 is
 ${\cal O}(\lambda^2)$, and so is small for weak coupling.
 Hence, the definition (\ref{eq:single_ptl_sd}) shows that
 the spectral density in
 the weak coupling limit has sharp peaks near
 $k^0 = \pm E_k$, where the effective single
 particle energy $E_k$ satisfies
 the dispersion relation
 \begin{equation}
 E_k^2 = {\bf k}^2 + m_{\rm th}^2 + \Sigma_R(E_k,|{\bf k}|)
 \;.
 \end{equation}
 Near the peaks,
 the spectral density may be approximated by a combination of
 two
 Lorentzians,
 \begin{equation}
 \rho(k) =
 {1\over 2E_k}\,
 \left(
 {2\Gamma_k\over (k^0 {-} E_k)^2 + \Gamma_k^2}
 -
 {2\Gamma_k\over (k^0 {+} E_k)^2 + \Gamma_k^2}
 \right)\times\left(1 + {\cal O}(\lambda^2) \right)
 \;,
 \label{eq:Lorentzian_sd}
 \end{equation}
 where $\Gamma_k$ is the momentum dependent thermal width
 given by
 \begin{equation}
 \Gamma_k \equiv \Sigma_I(E_k,|{\bf k}|)/2E_k
 \;.
 \label{eq:thermal_width}
 \end{equation}
 The thermal width $\Gamma_k$ is always positive
 since $\Sigma_I(k)$,
 or equivalently the single particle spectral density,
 must be positive for positive frequencies.
 This can be easily seen from the relation between the spectral
 density
 and the Wightman function and the positivity of
 the (Fourier transformed) Wightman function
 $\langle\phi(x)\phi(0)\rangle$\cite{Fetter}.
 Note that altogether
 $\rho(k)$ has four poles at $k^0 = E_k{\pm}i\Gamma_k$ and
 $k^0 = -E_k{\pm}i\Gamma_k$.
 In terms of the single particle spectral density, the cut propagator
 is
 \begin{equation}
 S(k)
 =
 [1{+}n(k^0)]\,\rho(k)
 =
 { 2\,[1{+}n(k^0)]\,\Sigma_I(k)
	\over
   \left| k^2 + m_{\rm th}^2 + \Sigma(k) \right|^2
 }
 \;.
 \end{equation}

 For the uncut propagator given by Eq.~(\ref{eq:realtimeprop}),
 the frequency integral
 can be exactly carried out to yield
 (see appendix~\ref{app:propagators} for details)
 \begin{mathletters}
 \begin{eqnarray}
 \tilde{G}(k)
 & = & \displaystyle
 -i{1+n(k^0)\over k^2 + m_{\rm th}^2 + \Sigma(k)}
 + i{n(k^0) \over k^2 + m_{\rm th}^2 + \Sigma(k)^*}
 \nonumber\\
 & = & \displaystyle
 -i
 {
 k^2 + m_{\rm th}^2 + \Sigma_R(k)
 	\over
 \left| k^2 + m_{\rm th}^2 + \Sigma(k) \right|^2
 }
 +
 {\coth(k^0\beta/2) \over 2}\,\rho(k)
 \;.
 \label{eq:prop_a}
 \end{eqnarray}
 In the weak coupling limit, this becomes
 \begin{equation}
 \tilde{G}(k)
 =
 \left(
 -i{1+n(E_k) \over (E_k{-}i\Gamma_k)^2 - k^2_0}
 + i{n(E_k) \over (E_k{+}i\Gamma_k)^2 - k^2_0}
 \right)\times \left( 1 + {\cal O}(\lambda^2) \right)
 \;.
 \label{eq:prop_b}
 \end{equation}
 \label{eq:prop}
 \end{mathletters}%
 \hspace{-1ex}%
 The first term in Eq.~(\ref{eq:prop_b}) has poles
 at $k^0=\pm (E_k{-}i\Gamma_k)$, and the second term has poles at
 $k^0=\pm (E_k{+}i\Gamma_k)$,
 coinciding with the pole positions of the spectral density $\rho(k)$.
 \begin{figure}
 \setlength {\unitlength}{1cm}
\vbox
    {%
    \begin {center}
 \begin{picture}(0,0)
 \put(1.6,-0.5){(a)}
 \put(5.6,-0.5){(b)}
 \put(9.6,-0.5){(c)}
 \put(13.6,-0.5){(d)}
 \end{picture}
	\leavevmode
	\def\epsfsize	#1#2{0.30#1}
	\epsfbox {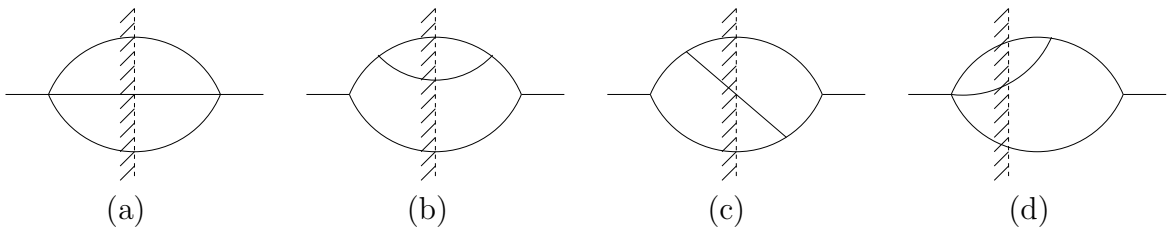}
    \end {center}
 \vspace{0.5cm}
 \caption{%
 Cut two-loop self-energy diagrams in a scalar
 $g\phi^3{+}\lambda\phi^4$ theory
 which produce a thermal width of order $\lambda^2 T$.
 }
 \label{fig:self_two_loop}
 }
 \end{figure}

 An important point to notice is
 that even though the statistical factor $n(k^0)$
 has a pole at $k^0 = 0$, both the cut propagator $S(k)$
 and the uncut propagator $\tilde{G}(k)$ are
 finite at zero frequency
 since the self-energy $\Sigma_I(k)$, which is an odd
 function of $k^0$, vanishes at $k^0 = 0$.
 Hence, although numerous factors of the statistical factors
 may appear
 in an expression for a diagram, one can be sure that there is
 no pole
 when loop frequencies approach zero.

 To determine the size of the thermal mass $m_{\rm th}$ and the
 thermal width $\Gamma_k$
 at high temperatures,
 the size of the one-loop (Fig.~\ref{fig:self_one_loop})
 and the two-loop (Fig.~\ref{fig:self_two_loop})
 self-energies at on-shell momenta must be known.
 At relativistic temperatures,
 $T
\,\, \vcenter{\hbox{$\buildrel{\displaystyle >}\over\sim$}} \,\,
 m_0$, the first one-loop
 diagram,
 Fig.~\ref{fig:self_one_loop}a,
 generates an ${\cal O}(\lambda T^2)$ contribution
 to the real part of the self-energy.
 Diagram \ref{fig:self_one_loop}b,
 with two cubic interaction vertices,
 is ${\cal O}(g^2 T^2/m_{\rm th}^2)$ which is at most ${\cal O}(\lambda
 T^2)$
 since by assumption
 $g={\cal O}(\sqrt{\lambda}m_0)$, and $m_{\rm th} \ge m_0$.
 At high temperature,
 the real part of diagram Fig.~\ref{fig:self_one_loop}c
 is ${\cal O}(g^2 \ln(T/m_{\rm th}))$ which is at most
 ${\cal O}(\lambda^2 T^2 \ln\lambda)$.  Hence, the diagram
 \ref{fig:self_one_loop}c does not contribute to the
 leading weak coupling behavior of the thermal mass
 correction.%
\footnote{%
	This estimate is for the external momentum of ${\cal O}(T)$.
	For a soft external momentum, the diagram
	\protect{\ref{fig:self_one_loop}}c
	is ${\cal O}(\lambda^{3/2}T^2)$.
        However, this is still ${\cal O}(\sqrt{\lambda})$ smaller
 than the
	diagrams Fig.~\protect{\ref{fig:self_one_loop}}a, b.
}
 Hence, the thermal mass is of order
 \begin{equation}
 m_{\rm th} = \sqrt{m_0^2 + {\cal O}(\lambda T^2)}
      \sim \sqrt{\lambda} T
 \;
 \end{equation}
 when
 $T
\,\, \vcenter{\hbox{$\buildrel{\displaystyle >}\over\sim$}} \,\,
 m_0/\sqrt{\lambda}$.

 The imaginary part of the self-energy receives an
 ${\cal O}(g^2 T^2)$
 contribution from the one-loop diagram \ref{fig:self_one_loop}c,
 but
 this contribution vanishes for on-shell external momenta since
 an
 on-shell excitation of mass $m_{\rm th}$ cannot decay into two on-shell
 excitations with the same mass.  (Nor can an on-shell excitation
 absorb
 the momentum of a thermal excitation and remain on-shell.)
 Hence, the dominant contribution to the on-shell imaginary part
 of the
 self-energy comes from the two-loop diagrams shown in
 Fig.~\ref{fig:self_two_loop}.
 At high temperature,
 these two-loop diagrams produce a ${\cal O}(\lambda^2 T^2)$
 imaginary part of the self-energy.%
\footnote{%
     Diagrams 3b and 3c for soft external on-shell momenta
     are ${\cal O}(g^4 T^2/m_{\rm th}^4)$,
     which is smaller than ${\cal O}(\lambda^2 T^2)$ by a factor
     of ${\cal O}(m_0^4/m_{\rm th}^4)$.
     Diagram 3d in the same limit is
     ${\cal O}(\lambda g^2 T^2/m_{\rm th}^2)$,
     which is smaller than ${\cal O}(\lambda^2 T^2)$ by a factor
     of ${\cal O}(m_0^2/m_{\rm th}^2)$.
     For hard external on-shell momenta of ${\cal O}(T)$,
     diagram 3a is strictly ${\cal O}(\lambda^2 T^2)$,
     while diagram 3b is
     ${\cal O}(g^4 T/m_{\rm th}^3) (< {\cal O}(\lambda^2 T^2))$
     due to near-collinear divergences cut-off by the mass,
     and diagrams 3c and 3d are
     ${\cal O}(g^4/T^2)$ and ${\cal O}(\lambda g^2)$, respectively.
     The explicit evaluation of diagram \ref{fig:self_two_loop}a,
     at zero external momentum, can be found in
     appendix~\ref{app:self_two_loop}.
}
 Consequently,
 at high temperature,
 the thermal width $\Gamma_k$,
 as defined in Eq.~(\ref{eq:thermal_width}),
 is ${\cal O}(\lambda^2 T)$ for hard (compared to $m_{\rm th}$)
 external on-shell momenta,
 and ${\cal O}(\lambda^{3/2} T)$ for soft on-shell momenta.

 \section{Classification of diagrams}
 \label{sec:classification}

 \subsection{Near on-shell singularities of cut diagrams}
 \label{subsec:on_shell_singularities}

\begin{figure}
 \setlength {\unitlength}{1cm}
\vbox
    {%
    \begin {center}
           \begin{picture}(0,0)
           \end{picture}
	\leavevmode
	\def\epsfsize	#1#2{0.4#1}
	\epsfbox{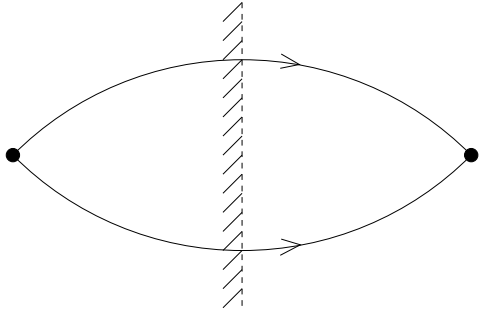}
    \end {center}
 \caption{%
 A typical one-loop cut diagram for the calculation of a Wightman
 function.
 The black dot at each end represents the external bilinear
 operator.
 }
 \label{fig:one_loop}
 }%
 \end{figure}

 Diagrams contributing to the spectral density of the stress
 tensor correlations function have two external vertices each of
 which connect to at least two propagators.
 For example,
 the shear viscosity requires evaluating the correlation function
 \begin{equation}
 \sigma_{\pi\pi}=\langle {\pi}_{lm} {\pi}^{lm} \rangle
 \;,
 \end{equation}
 where the traceless stress tensor,
 \begin{equation}
 {\pi}_{lm} \equiv
 \partial_l {\phi} \, \partial_m {\phi}
 -{\textstyle {1\over 3}}
 \delta_{lm}\partial_k {\phi}\, \partial^k {\phi}
 \;,
 \label{eq:pi_lm2}
 \end{equation}
 is quadratic in the scalar field.
 Naively, one would expect the dominant contribution to come from
 the one-loop diagram shown in Fig.~\ref{fig:one_loop}.
 However, in the zero momentum, small frequency limit,
 a finite temperature cut diagram
 such as this one can contain pairs of lines sharing the same
 loop momenta.
 As explained below, a near on-shell singularity appears
 \noindent
 wherever there is a product of two equal-momentum propagators.
 Since the thermal width that regulates these on-shell singularities
 is ${\cal O}(\lambda^2)$,
 the size of a diagram is no longer given simply by the number
 of explicit interaction vertices.%
\footnote{%
    In addition to the near on-shell singularities regulated
    by the thermal
    width, the soft and collinear
    singularities regulated by the thermal mass must be
    also considered at high temperatures.
    Fortunately,
    these soft and collinear singularities turn out not to affect
    the power counting in $\lambda$ presented in this section.
    Consequently,
    discussion of this point is deferred until
    appendix~\protect{\ref{app:soft_contributions}}.
}

 The infrared behavior of a cut diagram at non-zero temperature
 is more singular than at zero temperature.
 At zero temperature, lines in a diagram sharing the same loop
 momentum do not cause on-shell singularities
 because the poles in the frequency plane all
 reside on one side of the contour.
 However, at non-zero temperature, a propagator has poles on
 both sides of the contour, as can be seen in Eq.~(\ref{eq:prop_a}).
 Hence, products of free
 propagators sharing the same loop momentum
 contain poles pinching the contour, and thus produce an on-shell
 singularity.

 Inclusion of the finite thermal width,
 as in Eq.~(\ref{eq:single_ptl_sd}) and Eq.~(\ref{eq:prop_a}),
 regulates these on-shell singularities.
 The effect of these cut-off singularities
 may be illustrated by analyzing
 the would-be divergent part
 of the product of two propagators
 $\tilde{G}(k)\, \tilde{G}(k{+}\delta)$.  This product
 represents, for example,
 the two lines connected to the external vertex on the right
 side in
 \begin{figure}
 \setlength {\unitlength}{1cm}
\vbox
    {%
    \begin {center}
 \begin{picture}(0,0)
 \put(6.9,4.8){$k^0$}
 \put(0.5,1.2){$(-E_k{-}i\Gamma_k)$}
 \put(0.5,3.8){$(-E_k{+}i\Gamma_k)$}
 \put(4.5,1.2){$(E_k{-}i\Gamma_k)$}
 \put(4.5,3.8){$(E_k{+}i\Gamma_k)$}
 \end{picture}
	\leavevmode
	\def\epsfsize	#1#2{0.4#1}
	\epsfbox {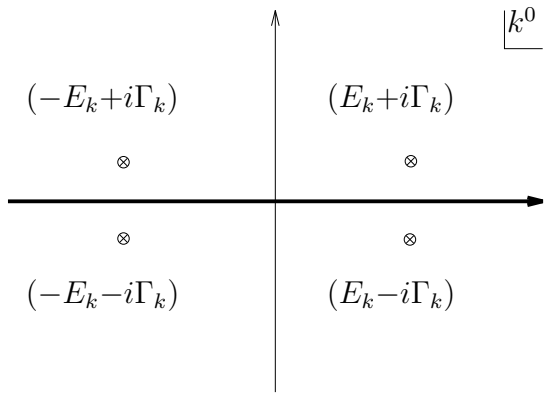}
    \end {center}
 \caption{%
 Nearly pinching poles in the product
 $G(k)^2$.
 The heavy line along the real axis represents the integration
 contour.
 }
 \label{fig:cmplx}
 }
 \end{figure}
 \noindent
 Fig.~\ref{fig:typical_cut}
 if the small external momentum leaving the vertex is $\delta$.

 As explained earlier, the propagator $\tilde{G}(k)$ has
 poles at $k^0=\pm (E_k{-}i\Gamma_k)$, and
 $k^0=\pm (E_k{+}i\Gamma_k)$.  Hence, when $\delta\to 0$,
 the product $\tilde{G}(k)\,\tilde{G}(k{+}\delta)$ contains
 poles separated by $\pm i\Gamma_k$ on opposite sides of
 the frequency contour, as illustrated in Fig.~\ref{fig:cmplx}.
 When the frequency integration is carried out,
 the contribution from these nearly pinching poles is
 ${\cal O}(1/\Gamma_k)={\cal O}(1/\lambda^2)$.
 Exactly the same argument applies to the case of two cut propagators,
 or the product of cut and uncut propagators.
 Hence, the product of any two equal-momentum propagators will
 contain nearly pinching poles.
 Consequently, a diagram with $m$ explicit
 interaction vertices and $n$ pairs of equal-momentum lines is
 potentially
 ${\cal O}(\lambda^m/\Gamma^n) = {\cal O}(\lambda^{m-2n})$.
 The naive expectation of one-loop dominance is not justified
 when $(2n{-}m) \geq 2$.

 The physical origin of the near infrared divergences
 caused by nearly pinching poles at non-zero temperature
 can be traced to the existence of on-shell thermal excitations.
 When a small momentum is introduced by an external operator,
 an on-shell thermal excitation can absorb the
 external momentum and become slightly off-shell.  The slightly
 off-shell particle may propagate a long time before it discharges
 the excess momentum and returns to the thermal distribution.
 Indefinite propagation of a stable on-shell excitation
 causes a divergence, since the amplitude is proportional to the
 infinite propagation time\cite{Coleman}.
 But at finite temperature,
 excitations cannot propagate indefinitely through the thermal medium
 without suffering collisions with other excitations.
 Hence, there are no stable excitations at non-zero temperature.
 If an excitation with momentum $k$ undergoes
 collisions at an average rate $1/\tau_k$, the contribution of that mode
 will be proportional to $\tau_k$, or the inverse of the width
 $\Gamma_k$.

 This may easily be seen explicitly in the product
 $\tilde{G}(k)\,\tilde{G}(k{+}\delta)$
 which contains the (nearly) singular piece,
 \begin{eqnarray}
 \left(
 \tilde{G}(k)\,
 \tilde{G}(k{+}\delta)
 \right)_{\rm pp}
 & \displaystyle = & \displaystyle
 \left(
 {
 -i[1{+}n(E_{k{+}\delta})]
	\over
 E_{k{+}\delta}^2
 {-} (k^0{+}\delta^0)^2
 {-} 2iE_{k{+}\delta}\Gamma_{k{+}\delta}
 }
 n(E_k)\rho(|k^0|, {\bf k})
 {+} ( k \leftrightarrow k{+}\delta )
 \right)_{\rm pp}
 \nonumber\\
 & \displaystyle & \displaystyle {} \times
 (1 + {\cal O}(\lambda^2))
 \;.
 \label{eq:pinched_prop}
 \end{eqnarray}
 Here the subscript ``pp'' indicates the pinching pole contribution.
 The spectral density with a Bose factor $n(E_k)$
 in Eq.~(\ref{eq:pinched_prop})
 may be interpreted as available
 phase space of the initial thermal particle.
 The rest
 may be interpreted as the Bose-enhanced amplitude for propagation
 of a particle after it has absorbed the soft momentum.
 When the thermal width is small compared to the average thermal
 energy,
 the single particle spectral density
 ({\it c.f.}~Eq.~(\ref{eq:Lorentzian_sd}))
 becomes sharply peaked near $k^0 = \pm E_k$.
 Near these peaks, the denominator in Eq.~(\ref{eq:pinched_prop})
 becomes ${\cal O}(E_k\Gamma_k)$.
 Hence, the contribution of
 $
 \left(
 \tilde{G}(k)\,
 \tilde{G}(k{+}\delta) \right)
 $
 contains an
 ${\cal O}(1/\Gamma_k)$ factor.

 \subsection{Classification}
 \label{subsec:classification}

 To simplify the presentation,
 the classification of $\lambda\phi^4$ diagrams will be examined
 first.
 The effect of adding an additional $g\phi^3$ interaction
 will be discussed afterwards.

 The classification of the diagrams is fairly straightforward.
 One only has to count the
 number of explicit interaction vertices in the diagram plus
 the number of equal-momentum pairs of lines as the
 external 4-momentum goes to zero.
 Since
 the thermal lifetime in $\lambda\phi^4$
 theory is ${\cal O}(1/\lambda^2)$,
 a finite temperature cut diagram with $m$ interaction
 vertices and
 $n$ two-particle intermediate states contributes at
 ${\cal O}(\lambda^{m{-}2n})$.
 For example, the one-loop diagram
 \begin{figure}
 \setlength {\unitlength}{1cm}
\vbox
    {%
    \begin {center}
 \begin{picture}(0,0)
 \end{picture}
	\leavevmode
	\def\epsfsize	#1#2{0.4#1}
	\epsfbox {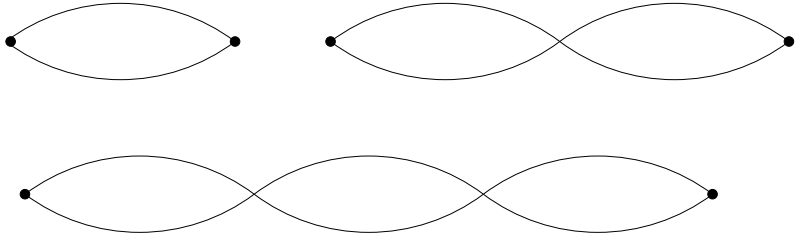}
    \end {center}
     \caption
 	{%
 	The first few chain diagrams in $\lambda\phi^4$ theory.
        Again, the black circles at each end represent bilinear
	external operators.
	 }
 \label{fig:chain}
     }
 \end {figure}
 \noindent
 in Fig.~\ref{fig:one_loop}
 has a single pair of lines with coincident momenta
 in zero external momentum and frequency limit.
 When cut, one line effectively forces the other line
 on-shell, and the contribution of the one-loop diagram in the zero
 momentum limit is ${\cal O}(1/\lambda^2)$.

 To determine what diagrams dominate
 in the calculation of a bilinear operator spectral density,
 one must examine
 which processes can scatter two particle intermediate states
 into two particle intermediate states.
 The minimal way of producing a two particle state from another
 two particle state is via a single elementary scattering.
 Diagrams in $\lambda\phi^4$ theory
 that consist of only these processes will be
 called ``chain'' diagrams.
 As illustrated in Fig.~\ref{fig:chain}, a chain diagram consists
 of  a series of one-loop bubbles.

 Adding each bubble to the chain introduces one additional factor
 of $\lambda$ from the interaction vertex and two inverse powers of
 $\lambda$ from the (nearly) pinching poles of the new bubble.
 Since the lowest order (one-loop) diagram is
 ${\cal O}(1/\lambda^2)$, a chain diagram with $n$ bubbles is
 potentially ${\cal O}(1/\lambda^{1+n})$.
 This suggests that
 the most significant contribution with a given
 number of interaction vertices
 would come from such chain diagrams.
 However,
 the contribution of each added bubble actually lacks a pinching
 pole
 contribution.
 Consequently,
 as will be shown shortly in section~\ref{subsec:chains},
 the net contribution of chain diagrams
 is to modify the contribution of the external vertex
 by a term of ${\cal O}(\lambda T^2)$.
 For the bulk viscosity,
 this correction is {\em not} negligible since
 an insertion of
 $\bar{\cal P}={\cal P}{-}v_{\rm s}^2\varepsilon$
 (where $v_{\rm s}^2$ is the speed of sound)
 produces an ${\cal O}(\lambda T^2)$ factor for typical loop momenta
 of
 ${\cal O}(T)$, as shown in section~\ref{subsec:inhomogeneous_terms}.
 For the shear viscosity, chain diagrams do not contribute at all
 since the angular integration over a single
 insertion of $\pi_{lm}$, $(k_l k_m{-}{1\over3}\delta_{lm}{\bf k}^2)$,
 vanishes due to rotational invariance.
 \begin{figure}
 \setlength {\unitlength}{1cm}
\vbox
    {%
    \begin {center}
 \begin{picture}(0,0)
 \put(0.8,3.5){$k$}
 \put(0.8,0.1){$k$}
 \put(3.0,3.5){$p$}
 \put(3.0,0.1){$p$}
 \put(1.0,1.8){$l$}
 \put(3.0,1.8){$l{-}k{+}p$}
 \put(6.0,2.8){$k$}
 \put(6.0,0.7){$k$}
 \put(10.0,3.3){$p$}
 \put(10.0,1.2){$p$}
 \put(7.1,0.5){$l$}
 \put(8.9,3.0){$l$}
 \put(8.2,1.9){$l{-}k{+}p$}
 \end{picture}
	\leavevmode
	\def\epsfsize	#1#2{0.4#1}
	\epsfbox {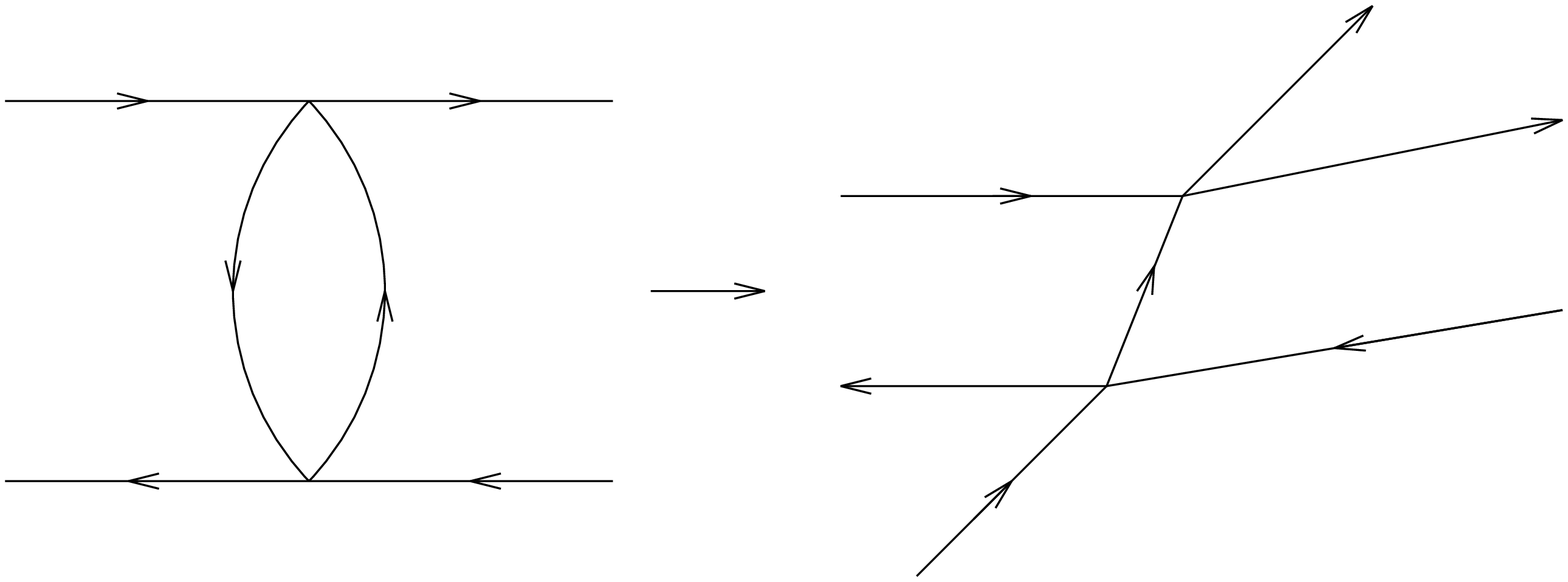}
    \end {center}
 \caption{%
 A diagrammatic representation of
 momentum exchange between two lines via one-loop process
 in the $\lambda\phi^4$ theory.
 When all momenta are on-shell,
 this process can be interpreted as a second order scattering
 that causes a transition between different two particle states.
 }
 \label{fig:ladd_exch}
 }
 \end{figure}

 The next most efficient way of
 causing a transition between different
 two particle states in $\lambda\phi^4$ theory
 is via a second order elastic scattering
 involving a spectator particle in the thermal medium,
 as illustrated in Fig.~\ref{fig:ladd_exch}.
 In this case, momentum is
 exchanged between two lines via a one-loop process as shown
 in the
 first diagram in Fig.~\ref{fig:ladd_exch}.
 When all momenta are on-shell,
 this process may be interpreted as
 a second order scattering involving a physical thermal particle
 with
 momentum $l$
 that causes a transition between two particle state
 with a common momentum $k$ and two particle state with a common
 momentum $p$.
 A diagrammatic representation of this process is shown
 in the second diagram in Fig.~\ref{fig:ladd_exch}.

\begin{figure}
 \setlength {\unitlength}{1cm}
\vbox
    {%
    \begin {center}
 \begin{picture}(0,0)
 \end{picture}
	\leavevmode
	\def\epsfsize	#1#2{0.4#1}
	\epsfbox {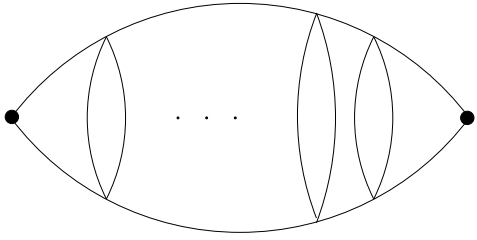}
    \end {center}
 \caption{%
 The planar ladder diagram with $N$ rungs
 in $\lambda\phi^4$ theory.
 The black dot at each end represents an insertion of
 an external operator.
 }
 \label{fig:ladder4}
 }
 \end{figure}

 Diagrams in $\lambda\phi^4$ theory consisting
 entirely of two parallel lines exchanging momenta
 via such one-loop diagrams will be called ``ladder'' diagrams, and are
 are illustrated in
 Fig.~\ref{fig:ladder4}.
 The one-loop sub-diagrams connecting the other two lines are
 the ``rungs'' of the ladder.
 All ladder diagrams
 contribute at the same order as the one-loop diagram
 ({\it i.e.,}~${\cal O}(1/\lambda^2)$) since
 each rung adds two more factors of $\lambda$ and one more
 ${\cal O}(1/\lambda^2)$ lifetime.
 Therefore, all ladder diagrams must be summed to evaluate
 the transport coefficients correctly.
 The explicit forms of these ladder diagrams
 will be examined more closely
 when the summation of all ladder diagrams is discussed
 in section~\ref{sec:summation}

%
%

 The presence of an additional cubic interaction generates
 one additional ``chain'' diagram and
 a set of simple ``ladder'' diagrams
 whose contribution potentially grows as more loops are added.
 The only ``chain'' diagram with only cubic
 interactions is the two loop

 \begin{figure}
 \setlength {\unitlength}{1cm}
\vbox
    {%
    \begin {center}
 \begin{picture}(0,0)
 \end{picture}
	\leavevmode
	\def\epsfsize	#1#2{0.5#1}
	\epsfbox {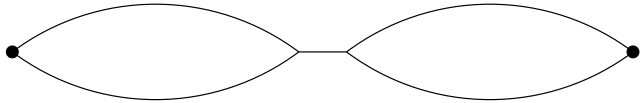}
    \end {center}
     \caption
 	{%
 	The two loop chain diagram in a scalar
	$g\phi^3$ theory.
 	}
 \label{fig:chain3}
     }
 \end {figure}
\begin{figure}
 \setlength {\unitlength}{1cm}
\vbox
    {%
    \begin {center}
 \begin{picture}(0,0)
 \end{picture}
	\leavevmode
	\def\epsfsize	#1#2{0.4#1}
	\epsfbox {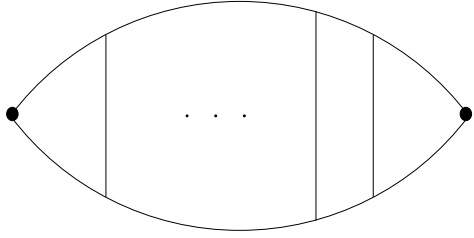}
    \end {center}
     \caption
 	{%
 	A ladder diagram with $N$ straight rungs in a scalar $g\phi^3$
 theory.
	}
 \label{fig:ladder3}
     }
 \end {figure}

 \noindent
 diagram illustrated in Fig~\ref{fig:chain3}.
 Other possible ``chain'' diagrams
 with more than two bubbles connected by single
 lines do not appear because they are a part of the resummed
 propagator.
 Again,
 for the shear viscosity, the two-loop diagram vanishes
 due to rotational invariance.
 For the bulk viscosity,
 as shown in section~\ref{subsec:chains},
 the contribution of this two-loop diagram is also to modify
 contribution of the $\bar{\cal P}$ vertex
 by a term of ${\cal O}(\lambda T^2)$ in addition to the
 modification from summing up $\lambda\phi^4$ chain diagrams.
 The set of diagrams that may potentially grow with the increasing
 number of loops is the set of $g\phi^3$ ``ladder'' diagrams
 with straight rungs, shown in Fig.~\ref{fig:ladder3}.
 Recall that $g = {\cal O}(\sqrt{\lambda}m_{\rm phys})$.
 Hence, superficially
 a ladder diagram with $n$ straight rungs could be
 ${\cal O}(1/\lambda^{n+2})$
 since there are $n{+}1$ factors of $1/\lambda^2$ coming from the
 $n{+}1$ pairs of equal momentum lines, and $2n$ factors of $g$
 (or equivalently, $n$ factors of $\lambda$) from
 the explicit interaction vertices.
 However,
 each straight rung actually contributes an ${\cal O}(g^4)$
 suppression rather than ${\cal O}(g^2)$ suppression,
 and hence all ladder diagrams with straight rungs can contribute at
 ${\cal O}(1/\lambda^2)$, the same as the one-loop diagram.

 To understand this suppression,
 first consider a ladder diagram with the cut running
 through all the straight rungs.
 When all loop momenta flowing through the side rails are
 \noindent
 forced on-shell by the pinching poles, the momenta flowing through
 the straight rungs are
 necessarily highly off-shell.
 Each cut rung contributes a factor of the spectral density
 \begin{equation}
 \rho(\underline{l}_1{+}\underline{l}_2) =
 {2 \Sigma_I(\underline{l}_1{+}\underline{l}_2)
 	\over
 |(\underline{l}_1{+}\underline{l}_2)^2 + m_{\rm th}^2|^2}
 \;,
 \label{eq:sd_estimate}
 \end{equation}
 where $\underline{l}_1$, $\underline{l}_2$
 are the on-shell 4-momenta flowing through the side rails sandwiching
 the rung.
 Recall that the imaginary part of the self-energy
 at an off-shell momentum is
 ${\cal O}(g^2) = {\cal O}(\lambda m_{\rm phys}^2)$.
 Hence,
 when the denominator is ${\cal O}(m_{\rm phys}^4)$,
 a cut rung is ${\cal O}(g^4/m_{\rm phys}^4)$, or ${\cal O}(\lambda^2)$.
 At temperatures comparable or smaller than the physical mass,
 $(T
 \,\,\vcenter {\hbox{$\buildrel{\displaystyle <}\over\sim$}} \,\,
 m_{\rm phys})$,
 the denominator in Eq.~(\ref{eq:sd_estimate}) is ${\cal O}(m_{\rm phys}^4)$
 since the typical size of loop momenta is ${\cal O}(T)$.
 Consequently, all ladder diagrams with straight cut rungs
 can contribute at ${\cal O}(1/\lambda^2)$ when
 $T
 \,\,\vcenter {\hbox{$\buildrel{\displaystyle <}\over\sim$}} \,\,
 m_{\rm phys}$.
 At $T = {\cal O}(m_{\rm phys}/\sqrt{\lambda})$,
 the denominator in Eq.~(\ref{eq:sd_estimate})
 can be
 ${\cal O}(m_{\rm phys}^4)={\cal O}(m_{\rm th}^4)$
 when the small loop momentum contribution cannot be ignored,
 which is the case when calculating the bulk viscosity.
 At much higher temperatures
 $(T
 \,\, \vcenter{\hbox{$\buildrel{\displaystyle >}\over\sim$}} \,\,
 m_{\rm phys}/\lambda)$,
 the contribution of a cut rung is at most
 ${\cal O}(g^4/m_{\rm th}^4) = {\cal O}(\lambda^4)$.
 Hence, the contribution of
 a ladder diagram containing such rungs may be ignored compared
 to the contribution of the one-loop diagram.

 This additional suppression would appear to be absent when there
 are uncut rungs.
 This is correct for individual diagrams with uncut rungs.
 However, as shown in the next
\noindent
 section,
 the real part of a rung cancels in the pinching pole approximation when
 all the cut diagrams associated with one original
 Feynman diagram are summed.
 Hence, after summation over all possible cuts,
 any straight rung may be regarded as ${\cal O}(\lambda^2)$.

 \begin{figure}
 \setlength {\unitlength}{1cm}
\vbox
    {%
    \begin {center}
 \begin{picture}(0,0)
 \end{picture}
	\leavevmode
	\def\epsfsize	#1#2{0.3#1}
	\epsfbox {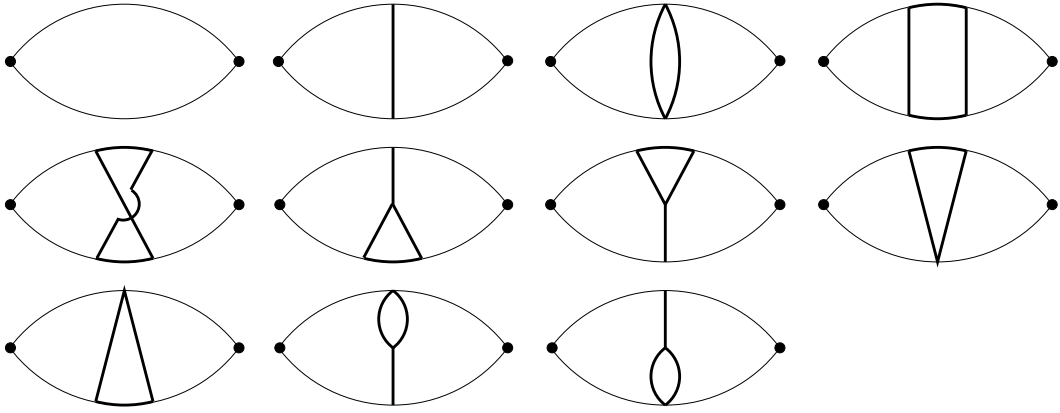}
    \end {center}
     \caption
 	{%
 One-, two- and three-loop
 diagrams contributing at $\Order(1/g^4)$.
 The thick lined sub-diagram are all $\Order(g^4)$ ``rungs''.
 Cut lines are not explicitly drawn.
	}
 \label{fig:lowest_ladder3}
     }
 \end {figure}

 The key result for the above estimate is that
 when the loop frequency integrations are carried out,
 the contribution of the sub-diagram
 sandwiched between pinching pole side rails
 (in this case, the straight rung) can be ${\cal O}(\lambda^2)$ for
 $T
 \,\,\vcenter {\hbox{$\buildrel{\displaystyle <}\over\sim$}} \,\,
 m_{\rm phys}/\sqrt{\lambda}$.
 Note that the sandwiched sub-diagram need not be restricted
 to the straight rung for the above estimate to hold.
 Substituting a straight rung with
 any of the other ``rungs'' shown in Fig.~\ref{fig:lowest_ladder3}
 would work just as well, since they all can be ${\cal O}(\lambda^2)$
 when
 $T
 \,\,\vcenter {\hbox{$\buildrel{\displaystyle <}\over\sim$}} \,\,
 m_{\rm phys}/\sqrt{\lambda}$ without further suppression.

\begin{figure}
 \setlength {\unitlength}{1cm}
\vbox
    {%
    \begin {center}
 \begin{picture}(0,0)
 \end{picture}
	\leavevmode
	\def\epsfsize	#1#2{0.4#1}
	\epsfbox {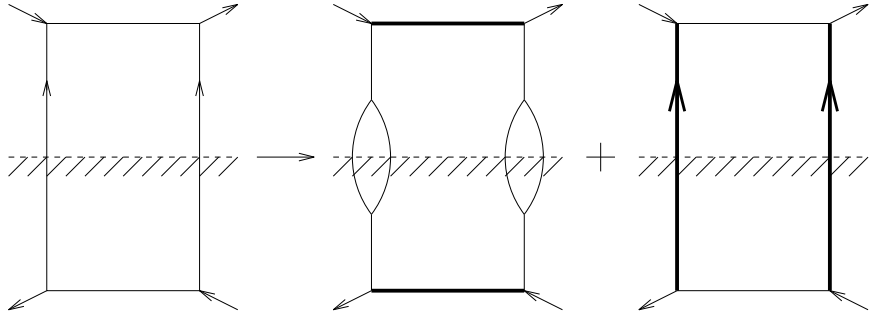}
    \end {center}
     \caption
 	{%
A box diagram which is a part of a ladder diagram with straight
 rungs.
The heavy lines implies that the corresponding momenta are put
 on-shell.
}
 \label{fig:box_rung}
     }
 \end {figure}

 One important complication is that, for straight $g\phi^3$ ladders,
 it is not sufficient to replace the product of propagators
 representing the side rails by their pinching pole part.  The
 non-pinching pole part,
 $\tilde{G}(k)\tilde{G}(k{+}\delta){-}
 (\tilde{G}(k)\tilde{G}(k{+}\delta))_{\rm pp}$,
 can also generate leading order contributions.
 Specifically,
 consider the box diagram shown in Fig.~\ref{fig:box_rung}.
 When the frequency integration is carried out,
 the residue of the pinching poles contained
 in the side rail propagators is ${\cal O}(\lambda^2)$
 due to four explicit factors of $g$ from the interaction vertices,
 and one ${\cal O}(1/\lambda^2)$ thermal lifetime
 compensated by two
 ${\cal O}(\lambda)$ cut propagators at off-shell momenta.
 This is not the only ${\cal O}(\lambda^2)$ contribution contained in
 the box diagram.
 Putting the two cut rungs on-shell also produces
 an ${\cal O}(g^4)={\cal O}(\lambda^2)$ contribution
 since no near-divergence cut propagator modifies the explicit
 factor of $g^4$.
 It will be convenient to regard
 the ${\cal O}(\lambda^2)$ non-pinching pole contribution
 of the box diagram as another elementary ``rung''
 \noindent
 which may be sandwiched between two pinching pole side rails.
 A more detailed examination of the non-pinching pole
 contribution from the box diagram is contained
 in section~\ref{sec:summation}
 where the summation of all ladder diagrams is discussed.

 \subsection{Higher order rungs in the calculation of the bulk viscosity}
 \label{subsec:inelastic_rungs}

 There are other higher-order ``rungs'' corresponding to
 processes more complicated
 than those shown in Fig.~\ref{fig:lowest_ladder3}.
 The processes corresponding to these ``rungs''
 contain more elementary scatterings than the rungs in
 Fig.~\ref{fig:lowest_ladder3} without the compensating pinching poles,
 and are sub-dominant as long as individual diagrams are compared.
 However, when an infinite number of diagrams are summed,
 the next order diagrams cannot be simply discarded without further
 analysis of the convergence of the sum of the leading order diagrams.

 For the shear viscosity calculation, no convergence problem arises.
 However, for the
 bulk viscosity calculation,
 the sum of the leading order part of the ladder diagrams diverges
 as shortly shown in section~\ref{sec:summation}.
 However, this is not a failure of the theory.
 As explained
 in section~\ref{subsec:qualitative},
 the bulk viscosity calculation must involve
 number-changing inelastic scattering processes.
 The leading order part of the simple ladder diagrams contains only
 the elastic scattering processes.
 Hence, it is no surprise that they cannot produce
 the correct
 leading order bulk viscosity.

 To calculate the leading order bulk viscosity,
 the next-to-leading order diagrams
 \begin{figure}
 \setlength {\unitlength}{1cm}
\vbox
    {%
    \begin {center}
 \begin{picture}(0,0)
 \end{picture}
	\leavevmode
	\def\epsfsize	#1#2{0.35#1}
	\epsfbox {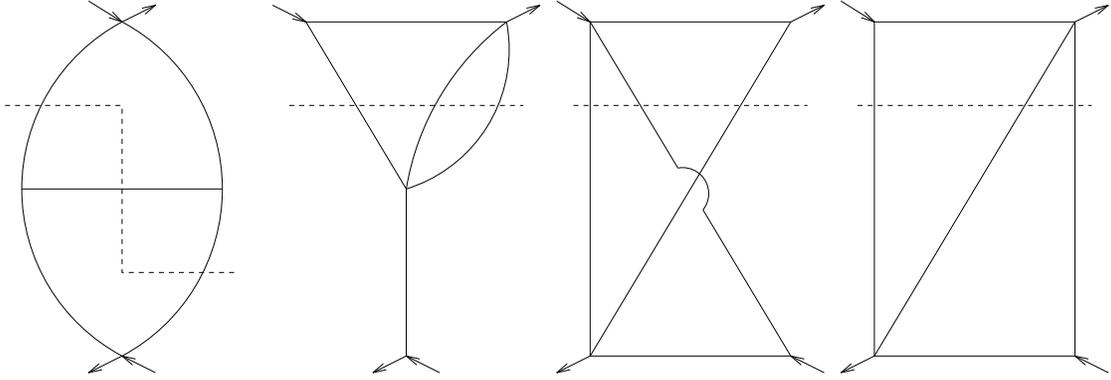}
    \end {center}
     \caption
 	{%
  Typical $g\phi^3{+}\lambda\phi^4$ theory
  $\Order(\lambda^2 g^2)$ rungs containing 2--3 scattering processes.}
  \label{fig:higher_rungs}
     }
 \end {figure}
 \noindent
 containing number-changing
 scattering processes must be included.
 The lowest order number
 changing process (hence the shortest relaxation time)
 in the $g\phi^3{+}\lambda\phi^4$
 theory is ${\cal O}(\lambda g)$ 2--3 scatterings.
 A few of such
 ``rungs'' containing these processes are illustrated in
 Fig.~\ref{fig:higher_rungs}.
 Other ${\cal O}(\lambda^2 g^2)$ rungs can be obtained by attaching
 one more line to the rungs in Fig.~\ref{fig:lowest_ladder3} in all
 possible ways consistent with the theory.
 Diagrams containing these rungs must be included in
 the bulk viscosity calculation in the
 $g\phi^3{+}\lambda\phi^4$ theory.

 For the pure $\lambda\phi^4$ theory, the lowest order number changing
 process is ${\cal O}(\lambda^2)$.  The ${\cal O}(\lambda^4)$
 rungs corresponding to these processes can be obtained by attaching
 two more lines to the rungs in Fig.~\ref{fig:lowest_ladder3}
 in all possible ways consistent with the $\lambda\phi^4$ theory.

 The rest of this section completes the classification of diagrams
 by showing how the chain diagrams modify the external vertex
 contribution.

 \subsection{Chain diagrams}
 \label{subsec:chains}

 Once again, for the sake of simplicity, $\lambda\phi^4$ diagrams
 are
 examined first.
 The analysis of the two-loop $g\phi^3$ chain diagram,
 diagrams with mixed $\lambda\phi^4$ and $g\phi^3$ bubbles,
 and the examination of chain diagrams with more complicated
 bubbles
 will follow.
 For a given number of interaction vertices, chain diagrams in
 $\lambda\phi^4$ theory, such as those in Fig.~\ref{fig:chain},
 \begin{figure}
 \setlength {\unitlength}{1cm}
\vbox
    {%
    \begin {center}
 \begin{picture}(0,0)
\put(-0.2,1.1){$q$}
\put(2.5,2.0){$l_1$}
\put(2.5,0.0){$q{-}l_1$}
\put(5.5,2.0){$l_2$}
\put(5.5,0.0){$q{-}l_2$}
\put(6.8,1.1){$q$}
\put(9.5,2.0){$l_1$}
\put(9.5,0.0){$q{-}l_1$}
\put(12.7,2.0){$l_2$}
\put(12.7,0.0){$q{-}l_2$}
 \end{picture}
	\leavevmode
	\def\epsfsize	#1#2{0.3#1}
	\epsfbox {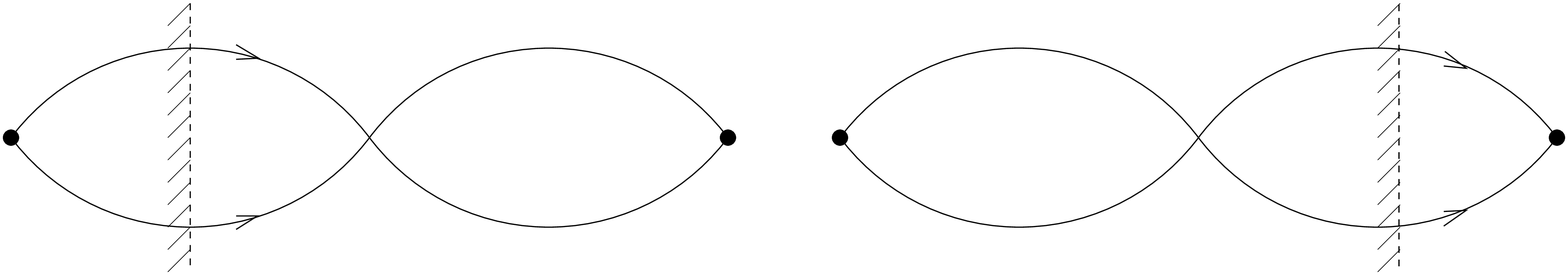}
    \end {center}
     \caption
 	{%
	Cut two-loop chain diagrams in the $\lambda\phi^4$ theory
	contributing to the Wightman function $\sigma_{AA}$.
 	}
 \label{fig:two_loop_chain}
     }
 \end {figure}
 \noindent
 contains the greatest number of pairs
 of the lines sharing the same loop momentum.
 A chain diagram with $n$ bubbles is potentially
 ${\cal O}(1/\lambda^{n+1})$
 because there are $n$ ${\cal O}(1/\lambda^2)$ thermal lifetimes
 and $(n{-}1)$ explicit factors of $\lambda$ from the interaction
 vertices.
 However,
 this is a severe over-estimate
 since the actual contribution of an added bubble
 lacks a pinching pole contribution.
 This is because (a) the discontinuity of a bubble vanishes
 in the zero external 4-momentum limit,
 and (b) the real part of a bubble does not contain pinching
 poles.
 For example, consider the two-loop chain diagrams, depicted in
 Fig.~\ref{fig:two_loop_chain},
 contributing to the calculation of the Wightman function
 $\sigma_{AA}^{\vphantom{x}}(q)$.
 Here, the external operator
 $\hat{A}$ may be any component of the stress-energy tensor,
 and is assumed to be even under a CPT transformation.

 The cut bubble is given by
 \begin{equation}
 L_A(q) \equiv {1\over 2} \int {d^4 l \over (2\pi)^4} \,
 		I_A(l,q{-}l)\, S(l)\, S(q{-}l)
 \;,
 \label{eq:cut_bubble}
 \end{equation}
 \hbox{and the uncut bubble in the unshaded region is}
 \begin{equation}
 C_A(q)  \equiv
	{-i\over 2}
	 \int {d^4 l \over (2\pi)^4} \,
         I_A(l,q{-}l)\, \tilde{G}(l)\, \tilde{G}(q{-}l)
 \;,
 \label{eq:uncut_bubble}
 \end{equation}
 \noindent
 where $q$ is the external 4-momentum, and
 $I_A(l,q{-}l)$ denotes the (polynomial)
 contribution from the external operator in such a manner that the
 contribution of $\phi^2$ $I_{\phi^2}(l,q{-}l) = 1$.

 Since the operator $\hat{A}$ is even under a CPT
 transformation, at zero external momentum
 $I_A(l,-l)$ is a real, even function of the loop momentum $l$.
 Hence,
 the sum of the two-loop chain diagrams
 in the zero external 4-momentum limit is
 \begin{eqnarray}
 \lim_{q^0\to 0}\,\lim_{{\bf q}\to 0}\,
 L_{AA}^{\hbox{\scriptsize{(2-loop)}}}(q)
 & \displaystyle = & \displaystyle
 4\lambda
 L_A(0)\,
 \left( C_A(0)+C_A(0)^* \right)
 \;
 \nonumber\\
 & \displaystyle = & \displaystyle
 8{\lambda} L_A(0)\,
 {\rm Re}\,C_A(0)
 \;.
 \label{eq:two_loop_chain}
 \end{eqnarray}
 In the same limit, the cut one-loop bubble
 $L_A(0)$ is ${\cal O}(1/\lambda^2)$ as before.
 To see that $\lambda{\rm Re}\,C_A(0)$
 does not exceed ${\cal O}(1)$, consider the following
 explicit form of the real part of an uncut bubble at zero momentum,
 \begin{eqnarray}
 {\rm Re}\,C_A(0)
 & \displaystyle = & \displaystyle
 {1\over 2}
 \int {d^4 l\over (2\pi)^4}\,
 I_A(l,-l)\,
 {\rm Re}\,\left(-i\tilde{G}^2(l)\right)
 \nonumber\\
 & \displaystyle = & \displaystyle
 {i\over 4}\int {d^4 l\over (2\pi)^4}\,
 I_A(l,-l)
 \coth(l^0\beta/2)
 \left(
 {1\over [l^2{+}m_{\rm th}^2{+}\Sigma(l)]^2 }
 -
 {1\over [l^2{+}m_{\rm th}^2{+}\Sigma(l)^*]^2}
 \right)
 \label{eq:Re_Ca}
 \;.
 \nonumber\\
 & \displaystyle & \displaystyle
 \end{eqnarray}
 Here, $I_A(l,-l) \sim l^2$ if $\hat{A}$ is a stress-energy tensor.
 Since the integrand does not contain pinching poles
 ({\it i.e.},~there is no $1/|l^2{+}m_{\rm th}^2{+}\Sigma(l)|^2$ term)
 no large lifetime factor appears when the frequency integration
 is performed.
 In appendix~\ref{app:chains},
 the real part of the uncut one-loop diagram is shown to be
 \begin{equation}
 {\rm Re}\,C_A(0) = {\cal O}(T^2)
 \;,
 \label{eq:Order_Re_Ca}
 \end{equation}
 using the fact that
 the integrand is appreciable only when $l$ is nearly on-shell.

 Individual higher order chain diagrams with more one-loop
 bubbles strung together may be analyzed in a similar manner.
 However,
 since chain diagrams form a geometric series,
 it is also straightforward
 to sum all cut chain diagrams with one-loop bubbles
 and examine the result of the summation.
 Of course, one can also
 perform the geometric sum first in imaginary-time,
 and then take the discontinuity of the result of summation.

 The summation of cut chain diagrams with one-loop bubbles
 is fairly simple.
 The only subtleties come from the
 cuts involved and the fact that there is an external operator
 at each end of a cut diagram.
 Due to the cuts, the equation for the resummed chain
 is a matrix equation instead of a single component linear equation.
 The presence of external operators implies that
 the bubbles at each end are not equivalent to the other bubbles.

 Since no additional difficulties than those already present
 in the two-loop calculation appear,
 performing the actual summation of the cut chain diagram
 is deferred to appendix~\ref{app:chains}.
 The result of the summation of all chain diagrams
 with one-loop bubbles is shown in appendix~\ref{app:chains}
 to be
 \begin{eqnarray}
 \lim_{q\to 0} \sigma_{AA}^{\rm chain}(q)
 & \displaystyle = & \displaystyle
 4L_{AA}(0)
 +
 {
 8\lambda\,{\rm Re}\,C_{A}(0)
		\over
 1 - \lambda {\rm Re}\,C_0(0)
 }
 L_A(0)
 +
 {
 4\left( \lambda\,{\rm Re}\,C_{A}(0) \right)^2
 		\over
 (1 - \lambda {\rm Re}\,C_0(0))^2
 }
 L_0(0)
 \nonumber\\
 & \displaystyle \equiv & \displaystyle
 4L_{\tilde{A}\tilde{A}}(0)
 \;,
 \label{eq:chain_sum}
 \end{eqnarray}
 where
 the finite temperature optical theorem
 \begin{equation}
 {\rm Im} \; C_0(q) = -{1\over2} \left( L_0(q) + L_0(-q) \right)
 \;,
 \label{eq:optical_thm}
 \end{equation}
 is used to simplify the result.
 (The optical theorem (\ref{eq:optical_thm})
 can be easily proven from
 Eq.~(\ref{eq:cut_bubble}) and Eq.~(\ref{eq:uncut_bubble}).)
 Here,
 $\sigma_{AA}^{\rm chain}(q)$ denotes the contribution
 of these chain diagrams to the correlation function $\sigma_{AA}(q)$,
 $L_{AA}(q)$ corresponds to the contribution of the one-loop
 diagram with the external operator $\hat{A}$ at both ends, and
 $L_A(q)$
 denotes the contribution of the one-loop diagram with $\hat{A}$
 at one
 end.  $L_0(q)$ and $C_0(q)$ are the cut
 and the uncut bubbles with $I_A=1$.
 The modified one-loop contribution $L_{\tilde{A}\tilde{A}}$
 contains the (modified) vertex contribution
 \begin{eqnarray}
 I_{\tilde{A}}(l,-l)
 & \displaystyle \equiv & \displaystyle
 I_A(l,-l)
 + {\lambda {\rm Re}\, C_A(0)
               \over
 1 - \lambda {\rm Re}\, C(0)}
 \nonumber\\
 & \displaystyle = & \displaystyle
 I_A(l,-l)
 + \lambda {\rm Re}\, C_A(0) \times
        \left( 1 +  {\cal O}(\sqrt{\lambda}) \right)
 \;,
 \label{eq:I_tilde_a}
 \end{eqnarray}
 where the estimate
 $\lambda{\rm Re}\,C(0) = {\cal O}(\sqrt{\lambda})$ is used.
 This estimate of $\lambda{\rm Re}\,C(0)$
 is justified in appendix~\ref{app:chains}.
 For the operator
 $\bar{\cal P}={\cal P}{-}v_{\rm s}^2\varepsilon$
 required for the bulk viscosity,
 $I_{\bar{\cal P}} = {\cal O}(\lambda T^2)$
 for a typical ${\cal O}(T)$ loop momentum, as shown in
 section~\ref{subsec:inhomogeneous_terms}.
 In the same section,
 the additional term
 $\lambda {\rm Re}\, C_{\bar{\cal P}}(0)$
 is also shown to be ${\cal O}(\lambda T^2)$.
 Hence, the correction term $\lambda{\rm Re}\,C_A(0)$
 in Eq.~(\ref{eq:I_tilde_a}) cannot be simply ignored.
 For the shear viscosity, ${\rm Re}\,C_{\pi}(0)$
 vanishes due to rotational invariance.
 Hence, no modification is needed in that case.

 When cubic interactions are added, the ``chain'' diagrams
 also include the two-loop diagram shown in Fig.~\ref{fig:chain3}
 where each bubble in the diagram now may be regarded as the
 sum of all
 $\lambda\phi^4$ chain diagrams.
 The sum of all chain diagrams in the $g\phi^3{+}\lambda\phi^4$
 theory
 is given by the sum of the $\phi^4$ chain result
 $L_{\tilde{A}\tilde{A}}$ (\ref{eq:chain_sum}) and this two-loop
 diagram.
 As shown in appendix~\ref{app:chains},
 a straightforward application of the cutting rules yields
 the sum of all chain diagrams as
 \begin{eqnarray}
 \lim_{q\to 0} \sigma_{AA}^{\hbox{\scriptsize{full-chain}}}(q)
 & \displaystyle = & \displaystyle
 4L_{\tilde{A}\tilde{A}}(0)
 -
 8{g^2\over m_{\rm th}^2}\,{\rm Re}\,C_{\tilde{A}}(0) L_{\tilde{A}}(0)
 +
 4{g^4\over m_{\rm th}^4}\,\left( {\rm Re}\,C_{\tilde{A}}(0) \right)^2
 L_0(0)
 \nonumber\\
 & \displaystyle \equiv & \displaystyle
 4L_{\cal{AA}}(0)
 \;,
 \label{eq:full_chain_sum}
 \end{eqnarray}
 where $L_{\cal{AA}}(0)$ contains
 the modified vertex contribution $I_{\cal A}$ given by
 \begin{eqnarray}
 I_{\cal{A}}(l,-l)
 & \displaystyle \equiv & \displaystyle
 I_{\tilde{A}}(l,-l)
 - {g^2 \over m_{\rm th}^2}{\rm Re}\,C_{\tilde{A}}(0)
 \nonumber\\
 & \displaystyle = & \displaystyle
 I_A(l,-l)
 + (\lambda - {g^2\over m_{\rm th}^2})\, {\rm Re}\, C_A(0) \times
        \left( 1 +  {\cal O}(\sqrt{\lambda}) \right)
 \;.
 \label{eq:chain_modified_Ia}
 \end{eqnarray}

 More complicated chain diagrams can be produced
 by including more complicated bubbles such
 as ladder diagrams.
 For these more complicated ``bubbles'',
 exactly the same argument given above will also apply
 provided that the generalized finite temperature optical theorem
 ({\it c.f.}~Eq.~(\ref{eq:optical_thm}))
 \begin{equation}
 {\rm Im}\, C_{\rm bubble}(q) =
 -{1\over 2} \left(
 		L_{\rm bubble}(q) + L_{\rm bubble}(-q)
 	   \right)
 \;
 \label{eq:optical_thm_bubble}
 \end{equation}
 holds for each bubble.
 However, unlike the imaginary part,
 the real part of higher order contributions,
 including ladder diagrams, to the bubble $C_A$ are
 suppressed compared to the
 real part of the one loop contribution.
 Hence they can be safely ignored.
 The generalized optical theorem
 (\ref{eq:optical_thm_bubble})
 can be inferred from the works of
 Kobes and Semenoff\cite{Semenoff}
 and will not be further discussed in this paper.

 \section{Summation of ladder diagrams}
 \label{sec:summation}

 As explained in section~\ref{subsec:qualitative},
 calculations of
 the shear and the bulk viscosities require different set of diagrams.
 More specifically, to evaluate the leading order shear viscosity,
 summation of only the leading order ladder diagrams is needed.
 Whereas, to evaluate
 the bulk viscosity, as shown in this section,
 ${\cal O}(\lambda^4)$ rungs must also be included.
 In this section, the leading order ladder summation for the shear
 viscosity is examined first.  More
 \begin{figure}
 \setlength {\unitlength}{1cm}
\vbox
    {%
    \begin {center}
 \begin{picture}(0,0)
 \put(1.7,7.63){$=$}
 \put(1.7,5.43){$=$}
 \put(1.7,3.18){$=$}
 \put(1.7,0.93){$=$}
 \put(-0.7,8.1){$k$}
 \put(-0.7,7.15){$q{-}k$}
 \put(-0.7,5.9){$k$}
 \put(-0.7,4.95){$q{-}k$}
 \put(-0.7,3.7){$k$}
 \put(-0.7,2.65){$q{-}k$}
 \put(-0.7,1.45){$q{-}k$}
 \put(-0.7,0.4){$k$}
 \put(3.0,7.63){$+$}
 \put(5.5,7.63){$+$}
 \put(5.5,5.43){$+$}
 \put(5.5,3.18){$+$}
 \put(5.5,0.93){$+$}
 \put(8.0,7.63){$+$}
 \put(8.0,5.43){$+$}
 \put(8.0,3.18){$+$}
 \put(8.0,0.93){$+$}
 \put(10.5,7.63){$+$}
 \put(10.5,5.43){$+$}
 \put(10.5,3.18){$+$}
 \put(10.5,0.93){$+$}
 \end{picture}
	\leavevmode
	\def\epsfsize	#1#2{0.27#1}
	\epsfbox {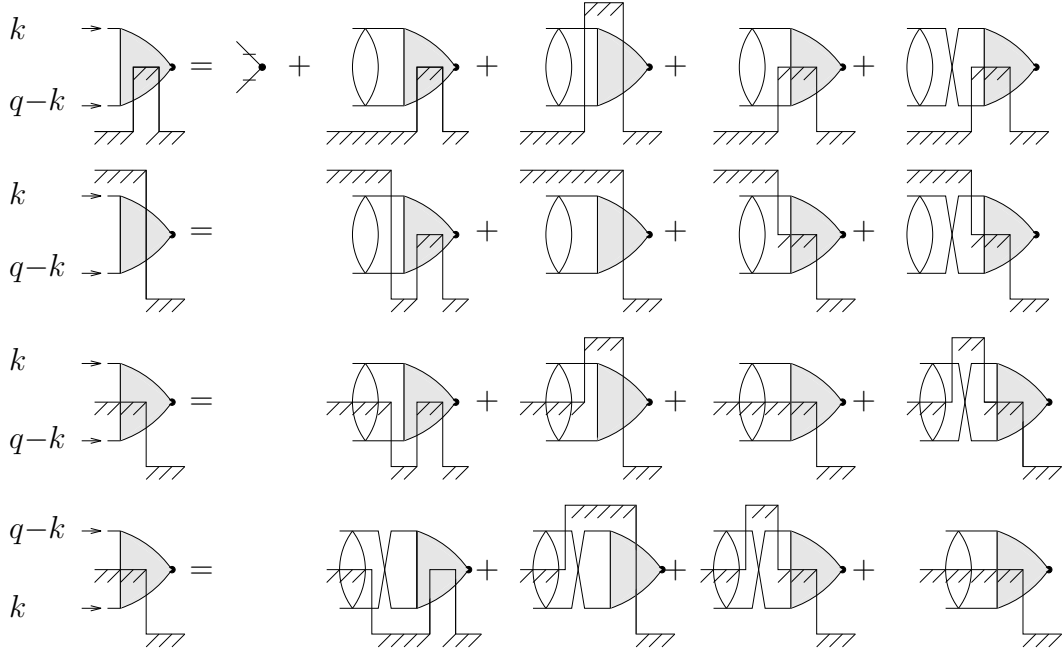}
    \end {center}
     \caption
 	{%
 The diagrammatic representation of the integral equation
 for the effective vertex function ${\cal D}_{\pi}(k,q{-}k)$.
 This corresponds to Eq.~({\protect\ref{eq:vertex_eq4}}).
     }
 \label{fig:vertex_eq4}
 }
 \end{figure}
 \noindent
 complicated analysis of summing the
 higher order contributions for the bulk viscosity follows.
 The results presented in this section are valid for all temperatures.

 \subsection{Ladder summation for the shear viscosity calculation
 in $\lambda\phi^4$ theory}
 \label{subsec:ladder_sum4}

 Cut ladder diagrams
 form a geometric series, and can be resummed
 by introducing a suitable effective vertex.
 Due to the various possible routings of
 the cut, the integral equation will involve a $4{\times}4$ matrix
 valued kernel.
 Hence, it is convenient to introduce an effective vertex
 ${\cal D}_{\pi}(k,q{-}k)$ which is a 4-component column vector.
 The subscript $\pi$ is a label for
 a component of the traceless part of the stress tensor.
 The resummed effective vertex satisfies the following linear
 integral equation
 \begin{equation}
 {\cal D}_{\pi}(k, q{-}k)
 =
 {\cal I}_{\pi}(k, q{-}k)
 +
 \int {d^4 p \over (2\pi)^4} \,
 {\cal M}(k{-}p) \,
 {\cal F}(p, q{-}p)
 {\cal D}_{\pi}(p, q{-}p)
 \;
 \label{eq:vertex_eq4}
 \end{equation}
 illustrated in Fig.~\ref{fig:vertex_eq4}.
 Here,
 ${\cal I}_\pi(k,q{-}k)$ is an inhomogeneous term representing the
 action of the bilinear operator $\hat \pi$ including the contribution
 of chain diagrams, ${\cal M}(k{-}p)$ is
 a $4{\times}4$ matrix representing the rungs of the ladder which
 consist of cut and uncut one-loop diagrams,
 and ${\cal F}(p,q{-}p)$ is a matrix
 representing the side rails of the ladder and
 consists of products of propagators.
 As shown in Fig.~\ref{fig:vertex_eq4}, the first component
 of ${\cal D}_{\pi}(k,q{-}k)$ corresponds to the effective
 vertex where momenta $k$ and $q{-}k$ enter vertices
 in the unshaded region.
 For the second component, $k$ and $q{-}k$ enters vertices in
 the shaded region.  In the third component,
 the momentum $k$ enters a vertex in the unshaded region while the
 momentum $q{-}k$ enters a vertex in the shaded region.
 The last component of ${\cal D}_{\pi}$ differs from the third component
 by changing $k$ to $q{-}k$, and vice versa.

 In a more symbolic form,
 the above equation can be compactly rewritten as
 \begin{equation}
 |{\cal I}_{\pi}\rangle
 =
 (1 - {\cal K})|{\cal D}_{\pi}\rangle
 \;,
 \label{eq:abstract_vertex_eq4}
 \end{equation}
 with the identification of the ``ladder kernel''
 \begin{equation}
 {\cal K} \equiv {\cal MF}
 \;.
 \end{equation}
 As is evident in Fig.~\ref{fig:vertex_eq4}, only the first component
 of the inhomogeneous term ${\cal I}_{\pi}$ is non-zero and given by
 $(k_lk_m{-}\delta_{lm}{\bf k}^2/3)$.
 Explicit expressions for ${\cal M}$ and ${\cal F}$ are given in
 appendix~\ref{app:ladder}.
 Note that all quantities depend on the
 external 4-momentum $q \equiv (\omega, {\bf q})$.

 The integral equation
 $|{\cal I}_{\pi}\rangle = (1{-}{\cal K})|{\cal D}_{\pi}\rangle$
 will be solvable only if any (left) zero modes of the kernel
 $(1{-}{\cal K})$ are orthogonal to the inhomogeneous term
 $|{\cal I}_{\pi}\rangle$.
 The operator $(1{-}{\cal K})$ does have four zero modes in the
 zero momentum, zero frequency limit.
 These four zero modes, denoted $\langle {\cal V}^{\mu} |$,
 are related to insertions of energy-momentum density $T^{\mu 0}$,
 and the existence of these zero modes is a direct consequence of
 energy-momentum conservation.  Explicit forms of these zero modes
 are shown in appendix~\ref{app:zero_modes}.
 Reassuringly,
 $|{\cal I}_{\pi}\rangle$ is orthogonal to the zero modes;
 this is also verified in appendix~\ref{app:zero_modes}.

 In terms of the resummed vertex $|{\cal D}_{\pi} \rangle$,
 the Wightman function of a pair of $\pi_{lm}$ is simply
 \begin{equation}
 \sigma_{\pi\pi}(q) =
 2\,
 \langle z_{\pi} | {\cal F} | {\cal D}_{\pi} \rangle
 \times (1+ {\cal O}(\lambda^2))
 \;,
 \end{equation}
 where
 $z_{\pi}$ represents the action of the operator $\pi_{lm}$
 in the same way ${\cal I}_{\pi}$ represents the action of the operator
 $\pi_{lm}$.%
\footnote{%
     The appropriate inner product is defined by
     \[
     \langle f | g \rangle \equiv
     \int {d^4 p \over (2\pi)^4}\,
     f(p)^{\dagger}\,g(p)
     \;.
     \]
}
 The only difference between $z_\pi$ and ${\cal I}_\pi$ is that
 $z_\pi$ has its only non-zero component in the second slot
 while ${\cal I}_\pi$ is non-zero in the first slot.
 The overall normalization constant 2 is chosen for convenience.

 In this notation,
 the shear viscosity $\eta$ is given by
 \begin{eqnarray}
 \eta & \displaystyle = & \displaystyle
        {\beta\over 10}\,
	\lim_{\omega\to 0}\, \lim_{|{\bf q}|\to 0}\,
	\langle z_{\pi} |{\cal F}| {\cal D}_{\pi} \rangle\times
	(1+{\cal O}(\lambda))
 \;.
 \end{eqnarray}
 \label{eq:ladder_viscosities}
 {}From now on, the external 4-momentum may simply be set to zero.

 In the limit of vanishing external momentum,
 the leading weak coupling behavior is generated by the (nearly)
 pinching pole contribution to the $p^0$ frequency integral.
 Hence, portions of the side rail matrix ${\cal F}$
 which do not contain pinching poles may be neglected.
 Examination of appendix~\ref{app:ladder} together with the explicit
 form of the cut and uncut single particle propagators shows
 that the leading order part of the remaining pinching pole part
 is%
\footnote{%
 	Actually, there is one more part of ${\cal F}$ that
 	contains pinching poles in the
	zero momentum, zero frequency limit.
	However, this part, denoted $hj^T$ in
	appendix~\ref{app:ladder},
	does not contribute to the leading order calculation for the
	following reason.
	First, $hj^T$ is orthogonal to the vertex parts
	since $z_A^Th=0$, and $j^T{\cal I}_{B}=0$.
 	Second, $hj^T$ is orthogonal to the rest of ${\cal F}$ in the
	sense that if $\bar{\cal F}\equiv{\cal F}{-}hj^T$
 	then $j^T\!{\cal M}\bar{\cal F}=0$ and $\bar{\cal F}{\cal M}h=0$.
	Hence, the $hj^T$ part
	does not affect the contributions of the
	ladder diagrams in
	$\sigma_{AB}^{(n)}
	=z_A^T\left({\cal FM}\right)^n\!{\cal F}{\cal I}_{B}$.
}
 \begin{equation}
 {\cal F}_{\rm pp}(p,-p)
 \equiv
 \bar{w}(p)u^T(p) \times
 [1{+}n(E_p)]\,n(E_p)\,
 {\rm sgn}(p^0)\,2\pi\delta(p_0^2{-}E_p^2){\bigg/}\Sigma_I(p)
 \;
 \label{eq:Fpp}
 \end{equation}
 where
 \begin{eqnarray}
 \bar{w}^T(p) & \displaystyle \equiv & \displaystyle
 (1,\; 1,\; (1{+}e^{-p^0\beta})/2,\; (1{+}e^{p^0\beta})/2) \,
 \;,
 \\
 u^T(p) & \displaystyle \equiv & \displaystyle
 (1,\; 1,\; (1{+}e^{p^0\beta})/2,\; (1{+}e^{-p^0\beta})/2)\,
 \;.
 \end{eqnarray}
 The leading order kernel is given by
 \begin{equation}
 {\cal K}_{\rm pp}
 \equiv
 {\cal M}_0 {\cal F}_{\rm pp}
 \;,
 \end{equation}
 where ${\cal M}_0$ contains one-loop rungs evaluated with free
 propagators, and the self-energy $\Sigma_I$ in ${\cal F}_{\rm pp}$
 contains only the contribution of the two-loop diagram calculated
 with the free propagators.  Since the factors of coupling constants
 from ${\cal M}_0$ and $\Sigma_I$ cancel each other,
 ${\cal K}_{\rm pp}$ is independent of $\lambda$ except for those
 contained in the thermal mass.
 Note that dropping non-pinching pole contributions reduces
 ${\cal F}$ to a rank one matrix.
 This allows one to greatly simplify the equation.

 For the change in the solution of the integral equation
 (\ref{eq:abstract_vertex_eq4}) caused by the replacement of
 ${\cal K}$ by ${\cal K}_{\rm pp}$ to be sub-leading in $\lambda$,
 the inhomogeneous term ${\cal I}_{\pi}$ must be orthogonal to the
 (left) zero modes of $(1{-}{\cal K}_{\rm pp})$
 as well as orthogonal to the original zero modes of $(1{-}{\cal K})$.
 Otherwise the reduced integral equation
 $|{\cal I}_{\pi}\rangle=(1{-}{\cal K}_{\rm pp})|{\cal D}_{\pi}\rangle$
 would be singular implying that the neglected part of ${\cal K}$
 could not be negligible.
 The issue of zero modes of $(1{-}{\cal K}_{\rm pp})$ does not
 arise when considering the size of an individual diagram
 as in section~\ref{sec:classification}, but rather reflects the
 convergence (or lack thereof) of the infinite series of ladder diagrams.
 Suppose the inhomogeneous term
 ${\cal I}_{\pi}$ had a non-zero projection onto a
 zero mode $y$.
 Then $({\cal K}_{\rm pp})^n y = y$, and
 all ladder diagrams would contain an identical
 ${\cal O}(1/\lambda^2)$ piece, $z_A^T{\cal F}_{\rm pp}y$,
 as a part of their pinching pole contribution.
 The infinite number of such terms would make the sum diverge.
 Hence, to produce a finite result, the inhomogeneous term must
 satisfy
 \begin{equation}
 \langle \bar{b}_{\mu,5}|{\cal I}_{\pi} \rangle = 0
 \;,
 \label{eq:inhom_cond}
 \end{equation}
 where $\langle \bar{b}_{\mu,5} |$ denotes
 the five zero modes of $(1{-}{\cal K}_{\rm pp})$ whose
 explicit forms are
 \begin{equation}
 \bar{b}_{\mu}(p)
 \equiv
 \langle \bar{b}_{\mu}|p \rangle
 =
 p_{\mu} \,[1{+}n(p^0)]\,S_{\rm free}(-p)\, u^T(p)
 \;,
 \label{eq:zero_modes}
 \end{equation}
 and
 \begin{equation}
 \bar{b}_5(p)
 \equiv
 \langle \bar{b}_5|p \rangle
 =
 {\rm sgn}(p^0) \,[1{+}n(p^0)]\,S_{\rm free}(-p)\, u^T(p)
 \;.
 \end{equation}
 Here, $\bar{b}_\mu$'s corresponds to the 4-momentum conservation, and the
 additional $\bar{b}_5$ corresponds to the particle number conservation.
 Of course the theory does not preserve the number of particles.
 However, the number changing scatterings are ${\cal O}(\lambda^4)$, and
 hence, do not contribute at the leading order.

 As a simple consequence of rotational invariance,
 the traceless stress operator involved in the calculation of
 the shear
 viscosity does satisfy Eq.~(\ref{eq:inhom_cond}).
 When $\bar{b}_i(k)$ is applied to ${\cal I}_{\pi}(k)$, it vanishes
 since rotational invariance requires that
 any rank 3 spatial tensor with two symmetric indices
 be a combination of $k_ik_lk_m$ and $k_i\delta_{lm}$.
 Applying $\bar{b}_0(k)$ or $\bar{b}_5(k)$ again results in zero
 because the angular integration over ${\cal I}_{\pi}(k)$ vanishes.

 The well-posed integral equation
 (\ref{eq:abstract_vertex_eq4})
 $|{\cal I}_{\pi}\rangle = (1{-}{\cal K}_{\rm pp})|{\cal D}_{\pi}\rangle$,
 can now be reduced,
 since the pinching pole kernel ${\cal F}_{\rm pp}$ (\ref{eq:Fpp})
 is
 a rank one matrix,
 by applying $u^T$ to both sides of the vector equation.
 The resulting linear integral equation is
 (dropping sub-leading corrections suppressed by
 ${\cal O}(\lambda)$),
 \begin{equation}
 I_{\pi}(k) =
 D_{\pi}(k) -
 \int{d^4 p\over (2\pi)^4}\,
 K_{\rm pp}(k,p)
 n(p^0)\,S_{\rm free}(p)
 \,{D_{\pi}(p)\over \Sigma_I(p)}
 \;,
 \label{eq:reduced_vertex_eq4_pp}
 \end{equation}
 where
 $I_{\pi}(k)$ is the first non-zero entry of
 $\langle k|{\cal I}_{\pi}\rangle$,
 the reduced effective vertex is
 \begin{equation}
 D_{\pi}(p) \equiv u^T(p)\langle p|{\cal D}_{\pi}\rangle
 \;,
 \end{equation}
 and the reduced integral kernel is
 \begin{eqnarray}
 K_{\rm pp}(k,p)
 & = & \displaystyle
 u^T(k)\, {\cal M}_0(k{-}p) \, \bar{w}(p)
 \nonumber\\
 & = & \displaystyle
 {1\over 2}\,
 (1{-}e^{-k^0\beta})\,L_0(k{-}p) \,(e^{p^0\beta}{-}1)
 \;.
 \label{eq:Kpp}
 \end{eqnarray}
 Here, the explicit form of the free cut particle propagator,
 \begin{equation}
 S_{\rm free}(p) =
 [1{+}n(p^0)]\,
 {\rm sgn}(p^0)\,2\pi\delta(p^2_0{-}E_p^2)
 \;,
 \label{eq:free_cut_prop}
 \end{equation}
 is used, and
 $L_0(k{-}p)$ is the cut rung given
 (in $\lambda\phi^4$ theory,) by
 \begin{equation}
 L_0(k{-}p)
 \equiv
 {\lambda^2\over 2}
 \int {d^4 l \over (2\pi)^4}\,
 S_{\rm free}(l{+}k{-}p)\, S_{\rm free}(-l)
 \;.
 \label{eq:cut_rung}
 \end{equation}
 Note that $K_{\rm pp}$ contains no reference to the real part
 of the
 uncut rung.
 When $u^T{\cal M}_0 \bar{w}$ is calculated, the real part of the uncut
 rung
 cancels.  Eq.~(\ref{eq:Kpp}) is obtained by expressing
 the remaining imaginary part of the uncut rung in terms
 of $L_0(k{-}p)$ with the help of the optical theorem
 (\ref{eq:optical_thm}).

 Due to the delta function present in
 the kernel, $p$ is an on-shell momentum.
 Also, since
 the leading weak coupling behavior of Wightman function is given by
 \begin{eqnarray}
 \sigma_{\pi\pi}^{\rm pp}(0)
 & \displaystyle = & \displaystyle 2 \langle z_{\pi} | {\cal F}_{\rm pp}
 | {\cal D}_{\pi}
 \rangle
 \nonumber\\
 & \displaystyle = & \displaystyle
 2\int{d^4 k\over(2\pi)^4}\, z_\pi^T(k)\, \bar{w}(k)
 n(k^0)\, S_{\rm free}(k)\,
 {D_{\pi}(k)\over \Sigma_I(k)}
 \;,
 \end{eqnarray}
 the final integral over $k$ will be also restricted to on-shell
 momenta.
 Hence, the reduced integral equation (\ref{eq:reduced_vertex_eq4_pp})
 need be solved for only on-shell momenta.

 To summarize,
 after summing all ladder diagrams in $\lambda\phi^4$ theory,
 the loop frequency integrals
 may be performed and the leading weak coupling behavior
 extracted from the pinching pole contribution.
 The resulting linear integral equation
 for the effective vertex reduces to
 a single component equation given explicitly by
 \begin{equation}
 I_{\pi}(\underline{k}) = D_{\pi}(\underline{k}) -
 \int {d^4 p\over (2\pi)^4}\,
 L_0(\underline{k}{-}p)\,
 {[1{+}n(p^0)] \over [1{+}n(\underline{k}^0)]}\,
 {\rm sgn}(p^0)\,
 2\pi\delta(p_0^2{-}E_p^2)\,
 {D_{\pi}(p) \over 2\Sigma_I(p)}\,
 \;,
 \label{eq:vertex_eq4_pp}
 \end{equation}
 where $\underline{k}$ is an on-shell momentum.

 The cut rung $L_0 (k{-}p)$ is easily shown to satisfy
 \begin{equation}
 L_0 (k{-}p) = e^{(k^0{-}p^0)\beta}\,L_0 (p{-}k)
 \;.
 \end{equation}
 Also, $\Sigma_I(p)$ is an odd function of $p^0$.
 Consequently,
 $D_{\pi}(-\underline{p})$ satisfies the same equation as does
 $D_{\pi}(\underline{p})$, provided
 $I_{\pi}(\underline{k})$ is an even function of $\underline{k}$.
 Hence,
 if $I_{\pi}(\underline{k})$ is an even function of $\underline{k}$, so is
 the solution $D_{\pi}(\underline{k})$.
 Since the energy-momentum tensor is even under CPT,
 the inhomogeneous terms for both the shear and bulk viscosities
 are even functions of the 4-momentum.

 In terms of the solutions of the reduced integral equation
 (\ref{eq:vertex_eq4_pp}), the shear viscosity is
 \begin{eqnarray}
 \eta
 & = & \displaystyle
 {\beta\over 10}\,
   \int{d^4 k\over (2\pi)^4}\,
	z_{\pi}^T(k)\,\bar{w}(k)\, n(k^0)\, S_{\rm free}(k)\,
	{D_{\pi}(k) \over \Sigma_I(k)}
 \nonumber\\
 & = & \displaystyle
 {\beta\over 10}\,
   \int{d^4 k\over (2\pi)^4}\,
	I_{\pi}(k)\, n(k^0)\,S_{\rm free}(k)\,
	{D_{\pi}(k)\over \Sigma_I(k)}
 \;,
 \label{eq:ladder_viscosities2}
 \end{eqnarray}
 neglecting sub-leading contributions suppressed by additional
 powers of $\lambda$.

 For the future use,
 we define the inner product of two functions of on-shell momentum as
 \begin{equation}
 ( f | g ) \equiv
 \int
 {d^4 l \over (2\pi)^4 \Sigma_I(l)}\,
 n(l^0)\,S_{\rm free}(l)\,
 f(l)^*\, g(l)
 \;.
 \label{eq:def_inner}
 \end{equation}
 In terms of this definition,
 the integral equation (\ref{eq:reduced_vertex_eq4_pp})
 can be expressed as
 \begin{equation}
 | I_{\pi} )
 =
 (1 - K_{\rm pp})| D_{\pi} )
 \;,
 \end{equation}
 whose five zero modes are
 \begin{eqnarray}
 (b_\mu|\underline{p})
 \equiv b_\mu(\underline{p})
 = \underline{p}_\mu\, \Sigma_I (\underline{p})
 \;,
 \label{eq:b_mu}
 \\
 \noalign{\hbox{and}}
 (b_5|\underline{p})
 \equiv b_5(\underline{p})
 = {\rm sgn}(\underline{p}^0)\, \Sigma_I (\underline{p})
 \;.
 \label{eq:b_5}
 \end{eqnarray}

 \subsection{Ladder summation for the shear viscosity calculation
 with an additional $g\phi^3$ interaction}
 \label{subsec:ladder_sum3}

\begin{figure}
 \setlength {\unitlength}{1cm}
\vbox
    {%
    \begin {center}
 \begin{picture}(0,0)
 \put(0.8,0.6){$k$}
 \put(2.2,3.6){$k$}
 \put(3.7,3.9){$k{-}l$}
 \put(3.7,0.0){$l{-}k$}
 \put(6.8,3.3){$p$}
 \put(6.8,0.8){$-p$}
 \put(3.1,2.4){${-}l$}
 \put(3.8,1.1){$l{-}k{+}p$}
 \end{picture}
	\leavevmode
	\def\epsfsize	#1#2{0.4#1}
	\epsfbox {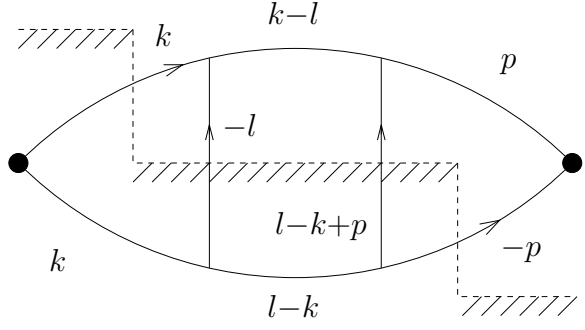}
    \end {center}
     \caption
 	{%
 A cut three loop ladder diagram in a scalar $g\phi^3$ theory.
 The black dot at each end represents an insertion of
 a bilinear external operator.
	}
 \label{fig:three_loop_ladder}
     }
 \end {figure}

 To start,
 consider ``simple'' ladder diagrams only containing the straight
 single line rungs,
 as illustrated in Fig.~\ref{fig:ladder3}.
 After summing these diagrams, including
 the contribution of the other required rungs will be easy.
 To sum these simple ladder diagrams, one again
 introduces an effective
 vertex ${\cal D}_{\pi}(k,q{-}k)$.
 Before performing any frequency integration,
 the effective vertex satisfies
 \begin{equation}
 {\cal D}_{\pi}(k, q{-}k)
 =
 {\cal I}_{\pi}(k, q{-}k)
 +
 \int {d^4 p \over (2\pi)^4} \,
 {\cal M}_{\rm line}(k{-}p) \,
 {\cal F}(p, q{-}p)
 {\cal D}_{\pi}(p, q{-}p)
 \;,
 \label{eq:vertex_eq3}
 \end{equation}
 where the elements of the matrix ${\cal M}_{\rm line}(k{-}p)$
 are simply cut and uncut single particle propagators.
 Before proceeding with the general analysis,
 it may be helpful to consider a typical example, such as
 the three-loop diagram in Fig.~\ref{fig:three_loop_ladder}.
 Applying the cutting rules,
 the contribution of this three-loop diagram
 (with zero external 4-momentum) is

 \begin{equation}
 \sigma_{\pi\pi}^{\hbox{%
\scriptsize{(Fig.\protect{\ref{fig:three_loop_ladder}})}}}(0)
 =
 2
 \int {d^4 k\over (2\pi)^4}\,{d^4 p\over (2\pi)^4}\,
 I_\pi (k)\,
 S(k)\,\tilde{G}(k)^*
 L_{\hbox{\scriptsize{full-box}}}(p,k)
 \,S(-p)\,\tilde{G}(p)
 I_{\pi}(p)
 \;,
 \label{eq:three_loop_ladder}
 \end{equation}
 where
 \begin{equation}
 L_{\hbox{\scriptsize full-box}}(p,k)
 =
 g^4\int {d^4 l \over (2\pi)^4}\,
 \tilde{G}(k{-}l)\, \tilde{G}(l{-}k)^* \,
 S(-l)\, S(l{-}k{+}p)
 \;
 \label{eq:box}
 \end{equation}
 is the one-loop box sub-diagram
 illustrated in Fig.~\ref{fig:box}.

 The size of the contributions from the (nearly) pinching poles
 in the complex $k^0$, $p^0$ and $l^0$ planes can be estimated
 as follows.
 Each pinch generates a thermal lifetime of ${\cal O}(1/\lambda^2)$.
 There are four ${\cal O}(\sqrt{\lambda}m_{\rm phys})$ factors of $g$.
 When all three loop momenta $k$, $p$ and $l$ are on-shell,
 the momenta flowing through the two cut propagators are well off-shell.
 An off-shell cut propagator is ${\cal O}(g^2)$ since it is
 proportional to the imaginary part of the
 ${\cal O}(g^2)={\cal O}(\lambda m_{\rm phys}^2)$ one-loop self-energy
 ({\it c.f.}~Eq.~({\ref{eq:single_ptl_sd})).
 Hence,
 when
 $T
 \,\,\vcenter {\hbox{$\buildrel{\displaystyle <}\over\sim$}} \,\,
 m_{\rm phys}$, the pinching pole contribution
 is $g^4\times{\cal O}(1/\lambda^6)\times{\cal O}(g^4)
 = {\cal O}(1/\lambda^2)$,
 or the same as the lowest-order one-loop diagram.
 Since the leading order shear viscosity is insensitive to small
 momentum contributions, at temperatures much greater than $m_{\rm phys}$,
 the cubic interaction becomes irrelevant compared to the quartic
 interaction.\footnote{%
       In contrast, the bulk viscosity is sensitive to small momentum
       contributions, and therefore the cubic interaction becomes negligible
       at much higher temperature $T \gg m_{\rm phys}/\sqrt{\lambda}$.
}

 Equivalently, when
 $T
 \,\,\vcenter {\hbox{$\buildrel{\displaystyle <}\over\sim$}} \,\,
 m_{\rm phys}$
 the pinching pole contribution of the box sub-diagram is
 ${\cal O}(g^8/\lambda^2) = {\cal O}(g^4)$,
 not ${\cal O}(g^4/\lambda^2)={\cal O}(1)$
 as one might have expected if the additional
 \begin{figure}
 \setlength {\unitlength}{1cm}
\vbox
    {%
    \begin {center}
 \begin{picture}(0,0)
\put(0.0,2.8){$-l$}
\put(3.3,2.8){$l{-}k{+}p$}
\put(1.5,4.3){$k{-}l$}
\put(1.5,-0.5){$l{-}k$}
 \end{picture}
	\leavevmode
	\def\epsfsize	#1#2{0.3#1}
	\epsfbox {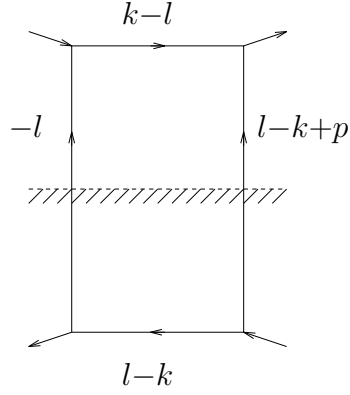}
    \end {center}
     \caption
 	{%
A box diagram which is a part of a ladder diagram with straight
 rungs.
	}
 \label{fig:box}
     }
 \end {figure}
 \noindent
 suppression from
 the off-shell self-energy were ignored.
 Consequently,
 the non-pinching pole contribution of the box diagram,
 which is also ${\cal O}(g^4)$ from
 the four explicit interaction vertices,
 is equally important as the pinching pole contribution.
 This complicates the treatment of these diagrams.

 The key observation of the above argument is that an off-shell
 straight cut rung is ${\cal O}(g^4/m_{\rm phys}^4) = {\cal O}(\lambda^2)$
 when
 $T
 \,\,\vcenter {\hbox{$\buildrel{\displaystyle <}\over\sim$}} \,\,
 m_{\rm phys}$.
 Hence, the leading weak coupling behavior
 of the non-pinching pole contribution is produced when
 the two cut propagators in the box are both on-shell.
 Otherwise, there will be additional
 suppression from the cut propagators which will
 make the contribution smaller than ${\cal O}(\lambda^2)$.
 Consequently, the contribution of the three-loop ladder diagram
 in Fig.~\ref{fig:three_loop_ladder}
 can be rewritten as
 \begin{eqnarray}
\lefteqn{
 \sigma_{\pi\pi}^{\hbox{%
\scriptsize{(Fig.\protect{\ref{fig:three_loop_ladder}})%
}}}(0)
 =}
 & \displaystyle & \displaystyle
 \nonumber\\
 & \displaystyle & \displaystyle \quad
 2\int {d^4 k\over (2\pi)^4}\,{d^4 p\over (2\pi)^4}\,
      {d^4 l\over (2\pi)^4}\,
 I_\pi (k)\,
 {\cal F}_{\rm pp}^{23}(k)\,
 L_{\rm line}(l{-}k)
 {\cal F}_{\rm pp}^{33}(l)\,
 L_{\rm line}(p{-}l)
 {\cal F}_{\rm pp}^{31}(p)\,
 I_{\pi}(p)
 \nonumber\\
 & \displaystyle & \displaystyle \quad\quad {}
 +
 2\int {d^4 k\over (2\pi)^4}\, {d^4 p\over (2\pi)^4}\,
 I_\pi (k)\,
 {\cal F}_{\rm pp}^{23}\,
 L_{\rm box}(p,k)
 {\cal F}_{\rm pp}^{31}\,
 I_\pi (p)
 \;,
 \label{eq:three_loop_ladder_pp}
 \end{eqnarray}
 where
 ${\cal F}_{\rm pp}^{ij}$ is the $(ij)$ element of the pinching
 pole
 side rail matrix ${\cal F}_{\rm pp}$,
 and the ``rungs'' between the pinching pole side rails are
 \begin{equation}
 L_{\rm box}(\underline{p},\underline{k})
 =
 g^4\,
 \int {d^4 l\over (2\pi)^4}\,
 \tilde{G}(k{-}l)\,\tilde{G}(l{-}k)^*\,
 S_{\rm free}(-l)\,
 S_{\rm free}(l{-}\underline{k}{+}\underline{p})\,
 \;,
 \end{equation}
 representing the ${\cal O}(\lambda^2)$ non-pinching pole
 contribution from the cut box sub-diagram
 \begin{figure}
 \setlength {\unitlength}{1cm}
\vbox
    {%
    \begin {center}
 \begin{picture}(0,0)
 \end{picture}
	\leavevmode
	\def\epsfsize	#1#2{0.5#1}
	\epsfbox {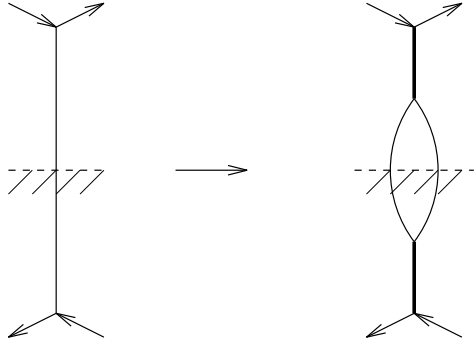}
    \end {center}
     \caption
 	{%
	Diagrammatic representation of a cut line rung corresponding to
	Eq.~(\protect{\ref{eq:L_line}}).  The heavy line in the unshaded
	region represents a retarded propagator $G_R$, and the heavy
	line in the shaded region represents an advanced propagator
	$G_A$.
	}
 \label{fig:line_rung}
     }
 \end {figure}
 \noindent
 (where the resummed
 cut propagators may be replaced by free cut propagators), and
 \begin{eqnarray}
 L_{\rm line}(\underline{l}{-}\underline{k})
 & \displaystyle = & \displaystyle
 g^2\,
 { [1{+}n(\underline{l}^0{-}\underline{k}^0)]\, 2\,
   \Sigma_I(\underline{l}{-}\underline{k})
                      \over
 \left| (\underline{l}{-}\underline{k})^2+m_{\rm th}^2
 +\Sigma(\underline{l}{-}\underline{k}) \right|^2
 }
 \nonumber\\
 \noalign{\smallskip}
 & \displaystyle = & \displaystyle
 g^2\,
 [1{+}n(\underline{l}^0{-}\underline{k}^0)]\, 2\,
 \Sigma_I(\underline{l}{-}\underline{k})\,
 G_R(\underline{l}{-}\underline{k})\, G_A(\underline{l}{-}\underline{k})
 \nonumber\\
 & \displaystyle = & \displaystyle
 {g^4\over \lambda^2}\, L(\underline{l}{-}\underline{k})
 G_R(\underline{l}{-}\underline{k})\, G_A(\underline{l}{-}\underline{k})
 \;,
 \label{eq:L_line}
 \end{eqnarray}
 representing a single cut straight rung as shown in
 Fig.~\ref{fig:line_rung}.
 The subscripts $R$ and $A$ here indicate the retarded and the
 advanced
 propagators ({\it c.f.}~Eq.~(\ref{eq:retarded_prop})).

 Applying the same argument above to other routings
 of the cut in the three-loop ladder,
 it is straightforward to see that
 the sum of all cut three-loop ladder diagrams has the form,
 \begin{eqnarray}
\lefteqn{
 \sigma_{\pi\pi}^{\hbox{\scriptsize{(3-loop)}}}(0)
 = }
 & \displaystyle & \displaystyle
 \nonumber\\
 & \displaystyle & \displaystyle \quad
 2\int {d^4 k\over (2\pi)^4}\,{d^4 p\over (2\pi)^4}\,
      {d^4 l\over (2\pi)^4}\,
 z_\pi^T(k)\,
 {\cal F}_{\rm pp}(k)\,
 {\cal M}_{\rm line}(k{-}l)
 {\cal F}_{\rm pp}(l)\,
 {\cal M}_{\rm line}(l{-}p)
 {\cal F}_{\rm pp}(p)\,
 {\cal I}_{\pi}(p)
 \nonumber\\
 & \displaystyle & \displaystyle \quad\quad {}
 +
 2\int {d^4 k\over (2\pi)^4}\, {d^4 p\over (2\pi)^4}\,
 z_\pi^T(k)\,
 {\cal F}_{\rm pp}(k)\,
 {\cal M}_{\rm box}(k,p)
 {\cal F}_{\rm pp}(p)\,
 {\cal I}_{\pi}(p)
 \;,
 \label{eq:three_loop_ladder_all_cut}
 \end{eqnarray}
 where a $4{\times}4$ matrix ${\cal M}_{\rm box}(k,p)$
 contains non-pinching pole contributions of
 cut and uncut box sub-diagrams.
 The previous expression (\ref{eq:three_loop_ladder_pp})
 is included since, as shown in appendix~\ref{app:ladder},
 the $(33)$ component of ${\cal M}_{\rm line}(k{-}l)$ is
 $L_{\rm line}(l{-}k)$, and
 the $(33)$ component of ${\cal M}_{\rm box}(k,p)$ is
 $L_{\rm box}(p{-}k)$.
 (Recall also that ${\cal I}_{\pi}=(I_{\pi},0,0,0)$
  and $z_\pi = (0,I_\pi,0,0)$.)
 Once again, merely replacing the side rail matrix ${\cal F}$
 by its pinching pole part ${\cal F}_{\rm pp}$,
 without changing the rung matrix ${\cal M}_{\rm line}$,
 is not sufficient to calculate the
 the leading weak coupling behavior of
 a simple ladder diagram; one must also include the box sub-diagram
 rung ${\cal M}_{\rm box}$.

 To sum all simple ladder diagrams,
 note that the first term in Eq.~(\ref{eq:three_loop_ladder_all_cut})
 can be regarded as the second iteration of the single-line kernel
 ${\cal M}_{\rm line}{\cal F}_{\rm pp}$,
 and the second term can be interpreted as the first
 iteration of the box kernel ${\cal M}_{\rm box}{\cal F}_{\rm pp}$.
 In exactly the same way, it is simple to deduce that
 the leading weak coupling behavior of a simple ladder diagram
 with
 $n$ straight rungs contains
 all possible sequences of
 $n{-}2m$ factors of single line rungs,
 ${\cal M}_{\rm line}{\cal F}_{\rm pp}$, and
 $m$ factors of box rungs,
 ${\cal M}_{\rm box}{\cal F}_{\rm pp}$,
 for all $m \le n/2$.
 Every such sequence can be interpreted as arising from the
 iteration of the integral equation
 \begin{equation}
 {\cal I}_{\pi}(k)
 =
 {\cal D}_{\pi}(k)
 -
 \int {d^4 p \over (2\pi)^4} \,
 \left(
 {\cal M}_{\rm line}(k{-}p) + {\cal M}_{\rm box}(k,p)
 \right)
 {\cal F}_{\rm pp}(p)
 {\cal D}_{\pi}(p)
 \;.
 \label{eq:vertex_eq3_pp}
 \end{equation}

 Because the pinching pole kernel ${\cal F}_{\rm pp}$ (\ref{eq:Fpp})
 is a rank one matrix, applying $u^T$ to both sides of
 Eq.~(\ref{eq:vertex_eq3_pp}) reduces the equation to
 \begin{equation}
 I_{\pi}(\underline{k})
 =
 D_{\pi}(\underline{k}) -
 \int {d^4 p\over (2\pi)^4}\,
 K_{\hbox{\scriptsize{simple-ladder}}}(k,p)
 n(p^0)\,S_{\rm free}(p)\,
 {D_{\pi}(p) \over \Sigma_I(p)}\,
 \;,
 \label{eq:reduced_vertex_eq3}
 \end{equation}
 where
 \begin{equation}
 K_{\hbox{\scriptsize{simple-ladder}}}(k,p)
 =
 {1\over 2}
 (1{-}e^{-k^0\beta})
 L_{\hbox{\scriptsize{simple-ladder}}}(k,p)
 (e^{p^0\beta}{-}1)
 \end{equation}
 with
 \begin{eqnarray}
 L_{\hbox{\scriptsize{simple-ladder}}}(k,p)
 & \displaystyle = & \displaystyle
 \int {d^4 l_1 \over (2\pi)^4}\, {d^4 l_2 \over (2\pi)^4}\,
 (2\pi)^4\delta(l_1{-}l_2{+}\underline{p}{-}\underline{k})\,
 S_{\rm free}(l_1)\, S_{\rm free}(-l_2)\,
 \nonumber\\
 & \displaystyle & \displaystyle \quad{}\times
 \left(
   {g^4\over 2}G_R(\underline{k}{-}\underline{p})\,
   G_A(\underline{k}{-}\underline{p})
 + {g^4\over 2}G_R(\underline{k}{-}l_1)\,G_A(\underline{k}{-}l_1)
 \right.
 \nonumber\\
 & \displaystyle & \displaystyle \qquad{}
 +
 \left.
 {g^4\over 2}G_R(\underline{k}{+}l_2)\,G_A(\underline{k}{+}l_2)
 \right)
 \;.
 \label{eq:L_simple_ladder}
 \end{eqnarray}
 In obtaining Eq.~(\ref{eq:reduced_vertex_eq3}),
 the following relation for the box rung is used,
 \begin{equation}
 u^T(k){\cal M}_{\rm box}(k,p)\bar{w}(p)
 =
 (1{-}e^{-k^0\beta})\,
 \left(
 {\cal M}_{\rm box}^{(44)}(k,p)
 {-}e^{k^0\beta}{\cal M}_{\rm box}^{(34)}(k,p)
 \right)
 (e^{p^0\beta}{-}1){\bigg /}2
 \;,
 \label{eq:u_M_box_w}
 \end{equation}
 together with an analogous relation
 for ${\cal M}_{\rm line}$ and $L_{\rm line}$.
 Verifying equations
 (\ref{eq:L_simple_ladder}) and (\ref{eq:u_M_box_w})
 is a straightforward but tedious
 exercise in the application of the generalized optical theorem.
 In short, to prove Eq.~(\ref{eq:u_M_box_w}), one must show that
 all cut box diagram contributions in $u^T{\cal M}_{\rm box}\bar{w}$
 other than the (44) and (34) contributions
 are canceled by the imaginary part of the uncut box diagram.
 The final expression (\ref{eq:L_simple_ladder}) in terms of
 the retarded and the advanced propagators
 results from the particular combination of (44) and (34) components
 in Eq.~(\ref{eq:u_M_box_w}).
 A sketch of the proof is given in appendix~\ref{app:ladder}.

 The above summation of simple ladder diagrams
 illustrates the general principle:
 To determine whether a diagram contributes to the leading weak
 coupling behavior, first carry out the frequency integrations.
 Then, if the contribution of each sub-diagram
 sandwiched between two pinching pole side rails
 is ${\cal O}(\lambda^2)$, the diagram as a whole will contribute
 to the leading weak coupling behavior.
 Hence, to identify {\em all} diagrams in the
 $g\phi^3{+}\lambda\phi^4$ theory contributing to the leading order
 behavior, one must identify all ${\cal O}(\lambda^2)$
 sub-diagrams (``rungs'') which may be sandwiched between two
 pinching pole side rails.
 In the simple ladder diagram, two of such ``rungs'' were
 identified, the single line ``rung'' and the box sub-diagram
 ``rung'', illustrated in Fig.~\ref{fig:rungs34} labeled (b)
 and (c).

 With cubic and quartic interactions, there are a total of 10
 different ${\cal O}(\lambda^2)$ ``rungs'' when
 $T
 \,\,\vcenter {\hbox{$\buildrel{\displaystyle <}\over\sim$}} \,\,
 m_{\rm phys}$
 as shown in Fig.~\ref{fig:rungs34}.
 These diagrams exhaust all possible
 ${\cal O}(\lambda^2)$, ${\cal O}(\lambda g^2)$, ${\cal O}(g^4)$
 rungs in $g\phi^3{+}\lambda\phi^4$ theory in those temperature range.
 The particular cuts shown in the figure correspond to
 the (44) components of the rung matrices.
 Consequently, the dominant set of ``ladder'' diagrams
 in $g\phi^3{+}\lambda\phi^4$ theory
 may be described as iterations of
 pinching pole side rails ${\cal F}_{\rm pp}$
 and a combined ${\cal O}(\lambda^2)$ rung matrix ${\cal M}_{\rm full}$,
 whose components are
 the sum of contributions of all possible cuts of the underlying
 diagrams of Fig.~\ref{fig:rungs34}.
 The summation of all such ladder diagrams
 is generated by the integral equation
 \begin{equation}
 {\cal I}_{\pi}(k)
 =
 {\cal D}_{\pi}(k)
 -
 \int {d^4 p \over (2\pi)^4} \,
 {\cal M}_{\rm full}(k,p)\,
 {\cal F}_{\rm pp}(p)
 {\cal D}_{\pi}(p)
 \;,
 \label{eq:vertex_eq_full_pp}
 \end{equation}
 where ${\cal M}_{\rm full}(k,p)$ consists of
 all cut and uncut ${\cal O}(\lambda^2)$ rungs
 in the $g\phi^3{+}\lambda\phi^4$ theory, {\it i.e.}, those in
 Fig.~\ref{fig:rungs34}.
 Once again, applying $u^T$ to both sides of
 Eq.~(\ref{eq:vertex_eq_full_pp}) reduces the equation
 \begin{figure}
 \setlength {\unitlength}{1cm}
\vbox
    {%
    \begin {center}
 \begin{picture}(0,0)
 \put(0.9,4){(a)}
 \put(3.4,4){(b)}
 \put(5.9,4){(c)}
 \put(8.4,4){(d)}
 \put(10.9,4){(e)}
 \put(0.9,-0.5){(f)}
 \put(3.4,-0.5){(g)}
 \put(5.9,-0.5){(h)}
 \put(8.4,-0.5){(i)}
 \put(10.9,-0.5){(j)}
 \end{picture}
	\leavevmode
	\def\epsfsize	#1#2{0.35#1}
	\epsfbox {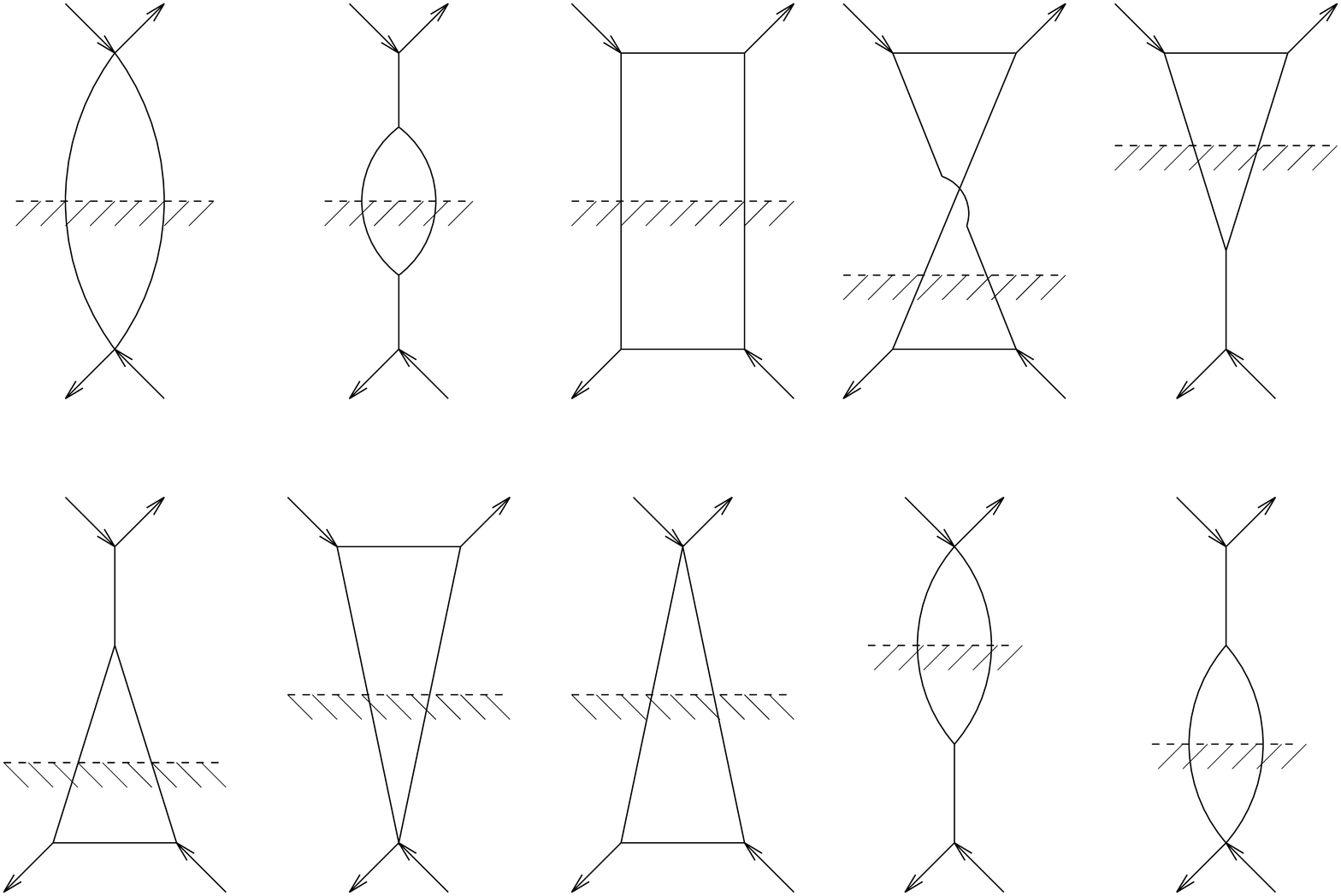}
    \end {center}
    \vspace{0.5cm}
     \caption
 	{%
	The $\Order(\lambda^2)$
	``rungs'' in the $g\phi^3{+}\lambda\phi^4$
	theory.  The cuts shown correspond to (44) components of the
	rung matrix.
	}
 \label{fig:rungs34}
     }
 \end {figure}
 \noindent
 to
 \begin{equation}
 I_{\pi}(\underline{k}) = D_{\pi}(\underline{k}) -
 \int {d^4 p\over (2\pi)^4}\,
       K_{\rm full}(\underline{k},p)\,
 n(p^0)\,
 S_{\rm free}(p)\,
 {D_{\pi}(p) \over \Sigma_I(p)}\,
 \;,
 \label{eq:reduced_vertex_eq_full}
 \end{equation}
 where the previous relation
 between the rung matrix ${\cal M}_{\rm full}$
 and its (44) and (34) components continues to hold,
 \begin{eqnarray}
 K_{\rm full}(k,p)
 & \displaystyle = & \displaystyle
 u^T(k){\cal M}_{\rm full}(k,p)\bar{w}(p)
 \nonumber\\
 & \displaystyle = & \displaystyle
 {1\over 2}
 (1{-}e^{-k^0\beta})\,
 L_{\rm full}(k,p)\,
 (e^{p^0\beta}{-}1)
 \label{eq:K_full}
 \\
 \noalign{\hbox{with}}
 L_{\rm full}(k,p)
 & \displaystyle = & \displaystyle
 \left(
 {\cal M}_{\rm full}^{(44)}(k,p)
 {-}e^{k^0\beta}{\cal M}_{\rm full}^{(34)}(k,p)
 \right)
 \;,
 \end{eqnarray}
 as a consequence of the generalized optical theorem.
 The proof of Eq.~(\ref{eq:K_full}) is
 discussed in appendix~\ref{app:ladder}.

 Noting that all the diagrams in
 Fig.~\ref{fig:rungs34} have two on-shell cut lines,
 a straightforward application of the cutting rules yields the
 sum of all the cut ``rungs'',
 \begin{eqnarray}
 L_{\rm full}(\underline{k},\underline{p})
 & \displaystyle = & \displaystyle
 \int {d^4 l_1 \over (2\pi)^4}\, {d^4 l_2 \over (2\pi)^4}\,
 S_{\rm free}(l_1)\, S_{\rm free}(-l_2)\,
 (2\pi)^4\delta(l_1{-}l_2{+}\underline{p}{-}\underline{k})
 \nonumber\\
 & \displaystyle & \displaystyle {}\times
 \bigg(
 {\lambda^2\over 2}
 +
 {g^4\over 2}G_R(\underline{k}{-}\underline{p})\,
 G_A(\underline{k}{-}\underline{p})
 +
 {g^4\over 2}G_R(\underline{k}{-}l_1)\,G_A(\underline{k}{-}l_1)
 \nonumber\\
 & \displaystyle & \displaystyle \quad {}
 +
 {g^4\over 2}G_R(\underline{k}{+}l_2)\,G_A(\underline{k}{+}l_2)
 \nonumber\\
 & \displaystyle & \displaystyle  {}
 +
 g^4\,{\rm Re}\,
 \left( G_R(l_1{+}\underline{p})\,G_A(l_1{-}\underline{k}) \right)
 +
 g^4\,{\rm Re}\,
 \left( G_R(l_1{+}\underline{p})\,G_A(l_1{-}l_2) \right)
 \nonumber\\
 & \displaystyle & \displaystyle  {}
 +
 g^4\,{\rm Re}\,
 \left( G_R(l_1{-}\underline{k})\,G_A(l_1{-}l_2 \vphantom{\underline{p}})
 \right)
 \nonumber\\
 & \displaystyle & \displaystyle  {}
 -
 \lambda g^2\,
 {\rm Re}\,G_R(l_1{+}\underline{p})
 -
 \lambda g^2\,
 {\rm Re}\,G_R(l_1{-}\underline{k})
 -
 \lambda g^2\,
 {\rm Re}\,G_R(l_1{-}l_2)
 \bigg)
 \label{eq:L_full_expanded}
 \\
 & \displaystyle = & \displaystyle
 {1\over 2}
 \int {d^4 l_1 \over (2\pi)^4}\, {d^4 l_2 \over (2\pi)^4}\,
 S_{\rm free}(l_1)\, S_{\rm free}(-l_2)\,
 (2\pi)^4\delta(l_1{-}l_2{+}\underline{p}{-}\underline{k})
 \nonumber\\
 & \displaystyle & \displaystyle \qquad{}\times
 \left|
   \lambda
   - g^2\,
   \left(
   G_R(l_1{+}\underline{p})+G_R(l_1{-}\underline{k})+G_R(l_1{-}l_2)
   \right)
 \right|^2
 \;.
 \label{eq:L_full}
 \end{eqnarray}
 Each term in (\ref{eq:L_full_expanded}) arises from specific
 diagrams
 of Fig.~\ref{fig:rungs34}.
 For example, consider the term
 $g^4\,{\rm Re}\,
  \left( G_R(l_1{+}\underline{p})\,G_A(l_1{-}\underline{k}) \right)$.
 This term corresponds to the ``cross'' diagram labeled as (d)
 in Fig.~\ref{fig:rungs34}, and redrawn in
 Fig.~\ref{fig:cross} with momentum labels.
 The cutting rules produce
 \begin{eqnarray}
 L_{\rm cross}(\underline{k},\underline{p})
 & \displaystyle = & \displaystyle
 {\cal M}_{\rm cross}^{(44)}(k,p)
 - e^{k^0\beta}
 {\cal M}_{\rm cross}^{(34)}(k,p)
 \nonumber\\
 & \displaystyle = & \displaystyle
 {g^4\over 2}
 \int {d^4 l_1 \over (2\pi)^4}\, {d^4 l_2 \over (2\pi)^4}\,
 (2\pi)^4\delta(l_1{-}l_2{+}\underline{p}{-}\underline{k})\,
 \nonumber\\
 & \displaystyle & \displaystyle
 \hspace{-1.5cm}
 {}\times
 \left(
 S_{\rm free}(l_1)\,S_{\rm free}(-l_2)\,\tilde{G}(\underline{p}{+}l_1)\,
 \tilde{G}(l_1{-}\underline{k})^*
 -
 e^{k^0\beta}\,
 S_{\rm free}(-l_1{-}\underline{p})\,S_{\rm free}(l_1{-}\underline{k})\,
 \tilde{G}(l_2)\,\tilde{G}(l_1)^*
 \right)
 \nonumber\\
 & \displaystyle = & \displaystyle
 {g^4\over 2}
 \int {d^4 l_1 \over (2\pi)^4}\, {d^4 l_2 \over (2\pi)^4}\,
 (2\pi)^4\delta(l_1{-}l_2{+}\underline{p}{-}\underline{k})\,
 \nonumber\\
 & \displaystyle & \displaystyle
 \hspace{-1.5cm}
 {}\times
 \left(
 S_{\rm free}(l_1)\,S_{\rm free}(-l_2)\,
 G_R(\underline{p}{+}l_1)\,G_A(l_1{-}\underline{k})
 {-}
 e^{k^0\beta}\,
 S_{\rm fee}(-l_1{-}\underline{p})\,S_{\rm free}(l_1{-}\underline{k})\,
 G_R(l_2)\,G_A(l_1)
 \right)
 \;,
 \nonumber\\
 & &
 \label{eq:L_cross}
 \end{eqnarray}
 where to obtain the last expression, the relation between propagators
 \begin{equation}
 \tilde{G}(k) = -iG_R(k)+S(-k) = -iG_A(k)+S(k)
 \;,
 \end{equation}
 and the fact that
 $S(l_1-\ud{k})$ and $S(\ud{p}+l_1)$ can be neglected
 since $l_1, l_2, \ud{p}, \ud{k}$ are all on shell
 are repeatedly used.
 Noting that the effective vertex $D_{\pi}(\underline{p})$ is an even
 function of
 $\underline{p}$, the sign of $\underline{p}$ in the second term of
 Eq.~(\ref{eq:L_cross}) may be changed
 inside the integral equation (\ref{eq:vertex_eq_full_pp}).
 Then changing the label $l_1 \to l_1{+}\underline{p}$,
 replacing full cut propagators by the free cut

 \begin{figure}
 \setlength {\unitlength}{1cm}
\vbox
    {%
    \begin {center}
 \begin{picture}(0,0)
 \put(-0.7,0.0){$-k$}
 \put(-0.7,5.0){$-k$}
 \put(4.4,0.0){$-p$}
 \put(4.4,5.0){$-p$}
 \put(5.8,0.0){$k$}
 \put(5.8,5.0){$k$}
 \put(10.4,0.0){$-p$}
 \put(10.4,5.0){$-p$}
 \put(1.6,5.2){$-p{-}l_1$}
 \put(1.6,-0.4){$l_1{-}k$}
 \put(3.6,4.0){$l_1$}
 \put(0.2,4.0){$-l_2$}
 \put(8.2,5.2){$l_2$}
 \put(8.2,-0.4){$l_1$}
 \put(5.8,4.0){$-l_1{-}p$}
 \put(9.8,4.0){$l_1{-}k$}
 \end{picture}
	\leavevmode
	\def\epsfsize	#1#2{0.4#1}
	\epsfbox {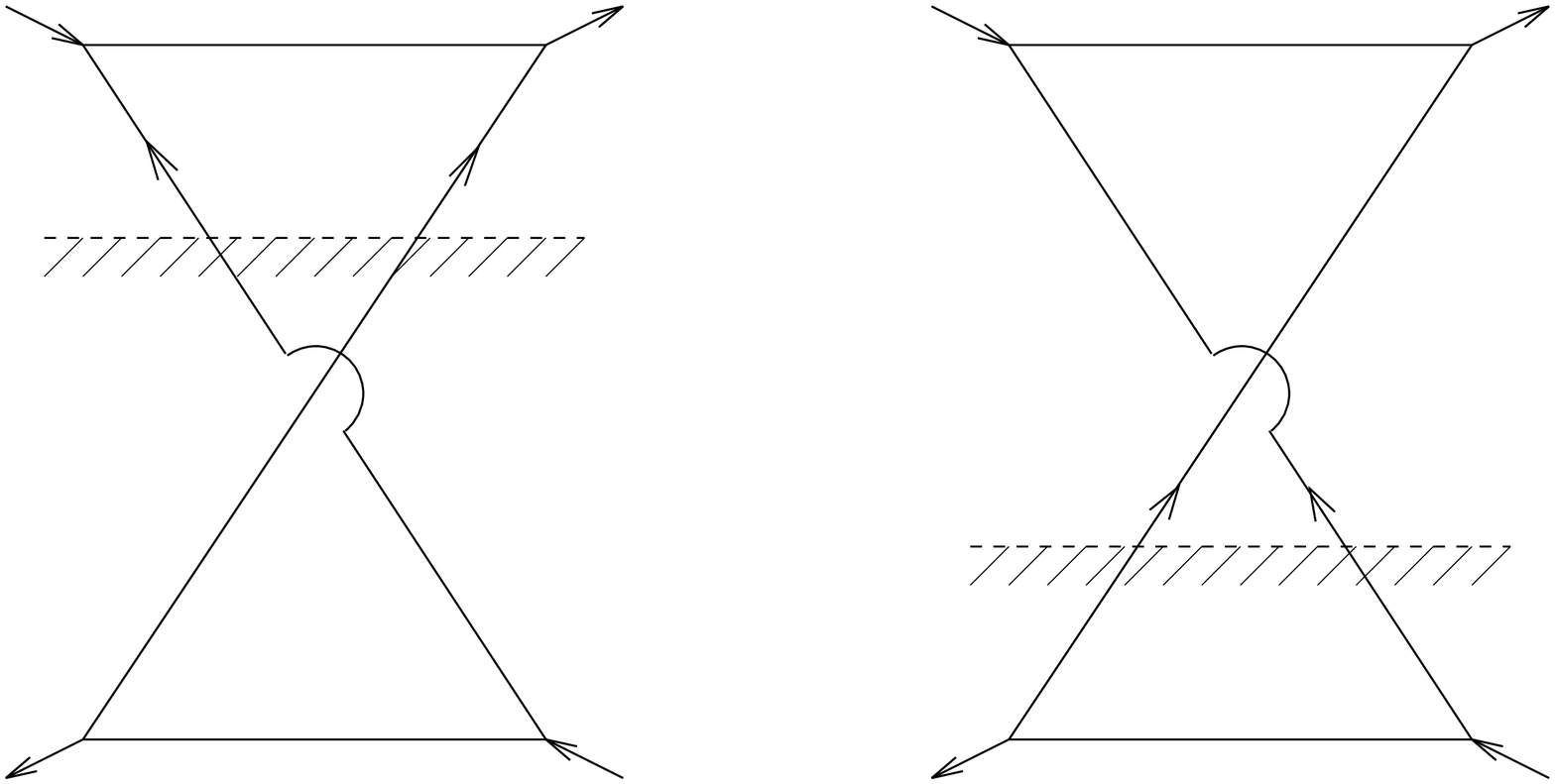}
    \end {center}
 \caption{%
 The (44) and (34) components of the ``cross'' sub-diagram rung.
 }
 \label{fig:cross}
 }
 \end{figure}
 \noindent
 propagators
 (and dropping sub-leading corrections suppressed by
 ${\cal O}(\sqrt{\lambda})$) yields,
 \begin{eqnarray}
 L_{\rm cross}(\underline{k},\underline{p})
 & \displaystyle = & \displaystyle
 g^4\,
 \int {d^4 l_1 \over (2\pi)^4}\, {d^4 l_2 \over (2\pi)^4}\,
 (2\pi)^4\delta(l_1{-}l_2{+}\underline{p}{-}\underline{k})\,
 \nonumber\\
 & \displaystyle & \displaystyle \qquad{}\times
 S_{\rm free}(l_1)\,S_{\rm free}(-l_2)\,
 {\rm Re}\left(
 G_R(l_1{+}\underline{p})\,G_A(l_1{-}\underline{k})
 \right)
 \;.
 \end{eqnarray}
 All other terms are produced similarly.

 The most important point to notice in Eq.~(\ref{eq:L_full})
 is that
 the various terms combine to produce the square of a single
 factor
 \begin{equation}
 {\cal T}(\underline{l}_1,\underline{p};\underline{l}_2,\underline{k})
 \equiv
   \lambda
   - g^2\,
   \left(
   G_R(\underline{l}_1{+}\underline{p})
  +G_R(\underline{l}_1{-}\underline{k})
  +G_R(\underline{l}_1{-}\underline{l}_2)
   \right)
 \;,
 \label{eq:almost_scattering_amp}
 \end{equation}
 which obviously resembles a tree level two-body ``scattering
 amplitude''.
 Strictly speaking, at non-zero temperature,
 one cannot define scattering amplitudes since no truly stable
 single particle excitations exist.
 However the expression
 (\ref{eq:almost_scattering_amp})
 may be regarded as an approximate scattering amplitude characterizing
 the dynamics of the finite temperature excitations on time scales
 short compared to their lifetime.
 The only difference between
 the result for $g\phi^3{+}\lambda\phi^4$ theory,
 and that for the $\lambda\phi^4$ theory,
 is the replacement of the constant tree level scattering amplitude
 $\lambda$ in the $\lambda\phi^4$ theory by the momentum dependent
 tree amplitude
 ${\cal T}(\underline{l}_1,\underline{p};\underline{l}_2,\underline{k})$.
 Note that ${\cal T}$ contains retarded propagators in place
 of the usual time-ordered propagators.
 At zero temperature, the scattering amplitude can be expressed
 both in terms of the time ordered Green function, or
 the retarded Green function\cite{Roman}.
 At non-zero temperature, it is
 retarded Green function which gives the correct amplitude.

 The arguments  of the propagators
 in Eq.~(\ref{eq:almost_scattering_amp})
 are all combinations of two on-shell momenta.
 A short exercise in kinematics shows that
 the 4-momentum squared of the sum of two on-shell momenta is
 always
 less than $-4 m_{\rm th}^2$,
 while the 4-momentum squared of the difference of
 two on-shell momenta is always greater than 0.
 Hence each propagator
 in (\ref{eq:almost_scattering_amp}) is bounded by
 $1/m_{\rm th}^2$ so that
 \begin{equation}
 |{\cal T}(\underline{l}_1,\underline{p};\underline{l}_2,\underline{k})|
 = {\cal O}(\lambda)+{\cal O}(g^2/m_{\rm th}^2) = {\cal O}(\lambda)
 \;,
 \end{equation}
 since $g^2=\Order(\lambda m_{\rm phys}^2)$ and
 $(m_{\rm phys}^2/m_{\rm th}^2)\le 1$.
 Consequently,
 the size of the effective vertex $D_{\pi}(\underline{p})$
 with the $g\phi^3{+}\lambda\phi^4$
 kernel does not differ (by powers of couplings) from
 the solution with only the $\lambda\phi^4$ kernel.
 Furthermore, formula (\ref{eq:ladder_viscosities2}) for the shear
 viscosity continue to hold in the $g\phi^3{+}\lambda\phi^4$ theory
 (with the self-energy now given by the full $g\phi^3{+}\lambda\phi^4$
 self-energy).
 Hence, in both theories, the shear viscosity is
 ${\cal O}(1/\lambda^2)$, only the numerical prefactors will differ.
 Note also that at temperatures $T \gg m_{\rm phys}$, the typical size
 of a propagator is ${\cal O}(1/T^2)$ and hence the contribution of
 the cubic interaction $g^2 G_R = {\cal O}(g^2/T^2)$ may be ignored
 compared to $\lambda$.

 For the comparison with the kinetic theory
 in the next section,
 it is helpful to note that
 the imaginary part of the on-shell two-loop self-energy
 $\Sigma_I(\underline{k})$
 ({\it c.f.}~Fig.~\ref{fig:self_two_loop}),
 used in Eq.~(\ref{eq:reduced_vertex_eq_full}),
 can be expressed in terms of
 the same ``rungs'' of Fig.~\ref{fig:rungs34} by
 joining two of the external lines by a cut propagator,
 \begin{equation}
 \Sigma_I(\underline{k})
 =
 {1\over 6}
 (1{-}e^{-\beta \underline{k}^0})
 \int{d^4 p\over (2\pi)^4}\,
 L_{\rm full}(\underline{k}{-}p)S_{\rm free}(p)
 \;.
 \label{eq:L_full_self}
 \end{equation}
 For instance, the contribution of the
 $\lambda\phi^4$ self-energy diagram can be obtained
 by attaching a cut line to the $\lambda\phi^4$ rung labeled
 (a) in
 Fig.~\ref{fig:rungs34} and dividing by 6.
 Since the diagram (a) has a symmetry factor of 2,
 this correctly reproduces the
 overall factor of $1/12$ associated with this self-energy diagram
 (6 from the symmetry factor for this two-loop diagram,
 and 2 from the relation between the
 imaginary part of a diagram and the discontinuity).
 The $g\phi^3$ theory two-loop self-energy diagram
 labeled (b) in Fig.~\ref{fig:self_two_loop}
 is similarly obtained by
 attaching a cut line to the rungs labeled (b) and (c)
 in Fig.~\ref{fig:rungs34} and dividing by 6.
 Since the diagram (b) has a symmetry factor of 2,
 and the box diagram (c) has a symmetry factor of 1,
 this again correctly produces the overall factor of
 $(1{+}1/2)/6=1/4$ associated with this self-energy diagram.
 All other two-loop self-energies can be reproduced from the
 rung
 diagrams in a similar manner.

 \subsection{Ladder summation for the bulk viscosity calculation in
 $\lambda\phi^4$ theory}
 \label{subsec:ladder_bulk}

 As explained in section~\ref{subsec:qualitative},
 when the volume of a system changes
 (or equivalently when the density changes),
 number-changing processes are
 ultimately responsible for restoring equilibrium.
 Hence, the calculation of the bulk viscosity must
 include the effect of changing particle number.
 The leading order equation
 \begin{equation}
 | {\cal I}_{\bar{\cal P}} \rangle
 =
 (1-{\cal M}_{\rm full}{\cal F}_{\rm pp})
 |{\cal D}_{\bar{\cal P}}\rangle
 \;,
 \label{eq:22bulk}
 \end{equation}
 where ${\cal M}_{\rm full}$ and
 ${\cal F}_{\rm pp}$ are given in Eq.~(\ref{eq:vertex_eq_full_pp}),
 contains only the effect of elastic binary scatterings,
 and therefore is not suitable for the calculation of the bulk
 viscosity.

 Mathematically,
 the integral equation (\ref{eq:22bulk}) is not well posed because
 only one of the consistency conditions
 $\langle \bar{b}_0|{\cal I}_{\bar{\cal P}}\rangle = 0$ and
 $\langle \bar{b}_5|{\cal I}_{\bar{\cal P}}\rangle = 0$
 can be satisfied by adjusting the value of a single
 free parameter $v^2$ in $\bar{\cal P}$.
 Since energy-momentum must be conserved, the condition
 $\langle \bar{b}_0|{\cal I}_{\bar{\cal P}}\rangle = 0$
 must be enforced by choosing
 \begin{eqnarray}
 v^2
 & \displaystyle = & \displaystyle
 {\langle \bar{b}_0 | {\cal I}_{\cal P}\rangle
                      \over
  \langle \bar{b}_0 | {\cal I}_{\varepsilon}\rangle}
 =
 {
 \displaystyle
 \int{d^3 {\bf k}\over (2\pi)^3}\, n(E_k)\,[1{+}n(E_k)]\,
 I_{\cal P}(k)
    \over
 \displaystyle
 \int{d^3 {\bf k}\over (2\pi)^3}\, n(E_k)\,[1{+}n(E_k)]\,
 I_{\varepsilon}(k)
 }
 \nonumber\\
 & \displaystyle = & \displaystyle
 {
 (\partial {\cal P}/ \partial T)
            \over
 (\partial {\varepsilon}/ \partial T)
 }
 = v_{\rm s}^2
 \;.
 \label{eq:cond_pressure}
 \end{eqnarray}
 Here,
 $I_{\cal P}(k)$, and $I_{\varepsilon}(k)$ represent the effect of
 the pressure and the energy density insertions {\em including} the
 contribution from chain diagrams
 ({\it c.f.}~Eq.~(\ref{eq:chain_modified_Ia})), and $v_{\rm s}$ is the
 speed of sound.\footnote{%
	 To see that the second expression does produce
	 the right speed of sound,
	 one must know the explicit forms of $I_{\cal P}$ and
	 $I_{\varepsilon}$ up to ${\cal O}(\lambda T^2)$.
	 Since these forms are
	 not essential to the present discussion, evaluation of
	 the inhomogeneous terms $I_{\cal P}$, $I_{\varepsilon}$
	 and the speed of sound $v_{\rm s}$ are deferred to
	 section~\ref{subsec:inhomogeneous_terms}.
}
 Note that the condition $v^2 = v_{\rm s}^2$
 is a familiar result from the Boltzmann equation.  In
 section~\ref{subsec:inhomogeneous_terms},
 it is shown that with this choice of $v_{\rm s}^2$,
 ${\cal I}_{\bar{\cal P}} \propto m_{\rm phys}^2$.
 The (leading order) bulk viscosity vanishes if the mass is zero
 since ${\cal I}_{\bar{\cal P}}=0$ in that case\cite{Weinberg}.

 When ${\cal O}(\lambda^4)$
 rungs are included in the ladder kernel,
 $\bar{b}_5$ is no longer a zero mode since the kernel now contains
 2--4 number-changing processes.
 Since the condition
 $\langle \bar{b}_5|{\cal I}_{\bar{\cal P}}\rangle=0$
 need no longer be satisfied,
 the resulting integral equation is well-posed, and hence
 the leading order bulk viscosity can be evaluated as follows.

 The integral equation for the effective vertex
 including ${\cal O}(\lambda^4)$ rungs may be written as
 \begin{equation}
 |{\cal I}_{\bar{\cal P}}\rangle
 =
 (1{-}{\cal K}_{\rm cons.}{-}\delta{\cal K}_{\rm ch.})
 |{\cal D}_{\bar{\cal P}}\rangle
 \;,
 \label{eq:bulk_integral_eq1}
 \end{equation}
 where ${\cal K}_{\rm cons.}$ is the number conserving part of the
 kernel, and $\delta{\cal K}_{\rm ch.}$ is the number changing part of
 the kernel.
 In terms of the solution $| {\cal D}_{\bar{\cal P}} \rangle$,
 the leading order bulk viscosity is given by
 \begin{equation}
 \zeta =
 \beta \langle {\cal I}_{\bar{\cal P}} |
 {\cal F} | {\cal D}_{\bar{\cal P}} \rangle
 \;.
 \end{equation}

 As will be shortly shown, $\bar{b}_5$ is no longer a zero mode of
 the kernel due to the number changing $\delta{\cal K}_{\rm ch.}$ part.
 However, $\bar{b}_5$ is still a zero mode of the number conserving part
 of the kernel so that
 \begin{equation}
 \langle \bar{b}_5 |
 (1{-}{\cal K}_{\rm cons.}{-}\delta{\cal K}_{\rm ch.})
 =
 -\langle \bar{b}_5 | \delta{\cal K}_{\rm ch.}
 \;.
 \end{equation}
 Since this vanishes as $\lambda\to 0$,
 the kernel
 $(1{-}{\cal K}_{\rm cons.}{-}\delta{\cal K}_{\rm ch.})$
 has a very small eigenvalue in the weak coupling limit.
 Hence, the solution of the integral equation
 (\ref{eq:bulk_integral_eq1})
 \begin{equation}
 |{\cal D}_{\bar{\cal P}}\rangle
 =
 { 1\over
   1-{\cal K}_{\rm cons.}-\delta{\cal K}_{\rm ch.}
 }
 |{\cal I}_{\bar{\cal P}}\rangle
 \;,
 \label{eq:bulk_sol}
 \end{equation}
 is totally dominated by the small eigenvalue component.
 To see this, the unit operator
 \begin{equation}
 {\bf 1} = \sum_{i} |f_i\rangle \langle f_i |
 \;,
 \end{equation}
 where $|f_i \rangle$'s are the eigenvectors of
 $(1{-}{\cal K}_{\rm cons.}{-}\delta{\cal K}_{\rm ch.})$
 with eigenvalues $\alpha_i$, may be inserted in Eq.~(\ref{eq:bulk_sol})
 to yield
 \begin{eqnarray}
 |{\cal D}_{\bar{\cal P}}\rangle
 & = & \displaystyle
 \sum_{i}
 {1\over \alpha_i}
 | f_i \rangle
 \langle f_i |{\cal I}_{\bar{\cal P}}\rangle
 \nonumber\\
 & = & \displaystyle
 {1\over \alpha_5}
 | f_5 \rangle
 \langle f_5 |{\cal I}_{\bar{\cal P}}\rangle
 \times (1 + {\cal O}(\lambda))
 \;.
 \label{eq:approx_bulk_sol}
 \end{eqnarray}
 In the last line, the eigenvector
 $| f_5 \rangle = | \bar{b}_5 \rangle  + {\cal O}(\lambda)$
 corresponds to the eigenvalue $\alpha_5$ which vanishes
 in the $\lambda \to 0$ limit.
 The leading order value of $\alpha_5$ can be obtained by
 \begin{eqnarray}
 \alpha_5
 & = & \displaystyle
 \langle f_5 |
 (1{-}{\cal K}_{\rm cons.}{-}\delta{\cal K}_{\rm ch.})
 | f_5 \rangle
 \nonumber\\
 & = & \displaystyle
 -\langle \bar{b}_5 | \delta {\cal K}_{\rm ch.} | \bar{b}_5 \rangle
  \times (1 + {\cal O}(\lambda))
 \;.
 \end{eqnarray}
 As is shortly shown, due to the statistical factors,
 the eigenvalue $\alpha_5 = {\cal O}(\lambda)$ even though it
 originates from ${\cal O}(\lambda^4)$ correction to ${\cal O}(\lambda^2)$
 rungs.

%
%

 Using the leading order solution (\ref{eq:approx_bulk_sol}),
 the leading order bulk viscosity is
 \begin{eqnarray}
 \zeta
 & = & \displaystyle
 \beta \langle {\cal I}_{\bar{\cal P}} |
 {\cal F} | {\cal D}_{\bar{\cal P}} \rangle
 \nonumber\\
 & = & \displaystyle
 -
 \beta
 {
 \langle {\cal I}_{\bar{\cal P}}|{\cal F}_{\rm pp} | \bar{b}_5 \rangle
 \langle \bar{b}_5 |{\cal I}_{\bar{\cal P}} \rangle
 \over
 \langle \bar{b}_5| \delta{\cal K}_{\rm ch.} |\bar{b}_5 \rangle
 }
 \times (1 + {\cal O}(\lambda))
 \;.
 \label{eq:leading_bulk}
 \end{eqnarray}
 At high temperature, the bulk viscosity
 explicitly contains $m_{\rm phys}^4$ provided by
 two factors of ${\cal I}_{\bar{\cal P}}$.

 To obtain the explicit form of the ladder kernel including
 ${\cal O}(\lambda^4)$ rungs, note that
 when the external operators are bilinear, any diagram contributing to
 the Wightman function calculation can be regarded as a ladder diagram.
 The side rail part, as before, consists of two-particle intermediate
 states, and the rung part consists of the two-particle irreducible
 sub-diagrams between two side rails.
 Hence, if the rung matrix ${\cal M}$ contains all possible
 rungs, and the propagators in ${\cal F}$ are the full propagators,
 the integral equation
 \begin{equation}
 {\cal D}_{\bar{\cal P}}(k, q{-}k)
 =
 {\cal I}_{\bar{\cal P}}(k, q{-}k)
 +
 \int {d^4 p \over (2\pi)^4} \,
 {\cal M}(k{-}p) \,
 {\cal F}(p, q{-}p)
 {\cal D}_{\bar{\cal P}}(p, q{-}p)
 \;
 \end{equation}

 \begin{figure}
 \setlength {\unitlength}{1cm}
\vbox
    {%
    \begin {center}
 \begin{picture}(0,0)
 \put(0.0,-0.5) {$l_2$}
 \put(1.0,-0.5) {$l_3$}
 \put(2.25,-0.5){$l_4$}
 \put(3.25,-0.5){$l_5$}
 \put(0.3,1.5)  {$l_1$}
 \put(3.1,1.5)  {$l_6$}
 \put(4.5,-0.5) {$l_2$}
 \put(5.3,-0.5) {$l_3$}
 \put(6.1,-0.5) {$l_4$}
 \put(7.6,-0.5) {$l_5$}
 \put(5.8,2.3)  {$l_1$}
 \put(6.5,2.3)  {$l_6$}
 \end{picture}
	\leavevmode
	\def\epsfsize	#1#2{0.4#1}
	\epsfbox {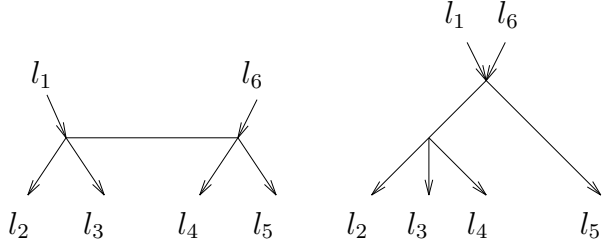}
    \end {center}
 \vspace{0.5cm}
     \caption
 	{%
 Diagrams corresponding to the $\Order(\lambda^{2})$
 2--4 scattering processes.
 Mirror images and different permutations of $l_i$'s are not
 explicitly shown.
 Combining any two of these diagrams,
 including the square of a diagram,
 and integrating over $l_i$ $(i=2,3,4)$
 produce $\Order(\lambda^4)$ rungs.
 Due to the 4-momentum conservation,
 $l_5 = l_1{+}l_6{-}l_2{-}l_3{-}l_4$.
}
 \label{fig:2_4_process}
     }
 \end {figure}
 \noindent
 may correspond to the sum of {\it all} diagrams contributing
 to the correlation function of the bilinear part of the operator
 $\bar{\cal P}$.
 The exact solution of this equation, of course, is impossible to obtain.
 However, as shown in Eq.~(\ref{eq:leading_bulk}),
 the integral equation need not be solved;
 only the leading order number changing part of the
 kernel and the zero modes of the number conserving part are needed to
 evaluate the leading order bulk viscosity.

 To extract the relevant terms in the kernel,
 it is convenient to include in the ``pinching pole part''
 the rungs up to and including ${\cal O}(\lambda^4)$ corrections
 all calculated with free propagators
 \begin{equation}
 {\cal K}_{\rm pp} \equiv
 ({\cal M}_0 + {\cal M}_1) \, {\cal F}_{\rm pp}
 \;.
 \end{equation}
 Here, ${\cal M}_0$ includes only the one-loop rung,
 and ${\cal M}_{1}$ includes
 ${\cal O}(\lambda^4)$
 rungs that can be obtained by adding two more lines to the
 diagrams in Fig.~\ref{fig:rungs34}, or equivalently,
 rungs that can be obtained by squaring the 2--4 amplitude shown in
 Fig.~\ref{fig:2_4_process}.
 ${\cal M}_1$ also contains various
 corrections to the simple one-loop rung shown
 in Fig.~\ref{fig:one_loop_corrected}.
 Among these rungs, only those containing a number changing process
 are important in calculating the leading order bulk viscosity.
 The form of the pinching pole side rail matrix
 ${\cal F}_{\rm pp}$ is the same as in Eq.~(\ref{eq:Fpp})
 except that the self-energy $\Sigma_I$ now includes
 contributions of ${\cal O}(\lambda^3)$ and ${\cal O}(\lambda^4)$
 diagrams
 shown in Fig.~\ref{fig:self_44}.
 With these definitions, the deviation from the ``pinching pole part''
 arises
 only from the side rails.
\begin{figure}
 \setlength {\unitlength}{1cm}
\vbox
    {%
    \begin {center}
 \begin{picture}(0,0)
 \end{picture}
	\leavevmode
	\def\epsfsize	#1#2{0.3#1}
	\epsfbox {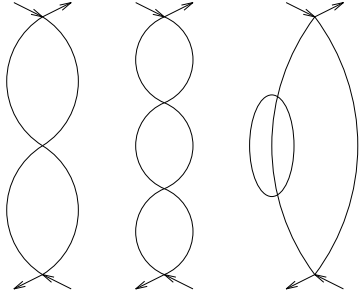}
    \end {center}
    \caption{%
$\Order(\lambda^3)$ and $\Order(\lambda^4)$ corrections to the simple
one-loop rung.  The first two diagrams
correspond to a part of the quartic vertex correction, the third one
corresponds to the finite width correction.
}
 \label{fig:one_loop_corrected}
    }%
\end {figure}
\begin{figure}
 \setlength {\unitlength}{1cm}
\vbox
    {%
    \begin {center}
 \begin{picture}(0,0)
 \end{picture}
	\leavevmode
	\def\epsfsize	#1#2{0.24#1}
	\epsfbox {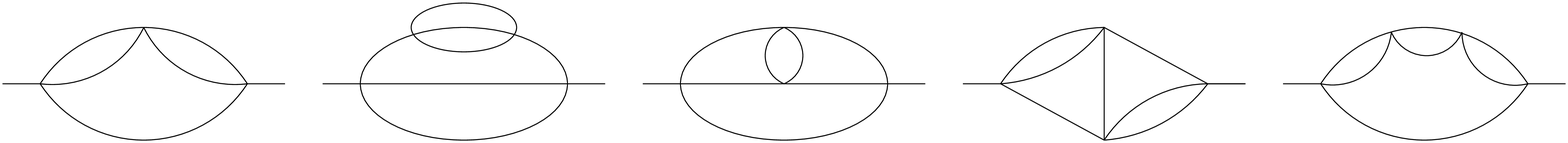}
    \end {center}
    \caption{%
$\Order(\lambda^3)$ and $\Order(\lambda^4)$ diagrams contributing to
to the imaginary part of the $\lambda\phi^4$ theory self-energy.
}
 \label{fig:self_44}
    }%
\end {figure}

 Rewriting the integral equation as
 \begin{equation}
 |{\cal I}_{\bar{\cal P}}\rangle
 = (1-{\cal K}_{\rm pp}-\delta {\cal K})|{\cal D}_{\bar{\cal P}} \rangle
 \;,
 \end{equation}
 multiplying it with $\delta {\cal K}$
 and adding the result to the original equation produce
 \begin{equation}
 (1+\delta{\cal K})|{\cal I}_{\bar{\cal P}}\rangle
 =
 ( 1 - {\cal K}_{\rm pp} - \delta {\cal K}\, {\cal K}_{\rm pp} )
 |{\cal D}_{\bar{\cal P}} \rangle
 +
 {\cal O}( (\delta {\cal K})^2 )
 \;.
 \label{eq:bulk_integral_eq}
 \end{equation}
 Here
 \begin{equation}
 \delta {\cal K}\, {\cal K}_{\rm pp}
 \equiv
 {\cal M}_0 \, \delta {\cal F}_{\rm pp} \,
 {\cal M}_0 \, {\cal F}_{\rm pp}
 \times (1 + {\cal O}(\lambda^2))
 \;,
 \end{equation}
 where $\delta {\cal F}_{\rm pp}$
 is the ${\cal O}(\lambda^2)$
 correction to the free single particle spectral density in
 ${\cal F}_{\rm pp}$.
 The non-pinching pole part does not contribute
 when sandwiched between two ${\cal M}_0$'s
 as shown in appendix~\ref{app:ladder}.

 The integral equation
 Eq.~(\ref{eq:bulk_integral_eq})
 is different from Eq.~(\ref{eq:bulk_integral_eq1})
 in two ways.
 First, the inhomogeneous term
 in Eq.~(\ref{eq:bulk_integral_eq})
 is not purely ${\cal I}_{\bar{\cal P}}$,
 and the kernel is not yet separated into the number conserving and
 the number changing part.
 Since only the leading order calculation is considered, the extra term
 in the inhomogeneous term is unimportant.
 For the separation of the number conserving and the number changing
 part,
 it is more convenient to evaluate directly
 the needed matrix element
 \begin{equation}
 \langle \bar{b}_5 | \delta {\cal K}_{\rm ch.} | \bar{b}_5 \rangle
 =
 - \langle \bar{b}_5 |
   ( 1 - {\cal K}_{\rm pp} - \delta {\cal K}\, {\cal K}_{\rm pp} )
   | \bar{b}_5 \rangle
 \;,
 \end{equation}
 rather than separating the two parts.

 Applying $u^T$ and using the same procedures as before,
 the integral equation (\ref{eq:bulk_integral_eq}) is reduced to
 \begin{eqnarray}
 I'_{\bar{\cal P}}(\underline{k})
 & = & \displaystyle
 D_{\bar{\cal P}}(\underline{k})
 -
 \int {d^4 l_1 \over (2\pi)^4}\,
 K_{\rm bulk} (\underline{k}, l_1)\,
 n(l_1^0)\,
 S_{\rm free}(l_1)\,
 {D_{\bar{\cal P}}(l_1) \over \Sigma_I(l_1)}\,
 \;,
 \label{eq:mod_bulk_eq}
 \end{eqnarray}
 where
 \begin{equation}
 I'_{\bar{\cal P}}(\underline{k})
 \equiv
 u^T(\underline{k}) (1+\delta {\cal K})
 {\cal I}_{\bar{\cal P}}(\underline{k})
 \;,
 \end{equation}
 and
 \begin{eqnarray}
 \lefteqn{K_{\rm bulk}(\underline{k}, l_1)\equiv
 -{1\over 2}\,
 (1{-}e^{-\underline{k}^0\beta})\,(1{-}e^{-l_1^0\beta})} & &
 \nonumber\\
 & \times & \displaystyle
 \bigg[
 {1\over 24}
 \int \prod_{i=2}^5
\left( {d^4 l_i \over (2\pi)^4}\, S_{\rm free}(-l_i) \right) \,
       (2\pi)^4\delta(\sum_{i=1}^5 l_i + \underline{k})\,
\left(
 \left|
 {\cal T}_{6}(\{ l_i \}, \underline{k})
 \right|^2
 + {T}_{3}(\{ l_i \}, \underline{k})
\right)
 \nonumber\\
 & + & \displaystyle
 \hphantom{\bigg[}
 {1\over 2}
 \int \prod_{i=2}^3
 \left( {d^4 l_i \over (2\pi)^4}\, S_{\rm free}(-l_i) \right) \,
 (2\pi)^4\delta(\sum_{i=1}^3 l_i + \underline{k})\,
 \left|
 {\cal T}_{4}(\{ l_i \},\underline{k})
 \right|^2
 \bigg]
 \;.
 \label{eq:mod_bulk_eq_K}
 \end{eqnarray}
 In the $\lambda\phi^4$ case, the scattering amplitude
 involving 4 particles, ${\cal T}_{4}( \{ l_i \}, \underline{k} )$,
 includes the lowest order amplitude $\lambda$, and
 ${\cal O}(\lambda^2)$ and ${\cal O}(\lambda^3)$ corrections.
 Since this part of the kernel
 conserves the particle number, the explicit form
 of ${\cal T}_4( \{ l_i \}, \underline{k})$ is not important in the bulk
 viscosity calculation.

 The lowest order scattering amplitude involving 6 particles is given by
 \begin{equation}
 {\cal T}_{6}( \{ l_i \}, \underline{k} )
 \equiv
 {\lambda^2}
 \sum \tilde{G}_R^{\rm free}(l_i{+}l_j{+}l_k)
 \;,
 \end{equation}
 where the sum is over all 10 different combinations of three momenta from
 the set $\{ l_i \}$.
\begin{figure}
 \setlength {\unitlength}{1cm}
\vbox
    {%
    \begin {center}
 \begin{picture}(0,0)
 \put(1,0.0){(a)}
 \put(4,0.0){(b)}
 \end{picture}
	\leavevmode
	\def\epsfsize	#1#2{0.3#1}
	\epsfbox {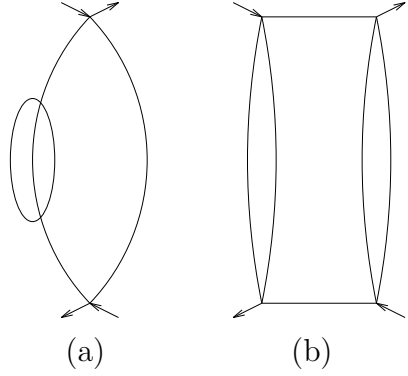}
    \end {center}
    \caption{%
 Diagrams corresponding to the squared terms in $|{\cal T}_6|^2$.
 Figure (a) corresponds to the finite width correction the one-loop
 rung, and figure (b) corresponds to the finite width correction to the
 pinching pole part of the side rail.
}
 \label{fig:squared_terms}

    }%
\end {figure}
\noindent
 The square $|{\cal T}_6|^2$ contains singular terms of the form
 $\left | \tilde{G}_R^{\rm free}(l_i{+}l_j{+}l_k) \right|^2$
 which produces an ill-defined product of delta functions.
 However, these products of delta functions are removed
 by an additional term $T_3$ given by
 \begin{eqnarray}
 {T}_{3}( \{ l_i \}, \underline{k} )
 & \displaystyle \equiv & \displaystyle
 {\lambda^4 \over 2}
 \sum
 \left( \tilde{G}_R^{\rm free}(l_i{+}l_j{+}l_k)
    {-} \tilde{G}_A^{\rm free}(l_i{+}l_j{+}l_k)
 \right)^2
 \nonumber\\
 & \displaystyle = & \displaystyle
 {\lambda^4 \over 2}
 \sum
 \rho_{\rm free}(l_i{+}l_j{+}l_k)^2
 \;.
 \end{eqnarray}
 The ${\cal O}(\lambda^4)$ rung diagrams corresponding
 to the terms in $|{\cal T}_6|^2$ can be obtained by
 attaching two more lines to the ${\cal O}(\lambda^2)$ rungs in
 Fig.~\ref{fig:rungs34} in all possible ways consistent with the
 $\lambda\phi^4$ theory.
 The prefactor $1/24$ accounts for the symmetry factors of the diagrams.

 To see the connection between the terms in $|{\cal T}_6|^2$
 and the diagrams, consider, for example, the 10 squared terms in
 $| {\cal T}_{6}( \{ l_i \}, \underline{k} )|^2 + T_3$.
 Using the symmetry in $l_i$'s under the integral, these terms can be
 re-expressed as
 \begin{eqnarray}
 T_{\rm sq}( \{ l_i \}, \underline{k} )
 & \displaystyle \equiv & \displaystyle
 4 \lambda^4
 \left(
 |\tilde{G}_R^{\rm free}(l_2{+}l_3{+}l_4)|^2
 +
 \rho_{\rm free}(l_2{+}l_3{+}l_4)^2/2
 \right)
 \nonumber\\
 & \displaystyle & \displaystyle {}
 +
 6 \lambda^4
 \left(
 |\tilde{G}_R^{\rm free}(l_1{+}l_2{+}l_3)|^2
 +
 \rho_{\rm free}(l_1{+}l_2{+}l_3)^2/2
 \right)
 \;.
 \end{eqnarray}
 The first term does not contain external momenta
 $l_1$ and $k$.
 Hence, the cut lines corresponding to $l_2,l_3,l_4$
 form a cut self-energy diagram
 (together with the 4-momentum conserving $\delta$-function)
 corresponding to
 the finite width correction to the one loop rung
 shown in Fig.~\ref{fig:squared_terms}a.
 The symmetry factor associated with this diagram is $1/6$.
 Factors in Eq.~(\ref{eq:mod_bulk_eq_K}) combine to yield $1/12$
 including an extra factor of $1/2$ from the relation of the form
 (\ref{eq:K_full}) used to obtain Eq.~(\ref{eq:mod_bulk_eq}).
 The remaining term containing $l_1,l_2,l_3$
 corresponds to the once iterated one-loop rung
 shown in Fig.~\ref{fig:squared_terms}b.
 The symmetry factor associated with this diagram is $1/4$.
 Factors in Eq.~(\ref{eq:mod_bulk_eq_K}) combine to yield $1/8$ again
 with the extra factor of $1/2$.

 Using the definition (\ref{eq:def_inner})
 the above integral equation can be expressed
 symbolically as
 \begin{equation}
 | I'_{\bar{\cal P}} )
 =
 (1 - K_{\rm bulk}) | D_{\bar{\cal P}} )
 \;.
 \end{equation}
 The self-energy $\Sigma_I$
 and the kernel $K_{\rm bulk}$ has the following relationship
 \begin{equation}
 \int {d^4 l \over (2\pi)^4}\, l^\mu \, n(l^0)\,S_{\rm free}(l)
 K_{\rm bulk}(l, \underline{p})
 =
 \underline{p}^\mu\, \Sigma_I (\underline{p})
 \;,
 \end{equation}
 which can be proven by a similar argument used
 to obtain Eq.~(\ref{eq:L_full_self}).
 From this relation, it is simple to see that the zero modes
 $b_{\mu}(\underline{l}) = \underline{l}_\mu\,\Sigma_I(\underline{l})$
 of the leading order ladder kernel $(1-K_{\rm pp})$
 are still the zero modes of the modified ladder kernel
 $(1 - K_{\rm bulk})$.
 However, a previous zero mode
 corresponding to number conservation,
 $b_5(\underline{l})
 = {\rm sgn}(\underline{l}^0)\, \Sigma_I (\underline{l})$
 is no longer a zero mode since
 \begin{equation}
 \int {d^4 l \over (2\pi)^4}\, {\rm sgn}(l^0) \,
 n(l^0) \, S_{\rm free}(l)
 K_{\rm bulk}(l, \underline{p})
 =
 {\rm sgn}(p^0)\, \Sigma_I (\underline{p})
 + \delta C(\underline{p})
 \;,
 \end{equation}
 or symbolically,
 \begin{equation}
 (b_5 | (1-K_{\rm bulk}) = -( \delta C |
 \;,
 \end{equation}
 where $\delta C(\underline{p})$ contains only the
 number-changing part of the kernel.
 Only the overlap $(\delta C| b_5)$ is needed to calculate the bulk
 viscosity since
 the leading order bulk viscosity is,
 ignoring corrections suppressed by
 ${\cal O}(\lambda)$,
 \begin{eqnarray}
 \zeta
 & = & \displaystyle
 \beta \langle {\cal I}_{\bar{\cal P}} |
 {\cal F} | {\cal D}_{\bar{\cal P}} \rangle
 =
 \beta ( I_{\bar{\cal P}} | D_{\bar{\cal P}} )
 \nonumber\\
 & = & \displaystyle
 \beta
 {
 ( I_{\bar{\cal P}}| b_5 ) ( b_5 |I'_{\bar{\cal P}} )
 		\over
 ( b_5| (1 - K_{\rm bulk}) | b_5 )
 }
 =
 - \beta
 {
 ( I_{\bar{\cal P}}| b_5 ) ( b_5 |I_{\bar{\cal P}} )
 		\over
 ( \delta C| b_5 )
 }
 \;.
 \label{eq:leading_bulk2}
 \end{eqnarray}
 Straight forward calculations yield
 \begin{equation}
 (b_5 | I_{\bar{\cal P}})
 =
 \int
 {d^3 {\bf l} \over (2\pi)^3 E_l}\,
 [1{+}n(E_l)]\,n(E_l)\,
 I_{\bar{\cal P}}(E_l,{\bf l})
 \;,
 \label{eq:b_5I_p}
 \end{equation}
 and
 \begin{eqnarray}
 -(\delta C | b_5)
 & = & \displaystyle
 ( b_5 |(1 - K_{\rm bulk})| b_5)
 \nonumber\\
 & = & \displaystyle
 2\int
 \prod_{i=1}^2 {d^3 {\bf l}_i \over (2\pi)^3}\,
 d\sigma_{12\to 3456}\, v_{12}\,
 \nonumber\\
 & & \displaystyle
 {}\quad \times
 n(E_1)\, n(E_2)\,
 [1{+}n(E_3)]\, [1{+}n(E_4)]\,
 [1{+}n(E_5)]\, [1{+}n(E_6)]\,
 \;,
 \label{eq:deltaCb_5}
 \end{eqnarray}
 where the differential scattering cross-section
 of 2 to 4 scatterings is given by \cite{Bjorken}
 \begin{equation}
 d\sigma_{12\to 3456}
 \equiv
 \prod_{i=3}^6 {d^3 {\bf l}_i \over (2\pi)^3 2 E_i}\,
 \left( |{\cal T}_6(\{\underline{l}_i\})|^2 \right)
 (2\pi)^4\delta( \underline{l}_1{+} \underline{l}_2{-} \underline{l}_3{-}
 \underline{l}_4
 		{-} \underline{l}_5{-} \underline{l}_6 )\,
      {\bigg /} (4 E_1 E_2 v_{12} 4!)
 \;,
 \label{eq:24_dsigma}
 \end{equation}
 where $v_{12}$ is the relative speed between ${\bf l}_1$ and
 $-{\bf l}_2$
 \begin{equation}
 v_{12} \equiv
 \left|
 { {\bf l}_1\over E_1 }
 +
 { {\bf l}_2\over E_2 }
 \right|
 \;.
 \end{equation}
 In Eq.~(\ref{eq:24_dsigma}),
 all underlined momenta have positive energy.
 In the 2--4 scattering amplitude,
 products of delta functions do not appear since
 combinations of on-shell momenta in the propagators
 cannot be on-shell due to kinematic constraints.
 Hence, the $T_3$ term is irrelevant here.

 Using Fermi's Golden rule\cite{Bjorken}, the expression
 $-(\delta C| b_5)$ in Eq.~(\ref{eq:deltaCb_5}) may be interpreted as
 (2 times) the total 2--4 reaction rate per volume in a thermal medium.
 Due to the statistical factors, ${\cal O}(m_{\rm th})$ momenta
 dominate the
 integral in the expression (\ref{eq:deltaCb_5}).  Hence, when
 $T = {\cal O}(m_{\rm phys}/\sqrt{\lambda})$,
 $(\delta C| b_5) = {\cal O}(\lambda^3 T^4)$.
 Then
 since $(b_5|I_{\bar{\cal P}}) = {\cal O}(m_{\rm phys}^2 T^2)
 = {\cal O}(\lambda T^4)$,
 the viscosity $\zeta = {\cal O}(\beta m_{\rm phys}^4 /\lambda^3) =
 {\cal O}(T^3/\lambda)$ which is ${\cal O}(\lambda)$
 smaller than the shear viscosity.
 Note that even at very high temperature, the bulk viscosity is a
 non-trivial function of the ${\cal O}(1)$ ratio $m_{\rm th}^2/\lambda T^2$.
 Hence, even at high temperature, the distinction between the thermal
 mass and the physical mass is important.
\begin{figure}
 \setlength {\unitlength}{1cm}
\vbox
    {%
    \begin {center}
 \begin{picture}(0,0)
 \put(0.0,-0.5) {$l_2$}
 \put(1.0,-0.5) {$l_3$}
 \put(1.7,-0.5) {$l_4$}
 \put(0.0,2.2) {$l_1$}
 \put(1.7,2.2) {$l_5$}
 \put(2.8,-0.5) {$l_2$}
 \put(3.3,-0.5) {$l_3$}
 \put(5.6,-0.5) {$l_4$}
 \put(2.5,1.5) {$l_1$}
 \put(6.2,1.5) {$l_5$}
 \put(7.1,-0.5) {$l_2$}
 \put(8.5,-0.5) {$l_3$}
 \put(9.8,-0.5) {$l_4$}
 \put(6.4,0.7) {$l_1$}
 \put(10.3,0.7) {$l_5$}
 \put(0.0,2.9) {$l_2$}
 \put(1.0,2.9) {$l_3$}
 \put(1.7,2.9) {$l_4$}
 \put(0.0,5.2) {$l_1$}
 \put(1.5,5.2) {$l_5$}
 \put(3.1,2.9) {$l_2$}
 \put(3.6,2.9) {$l_3$}
 \put(5.6,2.9) {$l_4$}
 \put(2.5,4.2) {$l_1$}
 \put(6.1,4.2) {$l_5$}
 \put(7.6,2.9) {$l_2$}
 \put(8.5,2.9) {$l_3$}
 \put(9.2,2.9) {$l_4$}
 \put(7.6,5.2) {$l_1$}
 \put(9.2,5.2) {$l_5$}
 \end{picture}
	\leavevmode
	\def\epsfsize	#1#2{0.4#1}
	\epsfbox {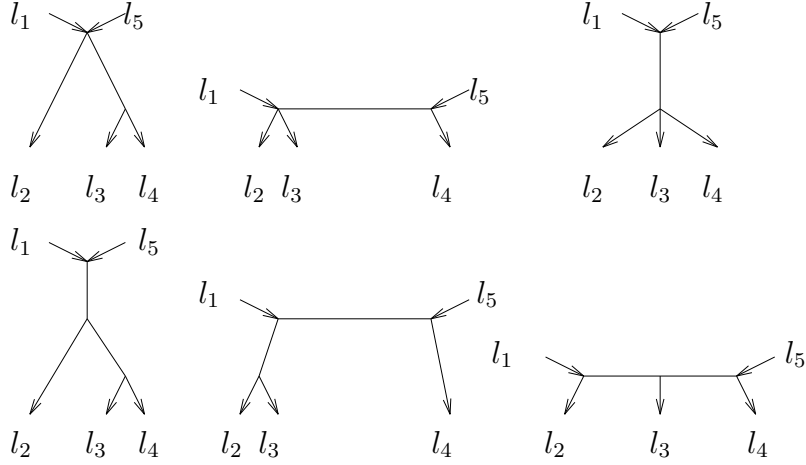}
    \end {center}
    \vspace{0.5cm}
    \caption{%
 Diagrams corresponding to the $\Order(\lambda^{3/2})$
 2--3 scattering processes.
 Mirror images and different permutations of $l_i$'s are not
 explicitly shown.
 Combining any two of these diagrams,
 including the square of a diagram,
 and integrating over $l_i$ $(i=2,3)$
 produce $\Order(\lambda^3)$ rungs.
 Due to the 4-momentum conservation,
 $l_4 = l_1{+}l_5{-}l_2{-}l_3$.
}
 \label{fig:23_scatterings}
    }%
\end {figure}

 \subsection{Ladder summation for the bulk viscosity calculation
 with an additional $g\phi^3$ interaction}
 \label{subsec:ladder_bulk34}

 In the $g\phi^3{+}\lambda\phi^4$ case,
 the lowest order number-changing process is
 ${\cal O}(\lambda g)={\cal O}(g^3)$ corresponding to 2 particles colliding
 to produce 3 particles.
 A few ${\cal O}(\lambda^3)$ rungs containing the effect of
 these scatterings are shown in Fig.~\ref{fig:higher_rungs}.
 Other rungs may be constructed by combining two diagrams
 among those in Fig.~\ref{fig:23_scatterings}
 corresponding to the 2--3 scattering processes, or equivalently
 adding one more line to the rungs in Fig.~\ref{fig:rungs34}
 in all possible ways consistent with the theory.

 With these rungs,
 the procedure used in the $\lambda\phi^4$ case can be again used
 to produce the reduced integral equation (\ref{eq:mod_bulk_eq})
 now with
 \begin{eqnarray}
 \lefteqn{K_{\rm bulk}(\underline{k}, l_1)\equiv
 -{1\over 2}\,
 (1{-}e^{-\underline{k}^0\beta})\,(1{-}e^{-l_1^0\beta})} & &
 \nonumber\\
 & \times & \displaystyle
 \bigg[
 {1\over 2}
 \int \prod_{i=2}^3
 \left( {d^4 l_i \over (2\pi)^4}\, S_{\rm free}(-l_i) \right) \,
 (2\pi)^4\delta(\sum_{i=1}^3 l_i + \underline{k})\,
 \left|
 {\cal T}_{4}(\{ l_i \},\underline{k})
 \right|^2
 \nonumber\\
 & + & \displaystyle
 \hphantom{\bigg[}
 {1\over 6}
 \int \prod_{i=2}^4
\left( {d^4 l_i \over (2\pi)^4}\, S_{\rm free}(-l_i) \right) \,
       (2\pi)^4\delta(\sum_{i=1}^4 l_i + \underline{k})\,
 \left|
 {\cal T}_{5}(\{ l_i \}, \underline{k})
 \right|^2
 \bigg]
 \;.
 \nonumber\\
 \end{eqnarray}
 The scattering amplitude
 involving 4 particles, ${\cal T}_{4}( \{ l_i \}, \underline{k} )$,
 includes the lowest order amplitude $\lambda$ and
 ${\cal O}(\lambda g)$ and ${\cal O}(\lambda^2)$ corrections.
 Since this part of the kernel conserves number, the explicit form
 of ${\cal T}_4( \{ l_i \}, \underline{k})$ is not important in evaluating
 the bulk viscosity.
 The lowest order scattering amplitude involving 5 particles is given by
 \begin{equation}
 {\cal T}_{5}( \{ l_i \}, \underline{k} )
 \equiv
 {\lambda g}
 \sum \tilde{G}_R^{\rm free}(l_i{+}l_j)
 - i {g^3}
 \sum \tilde{G}_R^{\rm free}(l_i{+}l_j)
 \tilde{G}_R^{\rm free}(l_m{+}l_n)
 \;,
 \end{equation}
 where the first
 sum is over all 10 different combinations of two members
 from the set $\{ l_i, \underline{k}\}$,
 and the second sum is over 15 different combinations of
 four members of the same set.
 Since all the momenta in the set $\{ l_i , \underline{k} \}$ are on-shell
 due to the delta function in $S_{\rm free}(-l_i)$,
 the combination $l_i{+}l_j$ cannot be on-shell.
 Hence, the ill-defined delta function products do not appear in
 $|{\cal T}_{5}( \{ l_i \}, \underline{k} )|^2$.  Also for the same reason,
 it makes no difference whether the retarded propagators are used or the
 time-ordered propagators are used.  For the sake of consistency, the
 retarded propagators are chosen here.
 The prefactor $1/6$ accounts for the symmetry factors of the
 diagrams.

 The self-energy $\Sigma_I$
 and the kernel $K_{\rm bulk}$ again has the following relationship
 \begin{equation}
 \int {d^4 l \over (2\pi)^4}\, l^\mu \, n(l^0)\,S_{\rm free}(l)
 K_{\rm bulk}(l, \underline{p})
 =
 \underline{p}^\mu\, \Sigma_I (\underline{p})
 \;,
 \end{equation}
 which can be proven by a similar argument used
 to obtain Eq.~(\ref{eq:L_full_self}).
 Again, the zero modes
 $b_{\mu}(\underline{l}) =
 \underline{l}_{\mu}\, \Sigma_I (\underline{l})$
 of the leading order ladder kernel $(1-K_{\rm pp})$
 are still the zero modes of the modified ladder kernel
 $(1 - K_{\rm bulk})$ the remaining zero mode
 $b_5(\underline{l}) =
 {\rm sgn}(\underline{l}^0)\, \Sigma_I (\underline{l})$
 is no longer a zero mode due to the number changing term in the kernel.
 The previous formula leading order bulk viscosity
 (\ref{eq:leading_bulk2})
 \begin{equation}
 \zeta
 =
 - \beta
 {
 ( I_{\bar{\cal P}}| b_5 ) ( b_5 |I_{\bar{\cal P}} )
 		\over
 ( \delta C| b_5 )
 }
 \;,
 \label{eq:leading_bulk2_34}
 \end{equation}
 still holds with the same $(b_5 | I_{\bar{\cal P}})$ in
 Eq.~(\ref{eq:b_5I_p}) but with
 \begin{eqnarray}
 -(\delta C | b_5)
 & = & \displaystyle
 ( b_5 |(1 - K_{\rm bulk})| b_5)
 \nonumber\\
 & = & \displaystyle
 {1\over2}\int
 \prod_{i=1}^2 {d^3 {\bf l}_i \over (2\pi)^3}\,
 d\sigma_{12\to 345}\, v_{12}\,
 n(E_1)\, n(E_2)\,
 [1{+}n(E_3)]\, [1{+}n(E_4)]\,
 [1{+}n(E_5)]\,
 \;.
 \label{eq:deltaCb_5_34}
 \end{eqnarray}
 Here the differential scattering cross-section
 of 2 to 3 scatterings is given by
 \begin{equation}
 d\sigma_{12\to 345}
 \equiv
 \prod_{i=3}^5 {d^3 {\bf l}_i \over (2\pi)^3 2 E_i}\,
 |{\cal T}_5(\{\underline{l}_i\})|^2
 (2\pi)^4\delta( \underline{l}_1{+} \underline{l}_2{-} \underline{l}_3{-}
 \underline{l}_4
 		{-} \underline{l}_5)\,
      {\bigg /} (4 E_1 E_2 v_{12} 3!)
 \;,
 \label{eq:23_dsigma}
 \end{equation}
 where $v_{12}$ is the relative speed between ${\bf l}_1$ and
 $-{\bf l}_2$.
 All underlined momenta in Eq.~(\ref{eq:23_dsigma})
 have positive energy.
 In the 2--3 scattering amplitude,
 the products of delta functions do not appear since
 combinations of momenta $\underline{l}_i{+}\underline{l}_j$
 in the propagators
 cannot be on-shell due to kinematic constraints.

 Again, the expression $-(\delta C|b_5)$ can be interpreted as
 (1/2 times) the total 2--3 reaction rate per volume.
 Due to the statistical factors,
 ${\cal O}(m_{\rm th})$ momenta dominate the integral.  Hence,
 when $T = {\cal O}(m_{\rm phys}/\sqrt{\lambda})$,
 $(\delta C|b_5) = {\cal O}(\lambda^{5/2} T^4)$.
 The bulk viscosity is then $\zeta = {\cal O}(T^3/\sqrt{\lambda})$
 which is ${\cal O}(\lambda^{3/2})$ smaller than the shear viscosity.
 Note that since the contribution of ${\cal O}(m_{\rm th})$ momenta
 dominates
 the integral, the bulk viscosity is a non-trivial function of the
 dimensionless ${\cal O}(1)$ ratio $g^2/\lambda m_{\rm th}^2$.
 Hence, even at high temperature, the distinction between the physical
 mass $m_{\rm phys}$ and the thermal mass $m_{\rm th}$ is important.
 As shown in \ref{subsec:inhomogeneous_terms} below,
 the only place where $m_{\rm phys}$ appears
 is in $I_{\bar{\cal P}}$.
 The propagators elsewhere must contain the thermal mass $m_{\rm th}$.

 \subsection{Inhomogeneous terms}
 \label{subsec:inhomogeneous_terms}

 The explicit forms of the inhomogeneous terms $I_{\pi}(\underline{k})$
 and
 $I_{\bar{\cal P}}(\underline{k})$ to leading order in weak coupling
 will be required in the following sections.
 For the shear viscosity, $I_{\pi}(\underline{k})$ corresponds to an
 insertion
 of the traceless stress tensor $\pi_{lm}$ given in
 Eq.~(\ref{eq:pi_lm}).  Hence,
 \begin{equation}
 I_{\pi}(\underline{k})
 =
 k_l k_m {-} {\textstyle{1\over 3}} \delta_{lm} {\bf k}^2
 \;.
 \end{equation}
 For the bulk viscosity, $I_{\bar{\cal P}}(\underline{k})$ corresponds
 to an
 insertion of the operator
 $\bar{\cal P}={\cal P}{-}v_{\rm s}^2\varepsilon$,
 and includes, as shown in section~\ref{subsec:chains},
 a one-loop ``renormalization'' contribution from chain diagrams.
 Since the standard form of the stress-energy tensor,%
\footnote{%
	Renormalization requires the counter term of the form
	$\delta T_{\mu\nu}
	= A (\partial_\mu\partial_\nu {-} g_{\mu\nu} \partial^2) \phi^2$
	where $A$ is a (infinite) constant\cite{Brown2}.
	However, this term does not concern us because its contribution
	to the inhomogeneous term is
	$\delta I_{\mu\nu} = (g_{\mu\nu} q^2 - q_\mu q_\nu)A$
	where $q_\mu$ is the external 4-momentum which is set to zero in
	the viscosity calculations.
}
 \begin{equation}
 {T}^{\mu\nu} =
 \partial^{\mu}{\phi}\partial^{\nu}{\phi}
 +
 g^{\mu\nu}{\cal L}
 \;,
 \label{eq:stress_energy_tensor}
 \end{equation}
 separates into the ``kinetic'' part,
 $\partial^\mu \phi \partial^\nu \phi$,
 and the Lagrangian part, $g^{\mu\nu}{\cal L}$,
 the inhomogeneous term for a pressure insertion
 may be separated into three parts,
 \addtocounter{equation}{1}
 $$
 I_{\cal P}
 =
 I_{\cal P}^{\rm kin.}
 +
 I_{\cal L}
 +
 I_{\cal P}^{\rm chain}
 \;,
 \eqno{(\theequation a)}
 $$
 $$
 I_{\varepsilon}
 =
 I_{\varepsilon}^{\rm kin.}
 -
 I_{\cal L}
 +
 I_{\varepsilon}^{\rm chain}
 \;,
 \eqno{(\theequation b)}
 $$
 with the obvious notations.
 The sign difference in the Lagrangian term is due to the
 metric $g_{\mu\nu}={\rm diag}(-1,1,1,1)$.

 The contribution from the ``kinetic'' part of ${\cal P}$
 and $\varepsilon$
 in the zero external momentum limit is simply
 \addtocounter{equation}{1}
 $$
 I_{{\cal P}}^{\rm kin.}(\underline{k})
 =
 {\textstyle{1\over 3}}{\bf k}^2
 \;,
 \eqno{(\theequation a)}
 $$
 $$
 I_{\varepsilon}^{\rm kin.}(\underline{k})
 =
 E_k^2
 \;.
 \eqno{(\theequation b)}
 $$
 To determine the contribution of a Lagrangian insertion,
 $I_{\cal L}$,
 it is convenient to rewrite the Lagrangian (\ref{eq:full_Lagrangian})
 using the equation of motion, as%
\footnote{%
	Alternatively, one can, of course,
	perform a straightforward diagrammatic
	analysis of the various terms arising
	from a Lagrangian insertion.}
 \begin{equation}
 {\cal L}
 =
 {1\over 2}E[\phi] + {g\over 2{\times}3!}\phi^3
         + {\lambda \over 4!}\phi^4
 \;,
 \label{eq:Lagrangian_decomposed}
 \end{equation}
 where
 \begin{equation}
 E[\phi]
 \equiv \phi {\partial \over \partial \phi}{\cal L}
 =
 \phi
 \left(
 \partial^2{-}m_0^2
 {-}{g\over 2!}\phi{-}{\lambda\over 3!}\phi^2 \right)
 \phi
 \;,
 \end{equation}
\begin{figure}
 \setlength {\unitlength}{1cm}
\vbox
    {%
    \begin {center}
 \begin{picture}(0,0)
 \end{picture}
	\leavevmode
	\def\epsfsize	#1#2{0.3#1}
	\epsfbox {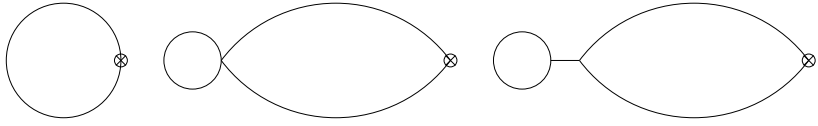}
    \end {center}
    \caption{%
      The lowest order diagrams contributing to
      $G_{\phi^2{\cal L}}$. Crossed circles indicate
      $\phi^2$ insertions.
}
 \label{fig:lagrange}
    }%
\end {figure}
\noindent
 is $\phi$ times the equation of motion.
 An insertion of the operator
 $E[\phi]$ in a time-ordered $N$-point correlation function
 simply produces an overall multiplicative factor\cite{Brown2},
 \begin{equation}
 G(x_1,\cdots,x_N; E[\phi(x)])
 =
 \sum_{a=1}^N\delta(x{-}x_a)\,G(x_1,\cdots,x_N)
 \;,
 \label{eq:E_insertion}
 \end{equation}
 provided an irrelevant disconnected contribution
 is suitably subtracted
 (most simply by using dimensional regularization).
 To evaluate $I_{\cal L}$,
 consider the correlation function of the Lagrangian ${\cal L}$
 with some other bilinear operator, such as $\phi^2$.
 In Euclidean space,
 \begin{eqnarray}
 G_{\phi^2 {\cal L}}(x,y)
 & \displaystyle \equiv & \displaystyle
 \langle {\cal T}(\phi^2(x)\,{\cal L}(y))\rangle
 \nonumber\\
 & \displaystyle = & \displaystyle
 {1\over 2}\langle {\cal T}(\phi^2(x)\,E[\phi(y)])\rangle
 +{g\over 2{\times}3!}
 \langle {\cal T}(\phi^2(x)\,\phi^3(y))\rangle
 +{\lambda \over 4!}
 \langle {\cal T}(\phi^2(x)\,\phi^4(y))\rangle
 \nonumber\\
 & \displaystyle = & \displaystyle
 \delta(x{-}y)\,
 \langle \phi^2(x) \rangle
 +{g\over 2{\times}3!}
 \langle {\cal T}(\phi^2(x)\,\phi^3(y))\rangle
 +{\lambda \over 4!}
 \langle {\cal T}(\phi^2(x)\,\phi^4(y))\rangle
 \;.
 \nonumber\\
 & \displaystyle & \displaystyle
 \end{eqnarray}
 The lowest order diagrams for $G_{\phi^2{\cal L}}$ are shown
 in Fig.~\ref{fig:lagrange}.
 The one-loop diagram in the figure is independent of
 external momentum.  Hence, when the external frequency is analytically
 continued and the discontinuity taken, the contribution of this
 simple one-loop diagram is zero.
 The other two two-loop diagrams represent insertions of
 half the ${\cal O}(\lambda T^2)$ one-loop self-energy.
 Hence, to lowest order in the weak coupling limit,
 a Lagrangian insertion merely produces a vertex factor of
 \begin{equation}
 I_{\cal L} = {1\over 2}\delta m_{\rm th}^2
 \;.
 \end{equation}

 To determine the chain diagram part of the inhomogeneous term,
 evaluation of
 ${\rm Re}\,C_{\cal P}(0)$ and ${\rm Re}\,C_{\varepsilon}(0)$
 is
 required ({\it c.f.}~Eq.~(\ref{eq:chain_modified_Ia})).
 The details of
 this evaluation are
 given in appendix~\ref{app:temperature_integrals}.
 The result, neglecting sub-leading corrections suppressed by
 ${\cal O}(\sqrt{\lambda})$, is simple,
 \addtocounter{equation}{1}
 $$
 I_{\cal P}^{\rm chain}
 =
 (\lambda-{g^2\over m_{\rm th}^2})\, {\rm Re}\,C_{\cal P}(0)
 =
 -{1\over 2}\delta m_{\rm th}^2,
 \eqno{(\theequation a)}
 $$
 $$
 I_{\varepsilon}^{\rm chain}
 =
 (\lambda-{g^2\over m_{\rm th}^2})\, {\rm Re}\,C_{\varepsilon}(0)
 =
 -{1\over 2}\delta m_{\rm th}^2
 \;.
 \eqno{(\theequation b)}
 $$

 Combining all three parts
 (and ignoring sub-leading corrections),
 the inhomogeneous terms for the pressure and
 energy density insertions are
 \addtocounter{equation}{1}
 $$
 I_{\cal P}(\underline{k})
 =
 {\textstyle{1\over 3}} {\bf k}^2
 \;,
 \eqno{(\theequation a)}
 $$
 $$
 I_{\varepsilon}(\underline{k})
 =
 E_k^2 - \delta m_{\rm th}^2
 =
 {\bf k}^2 + m_{\rm phys}^2
 \;,
 \eqno{(\theequation b)}
 $$
 where $m_{\rm phys}^2$
 is the physical {\em zero temperature} mass.
 Note that to lowest order $m_{\rm phys}^2$ is equivalent to
 $m_{\rm th}^2 - (T\partial m_{\rm th}^2/\partial T)/2$.

 Using these explicit forms of the inhomogeneous terms,
 the integrals in Eq.~(\ref{eq:cond_pressure}) can now be
 performed to show explicitly that the parameter $v^2$
 in ${\bar{\cal P}}={\cal P}{-}v^2\varepsilon$
 is equal to the speed of sound
 $v_{\rm s}^2 \equiv \partial{\cal P}/\partial\varepsilon$,
 \begin{eqnarray}
 v^2
 & \displaystyle = & \displaystyle
 \int{d^3 {\bf k}\over (2\pi)^3}\, n(E_k)\,[1{+}n(E_k)]\,
 {\textstyle{1\over 3}} {\bf k}^2
 {\bigg /}
 \int{d^3 {\bf k}\over (2\pi)^3}\, n(E_k)\,[1{+}n(E_k)]\,
     ({\bf k}^2{+}m_{\rm phys}^2)
 \nonumber\\
 & \displaystyle = & \displaystyle
 {1\over 3} - {5 \, m_{\rm phys}^2\over 12 \pi^2 T^2}
 + {\cal O}(\lambda^{3/2})
 =
 v_{\rm s}^2 + {\cal O}(\lambda^{3/2})
 \;.
 \end{eqnarray}
 The details of evaluating the speed of sound,
 and the various integrals involved
 are given in appendix~\ref{app:temperature_integrals}.
 Note that
 even though the energy $E_k$ is defined with the thermal mass
 $m_{\rm th}$,
 the speed of sound approaches $1/3$ as the zero temperature
 mass
 $m_{\rm phys}$ goes to zero, {\it not} as $m_{\rm th}$ goes to zero.
 In the massless limit, $m_{\rm phys}\to 0$,
 the stress-energy tensor is traceless due to
 scale invariance.%
\footnote{%
 	Scale invariance is, of course, broken quantum mechanically,
 and
	this leads to a trace anomaly.
	This implies that
	the relation between the pressure and the energy density is
	modified to
	$3{\cal P}{-}\varepsilon = \beta(\lambda)T^4/24^2$
	where
	$\beta(\lambda) = (T\partial \lambda/ \partial T)
	= 3\lambda^2/16\pi^2$ \cite{Brown}
	is the renormalization group $\beta$-function.
	(The explicit forms of ${\cal P}$ and ${\varepsilon}$
        are given in
	appendix~\protect{\ref{app:temperature_integrals}}.)
	Consequently, the speed of sound also receives
	an ${\cal O}(\lambda^2)$ correction of
	$\delta v_{\rm s}^2 = 5\beta(\lambda)/576\pi^2$.
 }
 In terms of equilibrium
 thermodynamic quantities this implies
 that $\varepsilon = 3 {\cal P}$, and the speed of sound is
 $v_{\rm s}^2 = 1/3$.

 Combining the results of $I_{\cal P}$, $I_{\varepsilon}$
 and $v_{\rm s}^2$, the inhomogeneous term for $\bar{\cal P}$
 insertion
 is (ignoring sub-leading corrections),
 \begin{eqnarray}
 I_{\bar{\cal P}}(\underline{k})
 & \displaystyle \equiv & \displaystyle
 I_{\cal P}(\underline{k})-v_{\rm s}^2 I_{\varepsilon}(\underline{k})
 \nonumber\\
 & \displaystyle =  & \displaystyle
 ({\textstyle 1\over 3} {-} v_{\rm s}^2){\bf k}^2
 - v_{\rm s}^2 m_{\rm phys}^2
 \nonumber\\
 & \displaystyle =  & \displaystyle
 {m_{\rm phys}^2 \over 3}
 \left(
  {A_0(m_{\rm th},T) {\bf k}^2 - A_2(m_{\rm th},T)
		\over
   A_2(m_{\rm th},T) + m_{\rm phys}^2 A_0(m_{\rm th},T)}
 \right)
 \nonumber\\
 & \displaystyle = & \displaystyle
 m_{\rm phys}^2\,
 \left( {5 {\bf k}^2 \over 12\pi^2 T^2} - {1\over 3} \right)
 \qquad ( m_{\rm phys}/T \ll 1 )
 \;,
 \label{eq:I_P_bar}
 \end{eqnarray}
 where
 \begin{equation}
 A_n(m_{\rm th},T)
 \equiv
 \int{d^3 {\bf p}\over (2\pi)^3}\, n(E_p)\,[1{+}n(E_p)]\, |{\bf p}|^n
 \;.
 \end{equation}
 Note that the inhomogeneous term is directly
 proportional to the zero temperature mass squared $m_{\rm phys}^2$.
 Hence,
 $I_{\bar{\cal P}}(\underline{k})$ vanishes in the massless limit,
 $m_{\rm phys} = 0$,
 and consequently, so does the leading order bulk viscosity
 (\ref{eq:ladder_viscosities2}) $\zeta$.%
\footnote{
 	This is another consequence of (classical) scale invariance.
 	Using the constitutive relation (\ref{eq:constitutive_rel}),
	the thermal average of the trace of stress-energy tensor may be
 	expressed as (ignoring higher derivative terms)
 	\[
        \langle T^{\mu}_{\mu} \rangle =
 	3{\cal P} - \varepsilon - \zeta \nabla{\cdot}{\bf u}
 	\;,
 	\]
        in the comoving (${\bf u}(x) = 0$) frame.
 	Classically,
	$T_\mu^\mu$ vanishes in the massless limit,
	and so does its equilibrium average $3{\cal P}{-}\varepsilon$.
	Hence, the leading order bulk viscosity
	$\zeta$ must also vanish in the massless limit\cite{Weinberg}.
        Again there are higher order corrections because
        scale invariance is broken quantum mechanically, and
        $3{\cal P}{-}\varepsilon = {\cal O}(\beta(\lambda))$.
	Consequently, in the massless limit,
        the bulk viscosity is $\zeta = {\cal O}(\lambda T^3)$.
  	Note that when $T\gg m_{\rm phys}/\lambda$,
 	the massless limit estimate
	${\cal O}(\lambda T^3)$
	can be larger than the massive theory estimate
	${\cal O}(m_{\rm phys}^4/\lambda^3 T)$.
}

 \section{Hydrodynamics and the Boltzmann equation}
 \label{sec:Boltzmann}

 Kinetic theory and the Boltzmann equation have traditionally
 been used to calculate transport properties of dilute many-body
 systems
 when (except during brief collisions)
 the underlying particles can be treated as classical particles
 with well defined position, energy, and momentum.

 For this classical picture of particle propagation to be valid,
 the mean free path must be large compared to
 the Compton wavelength of the underlying particle.
 At extremely high temperature, the mean free path scales as
 $1/T$.
 Hence, at high enough temperature,
 a Boltzmann equation describing the fundamental particles cannot
 be
 readily justified.
 However, as noted in previous sections,
 at high temperature the typical ${\cal O}(1/\lambda^2 T)$
 mean free path of thermal excitations
 is always larger than the ${\cal O}(1/\sqrt{\lambda}T)$
 Compton wavelength of slowly varying thermal excitations.
 Consequently,
 as will be discussed,
 a Boltzmann equation description of effective thermal excitations
 with
 a temperature dependent
 thermal mass and thermal scattering cross section
 can be consistent at any temperature.

 However, there is a fundamental complication when attempting
 to
 formulate a Boltzmann equation for effective excitations.
 In a non-equilibrium situation,
 the temperature may vary in space and time.
 Since the thermal mass depends on temperature,
 this implies that the mass of the effective excitations also
 varies in
 space and time.
 The correct treatment of this will be described shortly.
 As a warm up,
 first consider the usual Boltzmann
 equation with constant mass particles.
 It can be formulated from the statement
 that the rate of change in the comoving density of particles
 with an on-shell momentum $\underline{k}$ at position $x$
 equals the difference between the rates at which particles in
 this
 phase space region are generated or lost due to collisions,
 or\footnote{%
	Here the momentum ${\bf k}$ is the canonical momentum, not the
	kinetic momentum.  This is choice is necessary for the measure
	$d^3{\bf x}\, d^3{\bf k}$ to be invariant along particle
	trajectories.
}
 \begin{equation}
 \left(
    {\partial\over \partial t}
    + {{\bf k}\over E_k}{\cdot}\nabla
    + {F_{\rm ex}^{i} \over \gamma}\,
      {\partial\over \partial {k}^{i}}
 \right) f(x,\underline{k})
 =
 \left(
   {\partial f(x,\underline{k})\over \partial t}
 \right)_{\rm gain}
 -
 \left(
   {\partial f(x,\underline{k})\over \partial t}
 \right)_{\rm loss}
 \;.
 \label{eq:net_change}
 \end{equation}
 Here $f(x,\underline{k})$ is the single particle density function,
 $\left(\partial f(x,\underline{k})/ \partial t\right)_{\rm gain}$
 is the rate of increase of the density of particles with momentum
 $\underline{k}$ at $x$ due to collisions, and
 $\left(\partial f(x,\underline{k})/ \partial t\right)_{\rm loss}$
 is the corresponding
 rate at which particle density is lost.
 Here,
 $F_{\rm ex}^i$ represents whatever single particle external
 force
 or proper time derivative of 3-momentum,
 $F_{\rm ex}^i \equiv d k^i / d\tau$
 is present, and $\gamma$ is the usual $E_k/m$.
 In this section an underlined momentum signifies an on-shell
 momentum
 with only positive energy.
 The collision terms on the right hand side may be expressed
 in terms of scattering cross sections and distribution functions.
 For instance,
 the gain of particles with $\underline{k}$ at $x$
 due to scattering is given by
 \begin{equation}
 \left(
   {\partial f(x,\underline{k})\over \partial t}
 \right)_{\rm gain}
 =
 \sum_{\rm in,out}
	 \left|
   	    {\cal T}(\{\underline{p}_{\rm in}\};
	   	     \{\underline{p}_{\rm out}\},\underline{k})
         \right|^2\,
	          f_n(x; \{\underline{p}_{\rm in}\})\,
		  F_m(x; \{\underline{p}_{\rm out}\},\underline{k})
 \;.
 \label{eq:N_gain}
 \end{equation}
 Here,
 ${\cal T}$
 is a multi-particle scattering amplitude describing a process
 in which a particle with momentum $\underline{k}$
 is produced in the final state.
 The initial
 probability density to find $n$-particles
 with momenta $\{\underline{p}_{\rm in}\}$
 at position $x$ is given by the product of single particle densities
 \begin{equation}
 f_n(x; \{\underline{p}_{\rm in}\})
 =
 \prod_{\rm in}
 f(x,\underline{p}_{\rm in})
 \;
 \end{equation}
 under the assumption of molecular chaos.
 The factor $F_m(x; \{\underline{p}_{\rm out}\},\underline{k})$
 is the Bose or Fermi statistical factor for
 an $m$-particle final state with momenta $\{\underline{p}_{\rm out}\}$
 and $\underline{k}$, and is also given by a product of
 single particle statistical factors
 $[1{\pm}f(x,\underline{p})]$
 (with the upper sign for bosons, as considered here).
 Implicit in Eq.~(\ref{eq:N_gain}) is
 the assumption that the duration of scattering
 and the Compton wavelength of particles are
 short compared to the mean free time
 (and the scale of variation in the external force),
 so that the densities of particles participating in the scattering
 are accurately represented by the densities at a single position
 $x$.
 The loss rate
 $\left(\partial f(x,\underline{k})/ \partial t\right)_{\rm loss}$
 has a similar expression but with the particle with momentum
 $\underline{k}$ among the incoming particles.

 For simplicity, consider the $\lambda\phi^4$ theory.
 If the interaction strength is weak,
 only the first few terms in the sum in Eq.~(\ref{eq:N_gain}) are
 important.
 The included terms must contain the leading order
 number-changing scattering processes
 since as discussed in earlier sections,
 the bulk viscosity calculation requires such scatterings.
 In that case, Eq.~(\ref{eq:net_change}) combined with
 Eq.~(\ref{eq:N_gain}) becomes the basic relativistic Boltzmann
 equation,
 \begin{eqnarray}
 \lefteqn{
   \left( {\partial \over \partial t}
   + {{\bf k}\over E_k}{\cdot} \nabla
   + {F_{\rm ex}^{i} \over \gamma}\,
     {\partial\over \partial {k}^{i}}
   \right) f(x,\underline{k})  {d^3 {\bf k}_3\over (2\pi)^3}\,
 }
 & \displaystyle & \displaystyle
 \nonumber\\
 & \displaystyle & \displaystyle \qquad
 =
 \int_{123}
 {d^3 {\bf p}_3\over (2\pi)^3}\, {d^3 {\bf k} \over (2\pi)^3}\,
 d\sigma_{12\to3k}\, v_{12}\,
 \nonumber\\
 & \displaystyle & \displaystyle \qquad\qquad
 \left(
	f(x,\underline{p}_1)\,f(x,\underline{p}_2)\,
	F(x,\underline{p}_3)\,F(x,\underline{k})
	-
        F(x,\underline{p}_1)\,F(x,\underline{p}_2)\,
        f(x,\underline{p}_3)\,f(x,\underline{k})
 \right)
 \nonumber\\
 & \displaystyle & \displaystyle \qquad
 +
 2\int_{12345}
 {d^3 {\bf p}_1\over (2\pi)^3}\, {d^3 {\bf p}_2 \over (2\pi)^3}\,
 d\sigma_{12 \to 345k}\, v_{12}\,
 \nonumber\\
 & \displaystyle & \displaystyle \qquad\qquad
 \left(
	f(x,\underline{p}_1)\,f(x,\underline{p}_2)\,
	F(x,\underline{p}_3)\, F(x,\underline{p}_4)\,
	F(x,\underline{p}_5)\, F(x,\underline{k})\,
 \right.
 \nonumber\\
 & \displaystyle & \displaystyle \qquad\qquad\quad
 \left.
	-
        F(x,\underline{p}_1)\,F(x,\underline{p}_2)\,
	f(x,\underline{p}_3)\, f(x,\underline{p}_4)\,
	f(x,\underline{p}_5)\, f(x,\underline{k})\,
 \right)
 \nonumber\\
 & \displaystyle & \displaystyle \qquad
 +
 \int_{12345}
 {d^3 {\bf p}_5\over (2\pi)^3}\,
 {d^3 {\bf k} \over (2\pi)^3}\,
 d\sigma_{5k \to 1234}\, v_{5k}\,
 \nonumber\\
 & \displaystyle & \displaystyle \qquad\qquad
 \left(
	f(x,\underline{p}_1)\,f(x,\underline{p}_2)\,
	f(x,\underline{p}_3)\,f(x,\underline{p}_4)\,
	F(x,\underline{p}_5)\,F(x,\underline{k})\,
 \right.
 \nonumber\\
 & \displaystyle & \displaystyle \qquad\qquad\quad
 \left.
       -
        F(x,\underline{p}_1)\,F(x,\underline{p}_2)\,
	F(x,\underline{p}_3)\,F(x,\underline{p}_4)\,
	f(x,\underline{p}_5)\,f(x,\underline{k})\,
 \right)
 \;,
 \label{eq:Boltzmann}
 \end{eqnarray}
 where, for a convenient presentation, the momentum space volume
 element ${d^3 {\bf k} / (2\pi)^3}$ is included in the formula,
 and a short hand notation
 $F(x,\underline{p}) \equiv [1{+}f(x,\underline{p})]$ is used.
 The subscripts of the integral signs indicate that ${\bf k}$ is not
 integrated.
 Here, $d\sigma_{12\to 3k}$ is the usual
 two-body differential cross section,
 \begin{equation}
 d\sigma_{12\to 3k} \equiv
 \prod_{i=1,2} {d^3 {\bf p}_i\over (2\pi)^3 2E_i}\,
 (2\pi)^4
 \delta(\underline{p}_1{+}\underline{p}_2
	{-}\underline{p}_3{-}\underline{k})\,
 \left| {\cal T}_4
 (\underline{p}_1,\underline{p}_2;\underline{p}_3,\underline{k}) \right|^2
 {\bigg/} (4E_{3}E_k v_{12} 2)
 \;,
 \label{eq:diff_cross}
 \end{equation}
 where $v_{12}$ is the relative speed between
 particles with momenta $\underline{p}_1$ and $-\underline{p}_2$,
 and the symmetry factor of 2 in the denominator arises from the
 non-distinguishability of the final particles.
 The 2--4 differential cross section $d\sigma_{12\to 3456}$ is
 given by Eq.~(\ref{eq:24_dsigma}).
 The factor of 2 difference in the second and the third term is due to
 the fact that the second term is symmetric in
 $\bf{p}_3,\bf{p}_4,\bf{p}_5$ while the third term is symmetric in
 $\bf{p}_1,\bf{p}_2,\bf{p}_3,\bf{p}_4$.

 The scattering interaction at a given point $x$ still conserves
 energy and momentum
 even in the presence of a (slowly varying)  external force which
 changes the 4-momentum of an excitation
 during the free flight between collisions.
 Hence, when multiplied by $\underline{k}^\nu$ and
 integrated over ${\bf k}$,
 the right hand side of the Boltzmann equation (\ref{eq:Boltzmann})
 vanishes.
 This implies for the left hand side
 \begin{eqnarray}
 0 & \displaystyle = & \displaystyle
 \int{d^3{\bf k}\over (2\pi)^3 E_k}\,
 \underline{k}^\nu \underline{k}^\mu
 \partial_{\mu} f(x,\underline{k})\,
 +
 m\int{d^3{\bf k}\over (2\pi)^3 E_k}\,
 \underline{k}^\nu \,
 F_{\rm ex}^i\,
 {\partial \over \partial {k}^i}
 f(x,\underline{k})
 \nonumber\\
 & \displaystyle = & \displaystyle
 \partial_\mu
 \int{d^3{\bf k}\over (2\pi)^3 E_k}\,
 f(x,\underline{k})\, \underline{k}^\mu \underline{k}^\nu
 +
 m\int{d^3{\bf k}\over (2\pi)^3 E_k}\,
 \underline{k}^\nu \,
 {F}_{\rm ex}^i
 {\partial \over \partial {k}^i}
 f(x,\underline{k})
 \;.
 \label{eq:conservation_Boltz}
 \end{eqnarray}
 In the absence of any external force,
 this would become the local conservation
 equations for energy and momentum
 with the usual kinetic theory stress-energy tensor,
 \begin{equation}
 T^{\mu\nu}_{\rm kin.}(x) \equiv
 \int{d^3{\bf k}\over (2\pi)^3 E_k}\,
 \underline{k}^\mu \underline{k}^\nu f(x,\underline{k})
 \;.
 \end{equation}
 With an external force, one instead finds
 \begin{eqnarray}
 & \displaystyle
 \partial_{\mu}T^{\mu\nu}_{\rm kin.}(x)
 =
 {\cal S}_{\rm ex}^{\nu}(x)
 \;,
 & \displaystyle
 \\
 \noalign{\hbox{where the source ${\cal S}_{\rm ex}^\nu$
 is given by}}
 & \displaystyle
 {\cal S}_{\rm ex}^{\nu}(x) =
 -
 m\int{d^3{\bf k}\over (2\pi)^3 E_k}\,
 \underline{k}^\nu \,
 F_{\rm ex}^{i}
 {\partial \over \partial {k}^i}
 f(x,\underline{k})
 \;.
 \label{eq:S_mu}
 \end{eqnarray}

 Now consider the case of excitations with a space-time dependent
 mass,
 $m(x)$.
 Following the usual derivation of the equation of motion for
 a
 relativistic point particle, one finds
 \begin{equation}
 F_{\rm ex}^i(x)
 =
 {d\over d\tau}\,k^i
 =
 -\partial^i m(x)
 \;,
 \label{eq:dot_k}
 \end{equation}
 where $\tau$ is the proper time.
 Since the energy in this case,
 $E_k = \sqrt{{\bf k}^2 {+} m^2(x)}$, is space-time dependent,
 the partial derivatives in Eq.~(\ref{eq:conservation_Boltz})
 cannot be simply taken outside the integral.
 Including additional space-time derivatives of energy changes
 the external source term (\ref{eq:S_mu}) to
 \begin{eqnarray}
 {\cal S}_{\rm ex}^{\nu}(x)
 & \displaystyle = & \displaystyle
 \int{d^3{\bf k}\over (2\pi)^3}\,f(x,\underline{k})
 \partial_{\mu}
 \left(
 {\underline{k}^\mu \underline{k}^\nu \over E_k}
 \right)
 -
 m(x)\int{d^3{\bf k}\over (2\pi)^3 E_k}\,
 \underline{k}^\nu \,
 F_{\rm ex}^{i}
 {\partial \over \partial {k}^i}
 f(x,\underline{k})
 \nonumber\\
 & \displaystyle = & \displaystyle
 -{1\over 2}
 \left(  \partial^{\nu} m^2(x)  \right)
 \int{d^3{\bf k}\over (2\pi)^3 E_k}\,
 f(x,\underline{k})
 \;.
 \end{eqnarray}

 Finally, consider the situation of real interest where the mass
 of the
 effective excitations depends on the presence of other excitations
 in
 the medium.
 In the particular case of scalar interactions, relevant for
 $\lambda\phi^4$ theory, the effective mass $m(x)$
 (to the lowest order) will be
 \begin{eqnarray}
 m^2(x)
 & \displaystyle = & \displaystyle
 m_{\rm phys}^2
 +
 \delta m^2(x)
 \;,
 \\
 \noalign{\hbox{where}}
 \delta m^2(x)
 & \displaystyle \equiv & \displaystyle
 {{\lambda}\over 2} \int{d^3{\bf k}\over (2\pi)^3 E_k}\,f(x,\underline{k})
 \;.
 \label{eq:delta_m_Boltz}
 \end{eqnarray}
 This is simply a rewriting of the one-loop result for the thermal
 mass
 $m_{\rm th}^2$ when $f(x,\underline{k})$ is identified as the usual Bose
 factor.   The $g\phi^3{+}\lambda\phi^4$ theory result can be obtained
 by replacing
 ${\lambda} \to \lambda{-}g^2/m^2(x)$.  From now on,
 $\delta m^2(x)$ is assumed to have the form shown in
 (\ref{eq:delta_m_Boltz}).
 In this case,
 one can define a modified energy-momentum tensor which satisfies
 the conservation equations ignoring ${\cal O}(\lambda^2)$ corrections,
 \begin{equation}
 T^{\mu\nu}(x)
 \equiv
 \int{d^3{\bf k}\over (2\pi)^3 E_k}\,
 f(x,\underline{k})\,
 \left(
 \underline{k}^\mu \underline{k}^\nu{+}
 {\textstyle{1\over 4}}g^{\mu\nu}\delta m^2(x)
 \right)
 \;.
 \label{eq:conserved_T}
 \end{equation}
 At first sight,
 it may seem surprising that a conserved stress-energy tensor
 can be
 defined when the effective mass is space-time dependent.
 However, the underlying scalar theory does conserve energy and
 momentum.
 Hence, when the stress-energy tensor is correctly defined,
 the result must still be a conserved tensor
 regardless of whether the system is in or out of
 equilibrium.
 Appendix~\ref{app:temperature_integrals} shows that
 the form of $T^{\mu\nu}$ in Eq.~(\ref{eq:conserved_T})
 is identical to the equilibrium expression of the
 field theory stress-energy tensor up to ${\cal O}(\lambda)$.
 Hence, the Boltzmann equation (\ref{eq:Boltzmann}), with
 $F_{\rm ex}^i (x) = -\partial^i m_{\rm th}(x)$,
 may be regarded as a kinetic theory description of effective
 temperature dependent excitations.

 \subsection{The hydrodynamic limit of the Boltzmann equation}
 \label{subsec:hydro_limit}

 Hydrodynamic excitations are arbitrarily long-lived, long-wavelength
 fluctuations that characterize
 near-equilibrium behavior of interacting fluids.
 Consider describing the relaxation of a many-body system after
 a small initial disturbance.
 In a few mean free times,
 virtually all particles will have suffered
 numerous collisions
 with other particles in the medium.
 Hence, fluctuations in most degrees of freedom will relax
 in a few mean free times.
 However,
 an excess of a locally conserved quantity cannot simply
 disappear locally;
 to smooth a long-wavelength
 fluctuation in a conserved quantity,
 the conserved quantity must
 be physically transported over a distance comparable
 to the wavelength.
 For an arbitrarily large wavelength,
 this will require an arbitrarily long time.
 Consequently,
 for times long compared to the mean free time,
 the relaxation of the system may be described solely
 in terms of very long-wavelength, long-lived
 fluctuations in locally conserved quantities.
 These are the hydrodynamic fluctuations of the system.
 For a simple fluid without an additional conserved charge,
 the only locally conserved quantities are the energy and momentum.

 To solve the Boltzmann equation (\ref{eq:Boltzmann})
 in this near-equilibrium hydrodynamic regime,
 the single particle density
 $f(x,\underline{k})$ may be expressed as a small perturbation
 away from a ``local equilibrium'' distribution
 \begin{equation}
 f(x,\underline{k})  = f^{(0)}(x,\underline{k})
	      + f^{(1)}(x,\underline{k})
 \;,
 \label{eq:expansion_f}
 \end{equation}
 where the local equilibrium distribution
 $f^{(0)}(x,\underline{k})$ is characterized by
 a local inverse temperature $\beta(x)$ and unit local 4-velocity
 $u_{\mu}(x)$ (satisfying $u_\mu u^\mu = -1$),
 \begin{equation}
 f^{(0)}(x,\underline{k})  =
 1/\left( e^{-\beta(x)u_{\mu}(x)\underline{k}^{\mu}} - 1 \right)
 \;.
 \end{equation}

 To make the decomposition (\ref{eq:expansion_f}) unique, one
 must
 specify four conditions which serve to define
 the choice of local temperature and velocity.
 The most convenient choice is the
 Landau-Lifshitz condition\cite{Landau,Groot},
 which fixes
 the local temperature $\beta(x)$ and the local 4-velocity $u_\mu(x)$
 by
 requiring that $f^{(0)}(x,\underline{k})$
 produces the complete energy flow of the fluid\cite{Groot}.
 This means that
 \begin{eqnarray}
 T^{\mu\nu}(x)u_{\nu}(x)
 & \displaystyle \equiv & \displaystyle
 T_{(0)}^{\mu\nu}(x)u_{\nu}(x)
 \nonumber\\
 & \displaystyle = & \displaystyle -\varepsilon(x) u^{\mu}(x)
 \;,
 \label{eq:LL_cond}
 \end{eqnarray}
 where $T^{\mu\nu}(x)$ is
 the conserved kinetic theory stress-energy tensor
 which is defined in Eq.~(\ref{eq:conserved_T}), and
 \begin{eqnarray}
 T_{(0)}^{\mu\nu}(x)
 & \displaystyle \equiv & \displaystyle
 \int {d^3 {\bf k}\over (2\pi)^3 E_k}\,
 f^{(0)}(x,\underline{k})\,
 \left(
 \underline{k}^\nu \underline{k}^\mu
 {+} {\textstyle{1\over 4}} g^{\mu\nu}\delta m^2(x)
 \right)
 \nonumber\\
 & \displaystyle = & \displaystyle
 (\varepsilon(x){+}{\cal P}(x))u^{\mu}(x) u^{\nu}(x)
 + {\cal P}(x)g^{\mu\nu}
 \;,
 \label{eq:lowest_stress}
 \end{eqnarray}
 is the ``local equilibrium''
 contribution to the stress-energy tensor,
 characterized by an energy density $\varepsilon(x)$
 and pressure ${\cal P}(x)$.
 In other words, the correction
 $f^{(1)}(x,\underline{k})$
 to the distribution is required to make a vanishing contribution
 to
 $T^{\mu\nu}u_{\nu}$.

 By itself,
 the local equilibrium density function $f^{(0)}(x,\underline{k})$
 is not a solution of the Boltzmann equation.
 It makes the binary collision term in the Boltzmann equation
 (\ref{eq:Boltzmann}) vanish, while the convective derivative
 on the
 left hand side is non-zero
 (unless $\varepsilon$ and $u_\mu$ are constant).
 However, if $\varepsilon(x)$ and $u_\mu(x)$ vary on macroscopic
 scales,
 the size of the derivative
 $|\nabla f^{(0)}(x,\underline{k})|/f^{(0)}(x,\underline{k})$
 will be small compared to any microscopic inverse length scale.
 Hence, by adding a correction $f^{(1)}$
 to the local equilibrium distribution function,
 one may find a solution to the Boltzmann equation in which
 the size of the correction $f^{(1)}(x,\underline{k})$
 is small,
 \begin{equation}
 \left| f^{(1)}(x,\underline{k}) \right|
 \sim
 l_{\rm free}|\nabla f^{(0)}(x,\underline{k})|
 \ll \left|f^{(0)}(x,\underline{k})\right|
 \;.
 \label{eq:size_of_correction}
 \end{equation}
 Here $l_{\rm free} \sim 1/\bar{n}\bar{\sigma}$ is the mean free
 path.

 To produce an equation for
 the first order correction $f^{(1)}(x,\underline{k})$,
 it is convenient to choose the frame where
 ${\bf u}(x)=0$ at some particular position $x$, so that
 \begin{equation}
 f^{(0)}(x,\underline{k})
 =
 n(x,E_k)
 = {1\over e^{\beta(x)E_k} - 1}
 \;,
 \label{eq:local_eq}
 \end{equation}
 and to express the first order correction as
 \begin{equation}
 f^{(1)}(x,\underline{k})
 =
 - n(x,E_k)\,[1{+}n(x,E_k)]\,\phi(x,\underline{k})
 \;.
 \label{eq:first_correction}
 \end{equation}
 where $\phi(x,\underline{k})$ is a slowly varying function
 to be determined by the (linearized) Boltzmann equation.
 Also, the lowest order temperature dependent part of the mass
 can be
 evaluated as
 \begin{eqnarray}
 \delta m^2(x)
 & \displaystyle = & \displaystyle
 \lambda
 {1\over 2}\int{d^3{\bf k}\over (2\pi)^3 E_k}\,
 n(x,E_k)
 \nonumber\\
 & \displaystyle = & \displaystyle
 \lambda
 {1\over 24} T^2(x)
 \label{eq:explicit_delta_m}
 \;
 \end{eqnarray}
 ignoring sub-leading terms.

 By equating the derivatives of the local equilibrium density
 function
 $f^{(0)}(x,\underline{k})$ from the left hand side of
 Eq.~(\ref{eq:Boltzmann})
 with the terms linear in $\phi(x,\underline{k})$ from the collision
 term,
 the following equation for $\phi(x,\underline{k})$ in the ${\bf u}(x)=0$
 frame is obtained,
 \begin{eqnarray}
         I(x,\underline{k}) & \displaystyle = & \displaystyle
	   {1\over 4}\,\int
	    \prod_{i=1}^3 {d^3 {\bf p}_i \over 2E_{i}(2\pi)^3}\,\,
	     \left|
	     {\cal T}_4
	     (\underline{p}_1,\underline{p}_2;\underline{p}_3,\underline{k})
	     \right|^2\,
	     (2\pi)^4 \delta(\underline{p}_1 {+} \underline{p}_2
	     {-} \underline{p}_3 {-} \underline{k})\,
 \nonumber\\
	& \displaystyle & \displaystyle \qquad{}\times
	       [1{+}n(x,E_{1})]\,[1{+}n(x,E_{2})]\,
	       n(x,E_{3})/ [1{+}n(x,E_k)] \,
 \nonumber\\
   	& \displaystyle & \displaystyle \qquad{}\times
	       (\phi(x,\underline{k})+\phi(x,\underline{p}_3)
	       -\phi(x,\underline{p}_2)-\phi(x,\underline{p}_1))
 \nonumber\\
	  & \displaystyle & \displaystyle {}
	  + {1\over 24}\,\int
	    \prod_{i=1}^5 {d^3 {\bf p}_i \over 2E_{i}(2\pi)^3}\,\,
	     \left|
	     {\cal T}_6(\{ \underline{p}_i \},\underline{k})
	     \right|^2\,
	     (2\pi)^4 \delta(\underline{p}_1 {+} \underline{p}_2
	     {-} \underline{p}_3 {-} \underline{p}_4 {-} \underline{p}_5
	     {-} \underline{k})\,
 \nonumber\\
	& \displaystyle & \displaystyle \qquad{}\times
	       [1{+}n(x,E_{1})]\,[1{+}n(x,E_{2})]\,
	       n(x,E_{3})\, n(x,E_{4})\, n(x,E_{5})/ [1{+}n(x,E_k)] \,
 \nonumber\\
   	& \displaystyle & \displaystyle \qquad{}\times
	       (\phi(x,\underline{k})+\phi(x,\underline{p}_5)
	       +\phi(x,\underline{p}_4)+\phi(x,\underline{p}_3)
	       -\phi(x,\underline{p}_2)-\phi(x,\underline{p}_1))
 \nonumber\\
	  & \displaystyle & \displaystyle {}
	  + {1\over 48}\,\int
	    \prod_{i=1}^5 {d^3 {\bf p}_i \over 2E_{i}(2\pi)^3}\,\,
	     \left|
	     {\cal T}_6(\{ \underline{p}_i \},\underline{k})
	     \right|^2\,
	     (2\pi)^4 \delta(\underline{p}_1 {+} \underline{p}_2
	     {+} \underline{p}_3 {+} \underline{p}_4 {-} \underline{p}_5
	     {-} \underline{k})\,
 \nonumber\\
	& \displaystyle & \displaystyle \hspace{-1.0cm}{}\times
	       [1{+}n(x,E_{1})]\,[1{+}n(x,E_{2})]\,
	       [1{+}n(x,E_{3})]\,[1{+}n(x,E_{4})]
	       \, n(x,E_{5})/ [1{+}n(x,E_k)] \,
 \nonumber\\
   	& \displaystyle & \displaystyle \qquad{}\times
	       (\phi(x,\underline{k})+\phi(x,\underline{p}_5)
	       -\phi(x,\underline{p}_4)-\phi(x,\underline{p}_3)
	       -\phi(x,\underline{p}_2)-\phi(x,\underline{p}_1))
 \;.
 \nonumber\\
 \label{eq:Chapman}
 \end{eqnarray}
 The inhomogeneous term on the left
 is a polynomial in momentum and derivatives of
 the flow velocity ${\bf u}(x)$,
 \begin{eqnarray}
 I(x,\underline{k})
 & \displaystyle = & \displaystyle
 \beta(x)
 \left(
     {\textstyle {1\over 3}}{\bf k}^2
     {-}v_{\rm s}^2(x)({\bf k}^2{+}m_{\rm phys}^2)
 \right)
 \nabla{\cdot}{\bf u}(x)
 \nonumber\\
 & \displaystyle & \displaystyle {}
 +
 {\beta(x)\over 2}\,
 \bigg(
 k_ik_j{-}{\textstyle {1\over 3}}\delta_{ij}{\bf k}^2
 \bigg) \,
 \bigg(
 \nabla_i u_j(x) {+} \nabla_j u_i(x)
 {-} {\textstyle {2\over 3}}\delta_{ij}\nabla{\cdot}{\bf u}(x)
 \bigg)
 \;.
 \end{eqnarray}
 Note that choosing the ${\bf u}(x)=0$ frame does not imply
 that the gradient at $x$,
 $\nabla_i u_j(x)$ are zero (but $\partial_{\mu} u^0(x) = 0$
 since $u_\mu u^\mu = -1$).
 Also note that
 $I(x,\underline{k})$ contains
 $m_{\rm phys}^2$
 {\it not} the thermal mass
 $m^2(x)
 = m_{\rm phys}^2 {+} \delta m^2(x)$.
 This agrees with the result of the previous
 section (\ref{eq:I_P_bar}) where the inhomogeneous term also
 lacked
 the thermal mass correction.
 In simplifying the left hand side (the inhomogeneous term),
 the equilibrium thermodynamic identity
 \begin{equation}
 dT/T = d{\cal P}/(\varepsilon{+}{\cal P})
 \;,
 \end{equation}
 for the local thermodynamic quantities,\footnote{%
	This is a direct consequence of the form of the local
	distribution function $f^{(0)}$ (\protect\ref{eq:local_eq}).
}
 and the lowest order energy-momentum conservation equations
 \addtocounter{equation}{1}
 $$
 {\partial\over \partial t} \varepsilon(x)
 =
 -(\varepsilon(x)+{\cal P}(x))\,
 \nabla{\cdot}{\bf u}(x)
 \eqno{(\theequation a)}
 $$
 $$
 {\partial \over \partial t} {\bf u}(x)
 =
 -\nabla {\cal P}(x)/(\varepsilon(x)+{\cal P}(x))
 \;,
 \eqno{(\theequation b)}
 $$
 in combination with the definition
 $v_{\rm s}^2 = (\partial {\cal P}/\partial \varepsilon)$
 are used to rewrite the time derivatives of $\beta(x)$ and ${\bf u}(x)$
 in terms of the spatial derivatives.

 Given the form of the inhomogeneous function $I(x,\underline{k})$,
 rotational invariance
 (in the ${\bf u}(x)=0$ frame)
 requires that $\phi(x,\underline{p})$ have the following form,
 \begin{eqnarray}
 \phi(x,\underline{p})
 & \displaystyle = & \displaystyle
 \beta(x)A(x,\underline{p})\nabla{\cdot}{\bf u}(x)
 \nonumber\\
 & \displaystyle & \displaystyle \hspace{-1.0cm} {}
 +
 {\beta(x)\over 2}
 \left(
 \hat{p}_i\hat{p}_j{-}{\textstyle {1\over 3}}\delta_{ij}
 \right) \,
 B(x,\underline{p})
 \left(
 \nabla_i u_j(x) {+} \nabla_j u_i(x)
 {-} {\textstyle {2\over 3}}\delta_{ij}\nabla{\cdot}{\bf u}(x)
 \right)
 \;,
 \label{eq:sol_phi}
 \end{eqnarray}
 where $\hat{\bf p}$ is the unit vector in the direction of ${\bf p}$.
 Here,
 $A(x,\underline{p})$ is the amplitude of the spin 0
 (divergence) perturbation,
 and $B(x,\underline{p})$
 is the spin 2 (shear) perturbation amplitude.
 The solution in any other frame is, of course,
 related to the given solution $\phi(x,\underline{k})$ by a Lorentz
 boost.
 The linearized Boltzmann equation Eq.~(\ref{eq:Chapman}) is
 completely
 local in position; the parameter $x$ is simply a label
 and henceforth will be omitted.

 The scalar process and the tensor processes
 decouple and can be studied separately.
 For the spin 0 component,
 the integral equation for $A$ is obtained by replacing
 $I(x,\underline{k})$ in Eq.~(\ref{eq:Chapman}) by
 \begin{equation}
 I_{\bar{\cal P}}(\underline{k}) =
 {\textstyle {1\over 3}}{\bf k}^2
 - v_{\rm s}^2 ({\bf k}^2{+}m_{\rm phys}^2)
 \;,
 \end{equation}
 and $\phi(x,\underline{p})$ by $A(\underline{p})$.
 For the the spin 2 component, Eq.~(\ref{eq:Chapman}) simplifies
 to the inhomogeneous linear integral equation
 \begin{eqnarray}
 k_lk_m {-} {\textstyle{1\over 3}}\delta_{lm} {\bf k}^2
	 & \displaystyle = & \displaystyle
	   {1\over 4}\,\int
	    \prod_{i=1}^3 {d^3 {\bf p}_i \over 2E_{i}(2\pi)^3}\,\,
	     (2\pi)^4 \delta(\underline{p}_1 {+} \underline{p}_2
	     {-} \underline{p}_3 {-} \underline{k})\,
	     |{\cal T}_4 (\underline{p}_1,\underline{p}_2;
	     \underline{p}_3,\underline{k})|^2\,
 \nonumber\\
	& \displaystyle & \displaystyle \qquad {}\times
	       [1{+}n(E_{1})]\, [1{+}n(E_{2})]\,
	       n(E_{3})/ [1{+}n(E_k)] \,
 \nonumber\\
   	& \displaystyle & \displaystyle \qquad {}\times
	      ( B_{lm}(\underline{k}) + B_{lm}(\underline{p_3})
	      - B_{lm}(\underline{p_2})- B_{lm}(\underline{p_1}) )
 \;,
 \label{eq:Chapman2}
 \end{eqnarray}
 where $B_{lm}(\underline{p}) \equiv
 (\hat{p}_l\hat{p}_m{-}{\textstyle {1\over 3}}\delta_{lm})B(\underline{p})$,
 since, as discussed earlier,
 the 2--4 scattering terms are unnecessary for the shear viscosity
 calculation.

 After solving these linear equations to find
 the first order correction $\phi(x,\underline{p})$,
 the viscosities can be
 evaluated by computing the first order correction to the stress-energy
 tensor and comparing it to the constitutive relation
 (\ref{eq:constitutive_rel}).
 One finds
 \begin{eqnarray}
 T_{(1)}^{\mu\nu}(x)
 & \displaystyle = & \displaystyle
 \int {d^3 {\bf k}\over (2\pi)^3 E_k}\,
 f^{(1)}(x,\underline{k})\,
 \left(
 \underline{k}^\mu \underline{k}^\nu
 {+}
 {\textstyle{1\over 4}}g^{\mu\nu}\delta m^2(x)
 \right)
 \nonumber\\
 & \displaystyle = & \displaystyle
 -\beta
 \int {d^3 {\bf k}\over (2\pi)^3 E_k}\,
 n(E_k)\,[1{+}n(E_k)]\,
 \left(
 \underline{k}^\mu \underline{k}^\nu
 {+}
 {\textstyle{1\over 4}}g^{\mu\nu}\delta m^2(x)
 \right)
 \nonumber\\
 & \displaystyle & \displaystyle \hspace{-1.5cm} {}
 \bigg(
      A(\underline{k})\nabla{\cdot}{\bf u}(x)
      + {1\over 2}
      \left(
         \hat{k}_i\hat{k}_j{-}{\textstyle {1\over 3}}\delta_{ij}
      \right)\,
      B(\underline{k})
      \left(
         \nabla_i u_j(x) {+} \nabla_j u_i(x)
         {-} {\textstyle {2\over 3}}\delta_{ij}\nabla{\cdot}{\bf
 u}(x)
      \right)
 \bigg)
 \;,
 \nonumber\\
 & \displaystyle & \displaystyle
 \label{eq:first_correction_stress}
 \end{eqnarray}
 and
 (using the lowest order result
 $\nabla_i T_{(0)}^{0j}(x)
 = \left( \varepsilon{+}{\cal P} \right)\nabla_i u^j(x)$)
 obtains
 \addtocounter{equation}{1}
 $$
 \eta
 =
 {\beta\over 15}
 \int {d^3{\bf k}\over (2\pi)^3 E_k}\,
 {\bf k}^2\,
 n(E_k)\,[1{+}n(E_k)]\,B(\underline{k})
 \;,
 \eqno{(\theequation a)}
 $$
 $$
 \begin{array}{lcl}
 \zeta
 & \displaystyle = & \displaystyle
 \beta
 \int {d^3{\bf k}\over (2\pi)^3 E_k}\,
 \left(
     {\textstyle{1\over 3}}{\bf k}^2\,
   {+}{\textstyle{1\over 4}}\delta m^2(x)
 \right)
 n(E_k)\,[1{+}n(E_k)]\,A(\underline{k})
 \\
 & \displaystyle = & \displaystyle
 \beta
 \int {d^3{\bf k}\over (2\pi)^3 E_k}\,
 \left(
 {\textstyle{1\over 3}}{\bf k}^2
 {-}v_{\rm s}^2 ( {\bf k}^2{+}m_{\rm phys}^2 )
 \right)
 n(E_k)\,[1{+}n(E_k)]\,A(\underline{k})
 \;.
 \end{array}
 \eqno{(\theequation b)}
 $$
 In the last expression, the Landau-Lifshitz condition
 (\ref{eq:LL_cond}) $T_{(1)}^{00}(x)=0$
 has been used to express the bulk viscosity
 in terms of the same source term that defines
 the amplitude $A(\underline{k})$ ({\it c.f.}~Eq.~(\ref{eq:Chapman2})).

 \subsection{Equivalence of the Boltzmann equation to the field theory
 result}
 \label{subsec:equivalence}

 The Boltzmann equation result
 for the shear viscosity calculation, Eq.~(\ref{eq:Chapman2}),
 and the formulae from the finite temperature field theory
 Eq.~(\ref{eq:reduced_vertex_eq_full})
 are remarkably similar.
 In fact, once the free cut propagators are used to force all momenta
 on-shell,
 Eq.~(\ref{eq:Chapman2}) will have exactly the same form as
 Eq.~(\ref{eq:reduced_vertex_eq_full}),
 {\em provided} that the thermal dispersion relation is used,
 and the scattering amplitude
 for temperature dependent effective excitations
 (\ref{eq:almost_scattering_amp}) is used in the collision term.

 With the identification of
 \begin{eqnarray}
 D_{\pi}(\underline{k})
 & \displaystyle \leftrightarrow & \displaystyle
 (\hat{k}_l \hat{k}_m {-} {\textstyle{1\over 3}}\delta_{lm})
 \Sigma_I^{\rm Boltz.}(\underline{k})B(\underline{k})
 \\
 \noalign{\hbox{and}}
 m_{\rm th}^2
 & \displaystyle \leftrightarrow & \displaystyle
 m_{\rm phys}^2 + \delta m^2
 \;,
 \end{eqnarray}
 the equation (\ref{eq:Chapman2})
 for the spin 2 amplitude $B(\underline{k})$ can be rewritten as
 \begin{equation}
 I_\pi = D_{\pi}(k)
 -
 (1{-}e^{-\beta E_k})\,
 \int {d^4 p\over (2\pi)^4}\,
 L_{\rm Boltz.}(\underline{k}{-}p) \,
 S_{\rm free}(p)\,
 {D_{\pi}(p) \over 2\Sigma_I^{\rm Boltz.}(p)}
 \;,
 \label{eq:almost_vertex_eq}
 \end{equation}
 where $S_{\rm free}(p)$ is the ``free'' phase space amplitude
 for an
 effective single particle thermal excitation,
 \begin{equation}
 S_{\rm free}(k) =
 {\rm sgn}(k^0)\, [1{+}n(k^0)]\, 2\pi\delta(k^2+m_{\rm th}^2)
 \;,
 \end{equation}
 and the kernel $L_{\rm Boltz.}$ and the ``self-energy''
 $\Sigma_I^{\rm Boltz.}$ are,
 \begin{eqnarray}
 L_{\rm Boltz.}(\underline{k},\underline{p})
 & \displaystyle = & \displaystyle
 {1\over 2}
 \int
 {d^4 p_2\over (2\pi)^4}\,
 {d^4 p_3\over (2\pi)^4}\,
 \left|
 {\cal T}_4 (\underline{p},p_2;p_3,\underline{k})
 \right|^2
 (2\pi)^4\delta(\underline{p}{+}p_2{-}p_3{-}\underline{k})
 \nonumber\\
 & \displaystyle & \displaystyle \qquad {}\times
 S_{\rm free}(p_2)\, S_{\rm free}(-p_3)\,
 \;,
 \label{eq:Boltzmann_kernel}
 \end{eqnarray}
 \begin{equation}
 \Sigma_I^{\rm Boltz.} (\underline{k})
 =
 {1\over 6}\,
 (1{-}e^{-\beta E_k})\,
 \int {d^4 p\over (2\pi)^4}
 L_{\rm Boltz.}(\underline{k}{-}p)
 S_{\rm free}(p)
 \;.
 \end{equation}
 These are identical to the kernel
 $L_{\rm full}(\underline{k},\underline{p})$ and
 the self-energy
 $\Sigma_I(\underline{k})$ in the previous section.
 To write the equation in terms of a single function
 $D_{\pi}(x,\underline{p})$,
 an exchange of labels $3 \leftrightarrow 1$, and $2 \leftrightarrow 1$
 has been used, together with the fact that the scattering amplitude
 squared $|{\cal T}_4|^2$ is symmetric under the time-reversal.
 The only subtlety
 in deriving Eq.~(\ref{eq:almost_vertex_eq}) is that it is written in
 terms of on-shell momenta with both positive and negative energies
 while the Boltzmann equation is written solely in terms of on-shell
 momenta with positive energies.
 The equivalence is possible because the negative energy contributions
 vanish due to the kinematic conditions enforced by energy
 conservation.

 With the use of the thermal mass in place of
 the zero temperature mass,
 Eq.~(\ref{eq:almost_vertex_eq})
 with the scattering amplitude
 \begin{equation}
 {\cal T}(\underline{p},\underline{p}_2;\underline{p}_3,\underline{k})
 =
 \lambda
 -
 g^2
 \left(
 G_R(\underline{p}_2{+}\underline{p}) +
 G_R(\underline{p}_2{-}\underline{p}_3) +
 G_R(\underline{p}_2{-}\underline{k})
 \right)
 \;,
 \label{eq:Boltz_scattering_amp}
 \end{equation}
 has exactly the same structure as
 Eq.~(\ref{eq:reduced_vertex_eq_full}).

 For the bulk viscosity,
 the equivalence can be shown by first noting that
 the right hand side of the integral equation (\ref{eq:Chapman})
 also has a near zero mode corresponding to letting
 $\phi(x,\underline{p})$ be a constant, or integrating over
 $d^3 {\bf k}\, n(E_k)[1{+}n(E_k)]/(2\pi)^3E_k$.
 In particular,
 if the identification
 \begin{equation}
 D_{\bar{\cal P}}(\underline{k})
 \leftrightarrow
 \Sigma_I^{\rm Boltz.}(\underline{k})A(\underline{k})
 \;,
 \end{equation}
 is made, $\phi(x,\underline{p})$'s
 in the right hand side of Eq.~(\ref{eq:Chapman}) are
 replaced by 2
 (to account for both the positive and the negative energy on-shell
 momenta), and integrated over
 $d^3 {\bf k}\, n(E_k)[1{+}n(E_k)]/(2\pi)^3E_k$,
 the result on the right hand side is identical to
 $(b_5|(1{-}K_{\rm bulk})|b_5) = -(\delta C|b_5)$
 in Eq.~(\ref{eq:deltaCb_5}).
 The integral equation (\ref{eq:Chapman}) itself is
 {\em not} strictly identical to Eq.~(\ref{eq:mod_bulk_eq}).
 Eq.~(\ref{eq:mod_bulk_eq}) includes additional pieces in the number
 conserving parts of the kernel.  However,
 the number-changing part of the two kernels in the equations
 (\ref{eq:mod_bulk_eq}) and (\ref{eq:Chapman}) are the same.

 The existence of the near-zero mode implies that
 the integral equation for the spin 0 component $A$ is again dominated
 by this near-zero mode component.
 By using the same arguments as in section~\ref{subsec:ladder_bulk},
 the leading order bulk viscosity can be again written in terms of
 the near-zero mode matrix element.
 Since $I_{\bar{\cal P}}(\underline{k})$ is an even function of
 $\underline{k}^0$,
 the definition of the inner product of two functions
 (\ref{eq:def_inner}) can be used to express the bulk viscosity as
 \begin{eqnarray}
 \zeta
 & \displaystyle = & \displaystyle
 \beta \int {d^3{\bf k}\over (2\pi)^3 E_k}\,
 I_{\bar{\cal P}}(\underline{k})\,
 n(E_k)\,[1{+}n(E_k)]\,A(\underline{k})
 \nonumber\\
 & = & \displaystyle
 -\beta
 {
 ( I_{\bar{\cal P}}| b_5 ) ( b_5 |I_{\bar{\cal P}} )
 		\over
 (\delta C | b_5 )
 }
 \;,
 \end{eqnarray}
 where
 $(b_5 | I_{\bar{\cal P}})$ and $( \delta C | b_5 )$
 are again given by Eq.~(\ref{eq:b_5I_p}) and Eq.~(\ref{eq:deltaCb_5}).

 As stated earlier,
 the Boltzmann equation for the fundamental particles
 ceases to be valid at temperatures
 high enough to make the mean free path smaller than the Compton
 wavelength of the underlying particle.
 Thus, it may be surprising to find
 that a form of Boltzmann equation remains valid
 at all temperatures, provided the mass parameter is
 interpreted as the mass of effective thermal excitation
 and a temperature dependent scattering amplitude is used.

 It is interesting to consider the effect of using the thermal
 mass and
 the thermal scattering amplitude
 in various temperature range.
 At low temperatures ($T{\ll}m_{\rm phys}$),
 the thermal correction to the mass
 ({\it c.f.}~Eq.(\ref{eq:delta_m_Boltz}))
 is completely negligible.
 Consequently, for the leading order calculation, thermal quantities
 are
 unnecessary and the viscosities may be calculated by using
 kinetic theory of the non-relativistic particles.
 If the temperature is in the range
 $m_{\rm phys}
 \,\,\vcenter {\hbox{$\buildrel{\displaystyle <}\over\sim$}} \,\,
 T {\ll} m_{\rm phys}/\sqrt{\lambda}$,
 most particles are highly energetic,
 but the thermal corrections to the mass and the scattering
 amplitude are negligible.
 Consequently, the viscosities at these temperatures can be calculated
 by
 the kinetic theory of relativistic particles with the zero temperature
 mass and the scattering amplitude.
 At very high temperatures,
 $T {\gg} m_{\rm phys}/\lambda$,
 all mass scales, including the cubic coupling constant,
 other than temperature may be ignored.
 Hence, both the shear and bulk viscosity may be calculated from
 the
 kinetic theory of massless excitations with only the quartic
 interaction.%
\footnote{%
	The power counting performed in this paper is also
	valid for the massless scalar theory
	since the excitation in this case develops non-zero
	thermal mass of ${\cal O}(\sqrt{\lambda}T)$ at non-zero
	temperature.
}

 The most interesting region is
 at intermediate temperatures
 $T = {\cal O}(m_{\rm phys}/\sqrt{\lambda})$.
 Since this is much larger than
 $m_{\rm phys}$,
 and the typical size of loop momenta at high
 temperature is ${\cal O}(T)$,
 one might expect that
 the replacement of zero temperature mass by
 the thermal mass should
 have a negligible effect.
 This is true for some observables, such as the shear viscosity.
 However,
 for the bulk viscosity
 the contribution from momenta of ${\cal O}(m_{\rm th})$
 is not negligible compared to the hard momentum
 contribution.

 To understand the behavior of the bulk viscosity,
 first note that the classical scale invariance requires
 the classical bulk viscosity to be proportional
 to $m_{\rm phys}^4/T$~\cite{Weinberg}.
 When $T = {\cal O}(m_{\rm phys}/\sqrt{\lambda})$,
 the effect of quantum mechanically broken scale invariance
 is negligible compared to the $m_{\rm phys}$.
 In a scalar $g\phi^3{+}\lambda\phi^4$ theory,
 the 2--3 amplitude for soft momenta is
 ${\cal T}_5 \sim {\cal O}(\lambda g/m_{\rm th}^2)$
 Hence,
 the expression $(\delta C|b_5)$ above is a non-trivial function of the
 dimensionless ${\cal O}(1)$ ratio $g^2/\lambda m_{\rm th}^2$.
 As discussed in section~\ref{subsec:ladder_bulk34},
 $(\delta C|b_5) = {\cal O}(\lambda^{5/2} T^4)$ in this temperature range.
 Consequently,
 \begin{equation}
 \zeta
 =
 {m_{\rm phys}^4 \over\lambda^{5/2} T}\,
 d_{\rm bulk}(g^2/\lambda m_{\rm th}^2)
 \times (1 + {\cal O}(m_{\rm th}/T))
 \;,
 \end{equation}
 where $d_{\rm bulk}$ is a dimensionless function of ${\cal O}(1)$.
 The coefficient function
 $d_{\rm bulk}$ cannot be calculated from the massless scalar
 theory.
 Thus, including the thermal correction to the mass
 is essential to calculate
 the correct leading weak coupling behavior of
 bulk viscosity when $T={\cal O}(m_{\rm phys}/\sqrt{\lambda})$.

 If the temperature is high enough,
 $T \gg m_{\rm phys}/\lambda$, the quantum scale anomaly dominates
 the effect of the physical mass term,
 and the leading weak coupling behavior
 of the bulk viscosity is identical to that of
 the massless theory with only the quartic interaction.
 Due to the scale anomaly,
 $(b_5|I_{\bar{\cal P}})({\bf k})$ is non-zero but proportional to the
 ${\cal O}(\lambda^2)$ $\beta$-function,
 $(b_5|I_{\bar{\cal P}})({\bf k})
 = {\cal O}(\beta(\lambda) {\bf k}^2)$.
 Then since
 $(\delta C|b_5) = {\cal O}(\lambda^3 T^4)$,
 $\zeta = {\cal O}(\lambda T^3)$ when $T \gg m_{\rm phys}/\lambda$.
 Note that at this temperature, ${\cal O}(\lambda T^3)$ is larger than
 ${\cal O}(m_{\rm phys}^4 /\lambda^3 T)$.

 In contrast,
 the shear viscosity is insensitive to the soft momentum contribution.
 For the typical momentum $k={\cal O}(T)$,
 $g^2\tilde{G}(k)={\cal O}(g^2/T^2) < {\cal O}(\lambda^2)$.
 Hence, in this case,
 the scattering amplitude is dominated by the quartic
 interaction term, or, ${\cal T} \sim \lambda$.
 Then, the dimensional analysis demands that
 \begin{equation}
 \eta
 =
 {T^3\over \lambda^2} d_{\rm shear}
 \;,
 \end{equation}
 where $d_{\rm shear}$ is a pure number of ${\cal O}(1)$ which
 {\em can} be calculated from
 the kinetic theory of massless excitations with only the quartic
 interaction.

 \section{Calculation of viscosities}
 \label{sec:viscosities}

 The purpose of this section is to apply the results of previous
 sections to the calculation of the bulk and the shear viscosities in
 the $\lambda\phi^4$ and the $g\phi^3{+}\lambda\phi^4$ theory.
 We begin with the bulk viscosity.  Recapping the result
 (\ref{eq:leading_bulk2}), the leading order bulk viscosity is given by
 \begin{eqnarray}
 \zeta
 & = & \displaystyle
 - \beta
 {
 ( I_{\bar{\cal P}}| b_5 ) ( b_5 |I_{\bar{\cal P}} )
 		\over
 ( \delta C| b_5 )
 }
 \;.
 \label{eq:leading_bulk_final_again}
 \end{eqnarray}
 Among the factors in Eq.~(\ref{eq:leading_bulk_final_again}),
 \begin{equation}
 (b_5 | I_{\bar{\cal P}})
 =
 \int
 {d^3 {\bf l} \over (2\pi)^3 E_l}\,
 [1{+}n(E_l)]\,n(E_l)\,
 I_{\bar{\cal P}}(E_l,{\bf l})
 \;
 \end{equation}
 reduces to a 1-$d$ integral over the magnitude of the momentum,
 and can be easily evaluated numerically.
 The denominator for the $\lambda\phi^4$ theory
 \begin{eqnarray}
 -(\delta C | b_5)
 & = & \displaystyle
 2\int
 \prod_{i=1}^2 {d^3 {\bf l}_i \over (2\pi)^3}\,
 d\sigma_{12\to 3456}\, v_{12}\,
 \nonumber\\
 & & \displaystyle {} \times
 n(E_1)\, n(E_2)\,
 [1{+}n(E_3)]\, [1{+}n(E_4)]\, [1{+}n(E_5)]\, [1{+}n(E_6)]\,
 \;,
 \end{eqnarray}
 with the differential cross section
 \begin{equation}
 d\sigma_{12\to 3456}
 \equiv
 \prod_{i=3}^6 {d^3 {\bf l}_i \over (2\pi)^3 2 E_i}\,
 |{\cal T}_6(\{\underline{l}_i\})|^2
 (2\pi)^4\delta( \underline{l}_1{+} \underline{l}_2{-} \underline{l}_3{-}
 \underline{l}_4 {-} \underline{l}_5{-} \underline{l}_6 )\,
      {\bigg /} (4 E_1 E_2 v_{12} 4!)
 \;
 \end{equation}
 has a complicated angle dependence through the scattering cross
 section ${\cal T}_6$.
 Using the energy-momentum conserving delta function and rotational
 invariance, 4 of the 18 dimensional integral involved in calculating
 $(\delta C| b_5)$ can be done.  The remaining 14 integrals involve 5
 integrations over the magnitudes of momentum, and 9 angle integrations.
 Due to rotational invariance, one of the solid angle integration can be
 trivially done reducing the expression to a 12-dim integral which must
 be evaluated numerically.
 The denominator for the $g\phi^3{+}\lambda\phi^4$ theory is similarly
 given but with the more complicated 2--3 scattering amplitude
 ({\it c.f.} Eq.~(\ref{eq:deltaCb_5_34})).  In this case, a 9-dim
 integral must be carried out numerically.

 In contrast to the bulk viscosity calculation,
 the shear viscosity calculation requires solving an integral equation.
 Here,
 the integral equations derived in section~\ref{sec:summation}
 for the effective vertices are
 further reduced to one dimensional integral equations,
 and the explicit form of the $\lambda\phi^4$ theory kernel is
 evaluated and briefly examined.
 The shear viscosity in terms of the effective vertex $D_\pi$
 is given by
 \begin{equation}
 \eta
 =
 {\beta\over 10}\,
 \int{d^4 k\over (2\pi)^4}\,
 I_{\pi}(k)\, n(k^0)\,S_{\rm free}(k)\,
 {D_{\pi}(k)\over \Sigma_I(k)}
 \;,
 \end{equation}
 where $D_{\pi}$ satisfies
 \begin{equation}
 I_{\pi}(\underline{k}) =
 D_{\pi}(\underline{k}) -
 (1{-}e^{-\underline{k}^0\beta})\int {d^4 p\over (2\pi)^4}\,
 L_{\rm full}(\underline{k},p)\,S_{\rm free}(p)\,
 {D_{\pi}(p) \over 2\Sigma_I(p)}\,
 \;.
 \label{eq:shear_formula}
 \end{equation}
 The kernel $L_{\rm full}$ is
 \begin{equation}
 L_{\rm full}(k,p)
 =
 {1\over 2}
 \int {d^4 l_1 \over (2\pi)^4}\,
 {d^4 l_2 \over (2\pi)^4}\,
 S_{\rm free}(l_1)\,S_{\rm free}(-l_2)\,
 (2\pi)^4\delta(l_1{-}l_2{+}\underline{p}{-}\underline{k})\,
 \left|{\cal T}
 (\underline{l}_1,\underline{p};\underline{l}_2,\underline{k})\right|^2
 \;,
 \label{eq:L_full_again}
 \end{equation}
 and involves the $g\phi^3{+}\lambda\phi^4$ tree level
 scattering amplitude
 \begin{equation}
 {\cal T}(\underline{l}_1,\underline{p};\underline{l}_2,\underline{k})
 =
   \lambda
   - g^2\,
   \left(
   G_R(\underline{l}_1{+}\underline{p})
  +G_R(\underline{l}_1{-}\underline{k})
  +G_R(\underline{l}_1{-}\underline{l}_2)
   \right)
 \;.
 \end{equation}
 The imaginary part of the self-energy $\Sigma_I(\underline{k})$ is
 \begin{equation}
 \Sigma_I(\underline{k})
 =
 {1\over 6}
 (1{-}e^{-\beta \underline{k}^0})
 \int{d^4 p\over (2\pi)^4}\,
 L_{\rm full}(\underline{k}{-}p)S_{\rm free}(p)
 \;,
 \label{eq:L_full_self_again}
 \end{equation}
 where the free cut propagator contains the thermal mass,
 \begin{equation}
 S_{\rm free}(p) =
 [1{+}n(p^0)]\,{\rm sgn}(p^0)\,(2\pi)^4\delta(p^2{+}m_{\rm th}^2)
 \;.
 \label{eq:S_free_again}
 \end{equation}
 The inhomogeneous term
 $I_{\pi}(\underline{k}) = k_lk_m{-}{1\over 3}\delta_{lm}{\bf k}^2$
 represents an insertion of the traceless stress tensor
 \begin{equation}
 {\pi}_{lm} \equiv
 \partial_l {\phi} \, \partial_m {\phi}
 -{\textstyle {1\over 3}}
 \delta_{lm}\partial_k {\phi}\, \partial^k {\phi}
 \;.
 \end{equation}

 Eq.~(\ref{eq:shear_formula}) is a set of 5 independent
 (due to the traceless symmetric spatial indices)
 3-dimensional (since all momenta are on-shell)
 linear integral equations.
 Using spatial rotational invariance,
 this may be further reduced to a single one dimensional
 linear integral equation.
 In section~\ref{sec:summation},
 the effective vertex
 $D_{\pi}(\underline{p})$ was shown to be an even function of the
 momentum.
 Isotropy then requires
 that
 $D_{\pi}(\underline{p})$ have the structure
 \begin{equation}
 D_{\pi}(\underline{p}) =
 (\hat{p}_l\hat{p}_m{-}{\textstyle{1\over 3}}\delta_{lm})
 D_{\rm shear}(|{\bf p}|)
 \;,
 \end{equation}
 where $\hat{\bf p}$ is the unit vector in direction of ${\bf
 p}$.
 Contracting Eq.~(\ref{eq:shear_formula}) with
 $\hat{k}_l \hat{k}_m$ and
 carrying out the frequency integration with the help of the
 on-shell
 $\delta$-function yields a single one-dimensional integral equation
 for $D_{\rm shear}(|{\bf p}|)$
 \begin{equation}
 D_{\rm shear}(|{\bf k}|)
 =
 {\bf k}^2 +
 {3 \over 16}
 \int {d|{\bf p}| \over (2\pi)^3}
      { {\bf p}^2 \over E_p^2 \Gamma_p^{\vphantom{x}}}\,
 n(E_p)\,[1{+}n(E_p)]\,
 N_{\rm shear}(|{\bf k}|,|{\bf p}|)\,
 D_{\rm shear}(|{\bf p}|)
 \;,
 \label{eq:1_d_integral_eq_shear}
 \end{equation}
 where the kernel $N_{\rm shear}$ is an angular average of the
 $3d$ kernel $L_{\rm full}$,
 \begin{eqnarray}
 \lefteqn{
 N_{\rm shear}(|{\bf k}|,|{\bf p}|) \equiv
 (1{-}e^{-\beta E_k})\,
 \int d\phi\,
 d\cos\theta \, (\cos^2\theta - {\textstyle {1\over 3}})
 } & &
 \nonumber\\
 & \displaystyle & \displaystyle
 {}\times
 \left[
  L_{\rm full}(E_k,{\bf k};E_p,{\bf p})\,(e^{\beta E_p}{-}1)
  -
  L_{\rm full}(E_k,{\bf k};-E_p,{\bf p})\,(1{-}e^{-\beta E_p})
 \right]
 \;.
 \label{eq:1_d_kernel_shear}
 \end{eqnarray}
 Here, $\theta$ is the angle between the vectors ${\bf k}$ and
 ${\bf p}$.
 As before, the on-shell thermal width $\Gamma_p$ is defined
 as
 \begin{equation}
 \Gamma_p \equiv {\Sigma_I(E_p,{\bf p}) \over 2 E_p}
 \;.
 \end{equation}
 Note that $N_{\rm shear}$ is symmetric,
 $N_{\rm shear}(|{\bf k}|,|{\bf p}|) =
  N_{\rm shear}(|{\bf p}|,|{\bf k}|)$,
 since $L_{\rm full}(E_k,{\bf k}; E_p,{\bf p})=
 e^{\beta(E_k{-}E_p)}\,L_{\rm full}(E_p,{\bf p}; E_k,{\bf k})$.

 Finally,
 in terms of the scalar function $D_{\rm shear}(|{\bf k}|)$,
 the shear viscosity is given by the simple integral
\begin{equation}
 \eta = {\beta\over 60 \pi^2}\,
 	\int
	d|{\bf k}|\,
 	{|{\bf k}|^4 \over E_k^2 \Gamma_k^{\vphantom{x}}}\,
 	n(E_k)\,[1{+}n(E_k)] \,D_{\rm shear}(|{\bf k}|)
 \;.
 \label{eq:eta_formula}
 \end{equation}

 The final one-dimensional integral equation
 (\ref{eq:1_d_integral_eq_shear}) must be solved numerically.
 Clearly, one must first evaluate the full $g\phi^3{+}\lambda\phi^4$
 theory kernels $N_{\rm shear}$
 (\ref{eq:1_d_kernel_shear}).
 With both cubic and quartic interactions,
 the full ``rung'' $L_{\rm full}$ (\ref{eq:L_full_again}),
 is too complicated to compute analytically.
 However, in a pure $\lambda\phi^4$ theory, the integral is reasonably
 straightforward and one finds,
 \begin{eqnarray}
 L^{\phi^4}_{\rm full}(\underline{k}{-}\underline{p})
 & \displaystyle = & \displaystyle
 {\lambda^2\over 2}
 \int {d^4 l \over (2\pi)^4}\,
 S_{\rm free}(-l)\,S_{\rm free}(l{+}\underline{k}{-}\underline{p})
 \nonumber\\
 \; & \displaystyle = & \displaystyle
 {
   \lambda^2
  [1{+}n(\underline{k}^0{-}\underline{p}^0)]
   \over 8\pi\beta
  |{\bf k}{-}{\bf p}|
}
 \nonumber\\
 \noalign{\smallskip}
 \; & \displaystyle \; & \displaystyle  {}\times
 \left\{\,
 \theta( (\underline{k}{-}\underline{p})^2 )
 \ln
 \left|
 {
 1 - \exp(-\beta r_{+}(\underline{k}{-}\underline{p}))
     \over
 1 - \exp(-\beta r_{-}(\underline{k}{-}\underline{p}))
 }
 \right|
 \right.
 \nonumber\\
 \noalign{\smallskip}
 \; & \displaystyle \; & \displaystyle
 \qquad {}
 +
 \left.
 \theta( -(\underline{k}{-}\underline{p})^2{-}4m_{\rm th}^2 )
 \ln
 \left|
 {
  \sinh(\beta r_{+}(\underline{k}{-}\underline{p})/2)
 	\over
  \sinh(\beta r_{-}(\underline{k}{-}\underline{p})/2)
 }
 \right|\,
 \right\}
 \;,
 \label{eq:explicit_cut_rung}
 \end{eqnarray}
 where
 \begin{equation}
 r_{\pm}(\underline{k}{-}\underline{p}) \equiv
 {1\over 2}
     \left(
 	|{\bf k}{-}{\bf p}|
	\sqrt{1{+}4m_{\rm th}^2/ (\underline{k}{-}\underline{p})^2}
  	\pm (\underline{k}^0{-}\underline{p}^0)
    \right)
 \;.
 \end{equation}
 The term
 involving $\theta((\underline{k}{-}\underline{p})^2)$ in
 (\ref{eq:explicit_cut_rung})
 represents the result of integrating over
 two on-shell delta functions with Bose factors
 $n(E_l)[1{+}n(E_{l+k-p})]$.
 These statistical factors indicate
 that this term describes the transfer of an
 incoming momentum $k{-}p$ to a thermal excitation with momentum
 $\underline{l}$,
 producing another on-shell excitation
 with 4-momentum $(E_{l+k-p},{\bf l}{+}{\bf k}{-}{\bf p})$
 with a stimulated emission factor of $[1{+}n(E_{l{+}k{-}p})]$.

 The second term involving
 $\theta(-(\underline{k}{-}\underline{p})^2{-}4m_{\rm th}^2)$
 describes the usual
 process of creating two propagating on-shell particles with
 total
 invariant mass
 larger than twice the mass of the initial single particle excitation.
 It differs from the zero temperature result,
 \begin{equation}
 \lim_{\beta \to \infty}
 L^{\phi^4}_{\rm full}(k{-}p)
 =
 \lambda^2
 {
 \theta(k^0{-}p^0)\,
 \theta( -(k{-}p)^2{-}4m_{\rm phys}^2 )
 }
 \sqrt{1{+}{ 4m_{\rm phys}^2 / (k{-}p)^2 }}
 / 16\pi
 \;,
 \end{equation}
 only because of the stimulated emission in the final state.

 To solve the integral equation (\ref{eq:1_d_integral_eq_shear})
 numerically, the magnitudes of momenta
 $|{\bf k}|$ and $|{\bf p}|$
 need to be discretized in order to turn the integral
 into a finite number of coupled linear equations.
 Given the explicit form of the $\lambda\phi^4$ theory kernel
 $L^{\phi^4}_{\rm full}(k{-}p)$,
 evaluating the
 coefficients of the linear equations requires
 numerically computing two 1-dimensional angle integrations;
 one
 for the angular averaged kernel
 $N^{\phi^4}_{\rm shear}(|{\bf k}|,|{\bf p}|)$,
 and the other for the self-energy $\Sigma^{\phi^4}_I(\underline{p})$
 ({\it c.f.}~Eq.~(\ref{eq:L_full_self_again})).
 In contrast, for the $g\phi^3{+}\lambda\phi^4$ theory calculation,
 evaluating each coefficient of the final matrix equation requires
 first computing two 2-dimensional angular integrations since
 $L_{\rm full}(k,p)$ is no longer just a function of $k{-}p$
 due to the non-trivial structure of the scattering amplitude.

 Although the angular integrations involved in $N^{\phi^4}_{\rm shear}$
 is too complicated to carry out analytically,
 some qualitative behaviors of the kernels can be easily found.
 One property of the kernels,
 important in carrying out numerical analysis, is that
 $N_{\rm shear}^{\phi^4}(|{\bf k}|,|{\bf p}|)$
 has a discontinuous first derivative across $|{\bf k}|=|{\bf p}|$.
 To see this, consider, for example, the expression of
 $N_{\rm shear}^{\phi^4}(|{\bf k}|,|{\bf p}|)$
 in Eq.~(\ref{eq:1_d_kernel_shear}).
 If the angle $\theta$ is defined to be the angle between two vectors
 ${\bf k}$ and ${\bf p}$, the azimuthal angle integration
 in Eq.~(\ref{eq:1_d_kernel_shear}) is trivial.
 For the $\cos\theta$ integration, it is convenient to change variable
 to $y\equiv |{\bf k}{-}{\bf p}|$.
 The Jacobian of this variable change cancels
 an explicit $1/|{\bf k}{-}{\bf p}|$ factor contained in
 $L^{\phi^4}$.
 Then Eq.~(\ref{eq:1_d_kernel_shear}) may be rewritten as
 \begin{equation}
 N_{\rm shear}^{\phi^4}(|{\bf k}|,|{\bf p}|)
 =
 \int^{|{\bf k}|{+}|{\bf p}|}_{\left||{\bf k}|{-}|{\bf p}|\right|}
 dy\,
 \bigg(
 F_-(E_k,E_p,y)
 -
 F_+(E_k,E_p,y)
 \bigg)
 \;,
 \end{equation}
 where
 $F_-(E_k,E_p,y)$ contains the logarithmic part multiplying
 $\theta((\underline{k}{-}\underline{p})^2)$ in $L^{\phi^4}_{\rm full}$
 with $(\cos^2\theta{-}{1\over 3})$,
 and $F_+(E_k,E_p,y)$ contains the logarithmic part multiplying
 $\theta(-4m_{\rm th}^2{-}(\underline{k}{-}\underline{p})^2)$
 in $L^{\phi^4}_{\rm full}$
 with $(\cos^2\theta{-}{1\over 3})$.

 Differentiating with respect to $|{\bf k}|$ yields,
 \begin{eqnarray}
 {\partial\over \partial |{\bf k}|}
 N_{\rm shear}^{\phi^4}(|{\bf k}|,|{\bf p}|)
 & \displaystyle = & \displaystyle
 \bigg(
 F_-(E_k,E_p,|{\bf k}|{+}|{\bf p}|)
 -
 F_+(E_k,E_p,|{\bf k}|{+}|{\bf p}|)
 \bigg)
 \nonumber\\
 & \displaystyle & \displaystyle {}
 -
 {\rm sgn}(|{\bf k}|{-}|{\bf p}|)
 \bigg(
 F_-(E_k,E_p,\left| |{\bf k}|{-}|{\bf p}| \right|)
 -
 F_+(E_k,E_p,\left| |{\bf k}|{-}|{\bf p}| \right|)
 \bigg)
 \nonumber\\
 & \displaystyle & \displaystyle {}
 +
 \int^{|{\bf k}|{+}|{\bf p}|}_{\left||{\bf k}|{-}|{\bf p}|\right|}
 dy\,
 {\partial\over \partial |{\bf k}|}
 \bigg(
 F_-(E_k,E_p,y)
 -
 F_+(E_k,E_p,y)
 \bigg)
 \;.
 \end{eqnarray}
 Note that in the second line the signature of
 $|{\bf k}|{-}|{\bf p}|$ explicitly appears as a result of
 differentiating the lower-limit of the integral.
 Since $F_{\pm}(E_k,E_p,\left| |{\bf k}|{-}|{\bf p}| \right|)$
 is non-zero in the $|{\bf k}|\to |{\bf p}|$ limit, this implies
 that the first derivative of
 $N_{\rm shear}^{\phi^4}(|{\bf k}|,|{\bf p}|)$
 has a discontinuity across $|{\bf k}|=|{\bf p}|$.

 \section{Numerical results}
 \label{sec:numeric}

 \subsection{Choices of parameters and discretization method}

 In this section,
 the 1-$d$ integral equation
 (\ref{eq:1_d_integral_eq_shear})
 reproduced here for convenience,
 \begin{equation}
 D_{\rm shear}(|{\bf k}|)
 =
 {\bf k}^2 +
 {3\lambda^2 \over 16}
 \int {d|{\bf p}| \over (2\pi)^3}
      { {\bf p}^2 \over E_p^2 \Gamma_p^{\vphantom{x}}}\,
 n(E_p)\,[1{+}n(E_p)]\,
 N_{\rm shear}(|{\bf k}|,|{\bf p}|)\,
 D_{\rm shear}(|{\bf p}|)
 \;,
 \end{equation}
 is numerically solved to obtain the shear viscosity in the
 $\lambda\phi^4$ theory, and the integrals involved
 in the bulk viscosity calculation
 \begin{equation}
 \zeta
 =
 - \beta
 {
 ( I_{\bar{\cal P}}| b_5 ) ( b_5 |I_{\bar{\cal P}} )
 		\over
 ( \delta C| b_5 )
 }
 \;,
 \end{equation}
 are numerically carried out.

 \begin{figure}
 \setlength {\unitlength}{1cm}
\vbox
    {%
    \begin {center}
 \begin{picture}(0,0)
 \put(-0.5,7.5){$\displaystyle \left|{dx\over d|{\bf p}|}\right|$}
 \put(11.0,-0.5){$|{\bf p}|$}
 \end{picture}
	\leavevmode
	\def\epsfsize #1#2{0.5#1}
	\epsfbox  {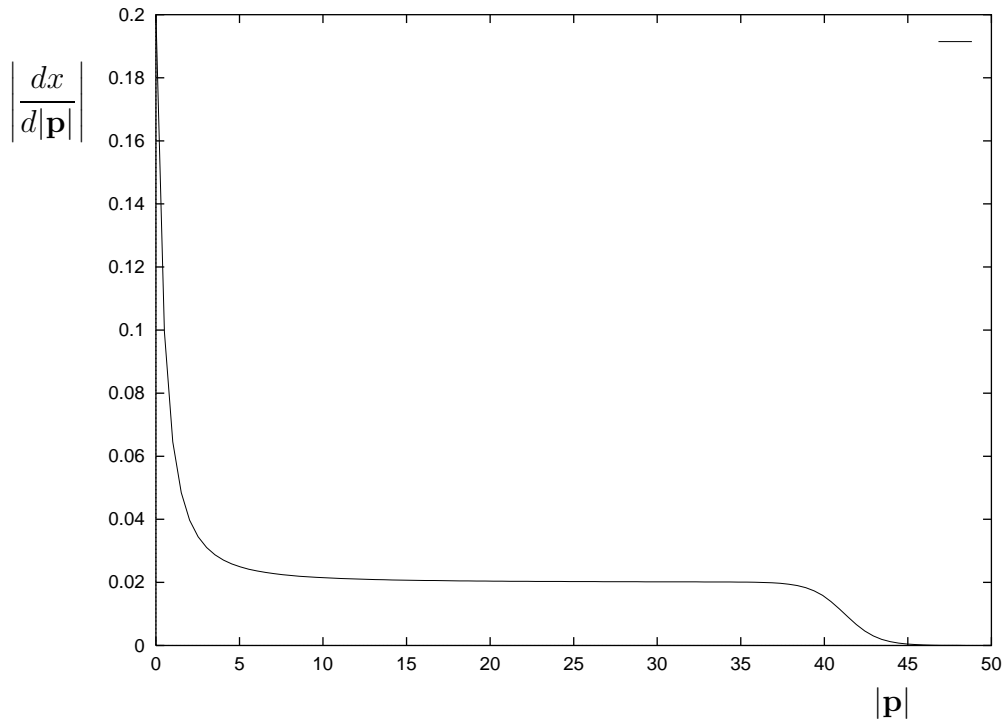}
    \end {center}
     \caption
 	{%
    A typical form of the Jacobian $|dx/d|{\bf p}||$.
    Here, the parameters are $M=50T$, $\mu = T$.
    }
 \label{fig:jacob}
 }
 \end{figure}
 \noindent
 First consider the shear viscosity calculation.
 By discretizing the magnitude of momenta,
 the integral equation can be turned
 into a finite set of linear equations which can be straightforwardly
 solved by computer.

 The discretization method chosen is
 the lowest order two-point Newton-Cotes formula
 \cite{Numerical},
 \begin{equation}
 \int_{x_0}^{x_{N+1}} dx \, f(x) =
 {3\over 2}(f(x_1) + f(x_N))
 +
 \sum_{i=2}^{N-1} f(x_i)
 + {\cal O}( (\Delta x)^2 )
 \;,
 \end{equation}
 where $\Delta x$ is the distance between two data points.
 The reason behind choosing this simple discretization is following:
 Due to the discontinuity in first derivatives of the kernel (a kink),
 second derivatives at $|{\bf p}| = |{\bf k}|$ is not well defined.
 Hence, in choosing a discretization method,
 higher order formulae are not necessarily more useful than the
 lowest order formula.

 To successfully implement the numerical analysis,
 a suitable parameterization must be chosen so that the improper integrals
 in the integral equations become proper ones.
 The parameterization used here is basically a logarithm of a Fermi
 distribution with chemical potential $M$
 \begin{eqnarray}
 \; & \displaystyle
 x = {\ln(
 	  e^{(M-f(|{\bf p}|))\beta } + 1
 	 )
      \over
      \ln(
 	  e^{M\beta} + 1
 	 )}
 & \displaystyle \;
 \label{eq:parametrization}
 \\
 \noalign{\hbox{where}}
 \; & \displaystyle
 f(|{\bf p}|) = |{\bf p}| - {c\, \mu^2 \over |{\bf p}| + \mu} + c\,\mu
 & \displaystyle
 \;.
 \end{eqnarray}
 The new parameter $x$ varies from $1$ to $0$ as
 $|{\bf p}|$ varies from $0$ to $\infty$.
 Constants $c$, $M$, and $\mu$ are adjustable parameters.

 The parameterization (\ref{eq:parametrization}) is chosen
 for the following two reasons:
 (a) To account for the soft-momentum contributions
 there must be enough data points near $|{\bf p}|=0$.
 (b) The kink in the kernel implies that
 the contribution from the momenta $|{\bf p}|{\sim}|{\bf k}|$
 cannot be ignored even for large $|{\bf k}|$.
 This implies that a discretization that
 sparsely samples large momentum values are not suitable
 since it will miss the non-negligible wiggles in those
 regions.  One must choose a parameterization that distributes data points
 more or less evenly in momentum space until the cut-off is reached.
 As shown in Fig.~\ref{fig:jacob}
 the Jacobian (which may be interpreted as the density of sampled
 points when $\Delta x$ is constant)
 \begin{equation}
 \left| {dx\over d|{\bf p}|} \right| =
 \beta {
 1+c\,\mu^2/( |{\bf p}| + \mu )^2
  \over
 (1+e^{(f(|{\bf p}|)-M)\beta}) \ln(1+e^{M\beta })
 }
 \;,
 \end{equation}
 shows that these two conditions are satisfied by the parameterization
 given in Eq.~(\ref{eq:parametrization}).
 The positive parameter $c$ controls the height at $|{\bf p}|=0$,
 and thus controls the percentage of data points sampled
 below $|{\bf p}|=\mu$,
 where $\mu$ is usually chosen to be ${\cal O}(m_{\rm th})$.
 In the present calculations, $c$ is chosen as
 \begin{equation}
 c = {2 (M-10 \mu)\over 9 \mu} \qquad (M > 10 \mu)
 \;.
 \end{equation}
 This particular choice of $c$ puts about ten percent of the
 total number of data points below $|{\bf p}|=\mu$.

 Once $|{\bf p}|$ exceeds $\mu$,
 the Jacobian is almost constant until $|{\bf p}|=M$ is reached.
 This implies that the data points are evenly distributed throughout
 the momentum range
 \begin{figure}
 \setlength {\unitlength}{1cm}
\vbox
    {%
    \begin {center}
 \begin{picture}(0,0)
 \put(-1.5,7.5){$D_{\rm shear}(|{\bf k}|)$}
 \put(11.0,-0.5){$|{\bf k}|/T$}
 \end{picture}
	\leavevmode
	\def\epsfsize  #1#2{0.45#1}
	\epsfbox {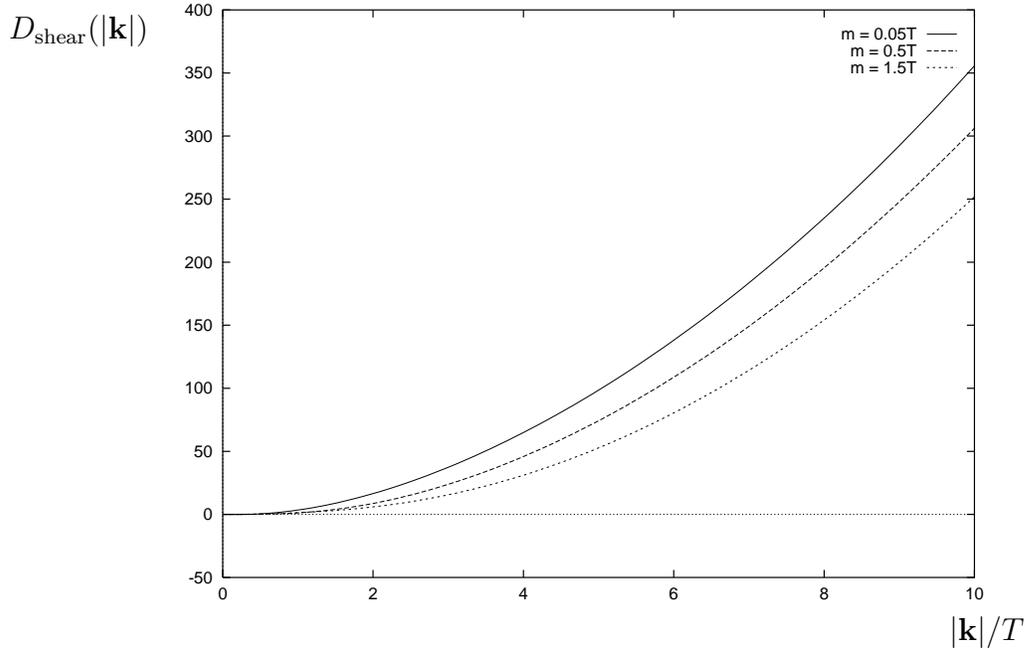}
    \end {center}
     \caption
 	{%
   Numerical solutions for $D_{\rm shear}(|{\bf k}|)$ for
   $\mth/T = 0.05, 0.5, 1.5$.
    }
 \label{fig:shear_sol}
 }
 \end{figure}
 \noindent
 $\mu
 \,\,\vcenter {\hbox{$\buildrel{\displaystyle <}\over\sim$}} \,\,
 |{\bf p}|
 \,\,\vcenter {\hbox{$\buildrel{\displaystyle <}\over\sim$}} \,\,
 M$.
 Since the Jacobian drops off very sharply once the momentum
 exceeds $M$, one can control the largest momentum sampled
 by choosing the value of $M$.
 Usually the largest momentum sampled is the size of $M$.
 Any $M$ much larger than any of the mass scales in the integrand
 will do.
 For the present calculation, the values of $M$ ranges from $30T$ to
 $100T$ as the mass increases.

 \subsection{Numerical results}

 For the shear viscosity, the value of the thermal mass $m_{\rm th}$
 are chosen to be $0.01T \leq m_{\rm th} \leq 3T$ to show the high
 temperature behavior of the shear viscosity.
 Typical forms of solutions for $D_{\rm shear}(|{\bf k}|)$ are shown in
 Fig.~\ref{fig:shear_sol} for $m_{\rm th}/T = 0.05, 0.5, 1.5$.
 The values of viscosity, extracted from the Newton-Cotes formula
 \begin{equation}
 \eta_{\infty} = \eta_{N} + a/N^2
 \;,
 \end{equation}
 with the number of data points $N=100,200,400,600$, are shown in
 Fig.~\ref{fig:shear_fit}.
 As expected, the shear viscosity rises as the mass is increased.
 In the non-relativistic limit of large
 \begin{figure}
 \setlength {\unitlength}{1cm}
\vbox
    {%
    \begin {center}
 \begin{picture}(0,0)
 \put(-1.0,7.5){$\eta\lambda^2/T^3$}
 \put(11.0,-0.5){$\mth/T$}
 \end{picture}
	\leavevmode
	\def\epsfsize  #1#2{0.45#1}
	\epsfbox {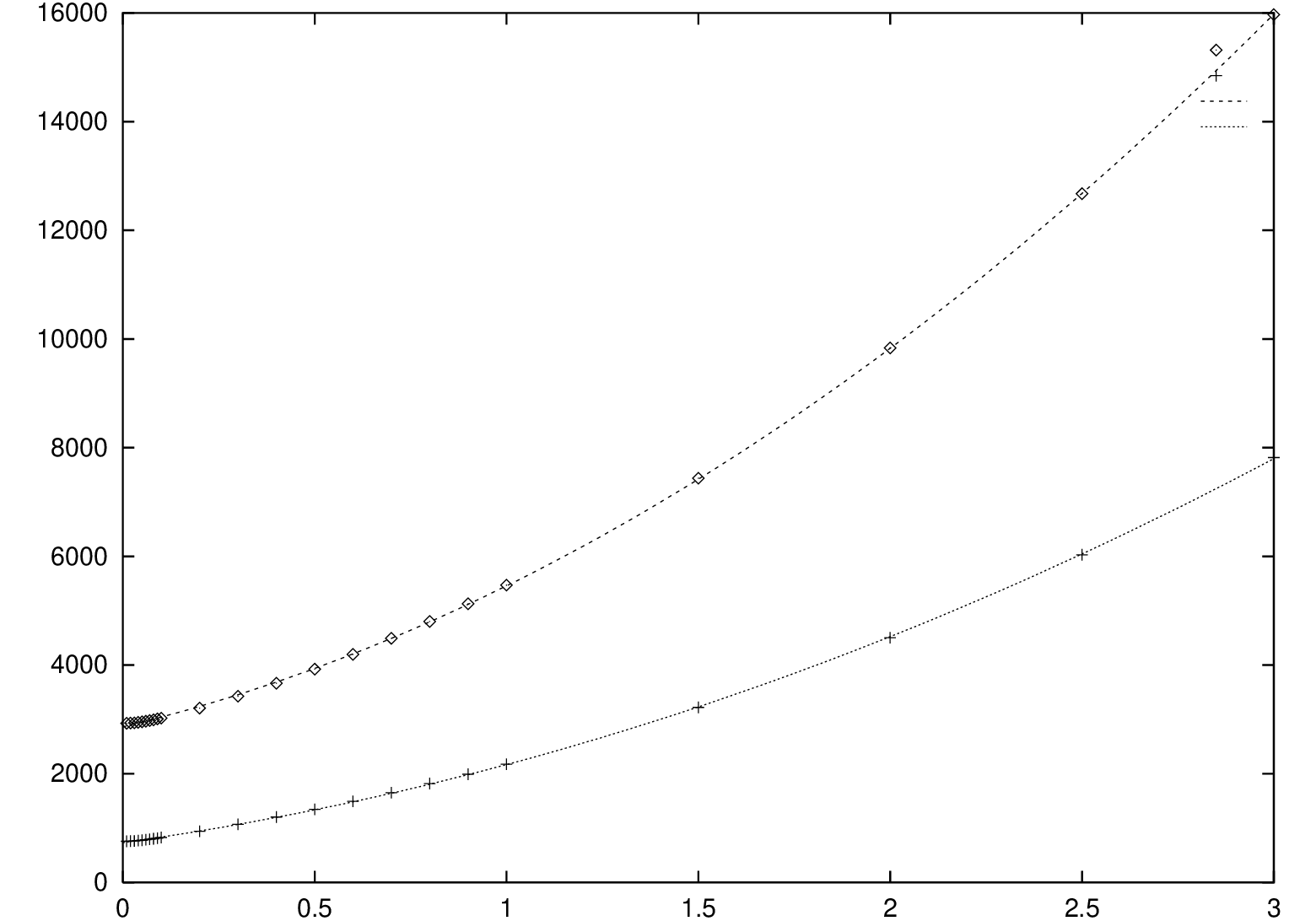}
    \end {center}
     \vspace{0.5cm}
     \caption
 	{%
 Upper curve represents numerical results for shear viscosity.
 Lower curve represents one-loop calculation result.
 The fit for the upper curve is
 $\eta(\mth/T)~=~2860 (T^3/\lambda^2)
 (1.0 + 0.596\mth/T + 0.310\mth^2/T^2)$.
 And the fit for the lower curve is
 $\eta_0(\mth/T)~=~733 (T^3/\lambda^2)
 (1.0 + 1.33\mth/T + 0.627\mth^2/T^2)$.
    }
 \label{fig:shear_fit}
 }
 \end{figure}
 \noindent
 $m_{\rm phys}/T$,
 as discussed in section~\ref{subsec:qualitative},
 the shear viscosity must rise as ${\cal O}( m_{\rm phys}^{5/2}  )$.
 In the ultra-relativistic or high temperature limit,
 the viscosity is nearly independent of the mass and ${\cal O}(T^3)$.
 In the high temperature limit, the result of resummation is about $4$
 times larger than the one-loop result alone.

 For the bulk viscosity,
 the numerical evaluation of the multiple integral involved in
 calculating  $(\delta C|b_5)$ was carried out by Monte-Carlo method.
 Typically, each point in the plot Fig.~\ref{fig:bulk_fit} is evaluated
 by about one million data points.
 As expected, the bulk viscosity behaves like $1/\lambda^3$
 for small values of $m_{\rm th}/T$ and rises sharply as the ratio
 increases.

 \section{Summary}
 \label{sec:summary}

 Hydrodynamic transport coefficients can be evaluated from first
 principles in a weakly coupled scalar field theory at arbitrary
 temperature.
 Using the diagrammatic rules derived
 \begin{figure}
 \setlength {\unitlength}{1cm}
\vbox
    {%
    \begin {center}
 \begin{picture}(0,0)
 \put(-1.5,7.5)
 {$\log\left(\displaystyle{\zeta\lambda^4T\over\mphys^4}\right)$}
 \put(11.0,-0.5){$\mth/T$}
 \end{picture}
	\leavevmode
	\def\epsfsize  #1#2{0.45#1}
	\epsfbox {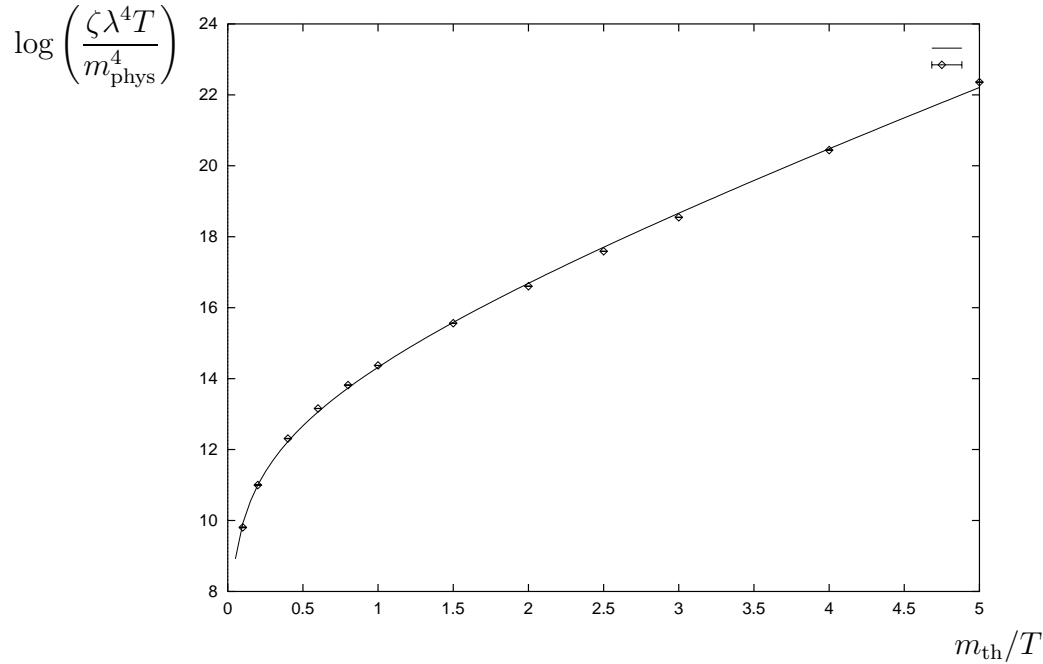}
    \end {center}
     \vspace{0.5cm}
     \caption
 	{%
 The plot of the numerical results for
 the bulk viscosities from resummed ladder diagrams.
 The fit is $\log(\zeta\lambda^4 T/m_{\rm phys}^4) =
 12.9 + 1.43(m_{\rm th}/T) + 1.36\log(m_{\rm th}/T)$.
 The error bar is typically about $1\%$ of the the value of $\zeta$.
    }
 \label{fig:bulk_fit}
 }
 \end{figure}
 \noindent
 in~\cite{Jeon}, it was shown that an infinite number of
 diagrams contribute to the leading weak
 coupling behavior of the viscosities.
 The dominant diagrams were identified by
 counting the powers of coupling constants,
 including those generated by near ``on-shell'' singularities
 cut-off by the single particle thermal lifetime.
 An infinite class of
 cut ``ladder'' diagrams were found to make
 the leading order contributions.
 The geometric series of cut ladder diagrams was summed
 by introducing a set of effective vertices, satisfying coupled
 linear integral equations.
 These equations were reduced to a
 single integral equation,
 which was then shown to be identical to the corresponding result
 obtained from a linearized Boltzmann equation describing effective
 thermal excitations with temperature dependent masses and scattering
 amplitudes.  The effective Boltzmann equation is valid even
 at very
 high temperature where the thermal lifetime and mean free path
 are
 short compared to the Compton wavelength of the underlying
 fundamental particles.

 Spatial isotropy allows one to reduce
 the dimension of the resulting integral equations
 \begin{figure}
 \setlength {\unitlength}{1cm}
\vbox
    {%
    \begin {center}
 \begin{picture}(0,0)
 \put(-1.5,7.5){$\left(\displaystyle{\zeta\lambda^4T\over\mphys^4}\right)$}
 \put(11.0,-0.5){$\mth/T$}
 \end{picture}
	\leavevmode
	\def\epsfsize  #1#2{0.45#1}
	\epsfbox {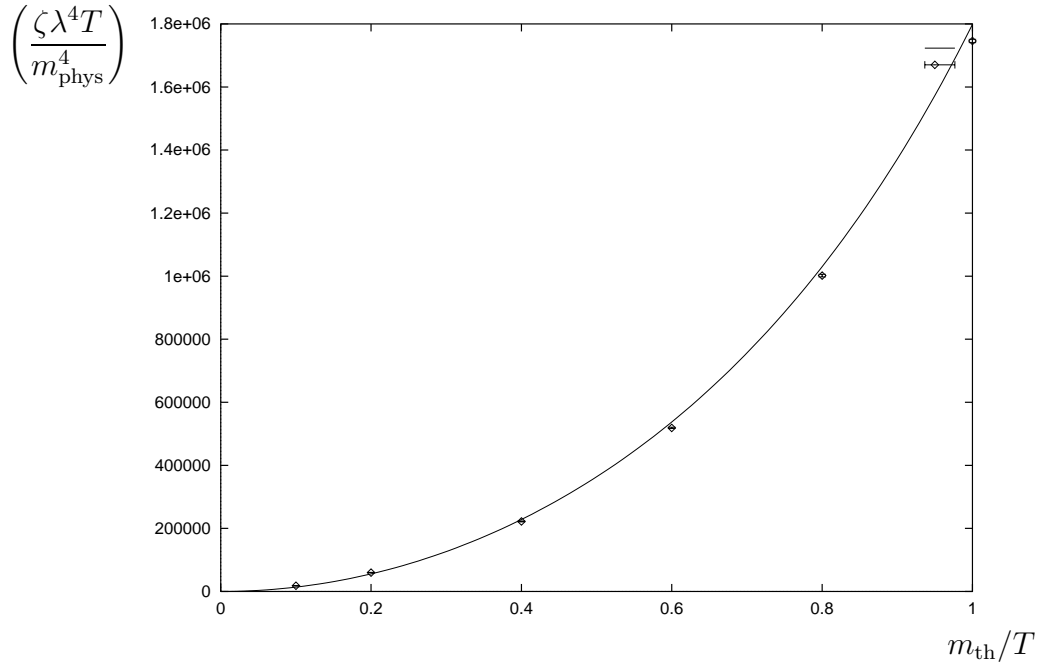}
    \end {center}
     \vspace{0.5cm}
     \caption
 	{%
 The plot of the numerical results for
 the bulk viscosities from resummed ladder diagrams for $m_{\rm th}\leq T$.
 The fit is
 $\zeta \lambda^4 T/m_{\rm phys}^4
 = 1.4\times 10^6 ((m_{\rm th}/T)^2 + 0.18 (m_{\rm th}/T)^4
 + 0.12 (m_{\rm th}/T)^6)$.
 Since $m_{\rm th}^2/T^2 \sim \lambda$, the bulk
 viscosity at high temperature behaves like $m_{\rm phys}^4/\lambda^3T$
 as expected.
 The error bar is typically about $1\%$ of the the value of $\zeta$.
    }
 \label{fig:bulk_fit_light}
 }
 \end{figure}
 \noindent
 to one-dimension,
 at which point they must be solved numerically.
 Numerical results for the viscosities
 in a scalar $\lambda\phi^4$ theory is reported.

 \acknowledgments{%
 Constant guidance and encouragement of L.~G.~Yaffe are greatly
 appreciated.}

 \appendix

 \section{Thermal Propagators}
 \label{app:propagators}

 The imaginary-time single particle propagator is
 \begin{equation}
 \tilde{G}_E({\bf k},i\omega_n) =
 {1 \over \omega^2_n + {\bf k}^2 + m_{\rm th}^2
 + \Sigma_E({\bf k},i\omega_n)}
 \;,
 \end{equation}
 where
 $\Sigma_E({\bf k},i\omega_n)$ is the full Euclidean self-energy,
 $\omega_n$ is the discrete frequency $2\pi n T$, and $m_{\rm th}$
 is the
 thermal mass
 containing ${\cal O}(\lambda T^2)$ thermal corrections.
 The Euclidean propagator has the spectral representation,
 \begin{equation}
 \tilde{G}_E({\bf k},i\omega_n) =
 \int {d\omega \over 2\pi}\,
 {\rho({\bf k},\omega) \over \omega - i\omega_n}
 \;.
 \end{equation}
 Hence,
 the single particle spectral density is obtained by
 analytically continuing $\tilde{G}_E$ in frequency and
 taking the discontinuity across the real axis,
 \begin{eqnarray}
 \rho(k) & \equiv & \displaystyle
 -i \left(
 \tilde{G}_E({\bf k}, k^0{+}i\epsilon)
 -
 \tilde{G}_E({\bf k}, k^0{-}i\epsilon)
 \right)
 \nonumber\\
 & = & \displaystyle
 {-i\over k^2 + m_{\rm th}^2 + \Sigma(k) }
 + {i\over k^2 + m_{\rm th}^2 + \Sigma(k)^* }
 \nonumber\\
 & = & \displaystyle
 { 2\Sigma_I(k)\over \left| k^2+m_{\rm th}^2+\Sigma(k) \right|^2 }
 \;,
 \end{eqnarray}
 where
 \begin{equation}
 \Sigma_E({\bf k},k^0{+}i\epsilon)
 \equiv
 \Sigma(k)
 \equiv
 \Sigma_R(k) -i \Sigma_I(k)
 \;,
 \end{equation}
 is the analytically continued Euclidean self-energy.

 The real-time propagator
 $\langle {\cal T} (\phi(x)\phi(0)) \rangle$
 used in the cutting rules is
 \begin{equation}
 \tilde{G} (k) \equiv
 \int {d\omega \over 2\pi} \,
 [ 1 {+} n(\omega) ] \,
 \rho(|{\bf k}|, \omega)
 \left(
 {
        2i\omega
         \over
 (k^0)^2 - (\omega - i\epsilon)^2
 }
 \right)
 \;.
 \end{equation}
 By changing the integration variable $\omega$ to $-\omega$ and
 adding the
 two expressions together, the real-time propagator can be re-expressed
 as
 \begin{eqnarray}
 \tilde{G}(k) & = & \displaystyle
 -{i\over 2}\,
 \int {d\omega \over 2\pi} \,
 \rho(|{\bf k}|, \omega)
 \left(
 {1\over \omega-k^0-i\epsilon} + {1\over \omega-k^0+i\epsilon}
 \right)
 +
 {1\over 2}\, \coth(\beta k^0/2)\, \rho(k)
 \nonumber\\
 & = & \displaystyle
 -{i\over 2}\,\left(
 \tilde{G}_E({\bf k},k^0{+}i\epsilon)
 + \tilde{G}_E({\bf k},k^0{-}i\epsilon)
 \right)
 +
 {1\over 2}\, \coth(\beta k^0/2) \, \rho(k)
 \nonumber\\
 & = & \displaystyle
 -i{1+n(k^0) \over k^2 + m_{\rm th}^2 + \Sigma(k)}
 +i{n(k^0) \over k^2 + m_{\rm th}^2 + \Sigma(k)^*}
 \;,
 \label{eq:uncut_prop_app}
 \end{eqnarray}
 where the fact that the spectral density $\rho(k)$ is an odd
 function
 of the frequency is repeatedly used.

 \section{Explicit form of the ladder kernels}
 \label{app:ladder}

 To analyze the zero momentum, small frequency limit of the viscosities,
 a detailed understanding of the structure of the ladder kernel
 ${\cal K}$ is needed.
 The $4{\times}4$ kernel is a product of two factor,
 ${\cal K}={\cal MF}$.
 First, consider the $\lambda\phi^4$ theory kernel.
 The $4{\times}4$ rung matrix is
 \begin{equation}
 {\cal M}(k{-}p)
 \equiv
 \left(
       \begin{array}{cccc}
        -iC(k{-}p) &    0         & 0         & 0 \\
           0       & iC(k{-}p)^*  & 0         & 0 \\
           0       &    0         & L(p{-}k)  & 0 \\
 	   0       &    0         & 0         & L(k{-}p)
       \end{array}
 \right)
 \;
 \end{equation}
 which has entries consisting of the uncut rung
 \begin{equation}
 C(k{-}p)
 \equiv
 -i{\lambda^2\over 2}
 \int {d^4 l \over (2\pi)^4}\,
 \tilde{G}(l{+}k{-}p)\, \tilde{G}(l)
 \label{eq:uncut_rung}
 \;,
 \end{equation}
 and the cut rung
 \begin{equation}
 L(k{-}p)
 \equiv
 {\lambda^2\over 2}
 \int {d^4 l \over (2\pi)^4}\,
 S(l{+}k{-}p)\, S(-l)
 \;.
 \label{eq:cut_rung_app}
 \end{equation}
 As before,
 $\tilde{G}(l)$ is the uncut propagator defined
 in Eq.~(\ref{eq:uncut_prop_app}) and
 $S(l) \equiv [1{+}n(l^0)]\rho(l)$ is the cut propagator.

 The side rail factor of the ladder kernel is given by
 \begin{equation}
 {\cal F}(p,q{-}p) =
 \left(
 	\begin{array}{lccc}
         \tilde{G}(p)\,\tilde{G}(q{-}p) & S(-p)\,S(p{-}q)  &
         \tilde{G}(p)\, S(p{-}q) & S(-p)\, \tilde{G}(q{-}p) \\
         S(p)\,S(q{-}p) & \tilde{G}(-p)^*\,\tilde{G}(p{-}q)^*
 &
 	 S(p)\, \tilde{G}(p{-}q)^* & \tilde{G}(-p)^*\, S(q{-}p) \\
         \tilde{G}(p)\, S(q{-}p) & S(-p)\, \tilde{G}(p{-}q)^*
 &
         \tilde{G}(p)\,\tilde{G}(p{-}q)^* & S(-p)\, S(q{-}p)
 \\
         S(p)\,\tilde{G}(q{-}p) & \tilde{G}(-p)^*\, S(p{-}q)
 &
         S(p)\, S(p{-}q) & \tilde{G}(-p)^*\,\tilde{G}(q{-}p)
       \end{array}
 \right)
 \;.
 \label{eq:side_rail_matrix}
 \end{equation}

 When the external momentum vanishes,
 the matrix ${\cal F}(p,-p)$ can be written as a sum of four
 outer products
 \begin{equation}
 {\cal F}(p,-p) =
 w(p)u^T(p) + h(p)j^T(p) + \kappa(p)\xi^T(p) + \mu(p)\zeta^T(p)
 \;
 \end{equation}
 where
 \begin{eqnarray}
 w^T(p) & \displaystyle \equiv & \displaystyle
 (1,\; 1,\; (1{+}e^{-p^0\beta})/2,\; (1{+}e^{p^0\beta})/2)\,
 [1{+}n(p^0)]\,n(p^0)\,{\rho(p) \over \Sigma_I(p)}\;,
 \\
 u^T(p) & \displaystyle \equiv & \displaystyle
 (1,\; 1,\; (1{+}e^{p^0\beta})/2,\; (1{+}e^{-p^0\beta})/2)\,
 \;,
 \\
 h^T(p) & \displaystyle \equiv & \displaystyle
 (0,\; 0,\; 1/4,\; -e^{p^0\beta}/4)\,
 {\rho(p) \over \Sigma_I(p)}\;,
 \\
 j^T(p) & \displaystyle \equiv & \displaystyle
 (0,\; 0,\; 1,\; -e^{-p^0\beta})\,
 \;,
 \\
 \kappa^T(p) & \displaystyle \equiv & \displaystyle
 (1,\; e^{-p^0\beta},\; e^{-p^0\beta},\; 1)\,
 [1{+}n(p^0)]^2/[p^2+m_{\rm phys}^2+\Sigma(p)]^2\;,
 \\
 \xi^T(p) & \displaystyle \equiv & \displaystyle
 (-1,\; -e^{-p^0\beta},\; -1,\; -e^{-p^0\beta})\,
 \;,
 \\
 \mu^T(p) & \displaystyle \equiv & \displaystyle
 (1,\; e^{p^0\beta},\; 1,\; e^{p^0\beta})\,
 n(p^0)^2/[p^2+m_{\rm phys}^2+\Sigma(p)^*]^2\;,
 \\
 \zeta^T(p) & \displaystyle \equiv & \displaystyle
 (-1,\; -e^{p^0\beta},\; -e^{p^0\beta},\; -1)\,
 \;.
 \end{eqnarray}

 In section~\ref{sec:summation} it is asserted that
 the $hj^T$ part is orthogonal to
 the inhomogeneous terms, and
 if $\bar{\cal F} \equiv {\cal F}{-}hj^T$, then
 $j^T{\cal M}\bar{\cal F}=\bar{\cal F}{\cal M}h =0$.
 Showing that $h(p)$ and $j(p)$ are orthogonal to
 the inhomogeneous terms ${\cal I}_A(p)$ and
 $z_A(p)$ is trivial because both $h(p)$ and $j(p)$ have vanishing
 first and second elements, while the only non-zero elements
 of
 ${\cal I}_A(p)$ and $z_A(p)$ are the first and the second ones,
 respectively.
 To show that $hj^T$ is orthogonal to
 ${\cal M}\bar{\cal F}$ and $\bar{\cal F}{\cal M}$,
 first note that $h(p)$ is orthogonal to $u(p)$,
 $\xi(p)$ and $\zeta(p)$,
 and that $j(p)$ is orthogonal to $w(p)$, $\kappa(p)$
 and $\mu(p)$.
 Hence, $\bar{\cal F}h=j^T\bar{\cal F}=0$.
 Due to the relation
 $L(k{-}p)=e^{(k^0{-}p^0)\beta}L(p{-}k)$,
 $h(p)$ and $j^T(p)$ are ``eigenvectors'' of the rung matrix,
 \begin{eqnarray}
 {\cal M}(k{-}p)h(p)
 & = & \displaystyle
 (0,0,L(p{-}k),-e^{p^0\beta}L(k{-}p))^T {\rho(p)\over 4\Sigma_I(p)}
 \nonumber\\
 & = & \displaystyle
 (0,0,1,-e^{k^0\beta})^T L(p{-}k){\rho(p)\over 4\Sigma_I(p)}
 \nonumber\\
 & \propto & \displaystyle
 h(k)
 \\
 \noalign{\hbox{and}}
 j^T(k){\cal M}(k{-}p)
 & = & \displaystyle
 (0,0,L(p{-}k),-e^{-k^0\beta}L(k{-}p))
 \nonumber\\
 & = & \displaystyle
 (0,0,1,-e^{-p^0\beta})L(p{-}k)
 \nonumber\\
 & \propto & \displaystyle
 j^T(p)
 \;.
 \end{eqnarray}
 Hence,
 $j^T\!{\cal M}\bar{\cal F}=\bar{\cal F}{\cal M}h =0$.

 In section~\ref{sec:summation}, the relation (\ref{eq:Kpp})
 for the $\lambda\phi^4$ theory rung,
 repeated here
 \begin{eqnarray}
 K_{\rm pp}(k,p)
 & = & \displaystyle u^T(k){\cal M}(k{-}p)w(p)
 \nonumber\\
 & = & \displaystyle
 (1{-}e^{-k^0\beta})\,L(k{-}p)
 \,S_{\rm free}(p){\bigg/}2\Sigma_I(p)
 \;,
 \label{eq:K_pp_app}
 \end{eqnarray}
 was important in simplifying
 the expression for the pinching pole contribution.
 This relation can be proven as follows.
 Let
 $\bar{w}(p) \equiv
 (1,1,(1{+}e^{-p^0\beta})/2,(1{+}e^{p^0\beta})/2)$.
 Then,
 \begin{eqnarray}
 u^T(k){\cal M}(k{-}p)\bar{w}(p)
 & = & \displaystyle
 2\,{\rm Im}\,C(k{-}p)
 \nonumber\\
 & & \displaystyle \quad{}
 +(1+e^{p^0\beta})(1+e^{-k^0\beta})
 \left(
 L(k{-}p) + e^{-(p^0{-}k^0)\beta}L(p{-}k)
 \right) {\bigg /}4
 \nonumber\\
 & = & \displaystyle
 (1{-}e^{-k^0\beta})L(k{-}p)(e^{p^0\beta}{-}1)/2
 \;,
 \label{eq:u_M_w_app}
 \end{eqnarray}
 where to obtain the last expression, the optical theorem
 \begin{equation}
 {\rm Im}\,C(k{-}p)
 =
 -(L(k{-}p)+L(p{-}k))/2
 \;,
 \end{equation}
 and a symmetry of the cut rung,
 \begin{eqnarray}
 L(p{-}k)
 & = & \displaystyle
 {\lambda^2\over 2}
 \int {d^4 l \over (2\pi)^4}\,
 S_{\rm free}(l{+}p{-}k)\, S_{\rm free}(-l)
 \nonumber\\
 & = & \displaystyle
 {\lambda^2\over 2}
 e^{(p^0{-}k^0)\beta}
 \int {d^4 l \over (2\pi)^4}\,
 S_{\rm free}(l{+}k{-}p)\, S_{\rm free}(-l)
 \nonumber\\
 & = & \displaystyle
 e^{(p^0{-}k^0)\beta} L(k{-}p)
 \;,
 \label{eq:L_rel}
 \end{eqnarray}
 are used.
 The second expression in Eq.~(\ref{eq:L_rel})
 is a consequence of
 the property of a cut propagator
 $S_{\rm free}(-k) = e^{-k^0\beta}S_{\rm free}(k)$.
 When combined with the remaining factors forming $w(p)$,
 this yields Eq.~(\ref{eq:K_pp_app}).
 Note that the real part of $C(k{-}p)$
 makes no contribution to the pinching pole part of the rung
 matrix.

 For the $g\phi^3$ rung matrix representing the straight single
 line
 rungs, exactly the same argument applies to yield
 \begin{eqnarray}
 K_{\rm line}(k,p)
 & = & \displaystyle u^T(k){\cal M}_{\rm line}(k{-}p)w(p)
 \nonumber\\
 & = & \displaystyle
 g^2\,(1{-}e^{-k^0\beta})\,S(k{-}p)
 \,S_{\rm free}(p){\bigg/}2\Sigma_I(p)
 \;,
 \end{eqnarray}
 where
 the rung matrix ${\cal M}_{\rm line}(k{-}p)$
 is now given by replacing $-iC$ with
 $-g^2\tilde{G}$, and $L$ with $g^2S$.
 And also in place of the optical theorem for the imaginary part
 of $C$,
 the straight rungs satisfy
 \begin{equation}
 {\rm Re}\,\tilde{G}(k) =
 (S(k)+S(-k))/2
 \;.
 \end{equation}

\begin{figure}
 \setlength {\unitlength}{1cm}
 \begin{picture}(0,0)
 \end{picture}
\vbox
    {%
    \begin {center}
	\leavevmode
	\def\epsfsize	#1#2{0.3#1}
	\epsfbox {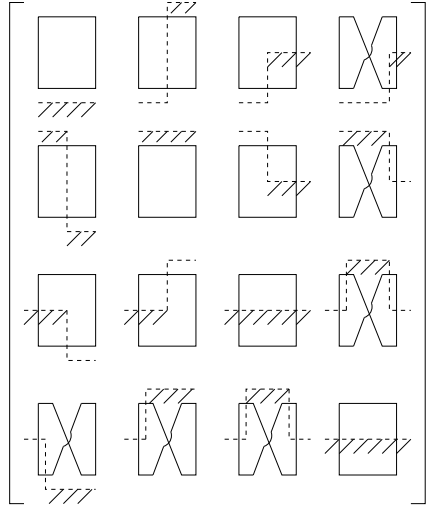}
    \end {center}
    \caption{%
 Diagrammatic representation of the rung matrix
 ${\cal M}_{\rm full-box}(k{-}p)$.
}
 \label{fig:box_mtx}
    }%
\end {figure}
 For the analogous relation for the box sub-diagram rung
 (\ref{eq:u_M_box_w}), reproduced
 in Eq.~(\ref{eq:u_M_box_w_again}) below,
 it is convenient to consider the ``rung'' matrix
 generated by (full) box sub-diagram
 \begin{equation}
 {\cal M}_{\hbox{\scriptsize{full-box}}}(k{-}p)
 \equiv
 \int{d^4 l\over (2\pi)^4}\,
 {\cal M}_{\hbox{\scriptsize{line}}}(k{-}l)
 {\cal F}(l,-l)
 {\cal M}_{\hbox{\scriptsize{line}}}(l{-}p)
 \;,
 \end{equation}
 illustrated in Fig.~\ref{fig:box_mtx}.
 The box sub-diagram rung
 ${\cal M}_{\hbox{\scriptsize{box}}}^{(ij)}(k{-}p)$
 is the non-pinching pole contribution of
 ${\cal M}_{\hbox{\scriptsize{full-box}}}^{(ij)}(k{-}p)$.
 The separation of the pinching pole contribution
 and non-pinching pole contribution can be made after contracting
 with
 $u$ and $w$.
 Since the line sub-diagram rung matrix is diagonal, contracting
 with
 $u(k)$ and $w(p)$ produces
 \begin{equation}
 u^T(k){\cal M}_{\hbox{\scriptsize{full-box}}}(k{-}p)\bar{w}(p)
 =
 \int {d^4 l\over (2\pi)^4}\,
 u_i(k){\cal M}_{\rm line}^{(ii)}(k{-}l)
 {\cal F}^{(ij)}(l,-l)
 {\cal M}_{\rm line}^{(jj)}(l{-}p)\bar{w}_j(p)
 \;.
 \label{eq:u_M_box_w_again}
 \end{equation}

 To simplify this expression, first, note that
 the elements of the rung matrix satisfy,
 \begin{mathletters}
 \begin{eqnarray}
 {\cal M}_{\hbox{\scriptsize{full-box}}}^{(14)}(\underline{k},\underline{p})
 & = & \displaystyle
 e^{-\underline{p}^0\beta}\,
 {\cal M}_{\hbox{\scriptsize{full-box}}}^{(13)}(\underline{k},\underline{p})
 \;,
 \\
 {\cal M}_{\hbox{\scriptsize{full-box}}}^{(24)}(\underline{k},\underline{p})
 & = & \displaystyle
 e^{-\underline{p}^0\beta}\,
 {\cal M}_{\hbox{\scriptsize{full-box}}}^{(23)}(\underline{k},\underline{p})
 \;,
 \\
 {\cal M}_{\hbox{\scriptsize{full-box}}}^{(41)}(\underline{k},\underline{p})
 & = & \displaystyle
 e^{\underline{k}^0\beta}\,
 {\cal M}_{\hbox{\scriptsize{full-box}}}^{(31)}(\underline{k},\underline{p})
 \;,
 \\
 {\cal M}_{\hbox{\scriptsize{full-box}}}^{(42)}(\underline{k},\underline{p})
 & = & \displaystyle
 e^{\underline{k}^0\beta}\,
 {\cal M}_{\hbox{\scriptsize{full-box}}}^{(32)}(\underline{k},\underline{p})
 \;,
 \\
 {\cal M}_{\hbox{\scriptsize{full-box}}}^{(43)}(\underline{k},\underline{p})
 & = & \displaystyle
 e^{(\underline{k}^0+\underline{p}^0)\beta}\,
 {\cal M}_{\hbox{\scriptsize{full-box}}}^{(34)}(\underline{k},\underline{p})
 \;,
 \\
 {\cal M}_{\hbox{\scriptsize{full-box}}}^{(33)}(\underline{k},\underline{p})
 & = & \displaystyle
 e^{(\underline{p}^0-\underline{k}^0)\beta}\,
 {\cal M}_{\hbox{\scriptsize{full-box}}}^{(44)}(\underline{k},\underline{p})
 \;,
 \end{eqnarray}
 \label{eq:rel_among_rungs}
 \end{mathletters}%
 \hspace{-0.5ex}%
 due to the corresponding property of the cut propagator
 $S(-p) = e^{-p^0\beta}S(p)$.
 Using the facts that
 the imaginary part of a uncut diagram  is given by
 the sum of all possible cut diagrams divided by a factor of
 $(-2)$
 ({\it c.f.}~Eq.~(\ref{eq:optical_thm})),
 one finds
 \begin{eqnarray}
 \lefteqn{%
 u^T(\underline{k})
 {\cal M}_{\hbox{\scriptsize{full-box}}}(\underline{k},\underline{p})
 \bar{w}(\underline{p}) =}
 & &
 \nonumber\\
 & & \displaystyle \qquad\qquad
 (1{-}e^{-\underline{k}^0\beta})
 \left(
 {\cal M}_{\hbox{\scriptsize{full-box}}}^{(44)}(\underline{k},\underline{p})
 -
 e^{k^0\beta}
 {\cal M}_{\hbox{\scriptsize{full-box}}}^{(34)}(\underline{k},\underline{p})
 \right)
 (e^{\underline{p}^0\beta}-1)
 {\bigg /}2
 \label{eq:u_M_box_w_final}
 \;.
 \end{eqnarray}
 Note that since the frequency integral is not yet carried out,
 this relation is valid for both the pinching pole contribution
 and
 for the non-pinching pole contribution.,

 For the full $g\phi^3{+}\lambda\phi^4$ theory rung matrix,
 the relations (\ref{eq:rel_among_rungs})
 can be again shown to hold.
 Hence, so does the relation (\ref{eq:K_full})
 \begin{equation}
 u^T(k){\cal M}_{\rm full}(k,p)w(p)
 =
 (1{-}e^{-k^0\beta})\,L_{\rm full}(k,p)
 \,S_{\rm free}(p){\bigg/}2\Sigma_I(p)
 \;,
 \end{equation}
 where
 \begin{equation}
 L_{\rm full}(k,p)
 =
 \left(
 {\cal M}_{\hbox{\scriptsize{full}}}^{(44)}(\underline{k},\underline{p})
 -
 e^{k^0\beta}
 {\cal M}_{\hbox{\scriptsize{full}}}^{(34)}(\underline{k},\underline{p})
 \right)
 \;.
 \end{equation}

 \section{Zero modes of ladder kernels}
 \label{app:zero_modes}

 The integral operator $(1{-}{\cal K})$ where ${\cal K}={\cal
 MF}$, has
 four zero modes as $\omega\to 0$ corresponding to the momentum-energy
 conservation.
 To see this, one must know the contribution of
 an momentum-energy density $T^{0\mu}$ insertion.
 The standard stress-energy tensor is given by
 \begin{equation}
 {T}^{\mu\nu} =
 \partial^{\mu}{\phi}\partial^{\nu}{\phi}
 +
 g^{\mu\nu}{\cal L}
 \;,
 \label{eq:stress_energy_tensor_again}
 \end{equation}
 where the Lagrangian is given in Eq.~(\ref{eq:full_Lagrangian}).
 The momentum density
 $T^{0i} = \partial^0\phi\partial^i\phi$ contains only the ``kinetic''
 part, and hence its contribution in the zero spatial momentum
 limit
 is simply
 \begin{equation}
 z^{i}(k,\omega{-}k)
 =
 (0,k^i(k^0{-}\omega/2),0,0)
 \;,
 \end{equation}
 where $k$ is the loop momentum flowing through the lines
 connected to $T^{0i}$ and $\omega$ is the external frequency.

 The energy density on the other hand contains both the kinetic
 part
 $\partial^0\phi\partial^0\phi$ and the Lagrangian part.
 The contribution from the kinetic part of $T^{00}$ in the zero
 spatial
 momentum limit is again simple,
 \begin{equation}
 z^{0}_{\rm kin.}(k,\omega{-}k)
 =
 (0,k^0(k^0{-}\omega/2),0,0)
 \;.
 \end{equation}
 The contribution of an Lagrangian insertion can be calculated
 by
 applying the method used in the main text
 ({\it c.f.}~Eq.~(\ref{eq:Lagrangian_decomposed})).
 The result up to ${\cal O}(\lambda^2)$ is
 \begin{eqnarray}
 z^{0}_{\Sigma}(k,\omega{-}k)
 & = & \displaystyle
 -
  (0,\, \bar{\Sigma} (k{-}\omega)^*,\,
   0,\, i[1{+}n(w{-}k^0)] \Sigma_I (w{-}k))
 \\
 \noalign{\hbox{where}}
 \bar{\Sigma} (k)^* & \equiv & \displaystyle
 {1\over 2}
 \left( [1{+}n(k^0)]\Sigma (k)^*
 - n(k^0)\Sigma (k) \right)
 \;,
 \end{eqnarray}
 is the uncut two-loop self-energy.

 Given the form of $z^{i}$ and $z^0=z^0_{\rm kin.}{+}z^0_{\Sigma}$,
 a tedious but straightforward calculation yields,
 \begin{eqnarray}
 {\cal V}^{\mu}(k, \omega{-}k)
 & \equiv & \displaystyle
 z^{\mu}(k, \omega{-}k)
 {\cal F}(k, \omega{-}k)
 \nonumber\\
 & \displaystyle = & \displaystyle
 g^{\mu}(k, \omega{-}k)/\omega
 -
 \int {d^4 p \over (2\pi)^4}\,
 g^{\mu}(p, \omega{-}p) \, {\cal M}(p{-}k)\,{\cal F}(k,\omega{-}k)/\omega
 \nonumber\\
 & \displaystyle = & \displaystyle
 g^{\mu}(1 - {\cal K})/\omega
 \;,
 \end{eqnarray}
 where, the row vector
 $g^{\mu}$ is given by
 \begin{eqnarray}
 g^{\mu}(k, \omega{-}k)
 & \displaystyle = & \displaystyle
 k^\mu(0, i(\tilde{G}(k{-}\omega)^* - \tilde{G}(k)^*),
           -iS(k), iS(\omega{-}k))
 \;.
 \end{eqnarray}

 Since $z^\mu {\cal F}$
 is finite as $\omega$ goes to zero,
 \begin{equation}
 \lim_{\omega\to 0}\, g^\mu (1 - {\cal K}) = 0
 \;.
 \end{equation}
 Hence, in the zero external momentum, and zero frequency limit,
 the operator $(1 - {\cal K})$ has four left zero modes given
 by
 $g^\mu$.%
\footnote{%
 	 The right zero modes of $(1-{\cal F}{\cal M})$
	 $f^\mu$ can be also obtained in an entirely similar way.
	 They are
	 \[
	 f^\mu(k,\omega{-}k) =
	 k^\mu (i(\tilde{G}(k){-}\tilde{G}(w{-}k)),0,-iS(w{-}k), iS(k))
	 \;.
	 \]
}
 Note that the first element of $g^\mu$ is always zero for any
 $\omega$.
 Hence, trivially, $g^\mu{\cal I}_A = 0$ for all $\omega$.
 This implies that the inhomogeneous term is orthogonal to the
 zero modes
 of the operator $\displaystyle\lim_{\omega\to 0}(1-{\cal K})$.

 Since ${\cal V}^{\mu}$ corresponds to
 an insertion of the energy-momentum tensor,
 these zero mode can be used to verify that correlation functions
 of
 the energy or momentum density
 vanish in the zero spatial momentum limit.
 Full correlation functions involving $T^{\mu 0}$
 must vanish as the momentum goes to zero
 since the conservation equation relates the time derivative
 of
 a conserved ``charge'' to the divergence of its current.
 Hence, for example, the correlation function of two momentum
 densities
 must behave like ${\bf k}^2$ in small momentum, finite
 frequency limit.
 In terms of the effective vertex, this implies that
 \begin{equation}
 {\cal V}^{\mu} {\cal D}_A = 0
 \;,
 \end{equation}
 for an arbitrary (non-singular) external operator $A$.
 Since, the effective vertex ${\cal D}_A$ can be expressed as
 \begin{equation}
 {\cal D}_A =
 {1\over 1-{\cal K}} {\cal I}_A
 \;,
 \end{equation}
 for momentum density correlation functions,
 \begin{eqnarray}
 {\cal V}^\mu {\cal D}_A
 & = & \displaystyle
 g^\mu (1-{\cal K}) {1\over 1-{\cal K}} {\cal I}_A
 \nonumber\\
 & = & \displaystyle
 g^\mu {\cal I}_A/\omega =0
 \;,
 \end{eqnarray}
 since as explained earlier
 $g^\mu {\cal I}_A = 0$.

 For the pinching pole part of the integral equation,
 \begin{equation}
 I_B(\underline{k}) = D_B(\underline{k}) -
 (1{-}e^{-\underline{k}^0\beta})
 \int {d^4 p\over (2\pi)^4}\,
       L_{\rm full}(\underline{k}{-}p)\,
 S_{\rm free}(p)\,
 {D_B(p) \over 2\Sigma_I(p)}\,
 \;,
 \end{equation}
 the (left) zero modes are
 \begin{eqnarray}
 \bar{b}_{\mu}(k)
 & \displaystyle = & \displaystyle
 k_{\mu}
 \,[1{+}n(k^0)]\,S_{\rm free}(-k)\,
 \;.
 \end{eqnarray}
 To verify this, note that $[1{+}n(k^0)]=1/(1{-}e^{-k^0\beta})$
 cancels
 the prefactor $(1{-}e^{-k^0\beta})$,
 and
 \begin{eqnarray}
 \int{d^4 k\over (2\pi)^4}\,
 k_{\mu}\,
 S_{\rm free}(-k)\,
 L_{\rm full}(k{-}p)
 & = & \displaystyle
 {1\over 2}
 \int{d^4 k\over (2\pi)^4}\,
 {d^4 l_1 \over (2\pi)^4}\, {d^4 l_2 \over (2\pi)^4}\,
 (2\pi)^4\delta(l_1{+}p{-}l_2{-}k)
 \left| {\cal T}(l_1,p;l_2,k) \right|^2
 \nonumber\\
 & & \displaystyle \qquad\quad{}\times
 k_{\mu}\,
 S_{\rm free}(-k)\,
 S_{\rm free}(l_1)\, S_{\rm free}(-l_2)\,
 \nonumber\\
 & = & \displaystyle
 {p_{\mu}\over 3}\,
 \int{d^4 k\over (2\pi)^4}\,
 S_{\rm free}(-k)\,
 L_{\rm full}(k{-}p)
 \nonumber\\
 & = & \displaystyle
 2p_\mu \Sigma_I(p)\,n(p^0)
 \;,
 \end{eqnarray}
 where to obtain the second line, the original expression is
 averaged
 with two equivalent expressions differing by
 the $k\leftrightarrow -l_1$, or $k\leftrightarrow l_2$
 labeling changes.
 The scattering amplitude is, as before,
 \begin{equation}
 {\cal T}(\underline{l}_1,\underline{p};\underline{l}_2,\underline{k})
 \equiv
 \left(
   \lambda
   - g^2\,
   \left(
   G_R(\underline{l}_1{+}\underline{p})
  +G_R(\underline{l}_1{-}\underline{k})
  +G_R(\underline{l}_1{-}\underline{l}_2)
   \right)
 \right)
 \;.
 \end{equation}
 Hence,
 when $\bar{b}_\mu(k)$
 is applied to the right hand side of the above integral
 equation, the two terms in the right hand side
 cancel each other exactly.

 \section{Stress-energy tensors and the speed of sound}
 \label{app:temperature_integrals}

 In this appendix,
 the equilibrium $\lambda\phi^4$ theory
 stress-energy tensor is calculated,
 including the ${\cal O}(\lambda T^4)$ correction.
 When the cubic interaction term is added to the Lagrangian,
 to this order, one only has to make the change
 \begin{equation}
 \lambda \to \lambda - {g^2\over m_{\rm th}^2}
 \;.
 \end{equation}
 In equilibrium, the stress-energy tensor is diagonal in the
 comoving
 frame,
 \begin{equation}
 \langle T^{\mu\nu} \rangle_{\rm eq}
 = {\rm diag}(\varepsilon, {\cal P}, {\cal P}, {\cal P})
 \;.
 \end{equation}
 Then due to the equilibrium thermodynamic identity
 \begin{equation}
 \varepsilon =
 T^2{\partial \over \partial T}
 \left( {1\over T}{\cal P} \right)
 \;,
 \label{eq:E_in_terms_of_P}
 \end{equation}
 only the pressure needs to be calculated.

 \begin{figure}
 \setlength {\unitlength}{1cm}
\vbox
    {%
    \begin {center}
 \begin{picture}(0,0)
 \end{picture}
	\leavevmode
	\def\epsfsize	#1#2{0.4#1}
	\epsfbox {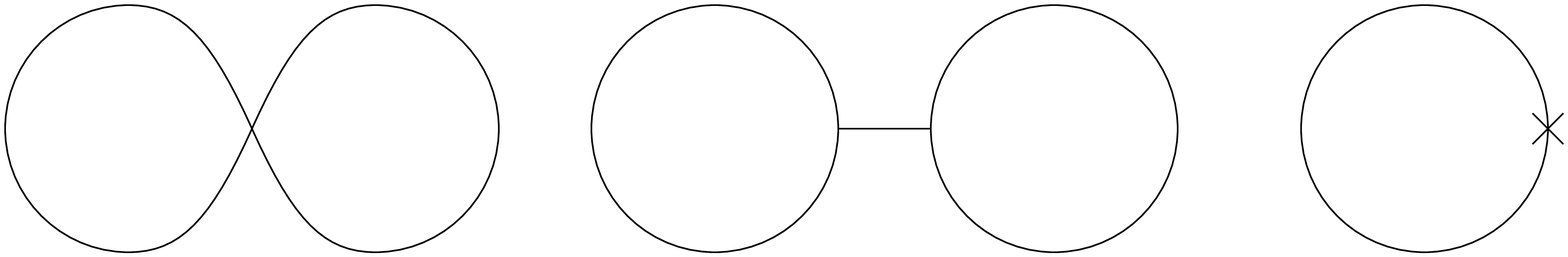}
    \end {center}
 \caption{%
 The lowest order connected vacuum diagrams for the thermal correction
 to the pressure in the $g\phi^3{+}\lambda\phi^4$ theory.
 The cross indicate an insertion of $\delta\mth^2\phi^2$.
 }
 \label{fig:pressure}
 }
 \end{figure}
 The easiest way to calculate
 the correction to the free-particle pressure
 is to sum the contribution of all connected vacuum graphs
 (see, for example, Ref.\cite{Kapusta}).
 To express the pressure in terms of the thermal mass,
 the Lagrangian may be rewritten as
 \begin{equation}
 -{\cal L} =
 {1 \over 2}
 \phi ( -\partial_{\tau}^2 - \nabla^2 + m_{\rm th}^2) \phi
 +
 {\lambda \over 4!} \phi^4
 - {\delta m_{\rm th}^2\over 2} \phi^2
 \;,
 \label{eq:thermal_Lagrangian}
 \end{equation}
 where
 $m_{\rm th}^2 = m_0^2{+}\delta m_{\rm th}^2$.
 To the lowest order, the zero temperature mass $m_0$ may be
 identified
 as the physical mass $m_{\rm phys}$.
 Then the first order correction to the free-particle pressure
 arises from the diagrams
 in Fig.~\ref{fig:pressure} where the term
 ${1\over 2}\delta m_{\rm th}^2\phi^2$ in the Lagrangian is treated
 as an
 additional interaction term.
 In Euclidean space, the free-particle pressure
 through ${\cal O}(\lambda)$ is,
 \begin{eqnarray}
 {\cal P}
 & = & \displaystyle
 T\sum_{n}\int{d^3 {\bf k}\over (2\pi)^3}\,
 {{\bf k}^2\over 3}
 {1 \over \omega_n^2 + {\bf k}^2 + m_{\rm th}^2}
 \nonumber\\
 & & \displaystyle {}
 - {\lambda \over 8}
 \left(
 T\sum_{n}\int{d^3 {\bf k}\over (2\pi)^3}\,
 {1 \over \omega_n^2 + {\bf k}^2 + m_{\rm th}^2}
 \right)^2
 \nonumber\\
 & & \displaystyle {}
 + {\delta m_{\rm th}^2 \over 2}
 T\sum_{n}\int{d^3 {\bf k}\over (2\pi)^3}\,
 {1 \over \omega_n^2 + {\bf k}^2 + m_{\rm th}^2}
 \;,
 \end{eqnarray}
 where $\omega_n$ is the usual discrete Euclidean frequency.

 Using standard techniques
 (see, for example, Ref.\cite{GrossPY}),
 the sum over $\omega_n$ can be separated into
 the zero temperature contribution
 and the non-zero temperature contribution.
 At high temperature,
 the (renormalized) ${\cal O}(m_{\rm phys}^4)$
 zero temperature contribution
 is negligible compared to the ${\cal O}(\lambda T^4)$
 thermal correction.
 Hence, to calculate the pressure through ${\cal O}(\lambda T^4)$,
 only the non-zero temperature contribution needs to be examined.
 For the one-loop diagrams in Fig.~\ref{fig:pressure},
 the non-zero temperature contribution
 is obtained by a simple replacement of
 \begin{equation}
 T\sum_{n=0}^{\infty}\,
 {1\over \omega_n^2+E_k^2} \to {n(E_k) \over E_k}
 \;,
 \end{equation}
 where $n(E_k)$ is the usual Bose factor, and the energy here
 is
 $E_k^2 \equiv {\bf k}^2{+}m_{\rm th}^2$.
 The thermal part of the pressure is then,
 \begin{equation}
 {\cal P}_{\rm th}
 =
 \int{d^3 {\bf k}\over (2\pi)^3 E_k}\, n(E_k)\,
 \left(
 {\textstyle{1\over 3}}{\bf k^2}
 +
 {\textstyle{1\over 4}}\delta m_{\rm th}^2
 \right)
 \;,
 \label{eq:P_th}
 \end{equation}
 where the fact that
 \begin{equation}
 \delta m_{\rm th}^2 =
 {\lambda \over 2}
 \int{d^3 {\bf k} \over (2\pi)^2 E_k}\,n(E_k)
 \;
 \label{eq:delta_mth_app}
 \end{equation}
 to lowest order is used to simplify the expression.

 To evaluate the pressure (\ref{eq:P_th}),
 consider first the
 the one-loop thermal correction to the mass (\ref{eq:delta_mth_app}).
 Since $\delta m_{\rm th}^2$ involves an explicit factor of $\lambda$,
 only
 the leading order term is needed.
 By setting the mass in the integrand to zero,
 the leading order contribution to the integral
 in Eq.~(\ref{eq:delta_mth_app}) can be evaluated as
 \begin{eqnarray}
 \delta m_{\rm th}^2
 & = & \displaystyle
 {\lambda \over 2}
 \int{d^3 {\bf k}\over (2\pi)^3 |{\bf k}|}\, n(|{\bf k}|)
 \nonumber\\
 & = & \displaystyle
 {\lambda\over 4\pi^2}
 \int_{0}^{\infty} d|{\bf k}|\, |{\bf k}|\,
 \sum_{s=1}^{\infty}\, e^{-s |{\bf k}| \beta}
 \nonumber\\
 & = & \displaystyle
 {\lambda \over 4\pi^2} T^2\,
 \sum_{s=1}^{\infty}\, {1\over s^2}
 = {\lambda \over 24} T^2
 \;,
 \label{eq:eval_thermal_one_loop}
 \end{eqnarray}
 neglecting
 sub-leading terms suppressed by ${\cal O}(\sqrt{\lambda})$.

 Next, consider the integral
 \begin{equation}
 {\cal P}_0 =
 \int{d^3 {\bf k}\over (2\pi)^3 E_k}\, n(E_k)\,
 {1\over 3}{\bf k}^2
 \;.
 \end{equation}
 The integrand depends only on the magnitude of the loop momentum
 $|{\bf k}|$.  Changing the integration variable
 $|{\bf k}|$ to $m_{\rm th}\sinh\theta$ yields
 \begin{eqnarray}
 {\cal P}_0
 & = & \displaystyle
 {m_{\rm th}^4\over 6\pi^2}
 \int_{0}^{\infty} d\theta \, \sinh^4\theta\,
 \sum_{s=1}^{\infty}\,e^{-s \beta m_{\rm th}\cosh\theta}
 \nonumber\\
 & = & \displaystyle
 {m_{\rm th}^4\over 2\pi^2}
 \sum_{s=1}^{\infty}\,
 \left(
 {T^2\over s^2 m_{\rm th}^2} K_2(s \beta m_{\rm th})
 +
 {1\over 8} K_0(s \beta m_{\rm th})
 \right)
 \;,
 \label{eq:pressure_0}
 \end{eqnarray}
 where $K_n(x)$ is the modified Bessel function of order $n$.
 expression in Eq.~(\ref{eq:pressure_0})
 is obtained by integrating by part, and
 using the standard expression for the Bessel
 functions\cite{math_table}.
 The leading and the next-to-leading
 terms of ${\cal P}_0$ in the small parameter
 $\beta m_{\rm th}$ can be calculated by
 using the small $x$ expansion of $K_2(x)$,
 \begin{equation}
 K_2(x) = {2\over x^2}
 		- {1\over 2} + {\cal O}(x\ln x)
 \;.
 \label{eq:bessel_2}
 \end{equation}
 Substituting Eq.~(\ref{eq:bessel_2}) into Eq.~(\ref{eq:pressure_0})
 and performing the elementary sums yield,
 \begin{equation}
 {\cal P}_0
 =
 {\pi^2 T^4 \over 90} - {m_{\rm th}^2 T^2 \over 24}
 + {\cal O}(T m_{\rm th}^3)
 \;.
 \label{eq:eval_P_0}
 \end{equation}
 All together, the thermal pressure up to ${\cal O}(\lambda T^4)$
 is
 \begin{eqnarray}
 {\cal P}_{\rm th}
 & = & \displaystyle
 {\pi^2 T^4 \over 90}
 - {m_{\rm th}^2 T^2 \over 24}
 + {T^2\over 48}\left( {\lambda T^2\over 24} \right)
 \nonumber\\
 & = & \displaystyle
 {\pi^2 T^4 \over 90}
 - {m_{\rm phys}^2 T^2 \over 24}
 - {T^2\over 48}\left( {\lambda T^2\over 24} \right)
 \;.
 \label{eq:eval_P_th}
 \end{eqnarray}
 {}From this, the thermal energy density
 (\ref{eq:E_in_terms_of_P}) can be easily calculated,
 \begin{eqnarray}
 \varepsilon_{\rm th}
 & = & \displaystyle
 3 {\cal P}_{\rm th}
 + {m_{\rm phys}^2 T^2 \over 12}
 \nonumber\\
 & = & \displaystyle
 \int{d^3 {\bf k}\over (2\pi)^3 E_k}\, n(E_k)\,
 \left( E_k^2 - {\textstyle{1\over 4}}\delta m_{\rm th}^2 \right)
 \;.
 \label{eq:eval_E_th}
 \end{eqnarray}
 To obtain the last expression,
 Eqs.~(\ref{eq:P_th}) and (\ref{eq:delta_mth_app}) are used.
 {}From the expression of the pressures (\ref{eq:P_th})
 and the energy density (\ref{eq:eval_E_th}),
 the thermal expectation of the  stress-energy tensor
 up to ${\cal O}(\lambda)$ can be compactly written as
 \begin{equation}
 \langle T^{\mu\nu} \rangle_{\rm eq}
 =
 \int{d^3 {\bf k}\over (2\pi)^3 E_k}\, n(E_k)\,
 \left(
 \underline{k}^{\mu}\underline{k}^{\nu}{+}
 {\textstyle{1\over 4}} g^{\mu\nu} \delta m_{\rm th}^2
 \right)
 \;.
 \end{equation}

 {}From the expressions (\ref{eq:eval_P_th})
 and (\ref{eq:eval_E_th}), the speed of sound
 $v_{\rm s}^2 =
 {\partial {\cal P}/\partial \varepsilon}$
 can now be straightforwardly calculated
 up to ${\cal O}(m_{\rm phys}^2/T^2)$,
 \begin{equation}
 v_{\rm s}^2
 =
 {(\partial {\cal P}/\partial T) \over
  (\partial \varepsilon /\partial T)}
 =
 {1\over 3} - {5 m_{\rm phys}^2 \over 12 \pi^2 T^2}
 \;.
 \end{equation}

 \section{Near Soft and Collinear Singularities}
 \label{app:soft_contributions}

 In this appendix, the soft momentum behavior of diagrams contributing
 to the calculation of the viscosities is briefly examined.
 In section~\ref{sec:classification}, it is asserted that
 the near soft singularities do not affect the
 power counting described in that section.
 Here a brief demonstration is presented.
 The temperature is assumed to satisfy $T \gg m_{\rm phys}$ so
 that
 $m_{\rm th}^2 = {\cal O}(\lambda T^2)$.

 To the leading order, the effect of adding one more
 rung ${\cal MF}$ is to provide the integrand one more factor
 of
 \begin{equation}
 u^T(\underline{k}){\cal M}_{\rm full}(\underline{k}{-}p)w(p)
 =
 (1{-}e^{-k^0\beta})\,L_{\rm full}(\underline{k}{-}p)\,[1{+}n(p^0)]\,
 S_{\rm free}(p){\bigg/}2\Sigma_I(p)
 \;,
 \label{eq:p_M_k}
 \end{equation}
 together with an additional integration over the 4-momentum
 $p$.
 Here $L_{\rm full}(\underline{k}{-}p)$
 consists of the cut diagrams such as those in Fig.~\ref{fig:rungs34}.
 In section~\ref{sec:classification}, this additional factor
 is regarded
 as ${\cal O}(1)$ since the inverse powers of $\lambda$ from the
 inverse of the self-energy is canceled by the explicit
 ${\cal O}(\lambda^2)$ scattering amplitude squared contained in
 $L_{\rm full}$.
 When the (on-shell) momenta $\underline{k}$ and $\underline{p}$ are soft,
 this argument could be upset if
 (a) the size of the self-energy $\Sigma_I(\underline{p})$ is {\em smaller}
 than ${\cal O}(\lambda^2 T^2)$,
 (b) the would-be soft singularities
 (factors of ${\cal O}(T/m_{\rm th})$) from the Bose factors
 are {\em not} compensated by the small momentum space volume,
 or
 (c) the non-pinching pole contribution of the side rail matrix
 is comparable in size to the pinching pole contribution.

 To see that none of these possibilities actually occur,
 first consider the size of the thermal ``scattering amplitude''
 which is contained in the expressions for
 $\Sigma_I(\underline{p})$ and $L_{\rm full}(\underline{k},\underline{p})$,
 \begin{equation}
 {\cal T}(\underline{l}_1,\underline{p};\underline{l}_2,\underline{k})
 \equiv
   \lambda
   - g^2\,
   \left(
   G_R(\underline{l}_1{+}\underline{p})
  +G_R(\underline{l}_1{-}\underline{k})
  +G_R(\underline{l}_1{-}\underline{l}_2)
   \right)
 \;.
 \label{eq:almost_scattering_amp_app}
 \end{equation}
 As before, the arguments  of the propagators
 in Eq.~(\ref{eq:almost_scattering_amp_app})
 are all combinations of two on-shell momenta.
 The 4-momentum squared of the sum of two on-shell momenta
 satisfies
 \begin{eqnarray}
 \left| ({\bf k}{\pm}{\bf p})^2 - (E_k{+}E_p)^2 \right|
 & = & \displaystyle
 2(E_kE_p \mp {\bf k}{\cdot}{\bf p} + m_{\rm th}^2)
 \nonumber\\
 & \ge & \displaystyle
 2(E_kE_p - |{\bf k}||{\bf p}| - m_{\rm th}^2)  + 4m_{\rm th}^2
 \nonumber\\
 & \ge & \displaystyle 4m_{\rm th}^2
 \;,
 \end{eqnarray}
 for all ${\bf k}$ and ${\bf p}$ since
 \begin{eqnarray}
 E_kE_p{-}|{\bf k}||{\bf p}|{-}m_{\rm th}^2
 & = & \displaystyle
 \left(
 (E_kE_p)^2 {-} ( |{\bf k}||{\bf p}| {+} m_{\rm th}^2 )^2
 \right)
 \left/(E_kE_p{+}|{\bf k}||{\bf p}|{+}m_{\rm th}^2) \right.
 \nonumber\\
 & = & \displaystyle
 m_{\rm th}^2(|{\bf k}|{-}|{\bf p}|)^2
 \left/(E_kE_p{+}|{\bf k}||{\bf p}|{+}m_{\rm th}^2) \right.
 \nonumber\\
 & \ge & \displaystyle 0
 \;.
 \end{eqnarray}
 Similarly, the 4-momentum squared of the difference of
 two on-shell momenta is
 \begin{eqnarray}
 \left| ({\bf k}{\pm}{\bf p})^2 - (E_k{-}E_p)^2 \right|
 & = & \displaystyle
 2(E_kE_p \mp {\bf k}{\cdot}{\bf p}  - m_{\rm th}^2)
 \nonumber\\
 & \ge & \displaystyle
 2(E_kE_p - |{\bf k}||{\bf p}| - m_{\rm th}^2)
 \nonumber\\
 & \ge & \displaystyle 0
 \;,
 \end{eqnarray}
 for all ${\bf k}$ and ${\bf p}$.
 Hence each propagator
 in (\ref{eq:almost_scattering_amp_app}) is bounded by
 $1/m_{\rm th}^2$
 for all on-shell momenta $\underline{l}_i$, $\underline{k}$ and
 $\underline{p}$, so that
 \begin{equation}
 |{\cal T}(\underline{l}_1,\underline{p};\underline{l}_2,\underline{k})|
 =
 {\cal O}(\lambda)+{\cal O}(g^2/m_{\rm th}^2) = {\cal O}(\lambda)
 \;,
 \label{eq:estim_T}
 \end{equation}
 since by assumption $g^2={\cal O}(\lambda m_{\rm phys}^2)$
 and $(m_{\rm phys}^2/m_{\rm th}^2)\le 1$.

 With this estimate of the size of $|{\cal T}|^2$,
 the size of the $g\phi^3{+}\lambda\phi^4$ theory cut rung,
 \begin{equation}
 L_{\rm full}(\underline{k},\underline{p})
 =
 {1\over 2}
 \int {d^4 l_1 \over (2\pi)^4}\, {d^4 l_2 \over (2\pi)^4}\,
 S_{\rm free}(l_1)\, S_{\rm free}(-l_2)\,
 (2\pi)^4\delta(l_1{-}l_2{+}\underline{p}{-}\underline{k})
 \left|
 {\cal T}(l_1,\underline{p};l_2,\underline{k})
 \right|^2
 \;,
 \end{equation}
 at soft $\underline{k}$ and $\underline{p}$ can be determined.
 When the external momenta $\underline{k}$ and $\underline{p}$ are
 ${\cal O}(m_{\rm th})$,
 the two $\delta$-functions in the cut propagators can be satisfied
 by ${\cal O}(m_{\rm th})$ loop momenta.
 Hence, the two momentum integrations
 over these two $\delta$-functions
 and the energy-momentum conserving $\delta$-function are ${\cal O}(1)$.
 Consequently, the size of
 $L_{\rm full}(\underline{k},\underline{p})$ at soft $\underline{k}$ and
 $\underline{p}$ is determined by the size of the scattering amplitude
 squared $|{\cal T}|^2 = {\cal O}(\lambda^2)$
 and two ${\cal O}(T/m_{\rm th})$ factors from the statistical factors.
 Hence, all combined,
 \begin{equation}
 L_{\rm full}(m_{\rm th})
 ={\cal O}(\lambda^2 T^2/m_{\rm th}^2)
 ={\cal O}(\lambda)
 \;,
 \end{equation}
 since $m_{\rm th}^2 = {\cal O}(\lambda T^2)$.

 Given this result, one may also estimate
 the imaginary part of the self-energy
 \begin{equation}
 \Sigma_I(\underline{p})
 =
 {1\over 6}
 (1{-}e^{-\underline{p}^0\beta})
 \int {d^4 k \over (2\pi)^4}\,
 S_{\rm free}(k)\,
 L_{\rm full}(\underline{p},k)
 \;,
 \end{equation}
 at soft external momenta $\underline{p}$.
 When $\underline{k}={\cal O}(m_{\rm th})$,
 the momentum integration together with $\delta$-function
 contribute a factor of ${\cal O}(m_{\rm th}^2)$.
 The ${\cal O}(T/m_{\rm th})$ Bose factor and the prefactor
 $(1{-}e^{-\underline{p}^0\beta})={\cal O}(m_{\rm th}/T)$ combined are
 ${\cal O}(1)$.
 Since $L_{\rm full}(\underline{p},\underline{k})$ at soft $\underline{k}$
 and $\underline{p}$ is
 ${\cal O}(\lambda)$,
 putting all terms together yields, for soft $\underline{p}$,
 \begin{equation}
 \Sigma_I(\underline{p})
 ={\cal O}(\lambda m_{\rm th}^2)
 ={\cal O}(\lambda^2 T^2)
 \;.
 \end{equation}
 Hence, $\Sigma_I(\underline{p})$ is ${\cal O}(\lambda^2 T^2)$ for both
 hard and
 soft $\underline{p}$.

 With all the ingredients at hand, the size of the soft momentum
 contribution to the additional rung
 $u^T(\underline{k}){\cal M}_{\rm full}(\underline{k}{-}p)w(p)$
 can be readily examined.
 When all the momenta involved are soft,
 the momentum integration combined with the $\delta$-function
 in the cut
 propagator provides a factor of ${\cal O}(m_{\rm th}^2)$.
 Once again, the prefactor $(1{-}e^{-\underline{p}^0\beta})$ combined
 with the
 Bose factor in the cut propagator is ${\cal O}(1)$.
 The cut rung is
 $L_{\rm full}(\underline{k},\underline{p}) = {\cal O}(\lambda)$,
 and the self-energy remains ${\cal O}(\lambda^2 T^2)$.
 Hence, all combined,
 the integration over
 $u^T {\cal M} w$ can be regarded as
 ${\cal O}(\lambda m_{\rm th}^2 /\lambda^2 T^2)
 = {\cal O}(1)$.
 Consequently, one can conclude that the power counting performed
 in
 section~\ref{sec:classification} is not altered by soft momentum
 contributions.

 For the near collinear singularities, note that the estimate
 for the
 scattering amplitude (\ref{eq:estim_T}) holds for all on-shell
 momenta.
 Hence, there is no large factor resulting from near collinear
 singularities and
 the power counting performed in section~\ref{sec:classification}
 is again not altered.

 \section{Details of chain diagram summation}
 \label{app:chains}

 \begin{figure}
 \setlength {\unitlength}{1cm}
\vbox
    {%
    \begin {center}
\begin{picture}(0,0)
 \put(0.0,5.6){$\sigma_{AA}=$}
 \put(4.3,5.6){$+$}
 \put(9.9,5.6){$+$}
 \put(-0.3,3.3){$+$}
 \put(7.6,3.3){$+$}
 \put(-0.3,0.9){$+$}
 \put(7.6,0.9){$+$}
\end{picture}
	\leavevmode
	\def\epsfsize	#1#2{0.3#1}
	\epsfbox {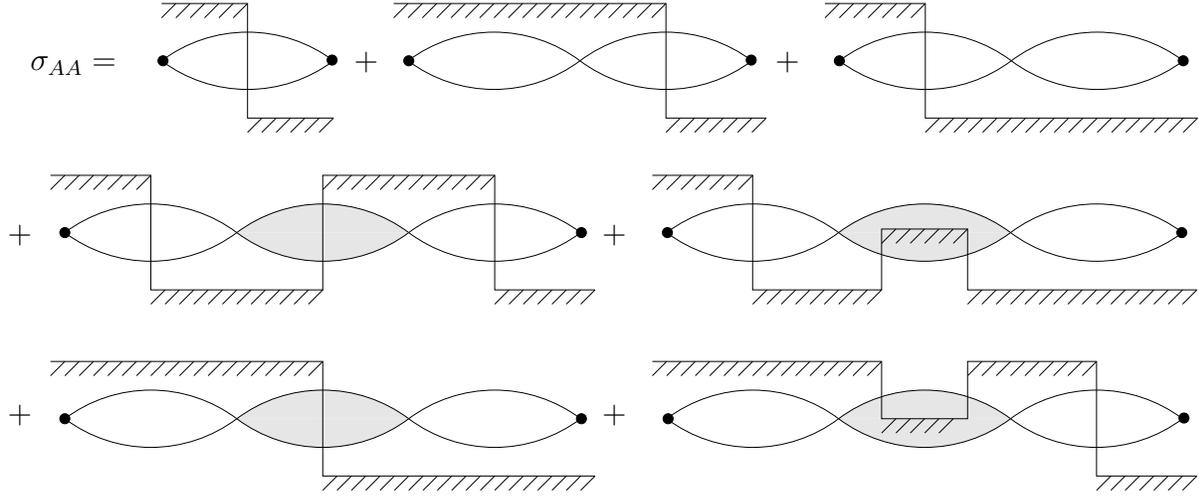}
    \end {center}
     \caption
 	{%
	Diagrammatic representation of the correlation function
	$\sigma_{AA}$.  Filled bubbles represent the sum of all cut
	chain diagrams.  The external operator $A$ is represented by
	black circles at each end.
 	}
 \label{fig:pi_chain}
     }
 \end {figure}
 For the sake of simplicity, $\lambda\phi^4$ diagrams are examined
 first.
 Diagrammatically,
 the sum of all $\lambda\phi^4$
 chain diagrams for the Wightman function
 $\sigma_{AA}^{\vphantom{x}}$ of a bilinear operator $\hat{A}$
 can be represented by Fig.~\ref{fig:pi_chain}.
 The filled bubble in Fig.~\ref{fig:pi_chain}
 with the ends of the cut lines on opposite sides
 represents sum of all cut chain diagrams with the same topology.
 This sum is denoted by
 $L_{\rm chain}(q)$.
 Similarly,
 the filled bubble with the ends of the cut on the
 same side represents the sum of all cut diagrams
 with the equivalent topology.  This sum is denoted by
 $C_{\rm chain}(q)$

 \begin{figure}
 \setlength {\unitlength}{1cm}
\vbox
    {%
    \begin {center}
 \begin{picture}(0,0)
 \put(2.6,0.9){$=$}
 \put(2.6,3.1){$=$}
 \put(5.5,0.9){$+$}
 \put(5.5,3.1){$+$}
 \put(10.5,0.9){$+$}
 \put(10.5,3.1){$+$}
 \end{picture}
	\leavevmode
	\def\epsfsize	#1#2{0.23#1}
	\epsfbox {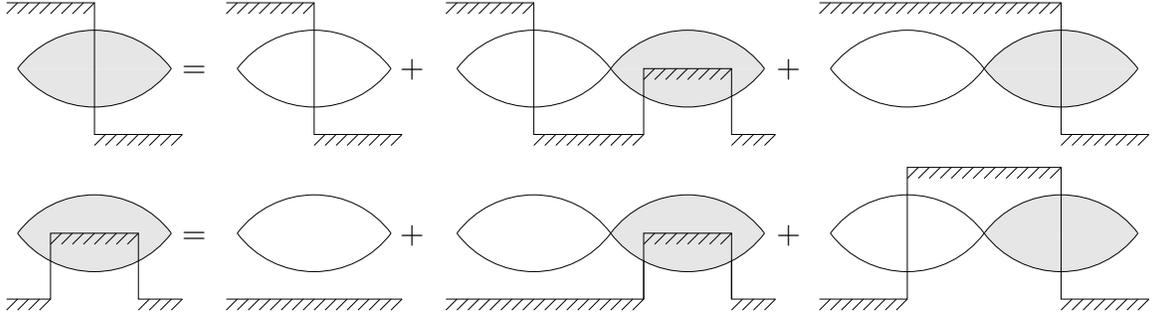}
    \end {center}
     \caption
 	{%
 Diagrammatic representation of the equations
 satisfied by the resummed cut
 and uncut bubbles.   In this diagram, the external operators
 at both ends
 are $\hat{\phi}^2$ rather than $\hat{A}$.   The sum of all chain
 diagrams for correlation function $\sigma_{AA}$ can be
 expressed in terms of the solutions of these equations.
 The empty cut bubble is denoted by $L(q)$, the empty uncut bubble
 $C(q)$.
 The filled bubble on the left hand side of the first diagram
 is denoted
 by $L_{\rm chain}(q)$, the filled bubble on the right hand side
 of the
 second diagram is $C_{\rm chain}(q)$.
 	}
 \label{fig:chain_sum}
     }
 \end {figure}
 The equations for
 $L_{\rm chain}(q)$ and $C_{\rm chain}(q)$ are
 \begin{equation}
 \left(
 \begin{array}{l}
 L_{\rm chain}(q)\\
 C_{\rm chain}(q)
 \end{array}
 \right)
 =
 \left(
 \begin{array}{l}
 L_0(q)\\
 C_0(q)
 \end{array}
 \right)
 +
 \lambda
 \left(
 \begin{array}{ll}
 C_0(q)^* & L_0(q) \\
 L_0(-q) & C_0(q)
 \end{array}
 \right)
 \left(
 \begin{array}{l}
 L_{\rm chain}(q)\\
 C_{\rm chain}(q)
 \end{array}
 \right)
 \;,
 \end{equation}
 as illustrated in Fig.~\ref{fig:chain_sum}.
 The cut bubble $L_0(q)$
 and the uncut bubble $C_0(q)$ are again defined by
 Eq.~(\ref{eq:cut_bubble}) and Eq.~(\ref{eq:uncut_bubble})
 with $I_A(l,q{-}l)=1$.

 The above matrix equation is easy to solve.  The solution is
 \begin{mathletters}
 \begin{eqnarray}
 L_{\rm chain}(q)
 & = & \displaystyle
 {
 		L_0(q)
 	        \over
  (1 - \lambda {\rm Re}\,C(q))^2
 +
  ({\textstyle {\lambda\over 2}})^2 (L(q)-L(-q))^2
 }
 \;
 \label{eq:cut_chain}
 \\
 \noalign{and}
 C_{\rm chain}(q)
 & = & \displaystyle
 {
  C(q)
  -
  \lambda( ({\rm Re}\,C(q))^2
  -
  {\textstyle{\lambda\over 4}}(L(q)-L(-q))^2
 		    \over
  (1 - \lambda {\rm Re}\,C(q))^2
  + ({\textstyle {\lambda\over 2}})^2 (L(q)-L(-q))^2
 }
 \label{eq:uncut_chain}
 \;,
 \end{eqnarray}
 \label{eq:solutions}
 \end{mathletters}%
 \hspace{-1ex}%
 where the finite temperature optical theorem
 \begin{equation}
 {\rm Im} \; C(q) = -{1\over2} \left( L(q) + L(-q) \right)
 \;,
 \label{eq:app_optical_thm}
 \end{equation}
 is used to obtain the form shown in (\ref{eq:solutions}).

 When cubic interactions are included, the ``chain'' diagrams
 also
 include cut two-loop diagrams shown in Fig.~\ref{fig:chain3}
 where the
 bubble in the diagram may be regarded as the sum of all $\lambda\phi^4$
 chain diagrams.
 Equivalently, the vertex contribution may be regarded as containing
 \begin{equation}
 I_{\tilde{A}}(l,-l)
 =
 I_A(l,-l)
 + \lambda {\rm Re}\, C_A(0) \times
        \left( 1 +  {\cal O}(\sqrt{\lambda}) \right)
 \;.
 \end{equation}
 A straight forward application of cutting rules then yields,
 \begin{eqnarray}
 \sigma_{AA}^{\hbox{\scriptsize{Fig.~\protect{\ref{fig:chain3}}}}}(0)
 & = & \displaystyle
 -4iL_{\tilde{A}}(0)\, \tilde{G}(0)\, C_{\tilde{A}}(0)
 +4iC_{\tilde{A}}(0)^*\, \tilde{G}(0)^*\, L_{\tilde{A}}(0)
 \nonumber\\
 & & \displaystyle {}
 + 4C_{\tilde{A}}(0)^*\, S(0)\, C_{\tilde{A}}(0)
 + 4L_{\tilde{A}}(0)\, S(0)\, L_{\tilde{A}}(0)
 \nonumber\\
 & = & \displaystyle
 -
 8{g^2\over m_{\rm th}^2}\,{\rm Re}\,C_{\tilde{A}}(0) L_{\tilde{A}}(0)
 +
 4{g^4\over m_{\rm th}^4}\,\left( {\rm Re}\,C_{\tilde{A}}(0) \right)^2
 L_0(0)
 \;,
 \end{eqnarray}
 where the optical theorem (\ref{eq:optical_thm}), and
 $S(0) = g^2L_0(0)/m_{\rm th}^4$, justified below, are used.

 For the remainder of this appendix,
 the estimates ${\rm Re}\,C_A(0) ={\cal O}(T^2)$,
 ${\rm Re}\,C_0(0)={\cal O}(1/\sqrt{\lambda})$,
 $g^2\tilde{G}(0)={\cal O}(g^2/m_{\rm th}^2)$, and
 $S(0) = {\cal O}(1/\lambda^2 T^2)$
 used in this appendix and section~\ref{subsec:chains}
 are examined.
 To estimate ${\rm Re}\,C_A(0)$,
 consider the following
 explicit form of the real part of an uncut bubble $C_A(0)$
 \begin{eqnarray}
 {\rm Re}\,C_A(0)
 & = & \displaystyle
 {1\over 2}
 \int {d^4 l\over (2\pi)^4}\,
 I_A(l,-l)\,
 {\rm Re}\,(-i\tilde{G}^2(l))\,
 \nonumber\\
 & = & \displaystyle
 {i\over 4}
 \int {d^4 l\over (2\pi)^4}\,
 I_A(l,-l)\,
 \coth(l^0\beta/2)\,
 \left(
 {1\over [l^2+m_{\rm th}^2+\Sigma(l)]^2 }
 -
 {1\over [l^2+m_{\rm th}^2+\Sigma(l)^*]^2}
 \right)
  \nonumber\\
 & = & \displaystyle
 {1\over 4}
 \left.
 {\partial \over \partial m^2}
 \int {d^4 l\over (2\pi)^4}\,
 I_A(l,-l)\,
 \coth(|l^0|\beta/2)\,
 2\pi\delta(l^2+m^2)
 \right|_{m^2 = m_{\rm th}^2}
 \times(1+{\cal O}(\lambda^2))
 \label{eq:app_Re_Ca}
 \;,
 \end{eqnarray}
 where to obtain the last line, the elementary relation
 \begin{equation}
 {\partial \over \partial x}
 \left( {1\over x+a} \right)
 = -{1\over (x+a)^2}
 \;,
 \end{equation}
 and the single particle spectral density in the weak coupling
 limit,
 \begin{eqnarray}
 \rho(l)
 & = & \displaystyle
 {i\over [l^2+m_{\rm th}^2+\Sigma(l)]}
 -
 {i\over [l^2+m_{\rm th}^2+\Sigma(l)^*]}
 \nonumber\\
 & = & \displaystyle
 \rho_{\rm free}(l)\,
 \times(1+{\cal O}(\lambda^2))
 \;,
 \end{eqnarray}
 are used.
 Note that the identification $m^2 = m_{\rm th}^2$ must be made {\em
 after}
 the derivative is taken.
 Again at high temperature, the zero temperature contribution
 is smaller
 than the thermal contribution.  Hence, the $\coth(\beta E_l/2)$
 factor
 in the integrand can be replaced by the Bose factor $2n(E_l)$
 to
 calculate the leading weak coupling behavior.

 For the pressure and the energy density insertions,
 \begin{eqnarray}
 {\rm Re}\, C_{\cal P}(0)
 & = & \displaystyle
 {1\over 2}
 \left.
 {\partial \over \partial m^2}
 \int {d^3 {\bf l} \over (2\pi)^3 E_l}\,
 n(E_l)\, {\textstyle{1\over 3}}{\bf l}^2\,
 \right|_{m^2 = m_{\rm th}^2}
 \nonumber\\
 & = & \displaystyle
 {1\over 2}
 \left.
 {\partial \over \partial m^2}
 {\cal P}_0
 \right|_{m^2 = m_{\rm th}^2}
 =
 -{T^2\over 48}
 \;,
 \end{eqnarray}
 and
 \begin{eqnarray}
 {\rm Re}\, C_{\varepsilon}(0)
 & = & \displaystyle
 {1\over 2}
 \left.
 {\partial \over \partial m^2}
 \int {d^3 {\bf l} \over (2\pi)^3 E_l}\,
 n(E_l)\, E_l^2\,
 \right|_{m^2 = m_{\rm th}^2}
 \nonumber\\
 & = & \displaystyle
 {1\over 2}
 \left.
 {\partial \over \partial m^2}
    \left( 3 {\cal P}_0 {+} {m^2 T^2\over 12} \right)
 \right|_{m^2 = m_{\rm th}^2}
 =
 -{T^2\over 48}
 \;,
 \end{eqnarray}
 ignoring higher order contributions.

 The expression for ${\rm Re}\,C_0(0)$ is also given by
 Eq.~(\ref{eq:Re_Ca}) by setting $I_A=1$.
 At high temperature, the leading ${\cal O}(T^2)$
 contribution to the integral in the last expression in
 Eq.~(\ref{eq:Re_Ca}) comes from loop momenta of ${\cal O}(T)$.
 However, when the derivative with respect to $m^2$ is taken,
 this is zero.
 The next largest contribution to the integral
 is ${\cal O}(m T)$ coming from loop momenta of ${\cal O}(m)$.
 When the mass derivative is taken and the identification
 $m\to m_{\rm th}$ made, this yields
 ${\rm Re}\,C_0(0) = {\cal O}(T/m_{\rm th}) ={\cal O}(1/\sqrt{\lambda})$.

 To estimate the size of propagators at zero momentum
 requires
 knowledge of the size of the self-energy at zero external
 momentum.
 The lowest order imaginary part of the off-shell self-energy
 comes from the $g\phi^3$ one-loop diagram shown
 in Fig.~\ref{fig:self_one_loop}.
 To estimate the size of this diagram, the propagators in the
 loop
 must be regarded as resummed propagators.
 Then the self-energy in the zero 4-momentum limit satisfies
 \begin{eqnarray}
 \lim_{q^0\to 0}\,\lim_{{\bf q}\to 0}\,
 2[1{+}n(q^0)]\,\Sigma_I(q)
 & = & \displaystyle
 g^2\,L_0(0)\times (1 + {\cal O}(\lambda))
 \nonumber\\
 & = & \displaystyle
 {\cal O}(g^2/\lambda^2) = {\cal O}(m_{\rm phys}^2/\lambda)
 \;,
 \end{eqnarray}
 where pinching pole approximation of $L(0)$ is again used.
 The above estimate for the self-energy implies that,
 \begin{eqnarray}
 \lim_{q^0\to 0}\,\lim_{{\bf q}\to 0}\, S(q) & = & \displaystyle
 \lim_{q^0\to 0}\,\lim_{{\bf q}\to 0}\,
 {2[1{+}n(q^0)]\Sigma_I(q)\over |q^2+m_{\rm th}^2+\Sigma(q)|^2 }
 \nonumber\\
 & = & \displaystyle
 { g^2 L_0(0) \over m_{\rm th}^4 }\times (1+{\cal O}(\sqrt{\lambda}))
 \nonumber\\
 & = & \displaystyle
     {\cal O}(g^2 / \lambda^4 T^4)
   = {\cal O}(m_{\rm phys}^2 / \lambda^3 T^4)
 \nonumber\\
 & = & \displaystyle
 {\cal O}(1/\lambda^3 T^2 \times (g^2/T^2))
 \le {\cal O}(1/\lambda^2 T^2)
 \;,
 \\
 \noalign{\hbox{and}}
 \lim_{q^0\to 0}\,\lim_{{\bf q}\to 0}\,
 g^2\,{\rm Re}\,\tilde{G}(q)
 & = & \displaystyle
 \lim_{q^0\to 0}\,\lim_{{\bf q}\to 0}\,
 {g^2\,[q^2 + m_{\rm th}^2 + \Sigma_R(q)]
       	\over
 |q^2 + m_{\rm th}^2 + \Sigma(q)|^2}
 \nonumber\\
 & = & \displaystyle
 {\cal O}(g^2/m_{\rm th}^2) =
 {\cal O}(\lambda m_{\rm phys}^2/m_{\rm th}^2) \le {\cal O}(\lambda)
 \;.
 \label{eq:soft_S}
 \end{eqnarray}

 \section{The imaginary part of two-loop self-energy}
 \label{app:self_two_loop}

 Using the cutting rules, the imaginary part of the two-loop self-energy
 can be expressed as
 \begin{equation}
 \Sigma_I^{\hbox{\scriptsize{two-loop}}}(\underline{q})
 \equiv
 {\lambda^2\over 12}
 (1{-}e^{-\beta E_q})\,
 {\cal S}(\underline{q})
 \;,
 \end{equation}
 where
 \begin{equation}
 {\cal S}(\underline{q})
 \equiv
 \int {d^4 l\over (2\pi)^4}\,{d^4 k \over (2\pi)^4}\,
 S_{\rm free}(l)\,
 S_{\rm free}(k{+}\underline{q})\,
 S_{\rm free}(-l{-}k)\,
 \;.
 \label{eq:cal_S_app}
 \end{equation}
 Using the on-shell $\delta$-functions
 free particle cut propagators
 \begin{equation}
 S_{\rm free}(l)
 =
 \sum_{\sigma_l =\pm 1}\,
 \sigma_l\, [1{+}n(\sigma_l E_l)]\,\pi\delta(l^0 - \sigma_l E_l){\bigg /}E_l
 \;,
 \end{equation}
 frequency integrations in Eq.~(\ref{eq:cal_S_app}) can be
 straightforwardly carried out to yield
 \begin{eqnarray}
 {\cal S}(\underline{q})
 & \displaystyle = & \displaystyle
 \sum_{\sigma = \pm 1}\,
 \sigma_{k+q} \sigma_{l+k} \sigma_{l}
 {1\over 8(2\pi)^5}
 \int
 {
 d^3 {\bf l} \, d^3 {\bf k}
	 \over
 E_l E_{k+q} E_{k+l}
 }\,
 [1{+}n(\sigma_{k+q}E_{k+q})]\,
 [1{+}n(\sigma_{l}E_{l})]\,
 \nonumber\\
 & \displaystyle & \displaystyle \quad\quad{}\times
 [1{+}n(\sigma_{k+l}E_{k+l})]\,
 \delta
 (E_q{-}\sigma_l E_l{-}\sigma_{k+q}E_{k+q}{-}\sigma_{k+l}E_{k+l})\,
 \;.
 \end{eqnarray}
 The argument of remaining $\delta$-function can be satisfied only when
 two of the $\sigma$ are $+1$ and the other one is $-1$.
 By suitably changing labels, the above then becomes
 \begin{equation}
 {\cal S}(\underline{q})
 =
 {3\over 8 (2\pi)^5}
 \int
 {
 d^3 {\bf l} \, d^3 {\bf k}
	 \over
 E_l E_{k+q} E_{k+l}
 }\,
 [1{+}n(E_{l})]\, [1{+}n(E_{k+q})]\, n(E_{k+l})\,
 \delta
 (E_q{+}E_{k+l}{-}E_{k+q}{-}E_{l})\,
 \;.
 \end{equation}

 To carry out remaining integrations, the angles between spatial vectors
 are defined as
 \begin{eqnarray}
 \cos \theta_k
 & \displaystyle \equiv & \displaystyle
 {{\bf k}{\cdot}{\bf q}/ |{\bf k}||{\bf q}|}
 \\
 \cos \theta_l
 & \displaystyle \equiv & \displaystyle
 {{\bf k}{\cdot}{\bf l}/ |{\bf k}||{\bf l}|}
 \;.
 \end{eqnarray}
 Changing variables to $E_{k+q}$ and $E_{k+l}$ with the Jacobians
 \begin{eqnarray}
 {d E_{k+q} / d\cos \theta_k}
 & \displaystyle = & \displaystyle
 {|{\bf k}||{\bf q}| / E_{k+q}}
 \\
 {d E_{k+l} / d\cos \theta_l}
 & \displaystyle = & \displaystyle
 {|{\bf k}||{\bf l}| / E_{k+l}}
 \;,
 \end{eqnarray}
 one finds
 \begin{eqnarray}
 {\cal S}(\underline{q})
 & \displaystyle = & \displaystyle
 {3\over 8(2\pi)^5 |{\bf q}|}
 \int
 dE_l \,d|{\bf k}|\,
 [1{+}n(E_{l})]\,
 \int_{E^-_{kl}}^{E^+_{kl}} dE_{k+l}\,
 \int_{E^-_{kq}}^{E^+_{kq}} dE_{k+q}\,
 [1{+}n(E_{k+q})]\,
 \nonumber\\
 & \displaystyle & \displaystyle \quad\quad{}\times
 n(E_{k+l})\,
 \delta (E_q{+}E_{k+l}{-}E_{k+q}{-}E_{l})\,
 \;,
 \end{eqnarray}
 where
 \begin{equation}
 E^{\pm}_{kl}
 \equiv
 \sqrt{(|{\bf k}|{\pm}|{\bf l}|)^2 + m_{\rm th}^2}
 \;,
 \end{equation}
 with analogous definitions for $E^{\pm}_{kq}$.
 Carrying $E_{k+q}$ and $E_{k+l}$ integrations
 amounts to figuring out the kinematic conditions.
 Straightforward calculation then yields
 \begin{eqnarray}
 \lefteqn{
 {\cal S}(\underline{q})
 =
 {3 T\over 32 \pi^3 |{\bf q}|}
 \int_{0}^{|{\bf q}|} d|{\bf k}|\,
 } & \displaystyle & \displaystyle
 \nonumber\\
 & \displaystyle & \displaystyle \quad{}\times
\left\{
 \int_{E_k}^{E_q} dE_l \,
 [1{+}n(E_{l})] [1{+}n(E_q{-}E_l)]\,
 \ln\left(
 e^{-\beta E_l}{-}e^{-\beta(E_q{+}E_k)}
	\over
 e^{-\beta E_l}(1{-}e^{-\beta E_k})
 \right)
\right.
 \nonumber\\
 & \displaystyle & \displaystyle \quad\;\; {}
 +
\left.
 \int_{E_q}^{\infty} dE_l \,
 [1{+}n(E_{l})] [1{+}n(E_q{-}E_l)]\,
 \ln\left(
 e^{-\beta E_q}(1{-}e^{-\beta E_k})
	\over
 e^{-\beta E_q}{-}e^{-\beta(E_l{+}E_k)}
 \right)
\right\}
 \;.
 \end{eqnarray}

 Changing variables from $E_l$ to $u \equiv e^{-\beta E_l}$,
 the above can be rewritten as
 \begin{eqnarray}
 \lefteqn{
 {\cal S}(\underline{q})
 =
 {3 T^2\over 32 \pi^3 |{\bf q}|} [1{+}n(E_q)]
 \int_{0}^{|{\bf q}|} d|{\bf k}|\,
 } & \displaystyle & \displaystyle
 \nonumber\\
 & \displaystyle & \displaystyle \quad{}\times
\left\{
 \int_{y}^{x} du \,
 \left(
 {1\over 1{-}u}+{1\over u{-}y}
 \right)
 \ln\left(
 {u {-} y x \over u(1{-}x)}
 \right)
\right.
 \nonumber\\
 & \displaystyle & \displaystyle \quad\;\; {}
 +
\left.
 \int_{0}^{y} du \,
 \left(
 {1\over 1{-}u}+{1\over u{-}y}
 \right)
 \ln\left(
 {y(1{-}x) \over y{-}u x}
 \right)
\right\}
 \;,
 \end{eqnarray}
 where
 \begin{eqnarray}
 y & \displaystyle \equiv & \displaystyle e^{-\beta E_q}
 \;,
 \\
 x & \displaystyle \equiv & \displaystyle e^{-\beta E_k}
 \;.
 \end{eqnarray}
 Another change of variable
 \begin{equation}
 z \equiv {u{-}y\over 1{-}u}
 \;,
 \end{equation}
 and the definition of dilogarithmic function,
 \begin{equation}
 Li_2(z)
 \equiv
 -\int_0^1 {du\over u}\,\ln(1{-}zu)
 =
 \sum_{n=1}^{\infty}\,{z^n\over n^2}
 \;,
 \end{equation}
 yield,
 \begin{eqnarray}
 \lefteqn{
 {\cal S}(\underline{q})
 =
 {3 T^2\over 32 \pi^3 |{\bf q}|}\, [1{+}n(E_q)]
 \int_{0}^{|{\bf q}|} d|{\bf k}|\,
 } & \displaystyle & \displaystyle
 \nonumber\\
 & \displaystyle & \displaystyle \quad{}\times
\left(
 \ln\left( {x\over y} \right)\,
 \ln\left( {1{-}y \over 1{-}x} \right)\,
 +
 2\ln^2 \left( {1{-}y\over 1{-}x} \right)\,
\right.
 \nonumber\\
 & \displaystyle & \displaystyle \quad{}
\left.
 +
 Li_2(y)
 +
 Li_2\left( { (1{-}xy)(x{-}y)\over x(1{-}y)^2 } \right)
 +
 Li_2\left( { x{-}y\over 1{-}y } \right)
 -
 Li_2\left( { x{-}y\over x(1{-}y) } \right)
\right)
 \;.
 \end{eqnarray}

 In the $|{\bf q}|\to 0$ limit, $x=y$.  Hence, immediately,
 \begin{equation}
 {\cal S}(m_{\rm th},0)
 =
 {3 T^2\over 32 \pi^3}\,[1{+}n(m_{\rm th})]\,
 Li_2(e^{-\beta m_{\rm th}})
 \;,
 \end{equation}
 and
 \begin{equation}
 \Sigma_I^{\hbox{\scriptsize{two-loop}}}(m_{\rm th},0)
 =
 {\lambda^2 T^2\over 2^7 \pi^3}\,
 Li_2(e^{-\beta m_{\rm th}})
 \;.
 \end{equation}
 When the temperature is high, $T\gg m_{\rm th}$, the identity
 \begin{equation}
 Li_2(x) = {\pi^2\over 6} -\ln(x)\ln(1{-}x) - Li_2(1{-}x)
 \;,
 \end{equation}
 yield\footnote{
 	Similar result was obtained by Parwani~\cite{Parwani}.
 }
 \begin{equation}
 \Sigma_I^{\hbox{\scriptsize{two-loop}}}(m_{\rm th},0)
 =
 {T^2 \lambda^2 \over 768 \pi}
 (1 + {\cal O}(m_{\rm th}/T  \ln (m_{\rm th}/T)))
 \;.
 \end{equation}
 When the temperature is low, $T \ll m_{\rm th}$,
 the dilogarithmic function $Li_2(e^{-\beta m_{\rm th}})
 = e^{-m_{\rm th}\beta}$,
 and
 \begin{equation}
 \Sigma_I^{\hbox{\scriptsize{two-loop}}}(m_{\rm th},0)
 =
 {\lambda^2 T^2 \over 128 \pi}\,
 e^{-m_{\rm th}\beta}\,
 \times (1 + {\cal O}(e^{-m_{\rm th}\beta}) )
 \;.
 \end{equation}

 In the opposite limit where $|{\bf q}|\to \infty$,
 the only term that survives in ${\cal S}$ is
 the first term in the bracket
 \begin{equation}
 \lim_{q\to \infty} {\cal S}(\underline{q})
 =
 -{3 T^2 \over 32 \pi^3}\,
 \int_{0}^{\infty} d|{\bf k}|\,
 \ln(1{-}e^{-\beta E_k})
 \;.
 \end{equation}
 In the high temperature limit, this yields
 \begin{eqnarray}
 \lim_{q\to \infty} {\cal S}(\underline{q})
 & \displaystyle = & \displaystyle
 {3 T^2\over 32 \pi^3}\,\left( {\pi^2\over 6} +
 {\cal O}(m_{\rm th}^2/T^2) \right)
 \;,
 \\
 \noalign{\hbox{and}}
 \lim_{q\to \infty}
 \Sigma_I^{\hbox{\scriptsize{two-loop}}}(\underline{q})
 & \displaystyle = & \displaystyle
 {T^2 \lambda^2 \over 768 \pi}
 (1 + {\cal O}(m_{\rm th}^2/T^2))
 \;.
 \end{eqnarray}
 Note that the leading order terms are the same.
 In the low temperature limit,
 \begin{eqnarray}
 \lim_{q\to \infty}
 \Sigma_I^{\hbox{\scriptsize{two-loop}}}(\underline{q})
 =
 {\lambda^2 \over 128 \pi}\,
 e^{-m_{\rm th}\beta}\, \sqrt{m_{\rm th} T^3 \over 2\pi^3}\,
 \times (1 + {\cal O}(e^{-m_{\rm th}\beta}) )
 \;.
 \end{eqnarray}

 \begin{figure}
 \setlength {\unitlength}{1cm}
\vbox
    {%
    \begin {center}
 \begin{picture}(0,0)
 \put(-1.0,7.5){$\tilde{\Sigma}_I(E_k)$}
 \put(11.0,-0.5){$|{\bf k}|$}
 \end{picture}
	\leavevmode
	\def\epsfsize  #1#2{0.5#1}
	\epsfbox {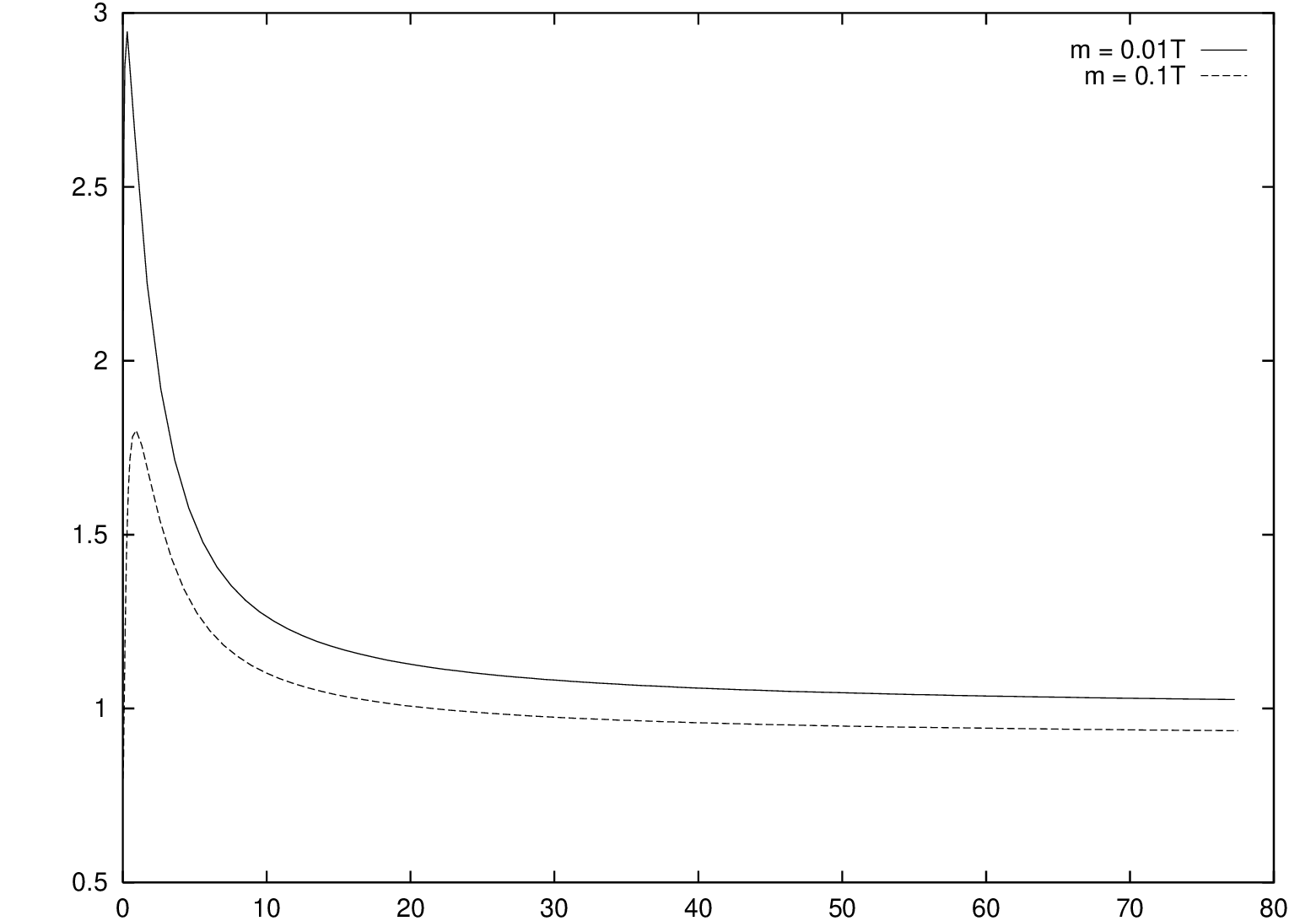}
    \end {center}
     \caption
 	{%
   Numerical results for
   $\tilde{\Sigma}_I(E_k) = 768\pi \Sigma_I(E_k)/\lambda^2 T^2$
   for $\mth/T = 0.01, 0.1$.
    }
 \label{fig:selflight}
 }
 \end{figure}

 \begin{figure}
 \setlength {\unitlength}{1cm}
\vbox
    {%
    \begin {center}
 \begin{picture}(0,0)
 \put(-1.5,7.5){$\tilde{\Sigma}_I(E_k)$}
 \put(12.0,-0.5){$|{\bf k}|$}
 \end{picture}
	\leavevmode
	\def\epsfsize  #1#2{0.5#1}
	\epsfbox {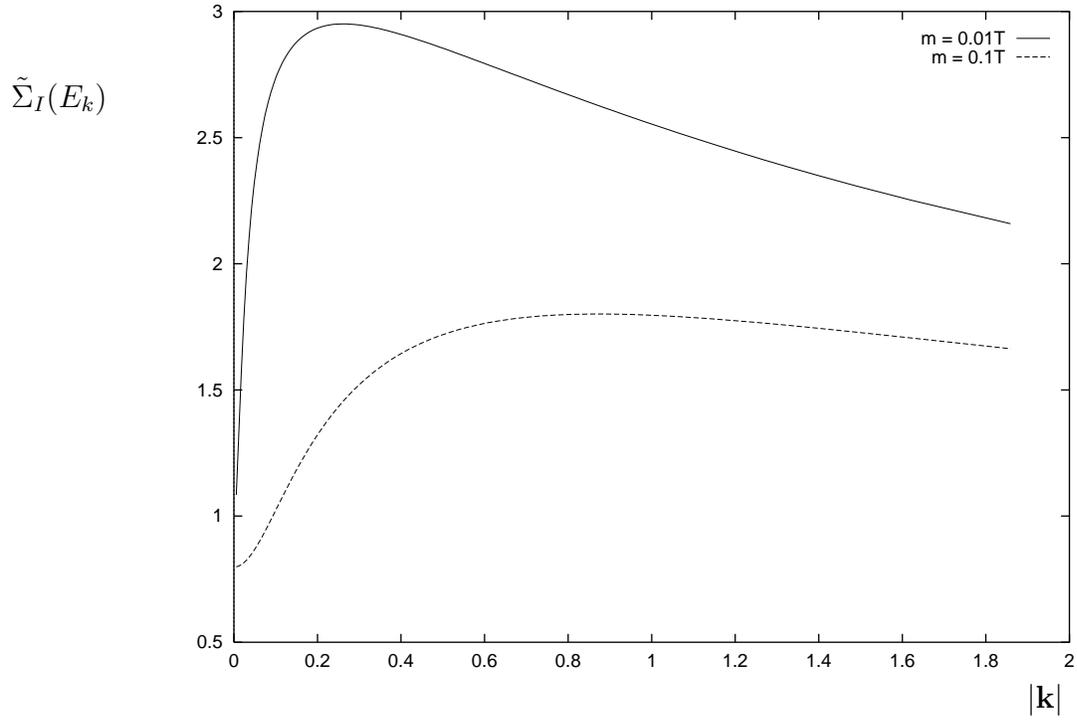}
    \end {center}
     \caption
 	{%
   $\tilde{\Sigma}_I(E_k) = 768\pi \Sigma_I(E_k)/\lambda^2 T^2$
   with $\mth/T = 0.01, 0.1$ for small values of $|{\bf k}|/T$.
    }
 \label{fig:light}
 }
 \end{figure}

 Numerical integration results shown in
 Fig.~\ref{fig:selflight} confirms this.
 A closer look at $\Sigma_I(\underline{q})$ with
 $m_{\rm th}/T=0.01,0.1$ for
 small values of $|{\bf k}|/T$ is presented in
 Fig.~\ref{fig:light}, and $\Sigma_I(\underline{q})$
 for $m_{\rm th}/T=1.0,2.0,3.0$
 is given in Fig.~\ref{fig:selfheavy}.
 Note that as the mass increases, the self-energy
 becomes flat throughout the momentum range.

\begin{figure}
 \setlength {\unitlength}{1cm}
\vbox
    {%
    \begin {center}
 \begin{picture}(0,0)
 \put(-1.0,7.5){$\tilde{\Sigma}_I(E_k)$}
 \put(11.0,-0.5){$|{\bf k}|$}
 \end{picture}
	\leavevmode
	\def\epsfsize  #1#2{0.5#1}
	\epsfbox {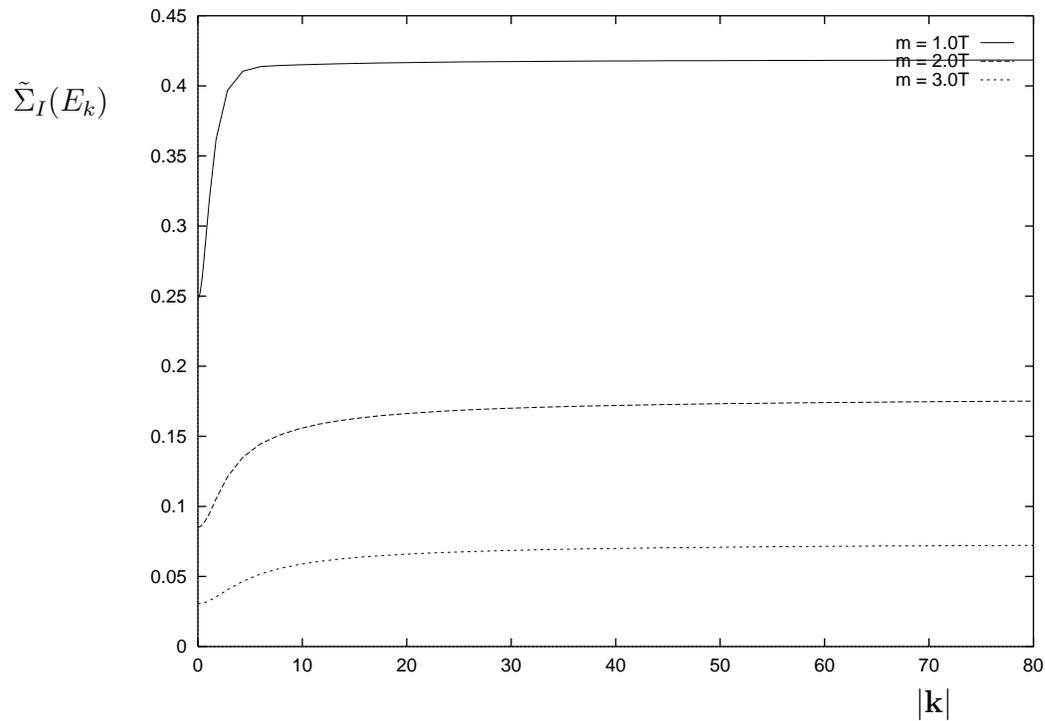}
    \end {center}
     \caption
 	{%
   Numerical results for
   $\tilde{\Sigma}_I(E_k) = 768\pi \Sigma_I(E_k)/\lambda^2 T^2$
   for $\mth/T = 1.0, 2.0, 3.0$.
    }
 \label{fig:selfheavy}
 }
 \end{figure}


 \end{document}